\documentclass[11pt]{article}

\usepackage{amssymb,amsmath,amsthm,amsopn,amsfonts}
\usepackage[T1]{fontenc}
\usepackage{ae,aecompl}
\usepackage{color}
\usepackage{colortbl}

\usepackage{algorithmic}
\usepackage{algorithm}

\usepackage{tabularx}
\usepackage{rotating}
\usepackage{array,multirow}

\usepackage{dutchcal} 

\usepackage{graphicx}
\usepackage[margin=20pt, justification=justified, font=small,labelfont=bf,labelsep=space]{caption}
\usepackage{float}
\usepackage{placeins}
\usepackage[percent]{overpic}  
\usepackage{morefloats}

\usepackage[normalem]{ulem}  
\usepackage{cancel}  

\usepackage{bm}
\usepackage{fancybox,layout,color}
\usepackage{fancyhdr}
\usepackage{scrtime}
\usepackage{setspace}
\usepackage{url}

\usepackage{mathtools}

\numberwithin{equation}{section}
\definecolor{purple}{RGB}{160,32,40}

\newtheorem{teo}{Theorem}[section]

\newtheorem{defi}[teo]{Definition}
\newtheorem{prop}[teo]{Proposition}

\newcommand{\R}{\ensuremath{\mathbb{R}} }
\newcommand{\dist}{\mathrm{dist}}

\newcommand{\Aff}{\mathrm{Aff}}

\DeclareMathOperator{\co}{\mathsf{co}}

\title{Compensated Convex Based Transforms \\for Image Processing and Shape Interrogation}

\author{\normalsize  
Antonio Orlando\thanks{CONICET, Departamento de Bioingenier\`ia,
Universidad Nacional de Tucum\'an, Argentina},
\, 
Elaine Crooks\thanks{Department of Mathematics, Swansea University,
Singleton Park, Swansea, SA2 8PP, UK}\, and Kewei Zhang\thanks{School  of Mathematical Sciences,
University of Nottingham, University Park, Nottingham, NG7 2RD, UK}
}

\date{ }

\begin{document}


\maketitle


\singlespacing
\pagestyle{fancy}
\fancyhead{}
\fancyfoot[OR,ER]{\tiny \today\,\, \thistime}
\cfoot{\thepage}


\begin{abstract}
	This paper reviews some recent applications of the theory of the compensated convex transforms or of 
	the proximity hull as developed by the authors to image processing and shape interrogation
	with special attention given to the Hausdorff stability and multiscale properties. 
	The paper contains also numerical
	experiments that demonstrate the performance of our methods compared to the state-of-art ones.
\end{abstract}


\medskip
{\bf 2000 Mathematics Subjects Classification number}:
90C25, 90C26, 49J52, 52A41, 65K10, 62H35, 14J17, 58K25, 53-XX,  65D17, 53A05,  26B25, 52B55, 65D18

\medskip
{\bf Email}:  aorlando@herrera.unt.edu.ar, e.c.m.crooks@swansea.ac.uk, kewei.zhang@nottingham.ac.uk  

\normalsize


\setcounter{equation}{0}
\section{Introduction}\label{Sec:Intro}
The compensated convex transforms were introduced in \cite{Zha08a,Zha08b} for the purpose of tight approximation of 
functions defined in $\R^n$ and their definitions 
were originally motivated by the translation method \cite{Mil90,Fir91,Gra96,Tar85}
in the study of the quasiconvex envelope in the vectorial calculus of variations
(see \cite{Dac08,Zha98} and references therein)
and in the variational approach of material microstructure \cite{Bal98,BJ87,BJ92}.
Thanks to their smoothness and tight approximation property, these transforms 
provide geometric convexity-based techniques for general functions  that yield  novel methods for identifying  singularities 
in functions \cite{ZOC15a, ZOC15b, ZCO15, ZCO16b} and  new tools for function and image interpolation and approximation 
\cite{ZCO16a, ZCO18}. 
In this paper we present some of the applications that have been tackled by this theory up to date.
These range from the detection of features in 
images or data \cite{ZOC15a, ZOC15b, ZCO16b}, to multi-scale  medial-axis extraction \cite{ZCO15}, to surface reconstruction from  
level sets, to approximation of scattered data and noise removal from images, to image inpainting \cite{ZCO16a,ZCO18}.

\medskip

Suppose $f:\mathbb{R}^n\to \mathbb{R}$ satisfies the following growth condition 
\begin{equation}\label{Eq:GrwthCond}
	f(x)\geq -A_1|x|^2-A_2\quad \text{ for any }x\in\mathbb{R}^n\,, 
\end{equation}
for some constants $A_1,\,A_2\geq 0$, then the quadratic lower compensated convex transform 
(lower transform for short) for a given $\lambda>A_1$ is defined in \cite{Zha08a} by
\begin{equation} \label{Eq:LwTr}
	C^l_\lambda(f)(x)=\co\left[\lambda|\cdot|^2+f\right](x)-\lambda|x|^2
	\qquad x\in\mathbb{R}^n,
\end{equation}
where $|x|$ is the Euclidean norm of $x\in \mathbb{R}^n$ and  $\co[g]$ the convex
envelope \cite{HL01,Roc70} of a function $g:\mathbb{R}^n\to \mathbb{R}$ bounded below.
Similarly, given $f:\mathbb{R}^n\to \mathbb{R}$ satisfying the growth condition 
\begin{equation}\label{Eq:GrwthCondUp}
	f(x)\leq A_1|x|^2+A_2\quad \text{ for any }x\in\mathbb{R}^n\,, 
\end{equation}
for some constants $A_1,\,A_2\geq 0$, the quadratic upper compensated convex transform 
(upper transform for short) for a given $\lambda>A_1$ is defined \cite{Zha08a} by
\begin{equation} \label{Eq:UpTr}
\begin{split}
	C^u_\lambda(f)(x)&=-C^l_\lambda(-f)(x)\\[1.5ex]
			 &=\lambda|x|^2-\co\left[\lambda|\cdot|^2-f\right](x)
	\qquad x\in\mathbb{R}^n\,.
\end{split}
\end{equation}

\begin{figure}[H]
\centerline{$\begin{array}{ccc}
	\includegraphics[width=0.24\textwidth]{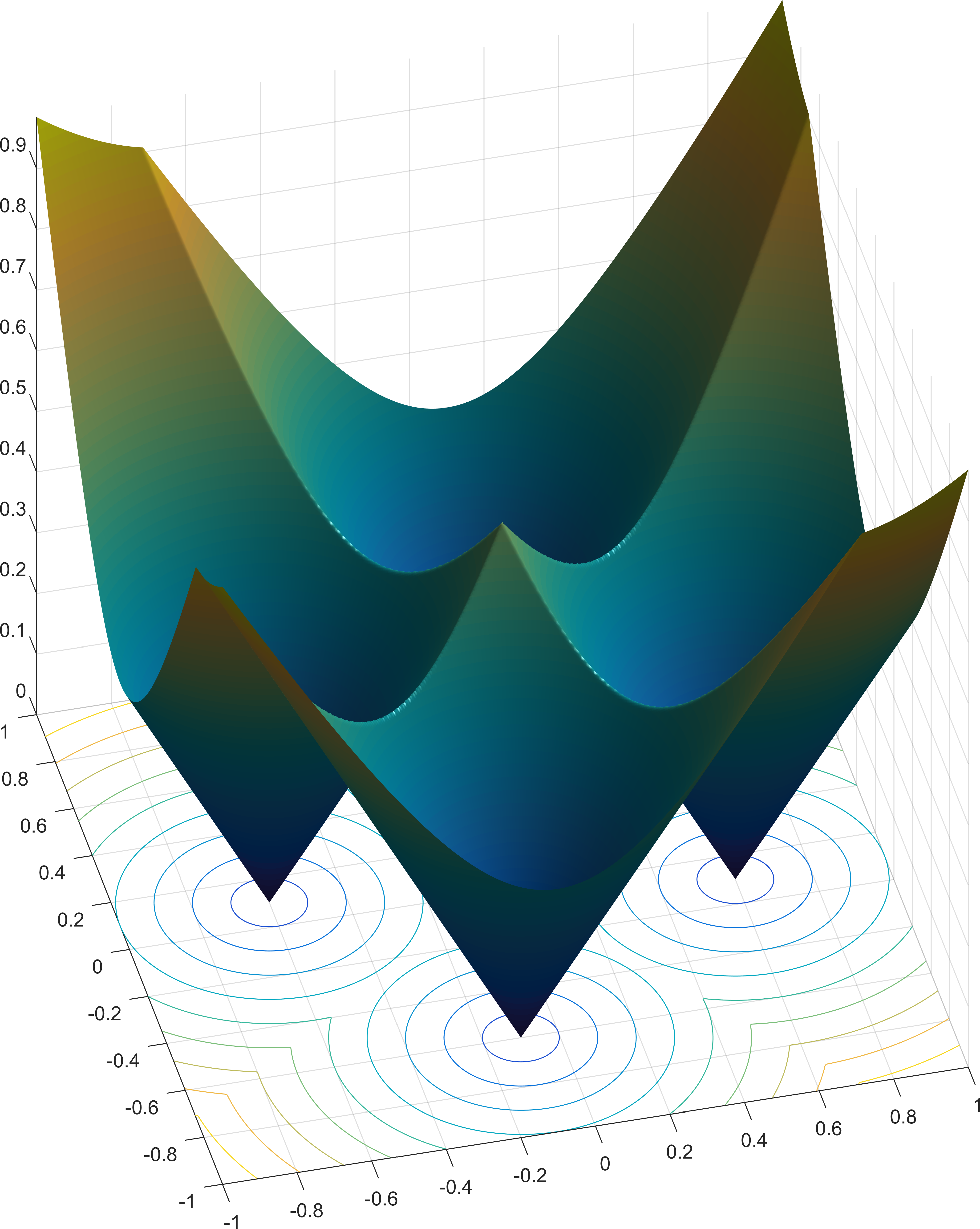}&
	\includegraphics[width=0.24\textwidth]{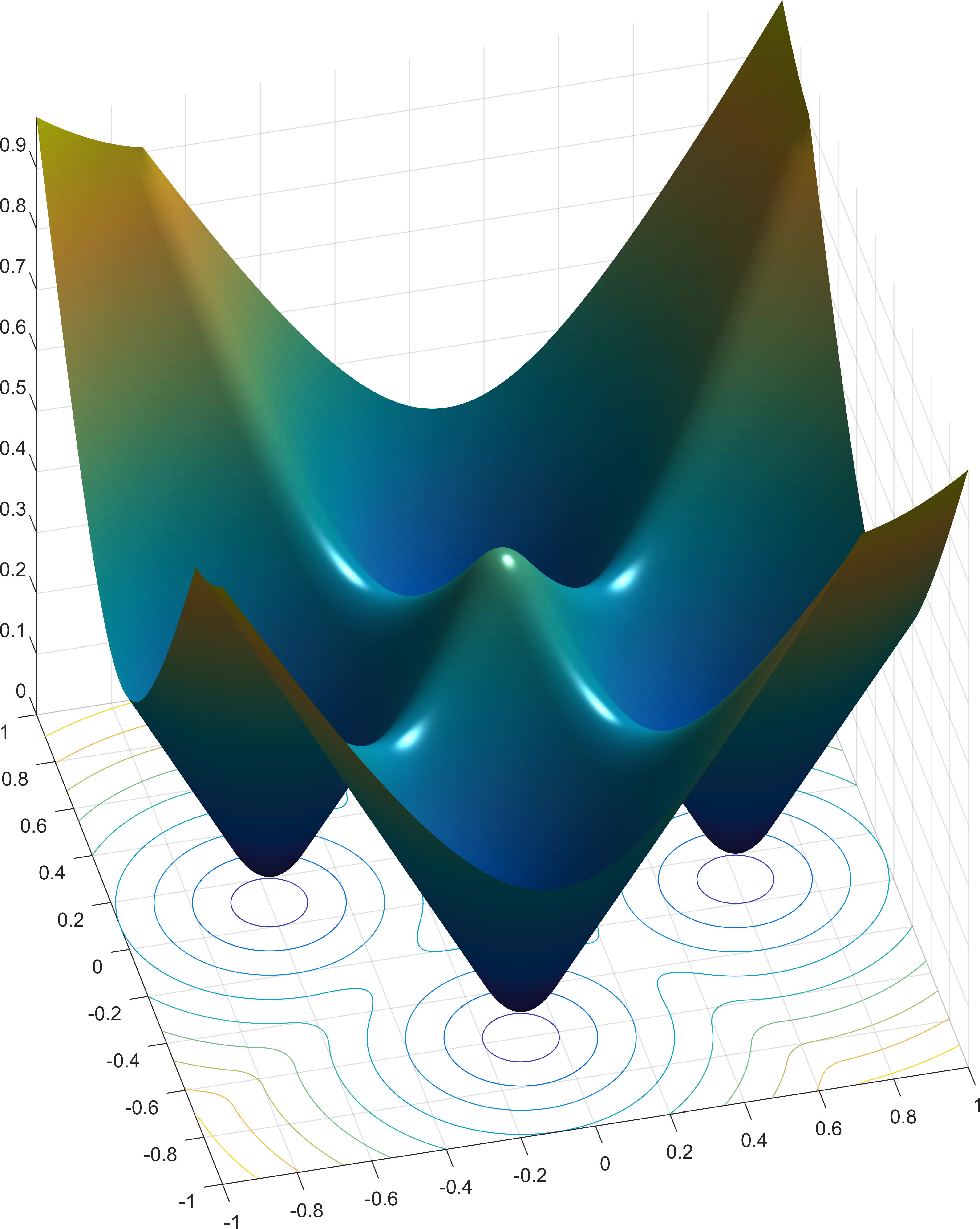}&
	\includegraphics[width=0.24\textwidth]{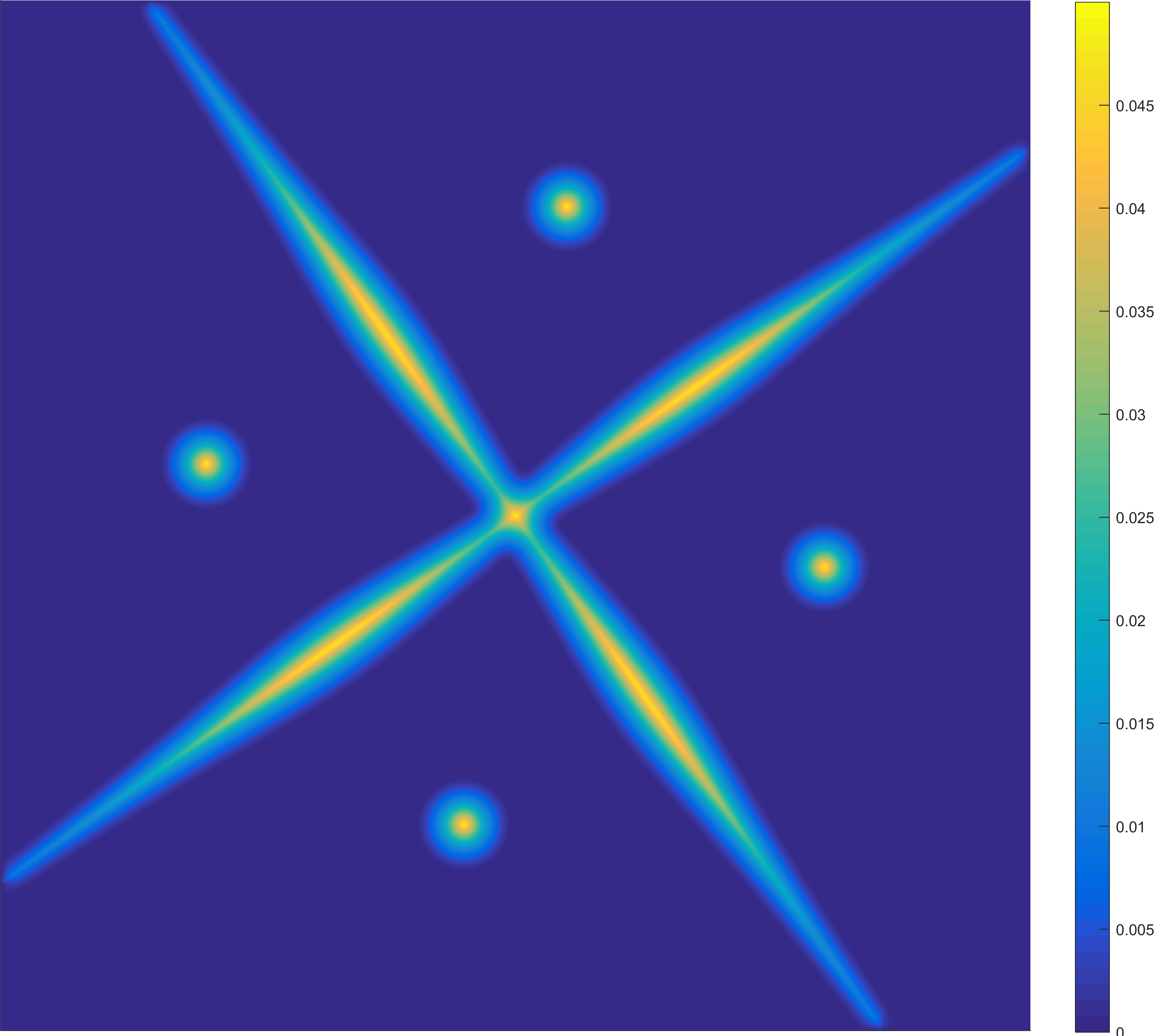}\\
	(a)&(b)&(c)
	\end{array}$}
\caption{\label{Fig:Smoothing}
Graph of $(a)$ a squared distance function $f$ to a four-point set,
$(b)$ its mixed transform $C_{\lambda}^u(C_{\lambda}^l(f))$ and $(c)$
the pointwise error $e=|f-C_{\lambda}^u(C_{\lambda}^l(f))|$.
}
\end{figure}

\medskip
It is not difficult to verify that if $f$ meets both \eqref{Eq:GrwthCond} and \eqref{Eq:GrwthCondUp}, 
for instance if $f$ is bounded, there holds 
\[
	C^l_\lambda(f)(x)\leq f(x)\leq C^u_\lambda(f)(x)\quad x\in\R^n\,,
\]
thus, the lower and upper compensated convex transforms are  $\lambda$-parametrised families of  transforms that
approximate $f$ from below and above respectively. Furthermore, they have smoothing effects and are tight approximations of $f$ 
in the sense that if $f$ is $C^{1,1}$ in a neighbourhood of $x_0$, there is a finite $\Lambda>0$, such
that $f(x_0)=C^l_\lambda(f)(x_0)$ (respectively, $f(x_0)=C^u_\lambda(f)(x_0)$ whenever
$\lambda\geq \Lambda$. This approximation property, which we refer to as tight approximation, is pivotal in the developments of the theory, because
it allows the transforms to be used for detecting singularities of functions 
by exploiting the fact that it is only when a point $x$ is close to a singularity point of $f$ we might find that the values of $C^l_{\lambda}(f)(x)$ and $C^u_{\lambda}(f)(x)$ might be different from that of $f(x)$  
\cite{ZOC15a}. Figure \ref{Fig:Smoothing} visualizes the smoothing and tight approximation of the mixed transform $C^u_{\lambda}(C^l_{\lambda}(f))$
of the squared distance function $f$ to a four-point set. Given the type of singularity of $f$, 
we apply the lower transform to $f$ which smooth the `concave'--like singularity followed by the upper transform
that smoothes the `convex'--like singularity of $C^l_{\lambda}(f)$ which are unalterated with respect to the original function $f$.
This can be appreciated by the graph of the 
pointwise error $e(x)=|f(x)-C_{\lambda}^u(C_{\lambda}^l(f))(x)|$ for $x\in\Omega$ which is zero everywhere 
but in a neighborhood of the singularities of $f$.

The transforms additionally satisfy the locality property that the values of $C^l_{\lambda}(f)$, $C^u_{\lambda}(f)$  
at  $x \in \R^n$ depend only on the values of $f$ in a neighbourhood of $x$, and are {translation invariant} 
in the sense that  $C^l_{\lambda}(f)$, $C^u_{\lambda}(f)$ are unchanged if the `weight' $|\cdot|^2$ in the formula \eqref{Eq:LwTr} 
and \eqref{Eq:UpTr} is replaced by $|\cdot-x_0|^2$ for any shift $x_0 \in \R^n$.  These  last two properties make the explicit calculation of  transforms tractable   
  for specific prototype functions $f$, which  facilitates the creation of dedicated 
extractors for a variety of different types of singularity using customised combinations of the transforms.

\medskip

\begin{figure}[H]
\centerline{$\begin{array}{cc}
	\includegraphics[width=0.35\textwidth]{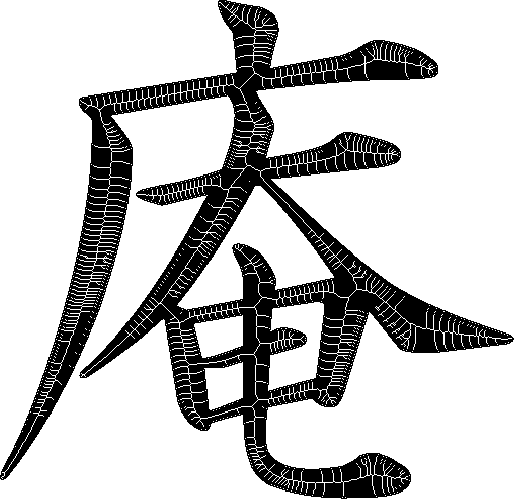}&
	\includegraphics[width=0.35\textwidth]{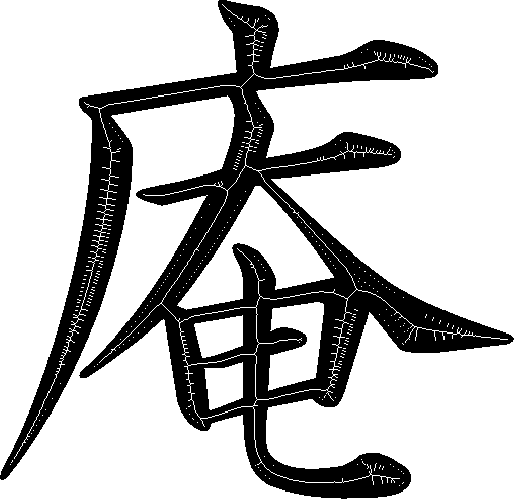}\\
	(a)&(b)
	\end{array}$}
\caption{\label{Fig:ScaleMMA}
Support of the multiscale medial axis map (suplevel set with level $t=10^{-8}\max_{x\in\R^2}\,{M_{\lambda}(\cdot;K)}$) 
with the `spurious' branches generated by pixelation of the boundary for $(a)$ $\lambda=1$
and  for  $(b)$ $\lambda=8$.}
\end{figure}

\begin{figure}[H]
\centerline{$\begin{array}{cc}
	\includegraphics[width=0.35\textwidth]{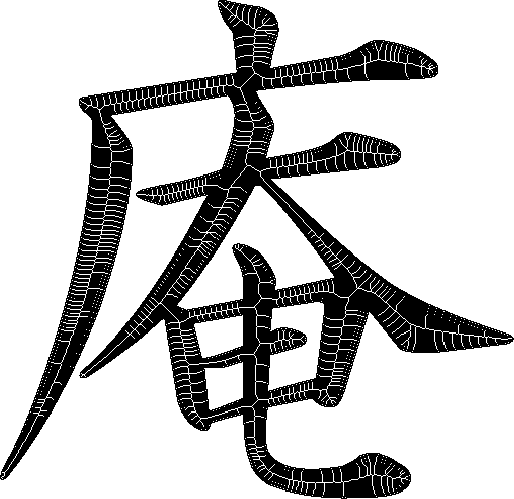}&
	\includegraphics[width=0.35\textwidth]{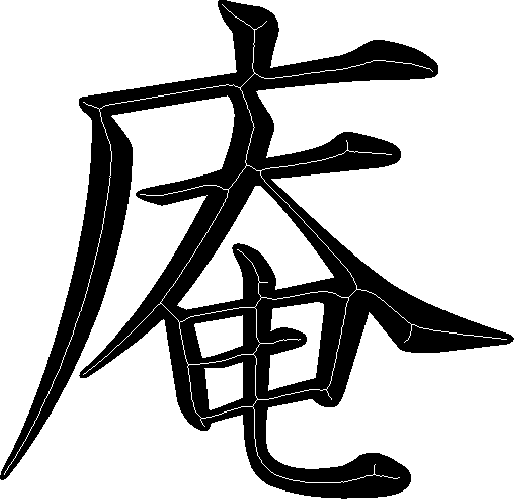}\\
	(a)&(b)
	\end{array}$}
\caption{\label{Fig:ScaleMMAHeights}
Selection of branches via the suplevel set of the multiscale medial axis map for $\lambda=1$ using different values of the threshold $t$,
$(a)$ $t=10^{-3}\max_{x\in\R^2}\,{M_{\lambda}(\cdot;K)}$ and 
$(b)$ $t=2\cdot 10^{-2}\max_{x\in\R^2}\,{M_{\lambda}(\cdot;K)}$.}
\end{figure}

These new geometric approaches enjoy key advantages over previous image and data processing techniques 
\cite{AK06,CS05,Sch15,VG15}. 
The curvature parameter $\lambda$  provides scales for features that allow users to select which size 
of feature they wish to detect, and the techniques are blind and global, in the sense that images/data 
are treated as a global object with no {\em a priori} knowledge required of, {\em e.g.,} feature location. 
Figure \ref{Fig:ScaleMMA} displays the $\lambda-$scale dependence in the case of the medial axis
where $\lambda$ is associated with the scale of the different branches whereas
Figure \ref{Fig:ScaleMMAHeights} shows the multiscale feature for given $\lambda$ associated with the height of the 
different branches of the multiscale medial axis map.

\medskip

\begin{figure}[H]
\centerline{$\begin{array}{cccc}
	\includegraphics[width=0.24\textwidth]{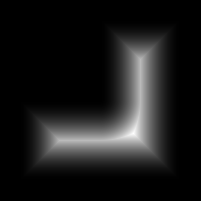}&
	\includegraphics[width=0.24\textwidth]{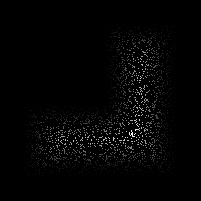}&
	\includegraphics[width=0.24\textwidth]{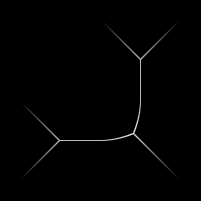}&
	\includegraphics[width=0.24\textwidth]{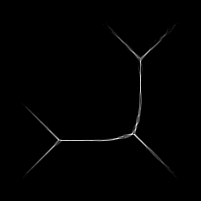}\\
	(a)&(b)&(c)&(d)
	\end{array}$}
\caption{\label{Fig:HausTent}
$(a)$ Image of $f(x,y)$;	$(b)$ Sampled image of $f(x,y)$ by random salt and pepper noise;
$(c)$ Stable ridges of $f(x,y)$;	$(d)$ Stable ridges from sampled image.
}
\end{figure}

Many of the methods  can also be shown  
to be stable under perturbation and different sampling techniques. Most significantly,  Hausdorff stability results 
can be rigorously proved for many of the methods. For example, the Hausdorff-Lipschitz continuity estimate \cite{ZOC15a} 
\[
	|C^u_{\lambda}(\chi_E)(x) - C^u_{\lambda}(\chi_F)(x)| \leq 2 \sqrt{\lambda} \dist_{\mathcal{H}}(E, F), \;\;\; x \in \R^n,
\]
shows that the upper transform $C^u_{\lambda}$ is Hausdorff stable  against sampling of geometric shapes defined by their characteristic functions.
Such stability is particularly important for the extraction of information when `point clouds' represent sampled domains. 
If a geometric shape is densely sampled, then from a human vision point of view, one can typically still identify geometric 
features of the sample and sketch its boundary. From the mathematical/computer science perspective, however, feature identification  
from sampled domains is challenging and usually methods are justified  only by either {\em ad hoc} arguments or numerical experiments.
Figure \ref{Fig:HausTent} displays an instance of this property where we show the edges of the continuous
nonnegative function 
$f(x,y)=\dist^2((x,y),\,\partial\Omega)$, with 
$(x,y)\in\Omega=([-1.5,\,1.5]\times[-1.5,\,1.5])\setminus([-1.5,\,0.5]\times[-1.5,\,-0.5])$, and 
of its sparse sampling $f\cdot\chi_A$ where $A\subset\Omega$ is a sparse set
(see Figure  \ref{Fig:HausTent}$(a)$, $(b)$ respectively). 
Due to the Hausdorff stability  of the stable
ridge transform, we are able to recover an approximation of the ridges from the sampled image 
(compare Figure \ref{Fig:HausTent}$(c)$, $(d)$).

\medskip

Via fast and robust numerical implementations of the transforms \cite{ZOC20}, this theory also gives rise to a highly-effective 
computational toolbox for applications. The  efficiency of the numerical computations  benefits  greatly from the locality property, 
which holds despite the global nature of the convex envelope itself. 

\medskip

Before we describe the applications of this theory, we provide next alternative
characterizations of the compensated convex transforms.

\subsection{Related areas: Semiconvex envelope}
Given the definitions \eqref{Eq:LwTr} and  \eqref{Eq:UpTr},
lower and upper compensated convex transforms can be considered as parameterized semiconvex and
semiconcave envelopes, respectively, for a given function.
The notions of semiconvex and semiconcave functions go back at least to Reshetnyak \cite{Res56} and have 
since been studied by many authors in different contexts 
(see, for example, \cite{Alb94,AAC92,CS04,LL86,Roc79,Via83}).
Let $\Omega\subseteq\R^n$ be an open set, we recall that a function $f:\Omega\to \mathbb{R}\cup\{+\infty\}$
is semiconvex  if there is a constant $C\geq 0$ such that
$f(x)=g(x)-C|x|^2$ with $g$ a convex function. More general weight functions,
such as $|x|\sigma(|x|)$, for example, are also used in the literature for defining more
general semiconvex functions \cite{Alb94,AAC92,ADT05,Rol79,Rol00}. 
Since general DC-functions (difference of convex functions) \cite{Har59} and
semiconvex/semiconcave functions are locally Lipschitz functions in their essential domains (\cite[Theorem 2.1.7]{CS04}),
Rademacher's theorem implies that they are  differentiable almost everywhere.
Fine properties for the singular sets of convex/concave and semiconvex/semiconcave functions have been
studied extensively \cite{AAC92,AC99,CS04} showing that the singular set of a semiconvex/semiconcave
function is rectifiable. By applying results and tools of the theory of compensated convex transforms, 
it is possible therefore to study how such functions can be effectively approximated by smooth functions, 
whether all singular points are of the same type, that is, for semiconcave (semiconvex) functions, whether all singular
points are geometric `ridge' (`valley') points, how singular sets can be effectively extracted beyond
the definition of differentiability and how the information concerning `strengths' of different
singular points can be effectively measured. These are all questions relevant to applications in 
image processing and computer-aided geometric design. An instance of this study, for example, has been 
carried out in \cite{ZCO15,ZCO16b} to study the singular set of the Euclidean squared-distance
 function $\dist^2(\cdot,\Omega^c)$ to the complement of a bounded open domain
 $\Omega\subset\mathbb{R}^n$ (called the medial axis \cite{Blu67} of the domain $\Omega$) 
 and of the weighted squared distance function \cite{OBSC00}.

\subsection{Related areas: Proximity hull}
Another characterization of the compensated convex transforms is in terms of the critical mixed Moreau envelopes, given that 
 \begin{equation}\label{Eq:CmpTrMo}
		C^l_\lambda(f)(x)=M^\lambda(M_\lambda(f))(x)\,,\qquad
		C^u_\lambda(f)(x)=M_\lambda(M^\lambda(f))(x)\,,
\end{equation}
where the Moreau lower and upper envelopes \cite{Mor65,Mor66} are defined, in our notation, respectively, by
\begin{equation}
\begin{array}{l}
	\displaystyle M_\lambda(f)(x)=\inf\{f(y)+\lambda|y-x|^2,\; y\in\mathbb{R}^n\} \,,\\[1.5ex]
	\displaystyle M^\lambda(f)(x)=\sup\{f(y)- \lambda|y-x|^2,\; y\in\mathbb{R}^n\}\,,
\end{array}
\end{equation}
with $f$ satisfying the growth condition \eqref{Eq:GrwthCond} and \eqref{Eq:GrwthCondUp}, respectively. 
Moreau envelopes play important roles in optimization, nonlinear analysis, 
optimal control and Hamilton-Jacobi equations, both theoretically and computationally \cite{CIL92,CS04,HL01,RW98}.
The mixed Moreau envelopes $M^\tau(M_\lambda(f))$ and $M_\tau(M^\lambda(f))$ coincide with the 
Lasry-Lions double envelopes $(f_{\lambda})^{\tau}$ and $(f^{\lambda})_{\tau}$ defined in \cite{LL86} 
by \eqref{Eq:LLLw} and \eqref{Eq:LLUp}, respectively, 
in the case of $\lambda=\tau$  and 
are also referred to in \cite{Str96,RW98} as
proximal hull and upper proximal hull, respectively. They have been extensively studied  
and used as approximation and smoothing methods of not necessarily convex functions \cite{AA93,CS04,PB13,Har09}. 
In particular,  in the partial differential equation literature, the focus of the study of the mixed Moreau 
envelopes $M^{\tau}(M_{\lambda}(f))$ and $M_{\tau}(M^{\lambda}(f))$ for the case $\tau>\lambda$ are known, under suitable growth conditions, 
 as the Lasry-Lions regularizations of $f$ of 
parameter $\lambda$ and $\tau$. In this case, the mixed Moreau envelopes are both $C^{1,1}$ functions \cite{AA93,CS04,LL86}. However, crucially they are not 
`tight approximations' of $f$, in contrast with our lower
and upper transforms $C^l_\lambda(f)(x)$ and $C^u_\lambda(f)(x)$ \cite{Zha08a}. 
Generalised $\inf$ and $\sup$ convolutions have also been considered, for instance in \cite{CS04,RW98}.
However, due to the way these regularization operators are defined, proof of mathematical
and geometrical results to describe how such approximations work has usually been challenging,
making their analysis and applications very difficult. As a result, the study of the proximal hull using the 
characterization in terms of the compensated convex transform would make them 
much more accesible and feasible for real world applications. 

\subsection{Related areas: Mathematical morphology}
Moreau lower and upper envelopes have also been employed in mathematical morphology in the 1990's \cite{Jac92,BS92},
to define greyscale erosion and dilation morphological operators,  
whereas the critical mixed Moreau envelopes $M^\lambda(M_\lambda(f))$ and $M_\lambda(M^\lambda(f))$ 
are greyscale opening and closing morphological operators  \cite{Ser82,Soi04,Shi09}.
In convex analysis, the infimal convolution of $f$ with $g$ is denoted as
$f\Box g$ and is defined as \cite{Roc70,BC17,CLSW98,RW98}
\[
	(f\Box g)(x)=\underset{y}{\inf}\, \{f(y)+g(x-y)\}\,.
\] 
This is closely related to the erosion of $f$ by $g$, 
given that 
\[
	(f\Box g)(x)=f(x)\ominus(-g(-x))\,.
\]
Thus if we denote by $b_{\lambda}(x)=-\lambda|x|^2$
the quadratic structuring function, introduced for the first time in \cite{Jac92,Boo92,JM96,Jac94,Jac95}, 
then with the notation
of \cite{Ser82,Soi04,Shi09,AGLM93}, we have
\begin{equation}
	\begin{array}{l}
		\displaystyle M_{\lambda}(f)(x)=\underset{y\in\mathbb{R}^n}{\inf}\{f(y)-b_{\lambda}(y-x)\}=: f\ominus b_{\lambda}\,,\\[1.5ex]
		\displaystyle M^{\lambda}(f)(x)=\underset{y\in\mathbb{R}^n}{\sup}\{f(y)+b_{\lambda}(y-x)\}=: f\oplus  b_{\lambda}
	\end{array}
\end{equation}
so that \eqref{Eq:CmpTrMo} can be written alternatively as 
\begin{equation} \label{Eq:OpClMatMrp}
	C_{\lambda}^l(f)=(f\ominus b_{\lambda})\oplus b_{\lambda}
	\quad\text{and}\quad
	C_{\lambda}^u(f)=(f\oplus b_{\lambda}) \ominus b_{\lambda}\,.
\end{equation}
The application of $M^\lambda(M_\lambda(f))$ and $M_\lambda(M^\lambda(f))$ in mathematical morphology  \cite{Ser82,Soi04,Shi09}, 
however, has not met with corresponding success, nor have  its properties been fully explored. 
This is in contrast with the r\^{o}le,  recognized since its introduction, that is played by paraboloid structuring functions
in defining morphological scale-spaces in image analysis \cite{Jac92,Boo92,SW16,Lin94,Lin11,MS87,MS87,WII99,Wei98}. 
For this and related topics concerning the
morphological scale-space representation produced by quadratic structuring functions, we refer to the pionering works \cite{Jac92,Boo92}.
Here, we would like only to observe that through identity \eqref{Eq:CmpTrMo}, we have a direct characterization
of the quadratic structuring based opening and closing morphological operators, either in terms of the convex envelope
(see \eqref{Eq:LwTr} and \eqref{Eq:UpTr}) or in terms of envelope from below/above with parabolas (see \eqref{Eq:LwTrEnvPrb} and \eqref{Eq:UpTrEnvPrb}). 
Such characterizations will allow us to derive various new
geometric and stability properties for the opening and closing morphological operators. 
Furthermore, when we apply compensated convex transforms to extract singularities from characteristic functions of compact
geometric sets, our operations can be viewed as the application of morphological operations devised for 
`greyscale images' to `binary images'. 
As a result, it might look not efficient to apply more involved operations for processing binary images, 
when in the current literature \cite{Ser82,Soi04,Shi09} there are `binary' set theoretic morphological 
operations that have been specifically designed for the tasks under examination.
Nevertheless, an advantage of adopting our approach is that the compensated convex transforms of characteristic functions 
are (Lipschitz) continuous, therefore 
applying a combination of transforms will produce a landscape of various levels (heights) that can be designed to highlight 
a specific type of singularity. We can then 
extract multiscale singularities by taking thresholds at different levels. In fact, the graphs of functions obtained by
combinations of compensated convex transforms contain much more geometric information than binary operations 
that produce simply a yes or no answer. Also, for `thin' geometric structures,
such as curves and surfaces, it is difficult to design `binary' morphological operations  to be 
Hausdorff stable.   


\subsection{Related areas: Quadratic envelopes}
From definition \eqref{Eq:LwTr}, it also follows that 
$C^l_\lambda(f)(x)$ is the envelope of all the quadratic functions with fixed quadratic term $\lambda |x|^2$ that
are less than or equal to $f$, that is,
\begin{equation} \label{Eq:LwTrEnvPrb}
	C^l_\lambda(f)(x)=\sup\left\{
				-\lambda|x|^2+\ell(x):\; -\lambda|y|^2+\ell(y)\leq f(y)\;\;\text{\rm for all }y\in\R^n
				\;\;\text{\rm and }\ell\;\;\text{\rm affine}\right \}\,,
\end{equation}
whereas from \eqref{Eq:UpTr} it follows that $C^u_\lambda(f)(x)$ is the envelope of all the quadratic functions with fixed quadratic term
$\lambda |x|^2$ that are greater than or equal to $f$, that is,
\begin{equation}\label{Eq:UpTrEnvPrb}
	C^u_\lambda(f)(x)=\inf\left\{
				\lambda|x|^2+\ell(x):\; f(y)\leq \lambda|y|^2+\ell(y)\;\;\text{\rm for all }y\in\R^n
				\;\;\text{\rm and }\ell\;\;\text{\rm affine}\right \} \,.
\end{equation}
This characterization was first given in \cite[Eq. (1.4)]{ZOC15a} and can be derived by noting that 
since the convex envelope of a function $g$ can be characterized as the pointwise supremum of the family of all the affine functions 
which are majorized by $g$, we have then
\begin{equation}\label{Eq:CharLwTR}
\begin{split}
		C_{\lambda}^{l}(f)(x)&=\co[f+\lambda|\cdot|](x)-\lambda|x|^2\\[1.5ex]
				     &=\sup{\ell(x):\,\ell(y)\leq f(y)+\lambda|y|^2\,\text{ for any }y\in\mathbb{R}^n}\\[1.5ex]
				     &=\sup{\ell(x)-\lambda|x|^2:\,\ell(y)-\lambda|y|^2\leq f(y)\,\text{ for any }y\in\mathbb{R}^n},
\end{split}
\end{equation}
which is \eqref{Eq:LwTrEnvPrb}. As stated before, \eqref{Eq:CharLwTR} can be in turn related directly 
to the Moreau's mixed envelope.
The characterization \eqref{Eq:LwTrEnvPrb} has been recently also reproposed by \cite{Car19} for the study of low-rank approximation 
and compressed sensing.

It is instructive to compare this characterization with \eqref{Eq:MorEnvParab} below about the Moreau envelopes.

\subsection{Outline of the Chapter}
The plan of the paper is as follows. 
After this general introduction, we will introduce relevant notation and recall 
basic results in convex analysis and compensated convex transforms in the next section. 
In Section \ref{Sec:Filters} we introduce the different compensated convex based transforms that
we have been developing. Their definition can be either motivated by a mere application of 
key properties of the basic transforms, namely the lower and upper transform, or by 
an \textit{ad--hoc} designed combinations of the basic transforms so to create a singularity 
at the location of the feature of interest. Section \ref{Sec:NumAlg} introduces some of the 
numerical schemes that can be used for the numerical realization of the compensated convex based transforms,
namely of the basic transform given by the
lower compensated convex transform. We will therefore describe the convex based and Moreau based 
algorithms, which can be both used according to whether we refer to the definition  \eqref{Eq:LwTr}
or the characterization \eqref{Eq:CmpTrMo} of the lower compensated convex transform.
Section \ref{Sec:NumExmpl} contains some representative applications of the
transformations introduced in this paper. More specifically, we will consider an application to shape interrogation
by considering the problem of identifying the location of intersections of manifolds represented by point clouds, 
and applications of 
our approximation compensated convex transform to the reconstruction of surfaces
using level lines and isolated points, image inpainting and salt \& pepper noise removal.  


\setcounter{equation}{0}
\section{Notation and Preliminaries}\label{Sec:Notat}
Throughout the paper $\mathbb{R}^n$ denotes the $n$-dimensional Euclidean space, whereas $|x|$ and 
$x\cdot y$ are the standard Euclidean norm and inner product respectively, for  $x,\, y\in \mathbb{R}^n$.
Given a non-empty subset $K$ of $\R^n$,
$K^c$  denotes the complement of $K$ in $\R^n$, i.e. $K^c=\R^n\setminus K$,  $\overline{K}$  its closure,
$\co[K]$  the convex hull of $K$, that is, the smallest (with respect to inclusion) 
convex set that contains the set $K$ and $\chi_K$ its characteristic function,  
that is, $\chi_K(x)=1$ if $x\in K$ and  $\chi_K(x)=0$ if $x\in K^c$. 
The Euclidean distance transform of a non-empty set $K\subset \mathbb{R}^n$ 
is the function that, at any point $x\in \mathbb{R}^n$, associates the Euclidean distance of 
$x$ to $K$, which is defined as $\inf\{|x-y|,\; y\in K\}$ and is
denoted as $\dist(x;\,K)$. Let $\delta>0$, the open
$\delta$-neighbourhood $K^\delta$ of $K$ is then defined by $K^\delta=\{x\in \mathbb{R}^n,\; \dist(x,\, K)<\delta\}$ and 
is an open set.
For $x\in \R^n$ and $r>0$, $B(x;\,r)$ indicates the open ball with center $x$ and radius $r$ whereas $S(x;\,r)$ denotes the 
sphere with center $x$ and radius $r$, that is, $S(x;\,r)=\partial B(x;\,r)$ is the boundary of $B(x;\,r)$.
The suplevel set of a function $f:\Omega\subseteq\mathbb{R}^n\to\mathbb{R}$ of level $\alpha$ is the set
\begin{equation}\label{Eq:SupLev}
	S_{\alpha}f=\{x\in\Omega:\,f(x)\geq \alpha\}\,,
\end{equation}
whereas the level set of $f$ with level $\alpha$ is also defined by \eqref{Eq:SupLev} with the inequality sign replaced by the 
equality sign. Finally, we use the notation $Df$ to denote the derivative of $f$.

Next we next list some  basic properties of compensated convex transforms. Without loss of generality, these properties
are stated mainly for the lower compensated convex transform given that it is then not difficult
to derive the corresponding results for the upper compensated convex transform using \eqref{Eq:UpTr}.
Only in the case $f$ is the characteristic function of a set $K$, i.e. $f=\chi_K$, we will refer 
explicitly to $C_{\lambda}^u(\chi_K)$ given that $C_{\lambda}^l(\chi_K)(x)=0$ for any $x\in\mathbb{R}^n$ if $K$ is, e.g., a finite set.
For details and proofs we refer to \cite{Zha08a,ZOC15a} and references therein, 
whereas for the relevant notions of convex analysis we refer to \cite{HL01,Roc70,BC17}.

\begin{defi}
Given a function $f:\mathbb{R}^n\to \mathbb{R}$ bounded below, the convex envelope $\co[f]$ is 
the largest convex function not greater than $f$. 
\end{defi}
This is a global notion. By  Carath\'eodory's Theorem \cite{HL01,Roc70}, we have
\begin{equation}
\begin{split}
	\co[f](x_0)=\inf_{\begin{subarray}{l}x_i\in \mathbb{R}^n\\
			i=1,\ldots,n+1\end{subarray}}\,\Big\{
			\sum_{i=1}^{n+1}\lambda_i f(x_i):&\,
			\,\sum_{i=1}^{n+1}\lambda_i=1,\,\sum_{i=1}^{n+1}\lambda_ix_i=x_0,\\[1.5ex]
				&\,\lambda_i\geq 0\,\,i=1,\ldots,n+1
			\Big\}\,,
\end{split}
\end{equation}
that is, the convex envelope of $f$ at a point $x_0\in\mathbb{R}^n$ depends on the values of 
$f$ on its whole domain of definition, namely $\mathbb{R}^n$ in this case.
We will however introduce also a local version of this concept
which will be used to formulate the locality property of the compensated convex transform 
and is fundamental for our applications. 

\begin{defi}
	Let $r>0$, $x_0\in\mathbb{R}^n$. Assume $f:B(x_0;\,r)\to\mathbb{R}$ to be bounded from below. 
	Then the value of the local convex envelope of $f$ at $x_0$ in $B(x_0;\,r)$ is defined by 
	\begin{equation}\label{Eq:LocCnvx}
	\begin{split}
		\co_{\overline{B}(x_0;\,r)}[f](x_0)=
		\inf_{\begin{subarray}{l}x_i\in B(x_0;\,r)\\
			i=1,\ldots,n+1\end{subarray}}\,\Big\{
			\sum_{i=1}^{n+1}\lambda_i f(x_i):&
			\,\,\sum_{i=1}^{n+1}\lambda_i=1,\,\sum_{i=1}^{n+1}\lambda_ix_i=x_0,\\[1.5ex]
			&\,\,\lambda_i\geq 0\,\,i=1,\ldots,n+1
			\Big\}\,.
	\end{split}
	\end{equation}
\end{defi}
Unlike the global definition, the infimum in \eqref{Eq:LocCnvx} is taken only over convex combinations
in $B(x_0;\,r)$ rather than in $\mathbb{R}^n$. 

As part of the convex analysis reminder, we also recall the definition of the Legendre-Fenchel transform.
\begin{defi}\label{Def:LT}
	Let $f:\mathbb{R}^n\to\mathbb{R}\cup\{+\infty\}$, $f\not\equiv +\infty$ and theer is an affine function minorizing $f$
	on $\mathbb{R}^n$. The conjugate (or Legendre-Fenchel transform) of $f$ is
	\begin{equation}\label{Eq:LFT}
			f^{\ast}:s\in\mathbb{R}^n\to f^{\ast}(s)=\sup_{x\in\mathbb{R}^n}\,\{ x\cdot s-f(x)  \}\,,
	\end{equation}
	and the biconjugate of $f$ is $(f^{\ast})^{\ast}$.
\end{defi}
We have then the following results.
\begin{prop}
	For $f$ satisfying the conditions of Definition \ref{Def:LT}, the conjugate $f^{\ast}$ is a lowersemicontinuous convex
	function and $(f^{\ast})^{\ast}$ is equal to the lowersemicontinuous convex envelope of $f$.
\end{prop}

Before stating the properties of interest of the compensated convex transforms, we describe the relationship between
the compensated convex transforms and other infimal convolutions. 

Let $f:\mathbb{R}^n\to \mathbb{R}$ 
satisfy \eqref{Eq:GrwthCond} and \eqref{Eq:GrwthCondUp}. As we have mentioned in the introduction, concepts closely related to the compensated convex transforms are the 
Lasry-Lions regularisations for parameters $\lambda$ and $\tau$ with $0<\tau<\lambda$, which are defined in \cite{LL86} as follows
\begin{equation}\label{Eq:LLLw}
\begin{split}
	(f_{\lambda})^{\tau}(x)	&=\sup_{y\in\R^n}\,\inf_{u\in\R^n}\,\big\{f(u)+\lambda|u-y|^2-\tau|y-x|^2\big\}	\\[1.5ex]
				&=M^{\tau}(M_{\lambda}(f))(x)\,,
\end{split}
\end{equation}
and 
\begin{equation}\label{Eq:LLUp}
\begin{split}
	(f^{\lambda})_{\tau}(x)	&=\inf_{y\in\R^n}\,\sup_{u\in\R^n}\,\big\{f(u)-\lambda|u-y|^2+\tau|y-x|^2\big\}	\\[1.5ex]
				&=M_{\tau}(M^{\lambda}(f))(x)\,.
\end{split}
\end{equation}
Both $(f_{\lambda})^{\tau}$ and $(f^{\lambda})_{\tau}$ approach $f$
from below and above respectively, as the parameters $\lambda$ and $\tau$ go to $+\infty$.
If $\lambda=\tau$, then $(f_{\lambda})^{\lambda}=M^{\lambda}(M_{\lambda}(f))$ is called proximal hull of $f$ whereas 
$(f^{\lambda})_{\lambda}=M_{\lambda}(M^{\lambda}(f))$ is refererd to as the upper proximal hull of $f$.
It is not difficult to verify that whenever $\tau>\lambda>0$ the following relation holds between the compensated convex transforms,
the Moreau envelopes and the Lary--Lions regularizations of $f$ \cite{Zha08a},
\begin{equation}
\begin{split}
		M_{\lambda}(f)(x)\leq M^{\lambda}(M_{\tau}(f))(x)\leq C_{\lambda}^l(f)(x)\leq f(x)\leq  C_{\tau}^u(f)(x)\leq
		M_{\lambda}(M^{\tau}(f))(x)&\leq M^{\tau}(f)(x)\\[1.5ex]
		&\text{for }x\in\R^n\,.
\end{split}
\end{equation}

Given $f:\mathbb{R}^n\to\mathbb{R}$, we recall also that the lower semicontinuous
envelope of $f$ is defined in \cite{HL01,Roc70} by 
\begin{equation}
	\underline{f}:\,x\in\mathbb{R}^n\,\mapsto\, \underline{f}(x)= \underset{y\to x}{\lim\inf}\,f(y)\,,
\end{equation}
and since there holds
\begin{equation}\label{Eq:EqLwTR}
	C_{\lambda}^l(f)(x)=C_{\lambda}^l(\underline{f})(x)\quad\text{ for }x\in\mathbb{R}^n\,\,,
\end{equation}
without loss of generality, in the following we can assume that the 
functions are lower semicontinuous.

The monotonicity and approximation properties of $C^l_\lambda(f)$ with respect to $\lambda$ is described by the following results. 
\begin{prop}\label{Sec2.Pro.MonLwTr}
Given $f:\mathbb{R}^n\to \mathbb{R}$ that satisfies \eqref{Eq:GrwthCond}, 
then for all $A_1<\lambda<\tau<\infty$, we have
	\begin{equation}
		C^l_\lambda(f)(x)\leq C^l_\tau(f)(x)\leq f(x)\qquad\text{for }x\in\R^n\,,
	\end{equation}
and, for $\lambda>A_1$
	\begin{equation}\label{Eq:Aprx}
		\lim_{\lambda\to\infty}\,C^l_\lambda(f)(x)=f(x)\qquad\text{for }x\in\R^n\,.
	\end{equation}
\end{prop}
The approximation of $f$ from below by $C^l_\lambda(f)$ given by \eqref{Eq:Aprx} can be better specified,
given that $C^l_\lambda(f)$ realizes a `tight' approximation of the function $f$ in the following sense 
(see \cite[Theorem 2.3$(iv)$]{Zha08a}).
\begin{prop}\label{Sec2.Pro.Tight}
	Let $f\in C^{1,1}(\overline B(x_0;\,r))$, $x_0\in\mathbb{R}^n$, $r>0$.
	Then for sufficiently large $\lambda>0$, we have that $f(x_0)=C^l_\lambda(f)(x_0)$.
	If the gradient of $f$ is Lipschitz in $\mathbb{R}^n$ with Lipschitz constant $L$, then $C_{\lambda}^l(f)(x)=f(x)$ for all 
	$x\in\mathbb{R}^n$ whenever $\lambda\geq L$.
\end{prop}
The property of `tight' approximation plays an important role in the definition of the transforms
introduced in Section \ref{Sec:Filters}. Related to this property is the density property of the lower
compensated transform established in \cite{ZOC15a} that can be viewed as a tight approximation 
for general bounded functions.
\begin{teo}\label{Thm.Dens}
	Suppose $f:\mathbb{R}^n\to\mathbb{R}$ is bounded, satisfying
	$|f(x)|\leq M$ for some $M>0$ and for all $x\in \mathbb{R}^n$.
	Let $\lambda>0$, $x_0\in\R^n$ and define $R_{\lambda,M}=(2+\sqrt{2})\sqrt{M/\lambda}$. 
	Then there are  $x_i\in \overline B(x_0;R_{\lambda,M})$, with $x_i\neq x_0$,
	and $\lambda_i\geq 0$ for $i=1,\ldots, n+1$, satisfying
	$\sum^{n+1}_{i=1}\lambda_i=1$ and
	$\sum^{n+1}_{i=1}\lambda_ix_i=x_0,$ such that
	\begin{equation*}
		C^l_\lambda(f)(x_i)=\underline{f}(x_i)\quad\text{\rm for }i=1,\ldots,n+1\,.
	\end{equation*}
\end{teo}
Since the lower transform satisfies 
\begin{equation*}
	C^l_\lambda(f)\leq \underline{f}\leq f\,,
\end{equation*}
if we consider the following set
\begin{equation*}
	T_l(f,\lambda)=\{x\in\mathbb{R}^n:\; C^l_\lambda(f)(x)=\underline{f}(x)\}\,,
\end{equation*}
as a result of Theorem \ref{Thm.Dens}, the set of points at which the lower compensated convex transform
equal the original function satisfies a density property, that is, 
the closed $R_{\lambda,M}$-neighbourhoods of $T_l(f,\lambda)$ covers $\mathbb{R}^n$. 
For any point $x_0\in\mathbb{R}^n$, the point $x_0$ is contained in the local convex hull
$\co\left[T_l(f,\lambda)\cap \bar B(x_0;R_{\lambda,M})\right]$. Furthermore, 
if $f$ is bounded and continuous, $T_l(f,\lambda)$ is exactly the set
of points at which $f$ is $\lambda$-semiconvex \cite{CS04}, i.e.,  points $x_0$ where 
\begin{equation*}
	f(x)\geq f(x_0)+\ell(x)-\lambda|x-x_0|^2\quad \text{\rm for all } x\in \mathbb{R}^n
\end{equation*}
with  $\ell$ an affine function
satisfying $\ell(x_0)=0$ and Condition \eqref{Eq:GrwthCond} holds for $f$.

A fundamental property for the appplications is the locality of the compensated convex transforms.
For a lowersemicontinuos function that is in addition bounded on any bounded set, 
the locality property was established for this general case
in \cite{Zha08a}. We next report its version for a bounded function which is relevant for the applications
to image processing and shape interrogation \cite{ZOC15a}.
\begin{teo}  
Suppose $f:\mathbb{R}^n\mapsto\mathbb{R}$ is bounded, satisfying $|f(x)|\leq M$ for some $M>0$ and 
for all $x\in \mathbb{R}^n$. Let $\lambda>0$ and $x_0\in\mathbb{R}^n$, then the following 
locality properties hold,
\begin{equation}\label{Eq:Loc} 
	\begin{array}{ll}
		\displaystyle C^l_\lambda(f)(x_0)= 
		\inf\Big\{\sum^{n+1}_{i=1}\lambda_i(f(x_i)+\lambda|x_i-x_0|^2),\;&
			\displaystyle  \lambda_i\geq 0,\;\sum^{n+1}_{i=1}\lambda_i=1,\;
			\sum^{n+1}_{i=1}\lambda_ix_i=x_0 \\[1.5ex]
		&\displaystyle |x_i-x_0|\leq R_{\lambda,M},\Big\}\,,
	\end{array}
\end{equation}
where $R_{\lambda,M}$ is the same as in Theorem \ref{Thm.Dens}.
\end{teo}
Since the convex envelope is affine invariant,  it is not difficult to realize that there holds
\begin{equation}
	C^l_\lambda(f)(x_0)=\co[\lambda|(\cdot)-x_0|^2+f](x_0)\quad\text{for }x_0\in\mathbb{R}^n\,
\end{equation}
thus condition \eqref{Eq:Loc} can be equivalently written as
\begin{equation}
	C^l_\lambda(f)(x_0)=\co_{\overline{B}(x_0;\,R_{\lambda,M})}\,[\lambda|(\cdot)-x_0|^2+f](x_0)\,.
\end{equation}
Despite the definition of $C^l_\lambda(f)$ involves the convex envelope of $f+\lambda|\cdot|^2$,
the value of the lower transform for a bounded function at a point depends on the values of the function 
in its $R_{\lambda,M}$-neighborhood. Therefore when $\lambda$ is large, the neighborhood
will be very small. If $f$ is globally Lipschitz, this result is a special case of Lemma 3.5.7 at p. 72
of \cite{CS04}.

The following property shows that the mapping $f\mapsto C_{\lambda}^l(f)$ is nondecreasing, 
that is we have

\begin{prop}\label{Sec2.Pro.OrdLwTr}
If $f\leq g$ in $\R^n$ and satisfy \eqref{Eq:GrwthCond}, then 
\begin{equation*}
	C^l_\lambda(f)(x)\leq C^l_{\lambda}(g)(x)\qquad\text{for }x\in\R^n\text{ and }
	\lambda\geq \max\{A_{1,f}, \,A_{1,g}\}\,.
\end{equation*}
\end{prop}

We conclude this section by stating some results on the 
Hausdorff stability of the compensated convex transforms. This is the relevant concept 
of stability we use to assess the change of the transformations with respect to 
perturbations of the set, thus it refers to the behaviour of the compensated convex transform of the 
characteristic functions of subsets $K$ of $\mathbb{R}^n$. We first state a result
that highlights the geometric structure of the upper transform of $\chi_K$.

\begin{teo} (Expansion Theorem) 
Let $E\subset \mathbb{R}^n$ be a non-empty set and let $\lambda>0$ be fixed, then
\begin{equation*}
	C^u_\lambda(\chi_E)(x)\quad\left\{\begin{array}{l} =1,\quad{\rm if}\, x\in \bar E,\\
	=0,  \quad{\rm if}\, x\in (\bar E^{1/\sqrt{\lambda}})^c,\\
	\in (0,\,1), \quad{\rm if}\, x\in E^{1/\sqrt{\lambda}}\setminus \bar E.
	\end{array}
	\right.
\end{equation*}
\end{teo}

Next, we recall  the definition of Hausdorff distance from \cite{AT04}.

\begin{defi} \label{Sec2.Def.HausDist}
	Let $E,\, F$ be non-empty subsets of $\R^n$. The Hausdorff distance
	between $E$ and $F$ is defined  by
	\begin{equation*}\label{Sec2.Eq.HausDist}
		\dist_{\mathcal{H}}(E,F)=\inf\left\{\delta>0: F\subset E^\delta\; \text{and}\;\, E\subset F^\delta\right\}.
	\end{equation*}
\end{defi}

This definition is also equivalent to saying that 
\[
	\dist_{\mathcal{H}}(E,F)=\max\Big\{\sup_{x\in E}\dist(x;\,F),\; \sup_{x\in F}	\dist(x;\,E) \Big\}\,.
\]

It is well-known and easy to prove that the Euclidean distance function $\dist(x,\, K)$ is 
Hausdorff-Lipschitz continuous in the sense that for given $K$ and $S\subset \mathbb{R}^n$ 
non-empty compact sets, we have
\begin{equation*}
	|\dist(x,\, K)-\dist(x,\, S)|\leq \dist_{\mathcal{H}}(K,S)\,.
\end{equation*}

In order to study the Hausdorff-Lipschitz continuity of the upper compensated convex transform 
of characteristic functions of compact sets, we introduce the distance 
based function $D^2_\lambda(x,\,K)$ defined by 
\begin{equation}\label{Eq:DistTr}
	D^2_\lambda(x,\,K)=\left(\max\left\{0,\,1-\sqrt{\lambda}\,\dist(x,\,K)\right\}\right)^2\,,\quad x\in\mathbb{R}^n\,.
\end{equation}
Clearly, we have
$0\leq D^2_\lambda(x,\,K)\leq 1$ in $\mathbb{R}^n$.  More precisely, we have
\begin{equation} 
	D^2_\lambda(x,\,K)\left\{\aligned
		&=1,\quad \text{if }\; x\in K,\\[1.5ex]
		&=0,\quad\text{if}\; \dist(x\, K)\geq \frac{1}{\sqrt{\lambda}},\\[1.5ex]
		&\in (0,\,1),\quad\text{if}\; 0<\dist(x,\, K)<\frac{1}{\sqrt{\lambda}}.
	\endaligned\right.
\end{equation}

Suppose $E,\, F \subset \mathbb{R}^n$ are two non-empty closed sets. It is, then, easy to see that
\begin{itemize}
	\item[$(i)$] if $E\subset F$,
		\begin{equation} 
			D^2_\lambda(x,\,E)\leq D^2_\lambda(x,\,F),\quad x\in\mathbb{R}^n;
		\end{equation}
	\item[$(ii)$] for $x\in \mathbb{R}^n$, if $E\cap \bar{B}(x,1/\sqrt{\lambda})\neq\varnothing$, then
	\begin{equation} 
		D^2_\lambda(x,\,E)=D^2_\lambda(x,\,E\cap \bar{B}(x,1/\sqrt{\lambda})).
	\end{equation}
\end{itemize}

For a given non-empty closed set $K$, by definition of the function $D^2_\lambda(x,\,K)$, we have 
\begin{equation*}
	0\leq \chi_K(x)\leq D^2_\lambda(x,\,K)\leq 1,\quad x\in\mathbb{R}^n\,.
\end{equation*}
The following result establishes the relationship between the upper transform of $\chi_K(x)$ and 
$D^2_\lambda(x,\,K)$ and it was established in \cite{ZOC15a}.

\begin{prop}\label{Prop.Up} 
Let $K\subset \mathbb{R}^n$ be a non-empty closed set and assume $\lambda>0$. Then, there holds
\begin{equation} 
	C^u_\lambda(\chi_K)(x)=C^u_\lambda(D^2_\lambda(\cdot,\,K))(x),\quad x\in\mathbb{R}^n\,.
\end{equation}
\end{prop}

The Hausdorff-Lipschitz continuity of $C^u_\lambda(\chi_K)(x)$ and 
$C^u_\lambda(D^2_\lambda(\cdot,\,K))(x)$ were also established in \cite{ZOC15a}.

\begin{teo}\label{Thm.HUp} 
Let $E,\, F\subset R^n$ be non-empty compact sets and let $\lambda>0$ be fixed, then for all $x\in \mathbb{R}^n$,
\begin{equation} 
	|D^2_\lambda(x, E)-D^2_\lambda(x, F)|\leq 2\sqrt{\lambda}\dist_{\mathcal{H}}(E,F),
\end{equation}
\begin{equation} 
	|C^u_\lambda(D^2_\lambda(\cdot,\,E))(x)-C^u_\lambda(D^2_\lambda(x,\,F))(x)|\leq 2\sqrt{\lambda}\dist_{\mathcal{H}}(E,F).
\end{equation}
Consequently,
\begin{equation} 
	|C^u_\lambda(\chi_E)(x)-C^u_\lambda(\chi_{F})(x)|\leq 2\sqrt{\lambda}\dist_{\mathcal{H}}(E,F).
\end{equation}
\end{teo}

\setcounter{equation}{0}
\section{Compensated convexity based transforms}\label{Sec:Filters}
The lower compensated convex transform \eqref{Eq:LwTr} and the upper compensated convex transform \eqref{Eq:UpTr}
represent building blocks for defining novel transformations 
to smooth functions, to identify singularities in functions, and to interpolate and approximate data.
For the creation of these transformations we follow mainly two approaches. One approach makes a direct use 
of the basic trasforms to single out singularities of the function or to smooth and/or approximate the function. 
By contrast, the other approach realises a suitably designed combination of the basic transforms 
that creates the singularity at the location of the feature of interest.
\subsection{Smoothing Transform}\label{Sec:Filters:St}
\begin{figure}[H]
\begin{center}$\begin{array}{cc}
	\includegraphics[width=0.35\textwidth]{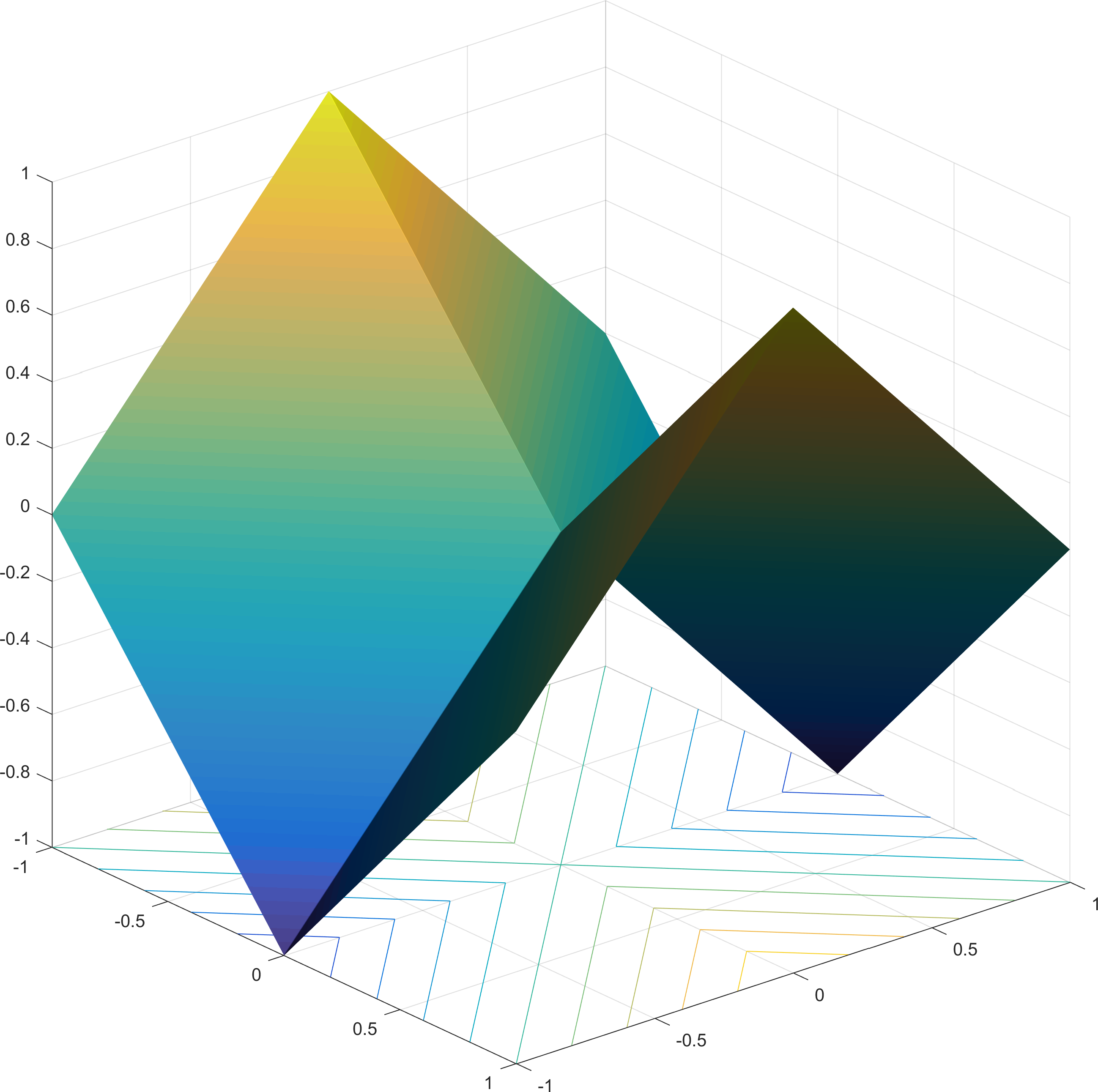}&
	\includegraphics[width=0.35\textwidth]{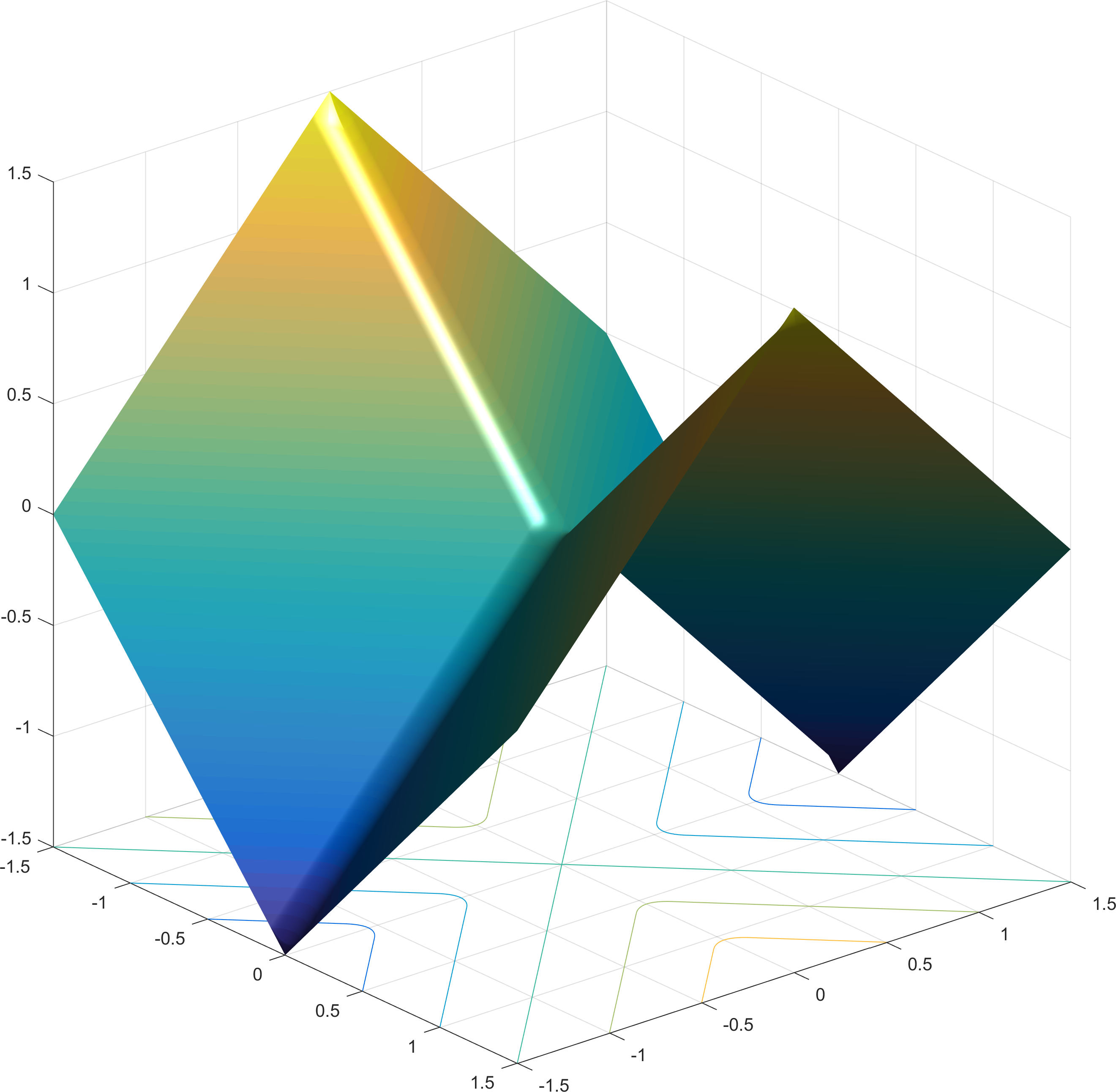}\\
	(a)&(b)\\
	\includegraphics[width=0.35\textwidth]{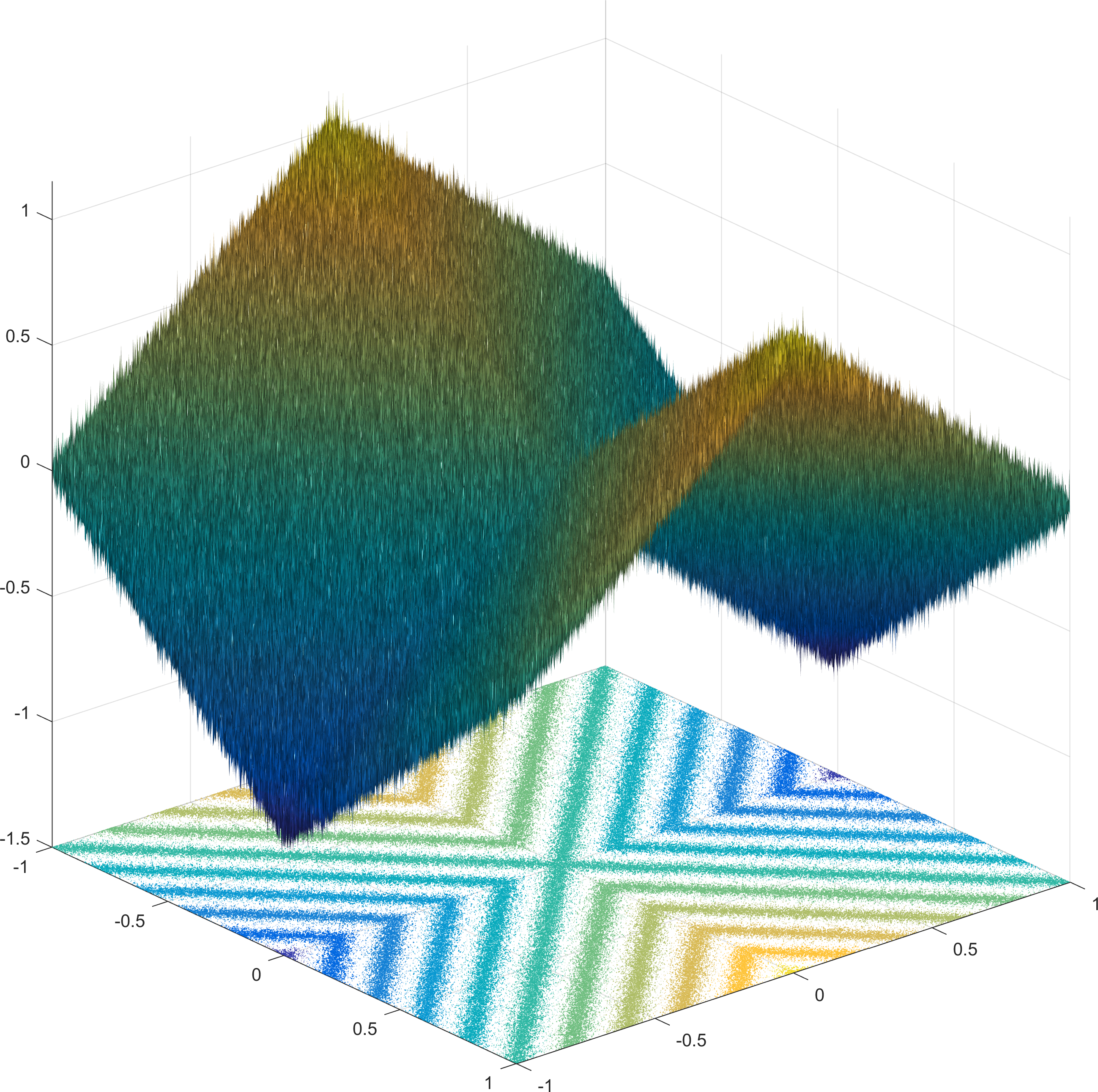}&
	\includegraphics[width=0.35\textwidth]{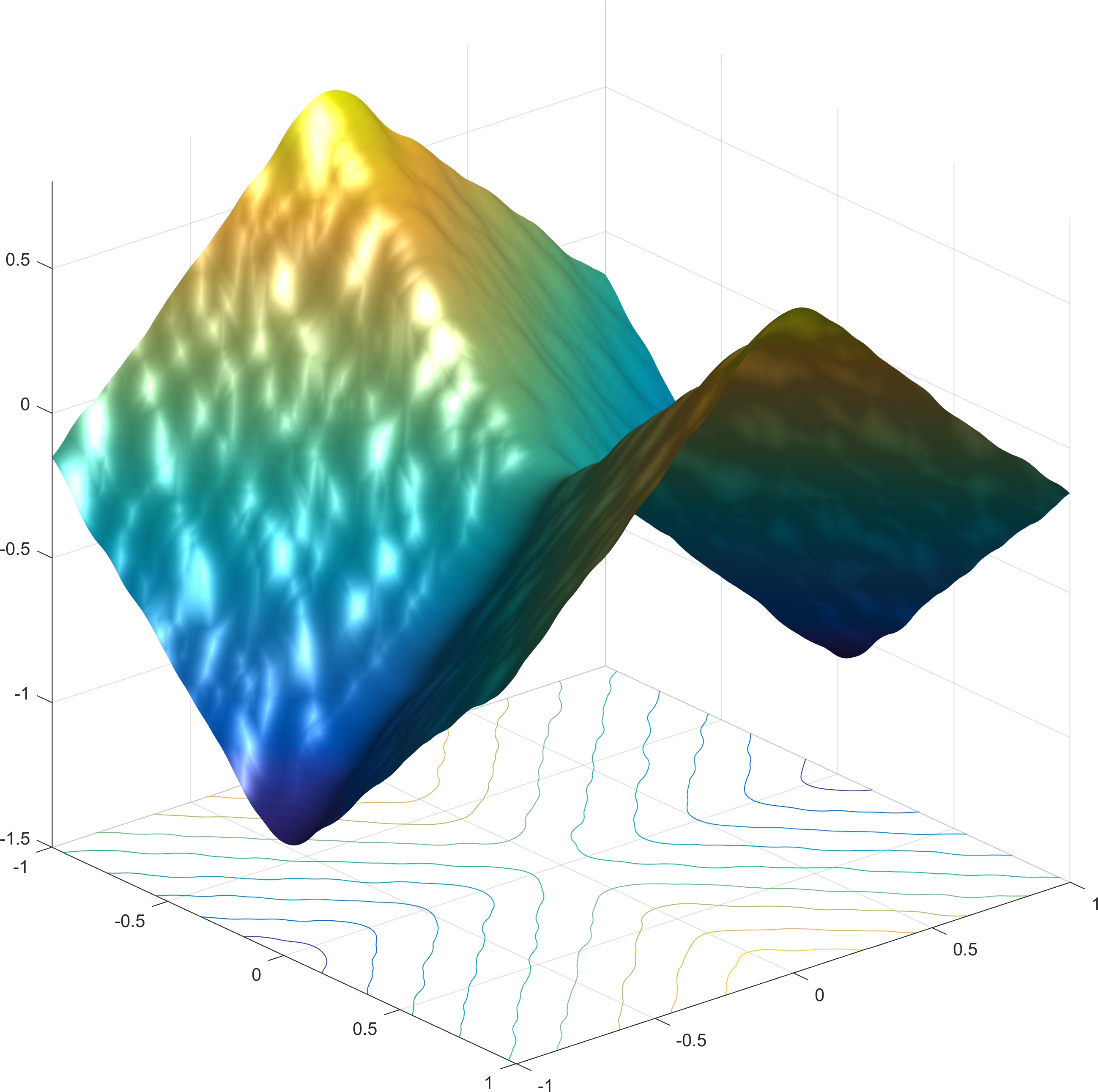}\\
	(c)&(d)
\end{array}$
\end{center}
   \caption{\label{Fig:SmoothFnct} 
   $(a)$ Input function $f(x,y)=|x|-|y|$;  
   $(b)$ Graph of $C_{\lambda,\tau}^{l,u}(f)$ for $\lambda=5$ and $\tau=5$;  
   $(c)$ Input function $f(x,y)+n(x,y)$ with $n(x,y)$ a bivariate normal distribution with mean value equal to $0.05$;  
   $(d)$ Graph of $C_{\lambda,\tau}^{l,u}(f+n)$ for $\lambda=5$ and $\tau=5$.}
\end{figure}

Let $f:\mathbb{R}^n\to\mathbb{R}$  satisfy a growth condition of the form
\begin{equation}\label{Eq.GrwthCond}
	|f(x)|\leq C_1|x|^2+C_2
\end{equation}
for some $C_1,C_2>0$, then given $\lambda,\,\tau>C_1$, we can define two (quadratic) mixed 
compensated convex transform as follows
\begin{equation}\label{Eq:MixTr}
	C_{\tau,\lambda}^{u,l}(f)(x):=C_{\tau}^u(C_{\lambda}^l(f))(x)\quad\text{and}\quad 
	C_{\lambda,\tau}^{l,u}(f)(x):=C_{\lambda}^l(C_{\tau}^u(f))(x),\quad x\in\mathbb{R}^n\,.
\end{equation}
From \eqref{Eq:UpTr}, we have that for every $\lambda,\tau>C_1$
\begin{equation}
	C_{\tau,\lambda}^{u,l}(f)(x)=-C_{\tau,\lambda}^{l,u}(-f)\,,
\end{equation}
hence properties of $C_{\tau,\lambda}^{l,u}(f)$ follow from those for $C_{\tau,\lambda}^{u,l}(f)$
and we can thus state appropriate results only for $C_{\tau,\lambda}^{u,l}(f)$. In this case, then, 
whenever $\tau,\lambda>C_1$ we have that $C_{\tau,\lambda}^{u,l}(f)\in C^{1,1}(\mathbb{R}^n)$.
As a result, if $f$ is bounded, then $C_{\tau,\lambda}^{u,l}(f)\in C^{1,1}(\mathbb{R}^n)$
and $C_{\tau,\lambda}^{l,u}(f)\in C^{1,1}(\mathbb{R}^n)$ for all $\lambda>0$ and $\tau>0$. This is important
in applications of the mixed transforms to image processing, because there the function representing the 
image takes a value from a fixed  range at each pixel point and so is always bounded. 
The regularizing effect of the mixed transform is visualized in Figure \ref{Fig:SmoothFnct}
where we display $C_{\lambda,\tau}^{l,u}(f)$ of the no-differentiable function 
$f(x,y)=|x|-|y|$, $(x,y)\in[-1,\,1]\times[-1,\,1]$ and of $f(x,y)+n(x,y)$
with $n(x,y)$ a bivariate normal distribution with mean value equal to $0.05$.
The level lines of $C_{\lambda,\tau}^{l,u}(f)$ and $C_{\lambda,\tau}^{l,u}(f+n)$
displayed in Figure \ref{Fig:SmoothFnct}$(b)$ and Figure \ref{}$(d)$, respectively,
are smooth curves.

Finally, as a consequence of 
the approximation result \eqref{Eq:Aprx} and likewise result for $C_{\tau}^u(f)$ (see Proposition \ref{Sec2.Pro.MonLwTr}) 
it is then to difficult to establish a similar approximation result also for the mixed transforms and verify that 
verify there are $\tau_j,\lambda_j\to\infty$ as $j\to\infty$ such that on every compact subset of 
$\mathbb{R}^n$, there holds
\begin{equation}
		C_{\tau_j}^u(C_{\lambda_j}^{l})(f)\to f \quad\text{uniformly as }j\to\infty\,.
\end{equation}
\subsection{Stable Ridge/Edge Transform}\label{Sec:Filters:SRt}
The ridge, valley and edge transforms introduced in \cite{ZOC15a} 
are basic operations for extracting geometric singularities. The key property is the tight approximation 
of the compensated convex transforms (see Proposition \ref{Sec2.Pro.Tight}) and the approximation to $f$ from 
below by $C_{\lambda}^l(f)$ and above by $C_{\lambda}^u(f)$, respectively.
\subsubsection{Basic transforms}\label{Sec:BasicRVE}
Let $f:\mathbb{R}^n\to\mathbb{R}$ satisfy the growth condition \eqref{Eq.GrwthCond}. The ridge 
$R_{\lambda}(f)$, the valley $V_{\lambda}(f)$ and the edge transforms $E_{\lambda}(f)$ of scale $\lambda>C_1$
are defined respectively by
\begin{equation}\label{Eq.RVEtransf}
	\begin{array}{l}
		   \displaystyle R_\lambda(f)=f-C^l_\lambda(f);\quad V_\lambda(f)=f-C^u_\lambda(f);\\[1.5ex]			
		   \displaystyle E_\lambda(f)=R_\lambda(f)-V_\lambda(f)=C^u_\lambda(f)-C^l_\lambda(f)\,.
	\end{array}	
\end{equation}
If $f$ is of sub-quadratic growth,
that is, $|f(x)|\leq A(1+|x|^\alpha)$ with $0\leq \alpha<2$, in particular
$f$ can be a bounded function, the requirement for $\lambda$ in \eqref{Eq.RVEtransf} is simply $\lambda>0$.

\begin{figure}[H]
\centerline{
	$\begin{array}{ccc}
		\includegraphics[width=0.30\textwidth]{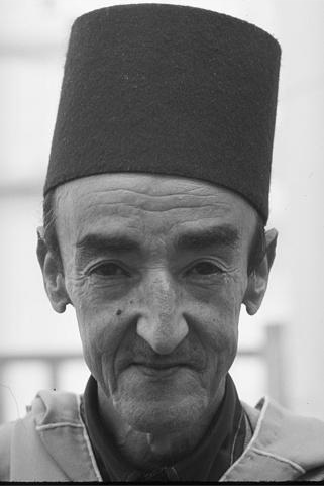}&
		\includegraphics[width=0.30\textwidth]{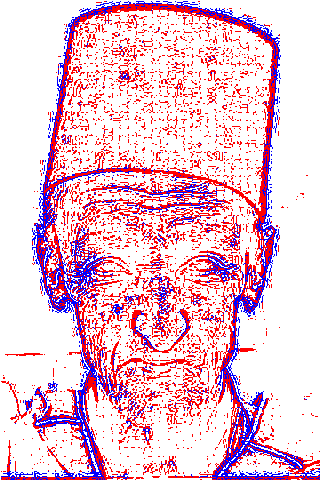}&
		\includegraphics[width=0.30\textwidth]{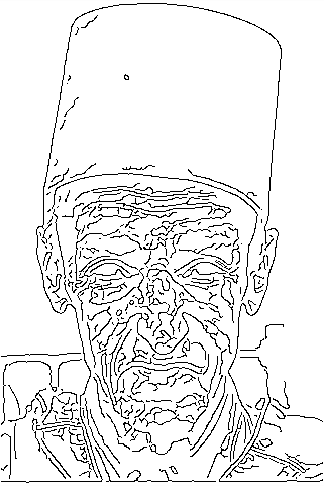}\\
		(a)&(b)&(c)
	\end{array}$
	}
\caption{\label{Fig:RVE}
	$(a)$ Input image; 
	$(b)$ Suplevel set of the ridge and valley transform with $\lambda=2.5$ and for the level equal 
	to $0.005\cdot\max{\left[R_{\lambda}(f)\right]}$
	and $0.005\cdot\max{\left[-V_{\lambda}(f)\right]}$, respectively;	$(c)$ Canny edges.}
\end{figure}

\begin{figure}[H]
\centerline{
	$\begin{array}{ccc}
		\includegraphics[width=0.33\textwidth]{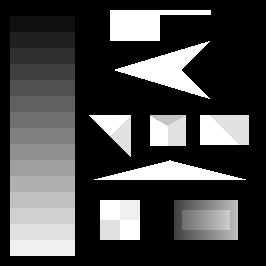}&
		\includegraphics[width=0.33\textwidth]{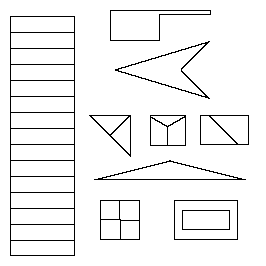}&
		\includegraphics[width=0.33\textwidth]{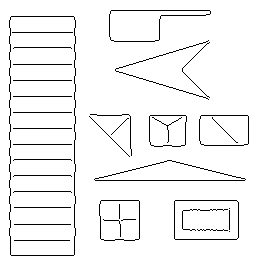}\\
		(a)&(b)&(c)
	\end{array}$
	}
\caption{\label{Fig:Susan}
	$(a)$ Input test image from \cite{SB97}; 
	$(b)$ Suplevel set of the ridge transform with $\lambda=0.1$ and for the level equal 
	to $0.004\cdot\max{\left[R_{\lambda}(f)\right]}$;
	$(c)$ Canny edges.}
\end{figure}

The ridge transform $R_\lambda(f)=f-C^l_\lambda(f)$  and
the valley transform  $V_\lambda(f)=f-C^u_\lambda(f)$ are non-negative and non-positive,
respectively, because of the ordering property of the compensated convex transforms
and their support set is disjoint to each other. In the applications,
we usually consider $-V_\lambda(f)$ to make the resulting function non-negative.
Figure \ref{Fig:RVE} displays the suplevel set of $R_\lambda(f)$ and $-V_\lambda(f)$
of the same level for a gray scale image $f$ compared to the Canny edge filter whereas 
Figure \ref{Fig:Susan} demonstrates on the test image used in \cite{SB97} the ability of 
$R_{\lambda}(f)$ to detect edges between different gray levels.

\begin{figure}[H]
\begin{center}$\begin{array}{cccc}
	\includegraphics[width=0.22\textwidth]{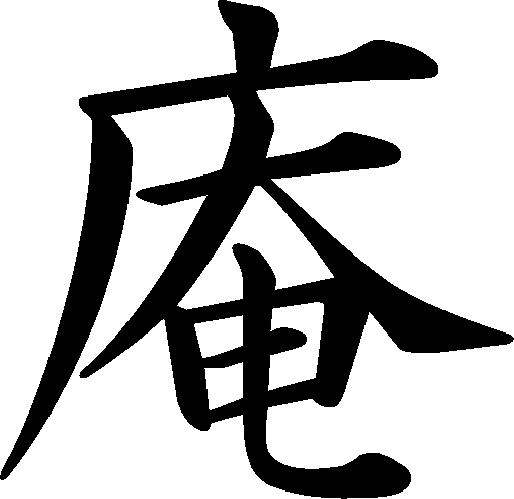}&
	\includegraphics[width=0.22\textwidth]{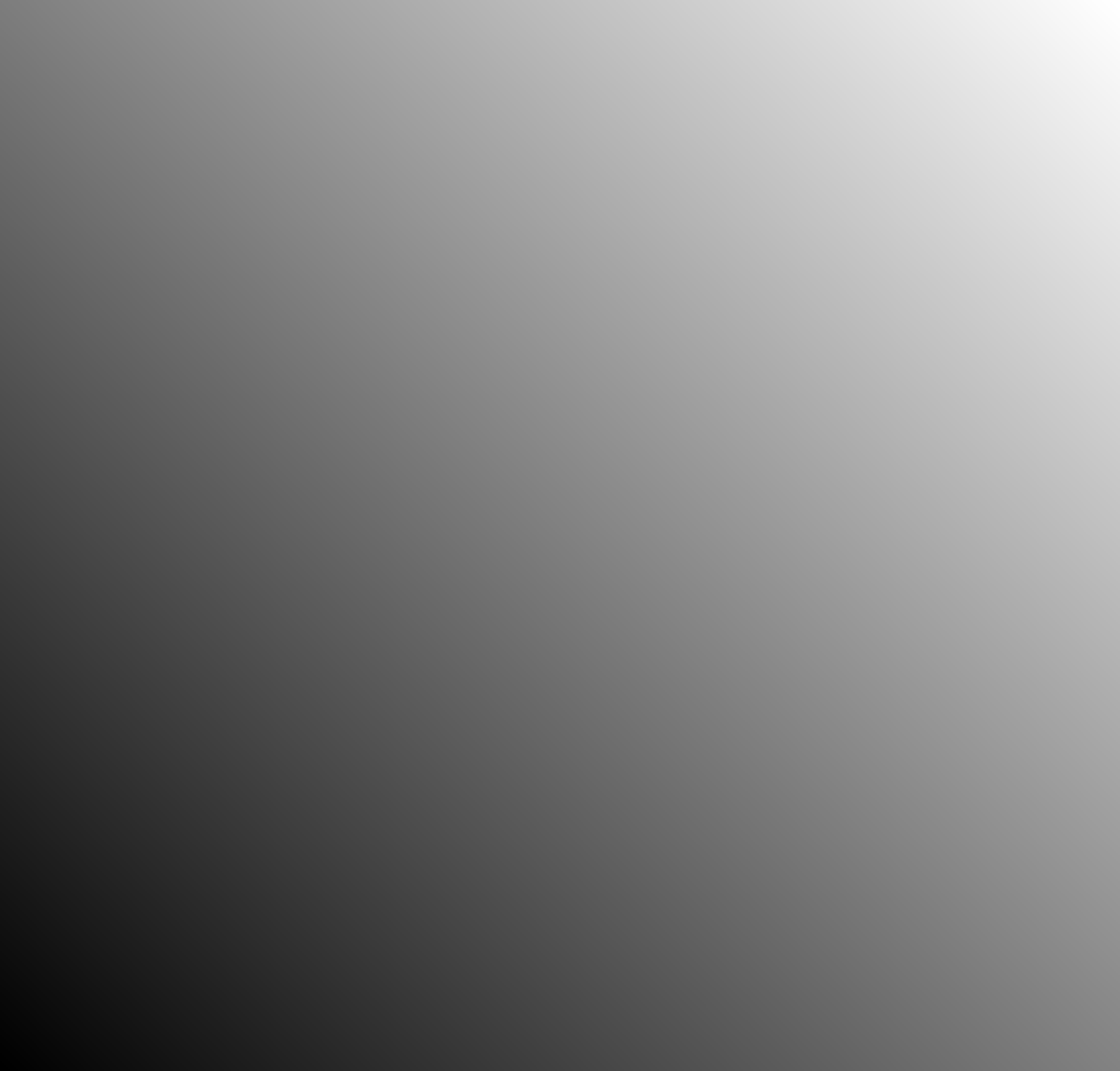}&
	\includegraphics[width=0.22\textwidth]{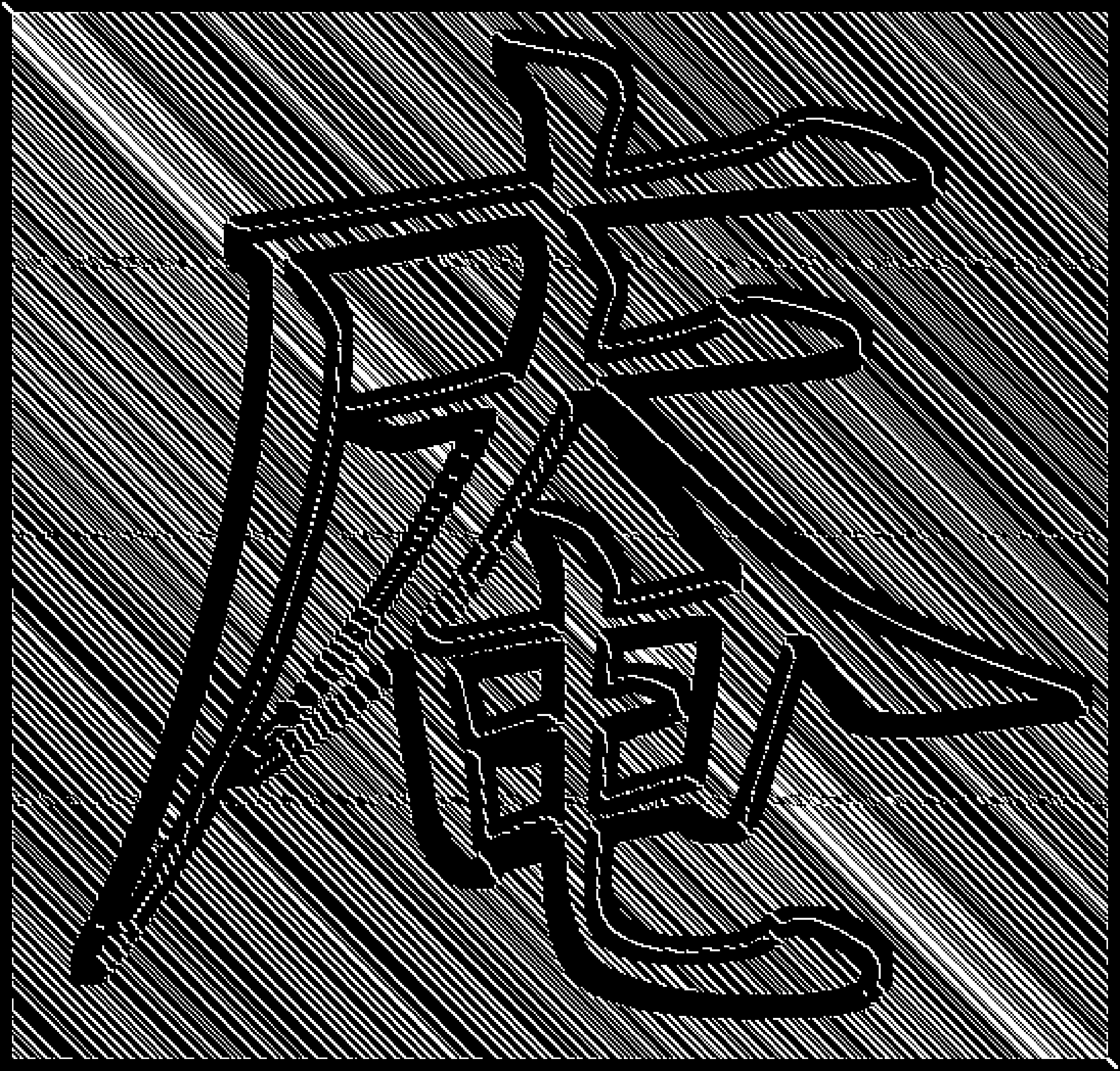}&
	\includegraphics[width=0.22\textwidth]{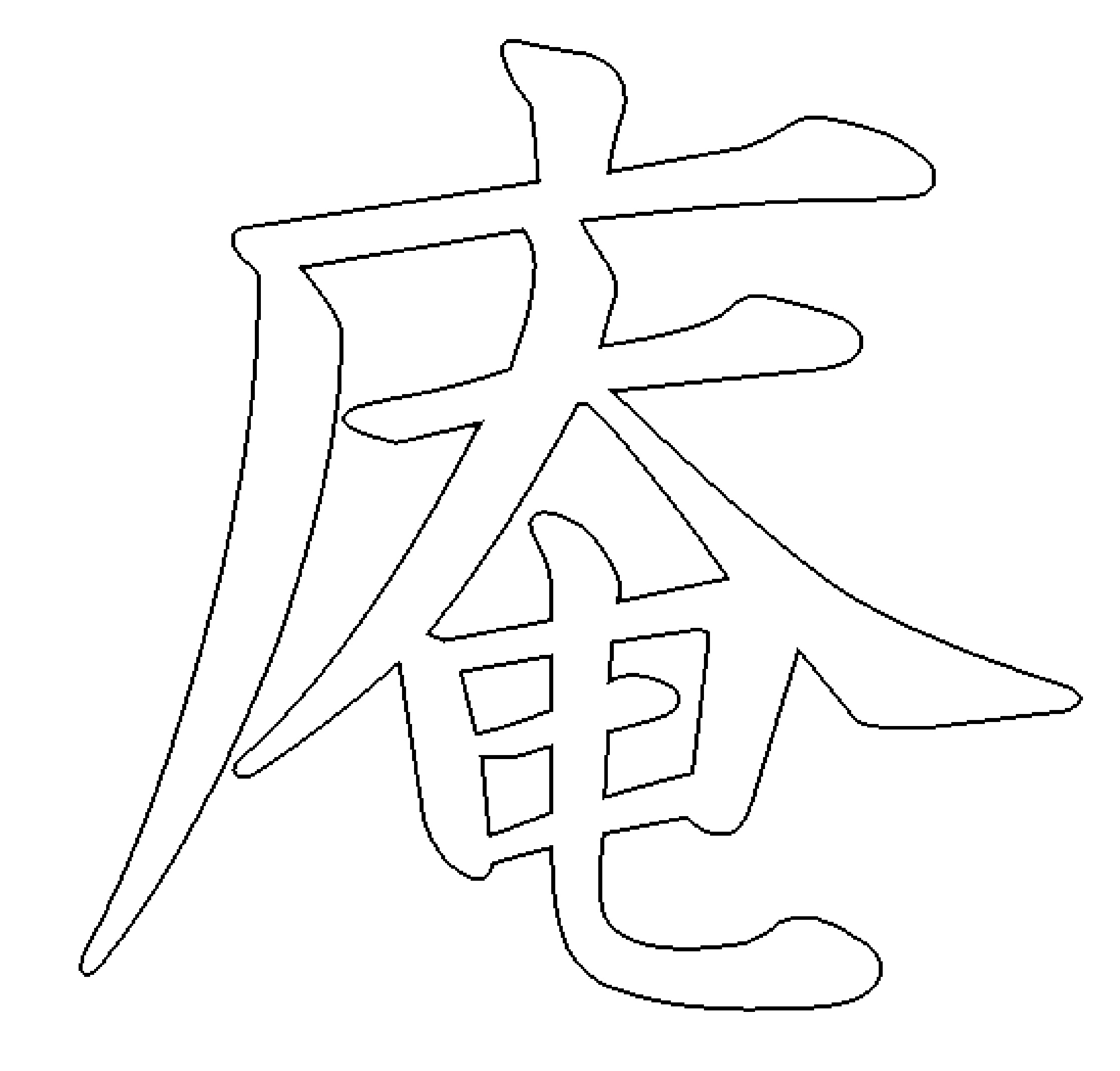}\\
	(a)&(b)&(c)&(d)
\end{array}$
\end{center}
   \caption{\label{Fig:PropRVEAfInv} $(a)$ A binary image $\chi$ of a Chinese character; $(b)$ Image $255\chi+\ell$ with $\ell=70(i-j)$
   for $1\leq i\leq 546 \,,1\leq j\leq 571$, i.e.
   the scaled characteristic
   function of the character plus an affine function;
   $(c)$ Edges extracted by Canny edge detector; $(d)$ Edges extracted by the edge transform
   $E_\lambda(f)$ with $\lambda=0.1$ after thresholding.}
\end{figure}

\begin{figure}[H]
\begin{center}$\begin{array}{ccc}
	\includegraphics[width=0.33\textwidth]{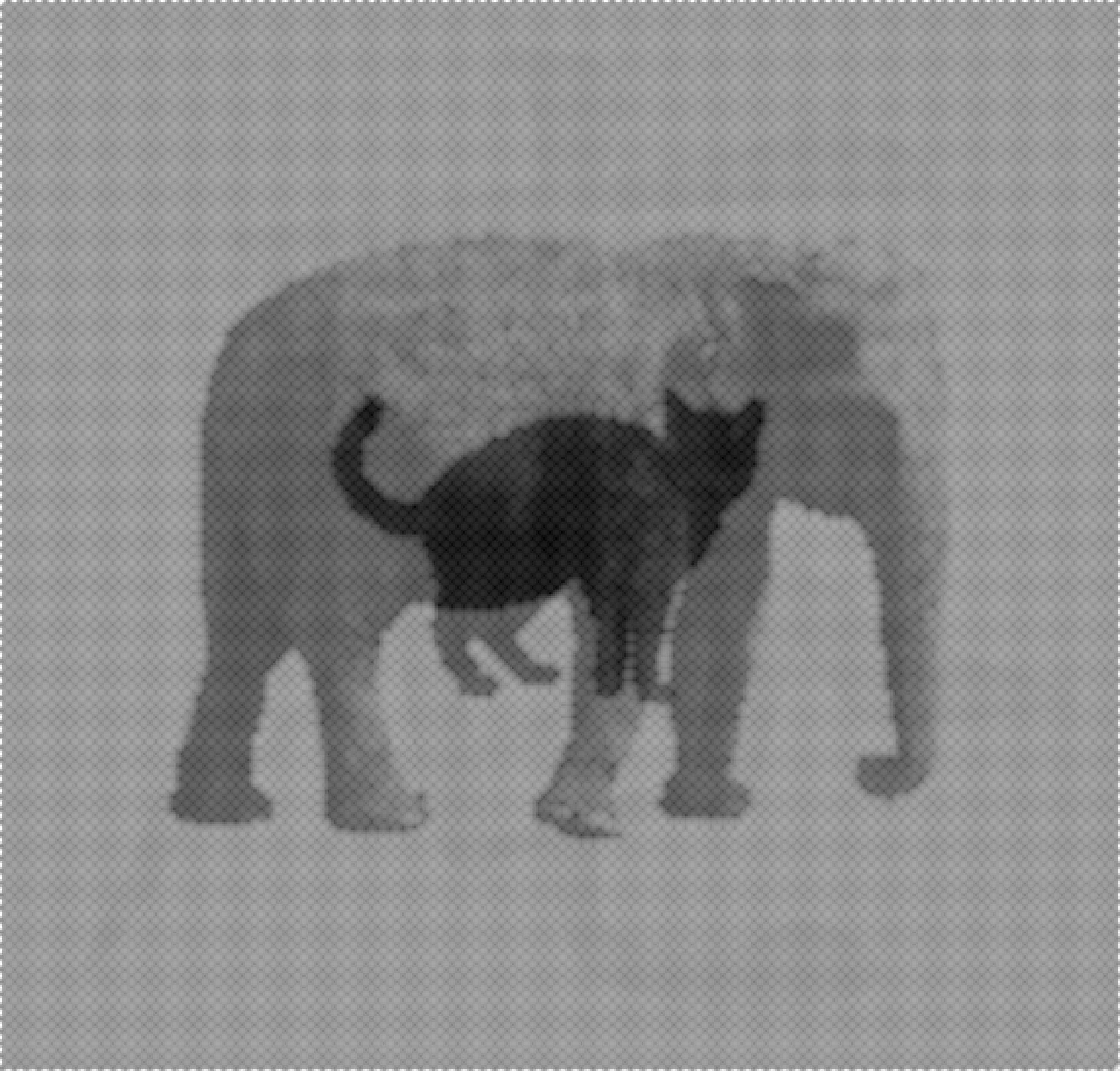}&
	\includegraphics[width=0.33\textwidth]{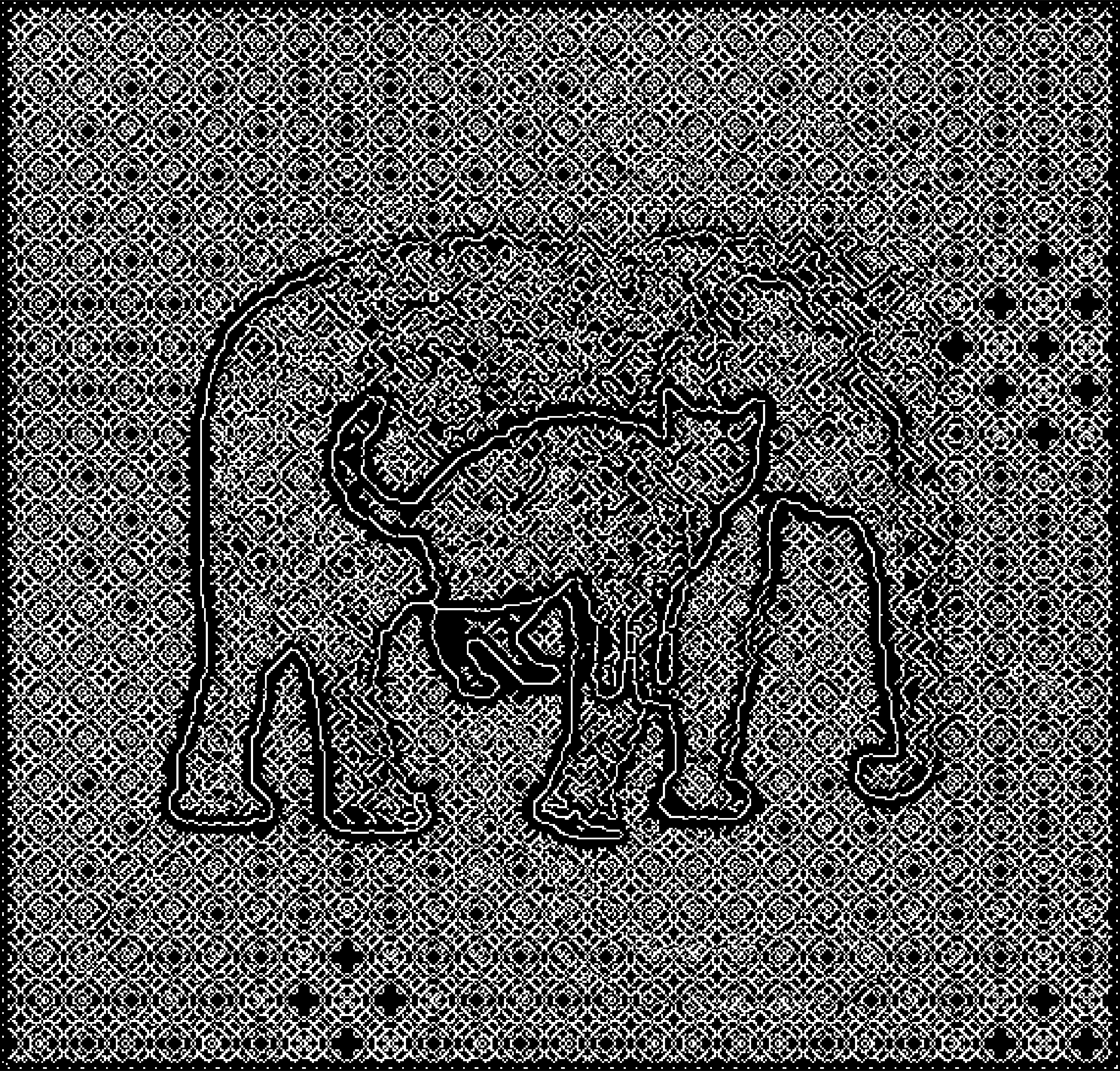}&
	\includegraphics[width=0.33\textwidth]{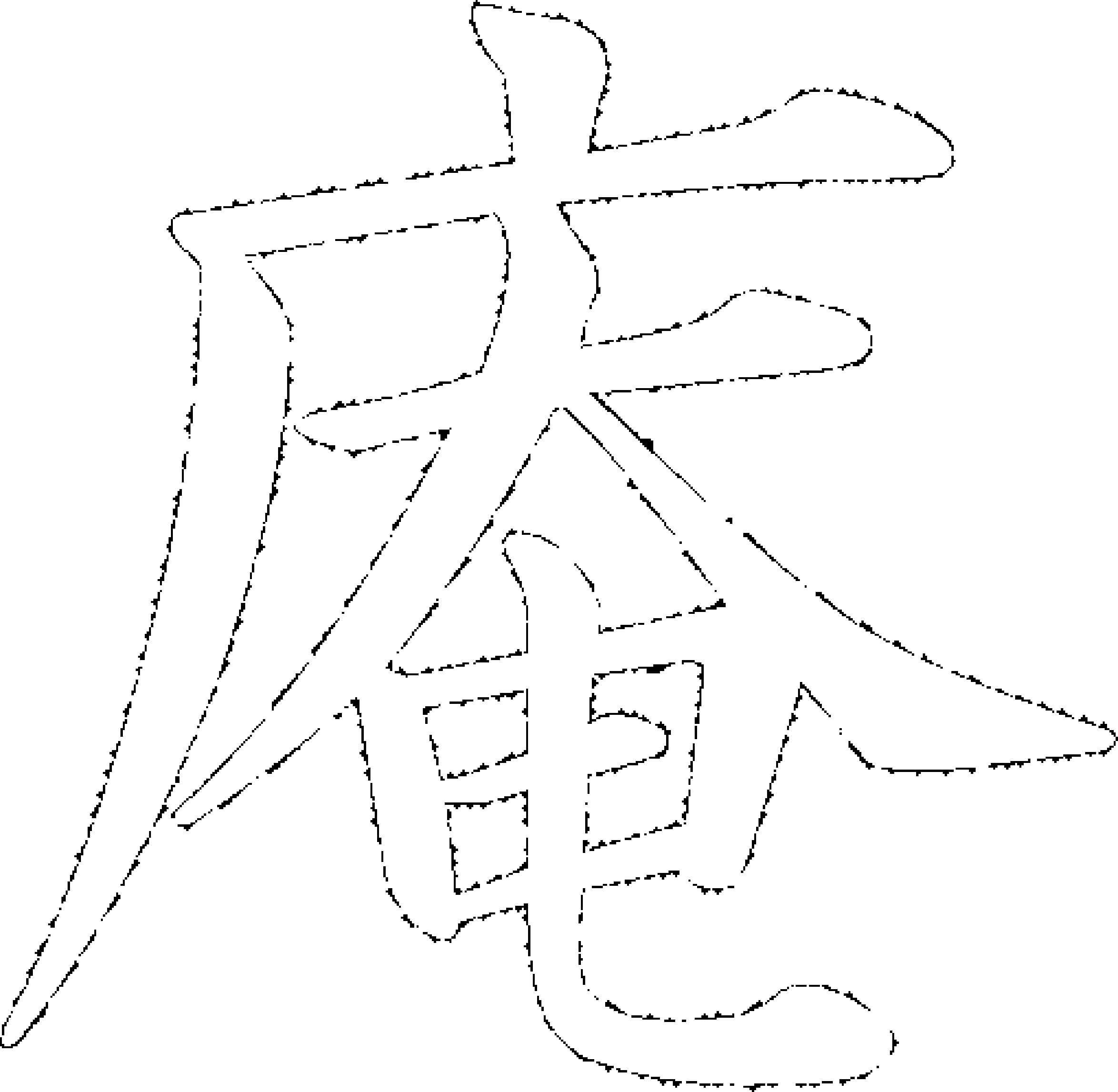}\\
	(a)&(b)&(c)
\end{array}$
\end{center}
   \caption{\label{Fig:PropRVEStab} $(a)$ A scaled binary image of a Chinese character perturbed by 
	a smooth image; $(b)$ Edges extracted by Canny edge detector; $(c)$ Edges extracted by the edge
	transform $E_\lambda(f)$ after thresholding. }
\end{figure}

The transforms  $R_\lambda(f)$ and $V_\lambda(f)$ satisfy the following properties
\begin{itemize}
	\item[(i)] The  transforms $R_\lambda(f)$ and $V_\lambda(f)$ are 
		invariant with respect to translation, in the sense that
		\begin{equation} \label{Eq.AfInv}
			R_\lambda(f+\ell)=R_\lambda(f)\quad\text{\rm and }\quad
			V_\lambda(f+\ell)=V_\lambda(f)
		\end{equation}
		for all affine functions $\ell\in \Aff(\mathbb{R}^n)$. Consequently, the edge transform
		$E_\lambda(f)$ is also invariant with respect to translation.
	\item[(ii)] The  transforms $R_\lambda(f)$ and $V_\lambda(f)$ are scale
		covariant in the sense that
		\begin{equation}
			R_\lambda(\alpha f)=\alpha R_{\lambda/\alpha}(f)\quad\text{\rm and }\quad
			V_\lambda(\alpha f)=\alpha V_{\lambda/\alpha}(f)
		\end{equation}
		for all $\alpha>0$. Consequently,
		the edge transform $E_\lambda(f)$ is also scale covariant.
	\item[(iii)] 	The transforms $R_\lambda(f)$, $V_\lambda(f)$ and $E_\lambda(f)$ are all 
		stable under curvature perturbations in the sense that
		for any $g\in C^{1,1}(\R^n)$ satisfying $|Dg(x)-Dg(y)|\leq \epsilon |x-y|$,
		if $\lambda>\epsilon$ then
		\begin{equation}\label{Eq.Stab}
		\begin{array}{l}
			\displaystyle R_{\lambda+\epsilon}(f)\leq R_\lambda(f+g)\leq R_{\lambda-\epsilon}(f);\quad
			\displaystyle V_{\lambda-\epsilon}(f)\leq V_\lambda(f+g)\leq V_{\lambda+\epsilon}(f);\\[1.5ex]
			\displaystyle E_{\lambda+\epsilon}(f)\leq E_\lambda(f+g)\leq E_{\lambda-\epsilon}(f).
		\end{array}
		\end{equation}
\end{itemize}
The numerical experiments depicted in Figure \ref{Fig:PropRVEAfInv} illustrate the affine invariance of the edge transform 
expressed by \eqref{Eq.AfInv} whereas Figure \ref{Fig:PropRVEStab} shows implications of the 
stability of the the edge transform under curvature perturbations according to \eqref{Eq.Stab}.
To get an insight on the geometric structure of the edge transform, it is informative to consider the case where $f$
is the characteristic function of a set. 
Let $\Omega\subset \R^n$ be a non-empty open regular set such that $\bar\Omega\neq \R^n$ 
and $\Gamma\subset\partial\Omega$, then for $\lambda>0$, we have that \cite{ZOC15a}
\begin{equation}\label{Eq.Thm.GeoRVT.3}
	E_\lambda(\chi_{\Omega\cup\Gamma})(x)\,
		\left\{\begin{array}{ll}
			= 0	& x\in	(\Omega^{1/\sqrt{\lambda}})^c\cup	
					\Omega\setminus (\Omega^c)^{1/\sqrt{\lambda}}\\[1.5ex]
			\in (0,\,1) & x\in \Omega^{1/\sqrt{\lambda}}\setminus\bar\Omega \cup
					(\Omega^c)^{1/\sqrt{\lambda}}\setminus \Omega^c	\\
			=1 & x\in\partial\Omega\,.
		\end{array}\right.
\end{equation}
Furthermore,  $E_{\lambda}(\chi_{\Omega\cup\Gamma})$ is continuous in $\R^n$ and, for $x\in \R^n$ there holds
\begin{equation}
	\underset{\lambda\to +\infty}{\lim}E_\lambda(\chi_{\Omega\cup\Gamma})(x) = \chi_{\partial\Omega}(x)\,,
\end{equation}	
that is, $\lambda$ controls the width of the neighborhood of $\chi_{\partial\Omega}$. As $\lambda\to\infty$, the support of
$E_{\lambda}(\chi_{\overline\Omega})$ shrinks to the support of $\chi_{\partial\Omega}$.

\begin{figure}[H]
\begin{center}$\begin{array}{cccc}
	\includegraphics[width=0.22\textwidth]{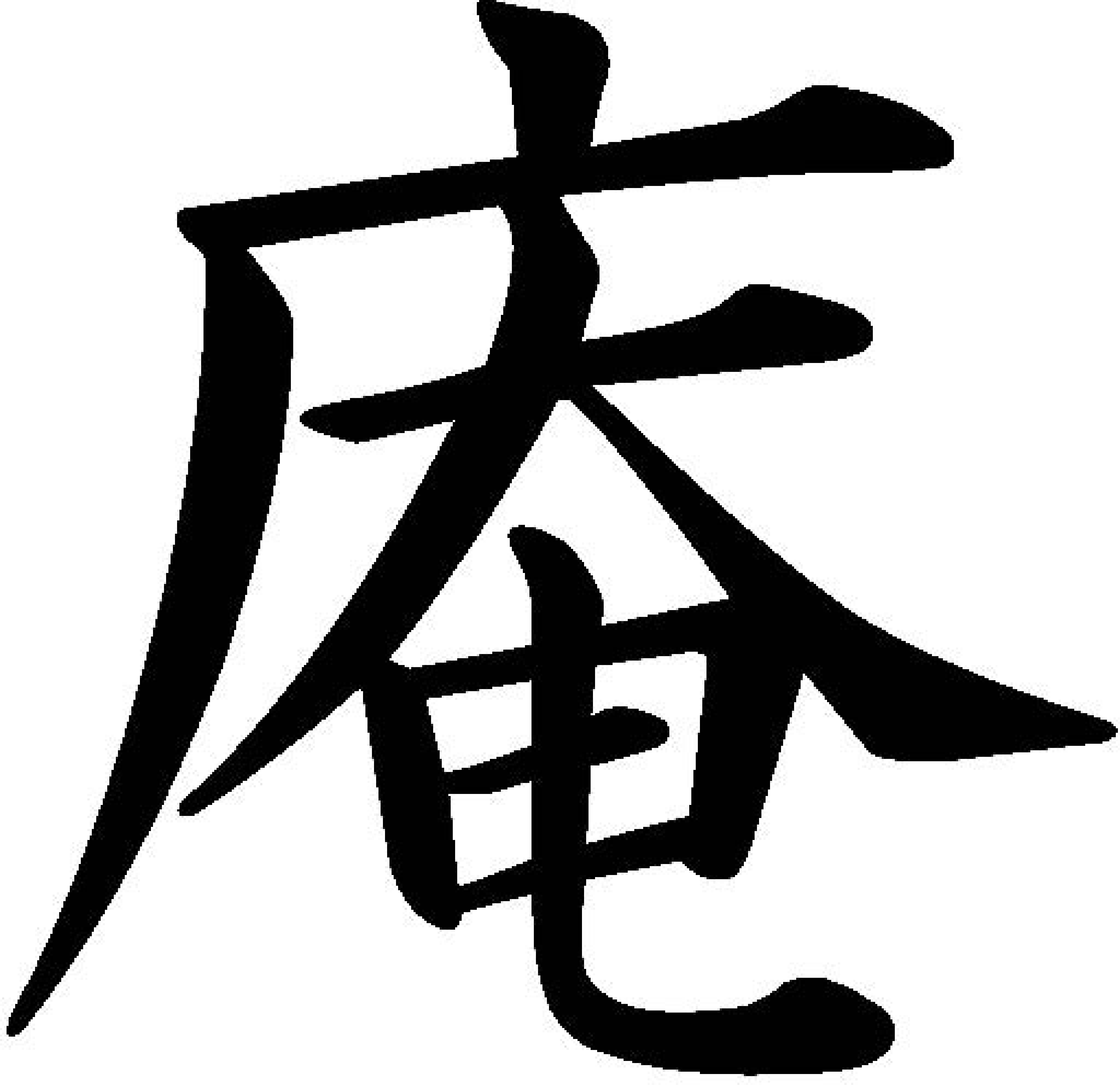}&
	\includegraphics[width=0.22\textwidth]{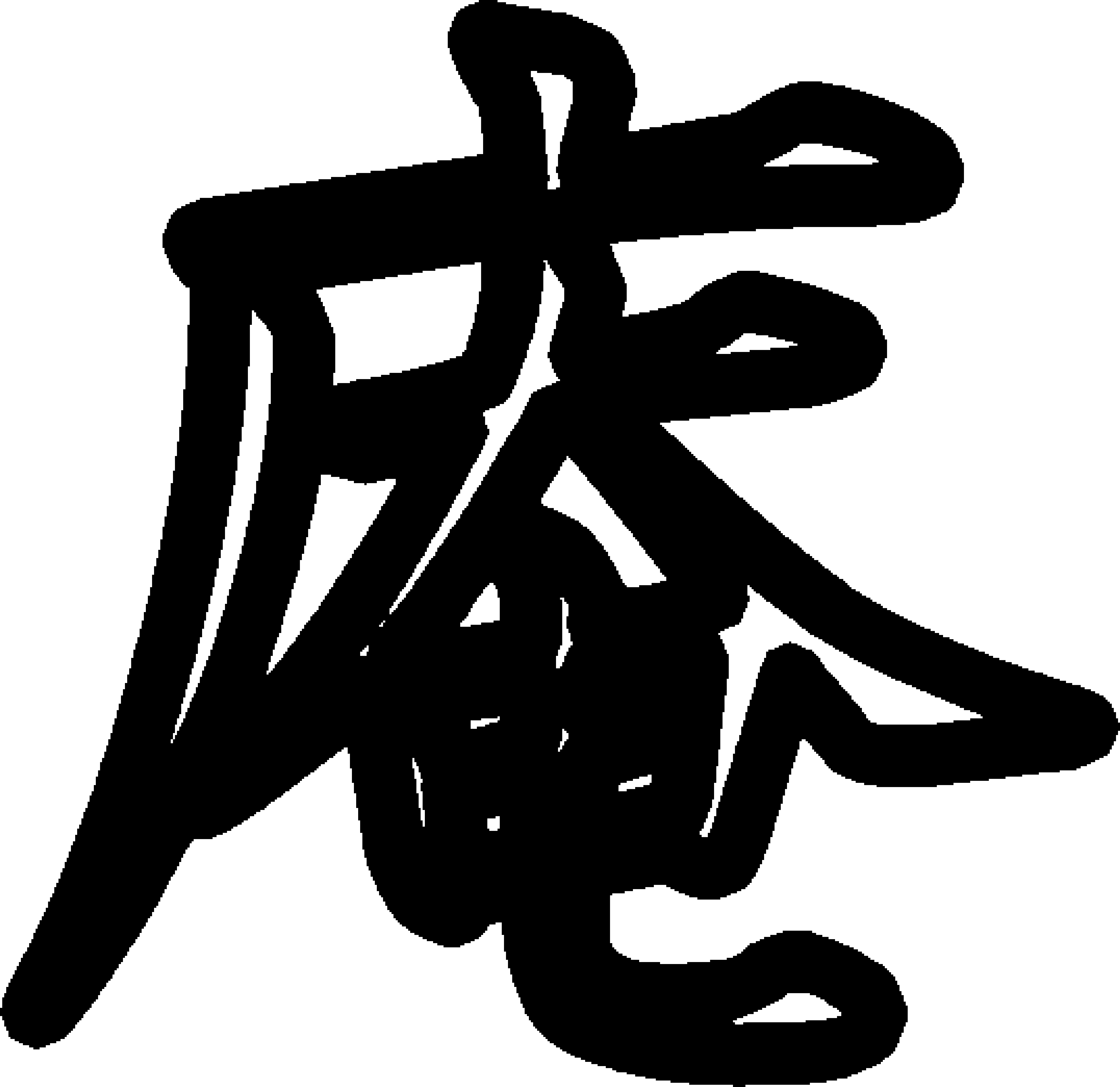}&
	\includegraphics[width=0.22\textwidth]{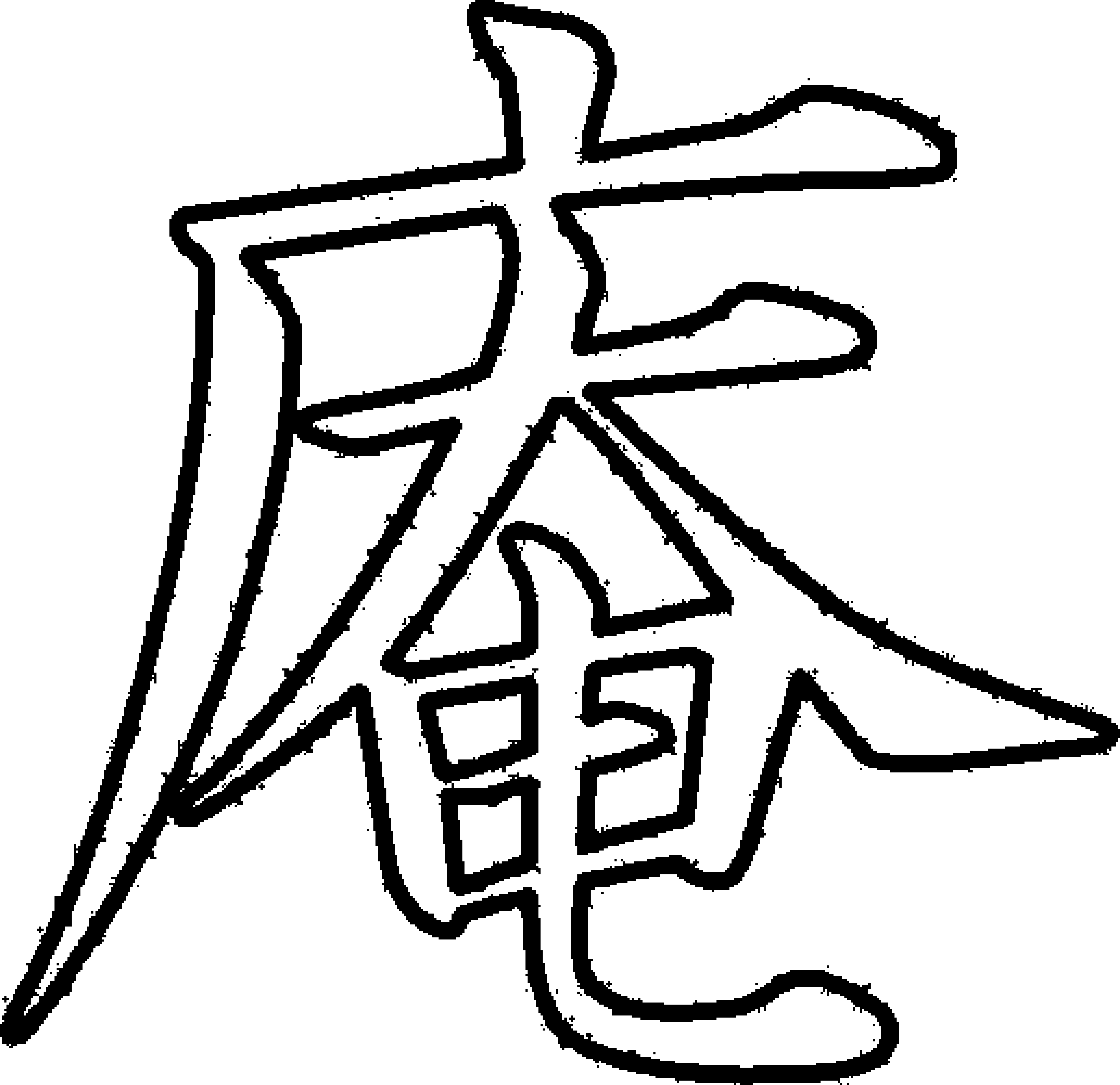}&
	\includegraphics[width=0.22\textwidth]{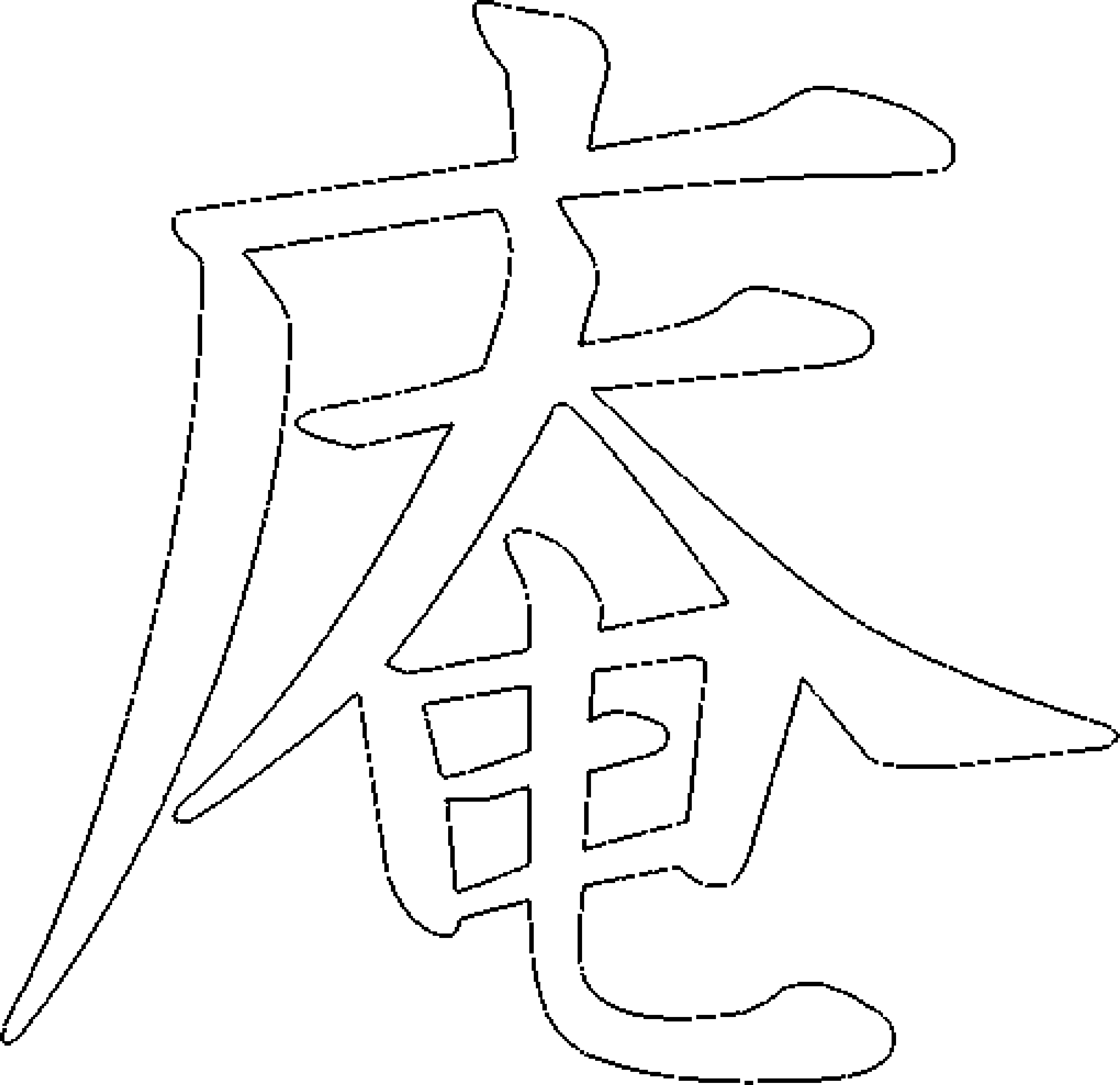}\\
	(a)&(b)&(c)&(d)
\end{array}$
\end{center}
   \caption{\label{Fig:EdgvsLmbd} Scale effect associated with $\lambda$ on the support of the edge transform of the $(a)$ image $f=255\cdot\chi$ of a Chinese character for 
   different values of $\lambda$: 
   $(b)$ $\lambda=1$;	$(c)$ $\lambda=10$;	$(d)$ $\lambda=100$.}
\end{figure}

\begin{figure}[H]
    \centerline{
    $\begin{array}{cccc}
		\includegraphics[width=0.23\textwidth]{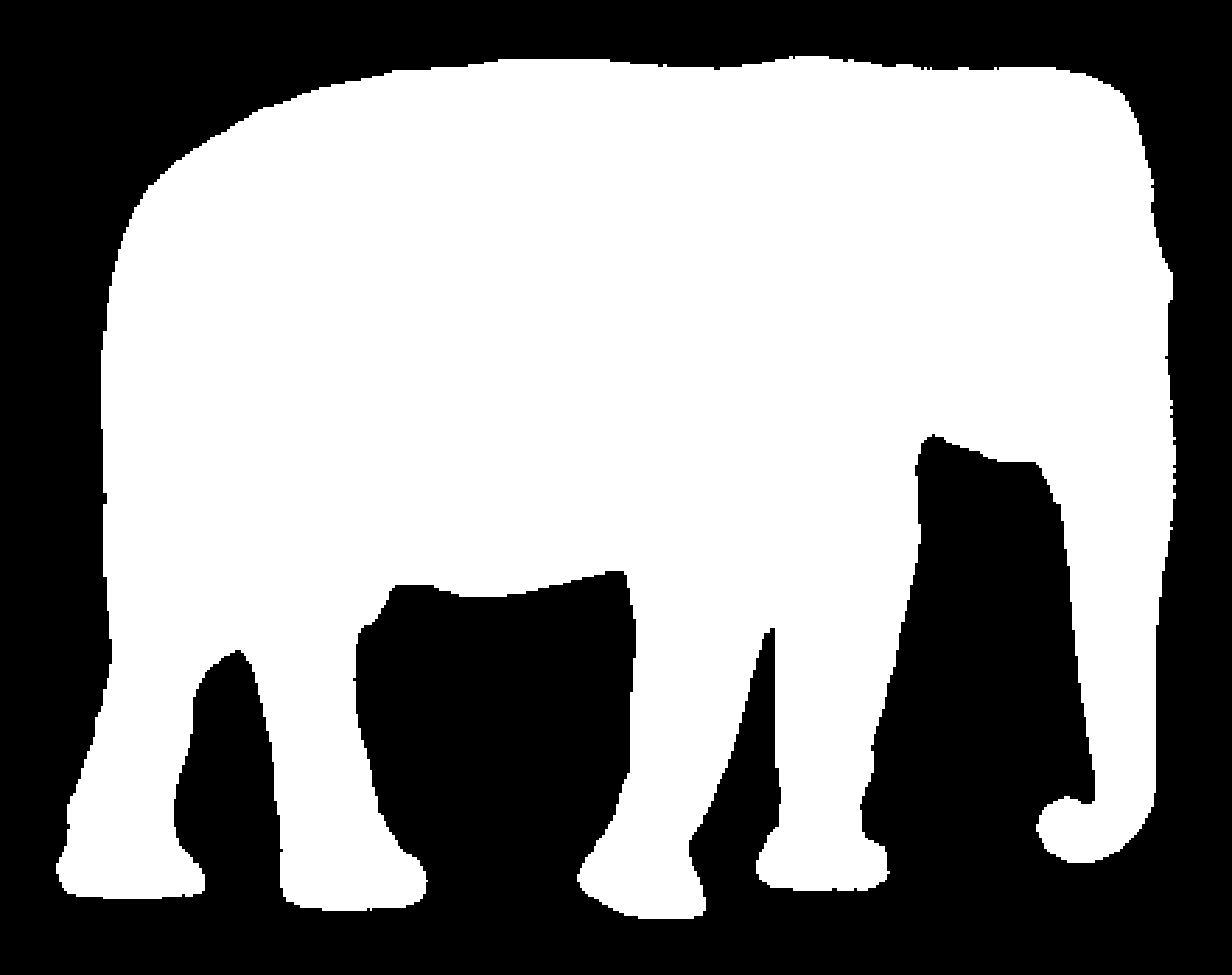}&
		\includegraphics[width=0.23\textwidth]{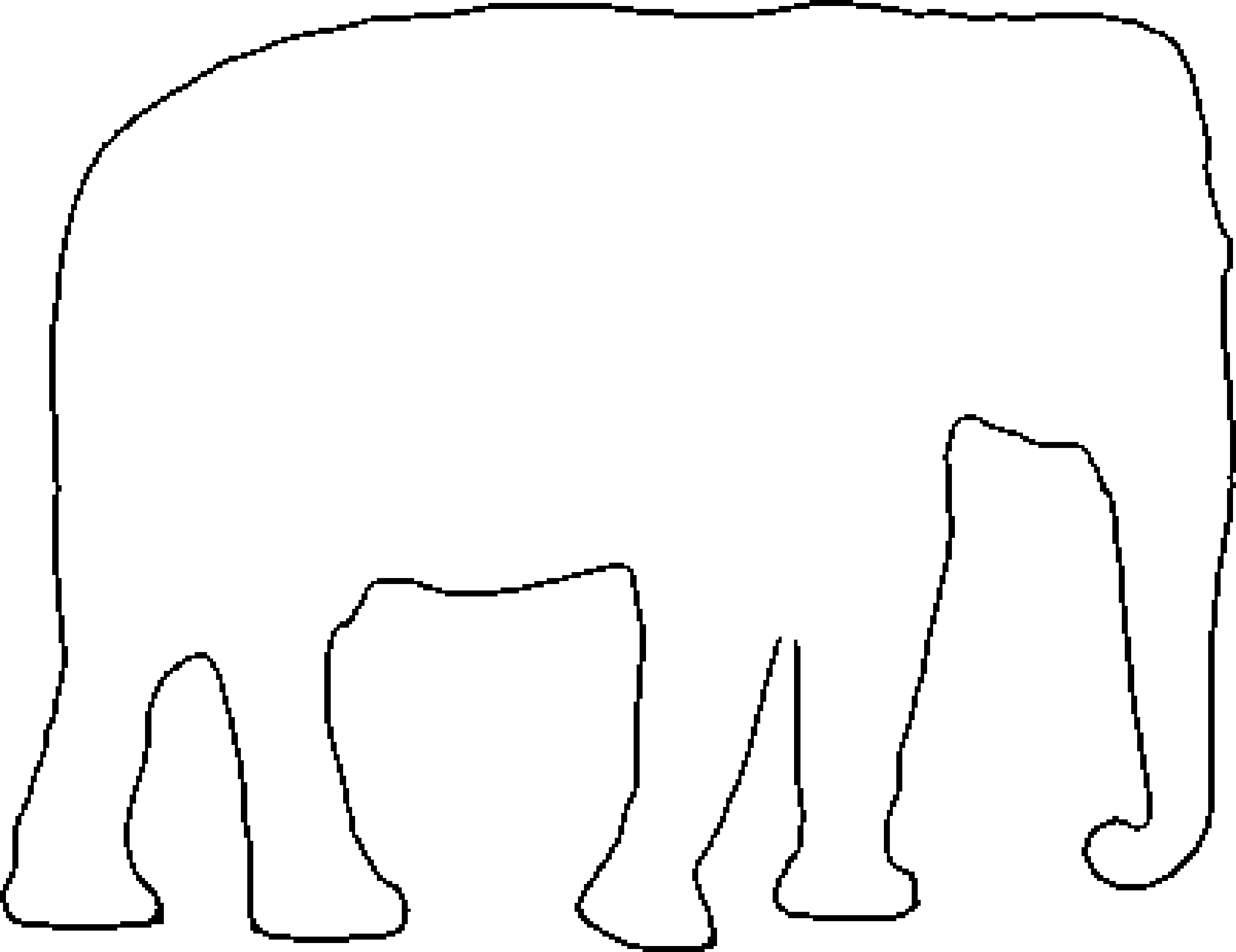}&
		\includegraphics[width=0.23\textwidth]{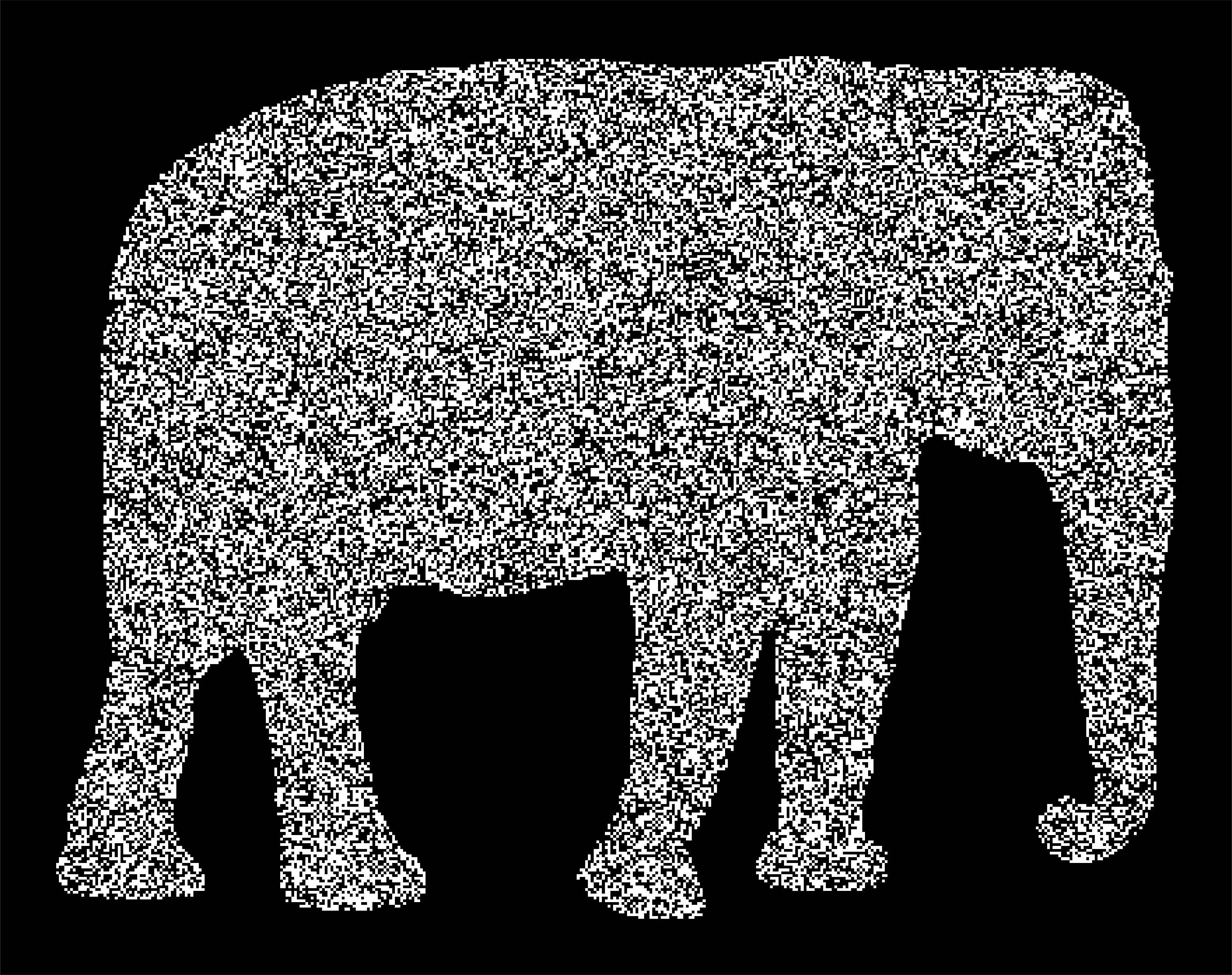}&
		\includegraphics[width=0.23\textwidth]{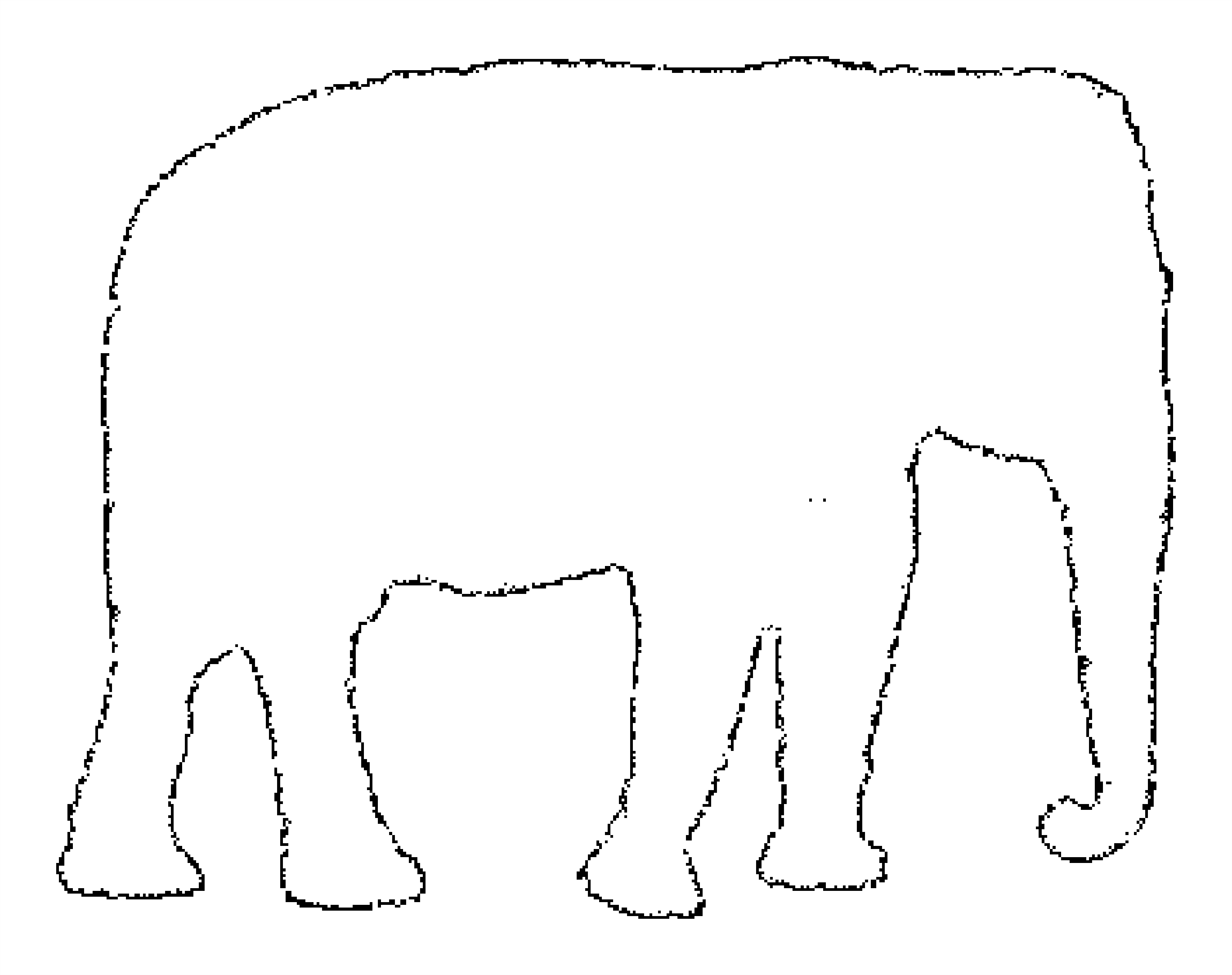}\\[1.5ex]
		(a)&(b)&(c)&(d)
		\end{array}$
}
\caption{\label{Sec3.Fig.Cat} $(a)$ Domain $E$ given by the image of an elephant displayed here as $1-\chi_E$; 
$(b)$ Boundary extraction using the stable ridge transform, $SR_{\lambda,\tau}(\chi_E))$, for $\lambda=0.1$ and $\tau=\lambda/8$; 
$(c)$ Domain $F$ obtained by randomly sampling $E$; 
$(d)$ Boundary extraction of the data sample after thresholding the stable ridge transform, $SR_{\lambda,\tau}(\chi_F))$,
computed for  $\lambda=0.1$ and $\tau=\lambda/8$.
}
\end{figure}

Figure \ref{Fig:EdgvsLmbd} illustrates the behaviour of $E_\lambda(\chi_{\overline\Omega})$ by displaying the support 
of $E_\lambda(\chi_{\overline\Omega})$ for different values of $\lambda$.
Since the original function $f$ is directly involved in the definitions of the ridge,
valley and edge transforms, the transforms \eqref{Eq.RVEtransf} are not Hausdorff stable
if we consider a dense sampling of the original function. It is possible nevertheless to establish 
a stable versions of ridge and valley transforms in the case that $f$ is the characteristic function 
$\chi_E$ of a non-empty closed set $E\subset\mathbb{R}^n$. For this result, it is fundamental the observation 
on the Hausdorff stability of the upper transform of the characteristic function $\chi_E$ of closed sets
which motivates the definition of stable ridge transform of $E$ as
\begin{equation}\label{Eq:StabRidge}
	\mathrm{SR}_{\tau,\lambda}(\chi_E)=C_{\lambda}^u(\chi_E)-C_{\tau}^l(C_{\lambda}^u(\chi_E))\,.
\end{equation}
For the ridge defined by \eqref{Eq:StabRidge} we have that 
if $E,\,F$ are non-empty compact subsets of $\R^n$, for $\lambda>0$ and $\tau>0$, 
then there holds
\begin{equation}\label{Sec3.Eq.StabRid} 
	|SR_{\lambda,\tau}(\chi_{E})(x)- SR_{\lambda,\tau}(\chi_{F})(x)|\leq 4\sqrt{\lambda}
	\dist_{\mathcal{H}}(E,F)	\quad(\text{for }x\in\R^n)\,.
\end{equation}
Figure \ref{Sec3.Fig.Cat} illustrates the meaning of \eqref{Sec3.Eq.StabRid}. Figure \ref{Sec3.Fig.Cat}$(a)$ displays
a domain $E$ represented by a binary image of a cat, $(c)$ shows a domain $F$ obtained by 
randomly sampling $E$, whereas $(b)$ and $(d)$ picture a suplevel set of the stable ridge transforms of the
respective characteristic functions.
Similarly to the Stable Ridge Transform of a non-empty compact subset $E$ of $\R^n$, we can then define the 
Stable Valley Transform of $E$ for $\lambda>\tau$ as
\[
	SV_{\lambda,\tau}(\chi_E)(x)=V_{\tau}(C_{\lambda}^u(\chi_E))(x)\quad x\in\R^n,\quad\lambda>\tau>0\,, 
\]
and the Stable Edge Transform  of $E$ for $\lambda>\tau$ as
\[
	SE_{\lambda,\tau}(\chi_E)(x)=E_{\tau}(C_{\lambda}^u(\chi_E))(x)\quad x\in\R^n,\quad\lambda>\tau>0\,.
\]
The condition $\lambda>\tau$ is invoked because it is not difficult to see that
\[
	C_{\tau}^u(C_{\lambda}^u(f)))=\left\{\begin{array}{ll}
						\displaystyle C_{\lambda}^u(f),   &\text{for }\lambda\leq \tau\\[1.5ex]	
						\displaystyle C_{\tau}^u(f),	     &\text{for }\lambda\geq \tau\,.
						\end{array}\right.
\]
Hence, if $\lambda\leq \tau$, we would get $SV_{\lambda,\tau}(\chi_E)(x)=0$ and
$SE_{\lambda,\tau}(\chi_E)(x)$ would simply equal to $SR_{\lambda,\tau}(\chi_E)(x)$. 

\subsubsection{Extractable corner points}
\begin{figure}[H]
    \centerline{\includegraphics[width=0.40\textwidth]{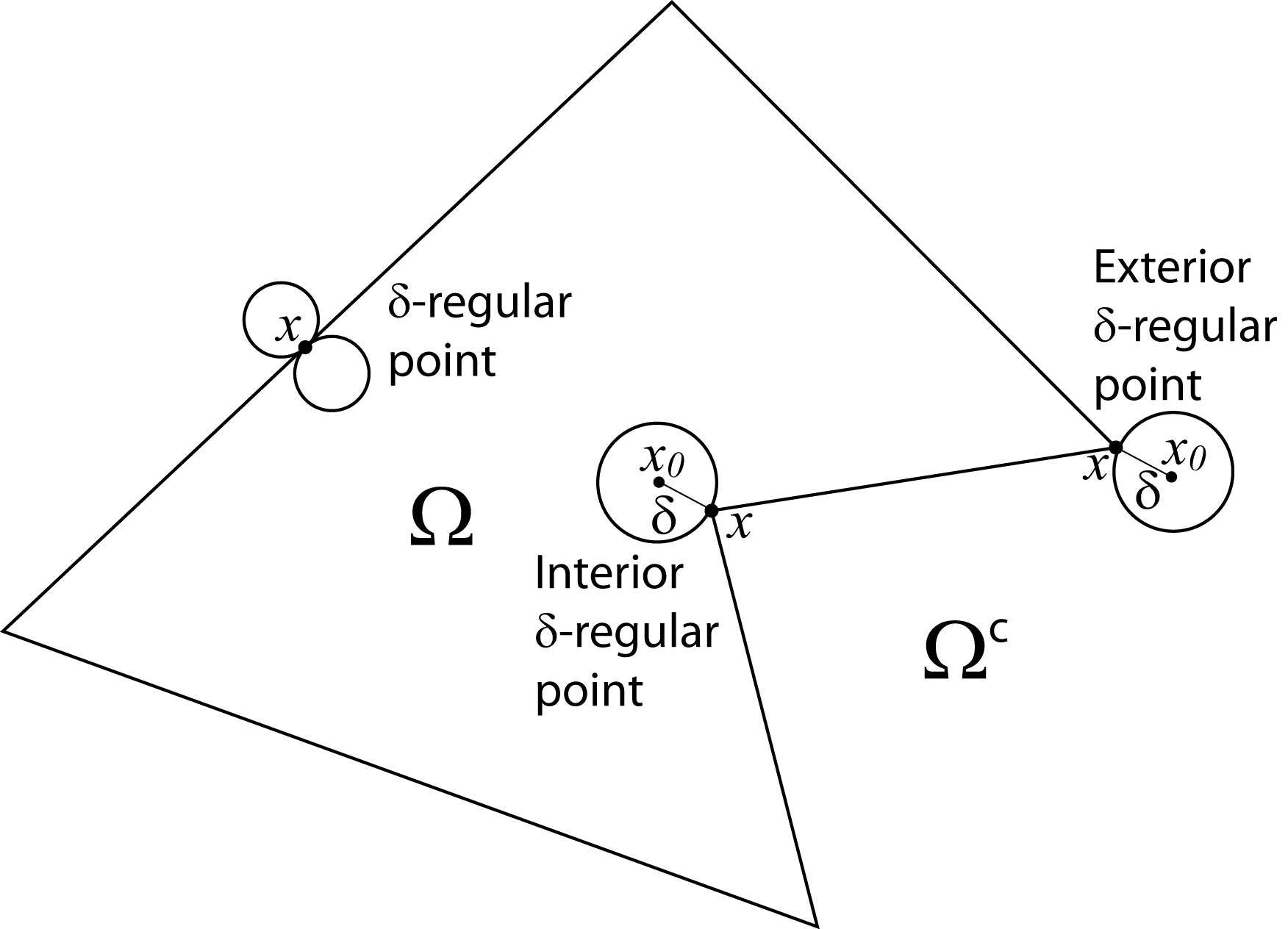}}
\caption{\label{Fig:RegPoints} Exterior and interior $\delta-$regular point of $\partial\Omega$.}
\end{figure}

\begin{figure}[H]
\centerline{
	$\begin{array}{ccc}
		\includegraphics[width=0.33\textwidth]{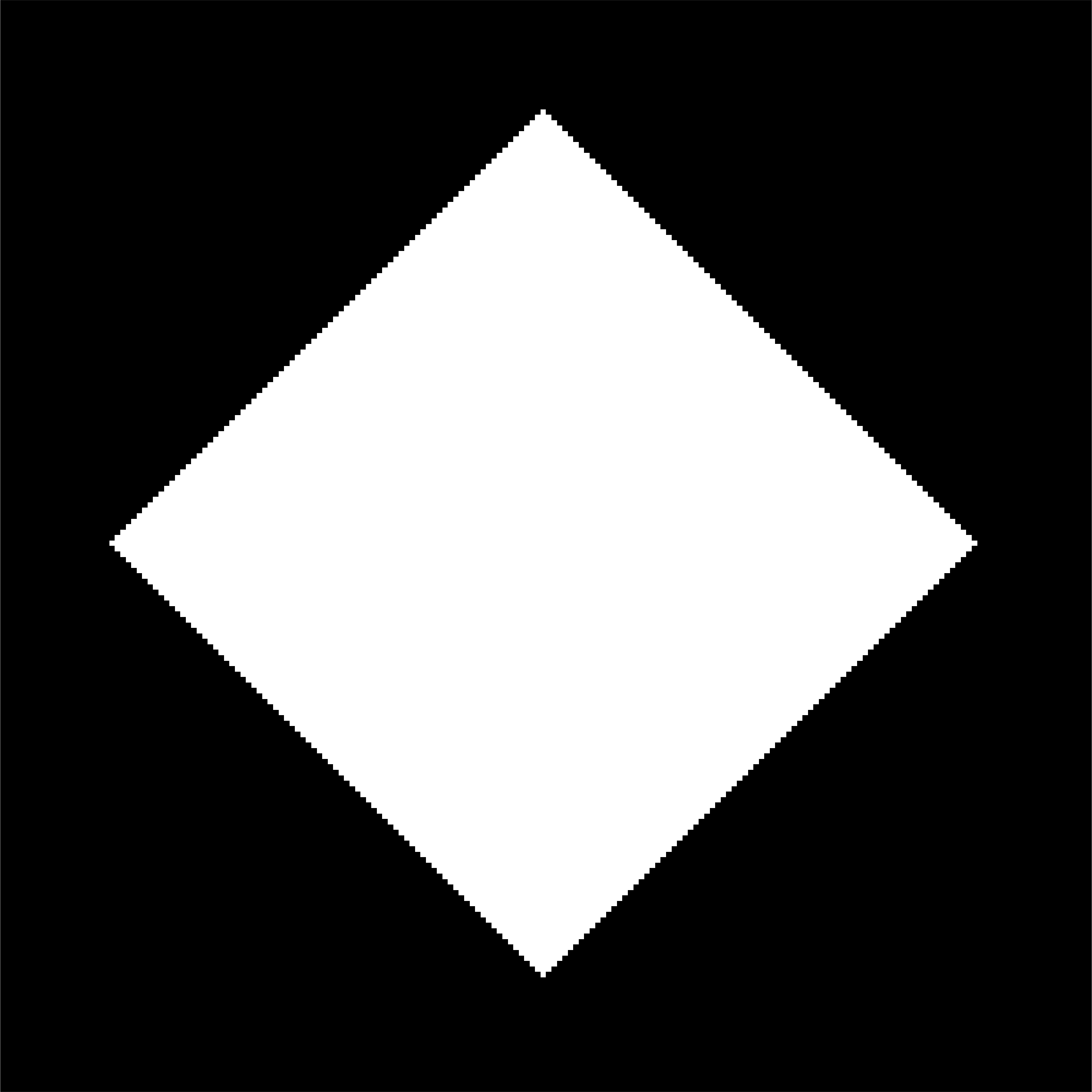}&
		\includegraphics[width=0.33\textwidth]{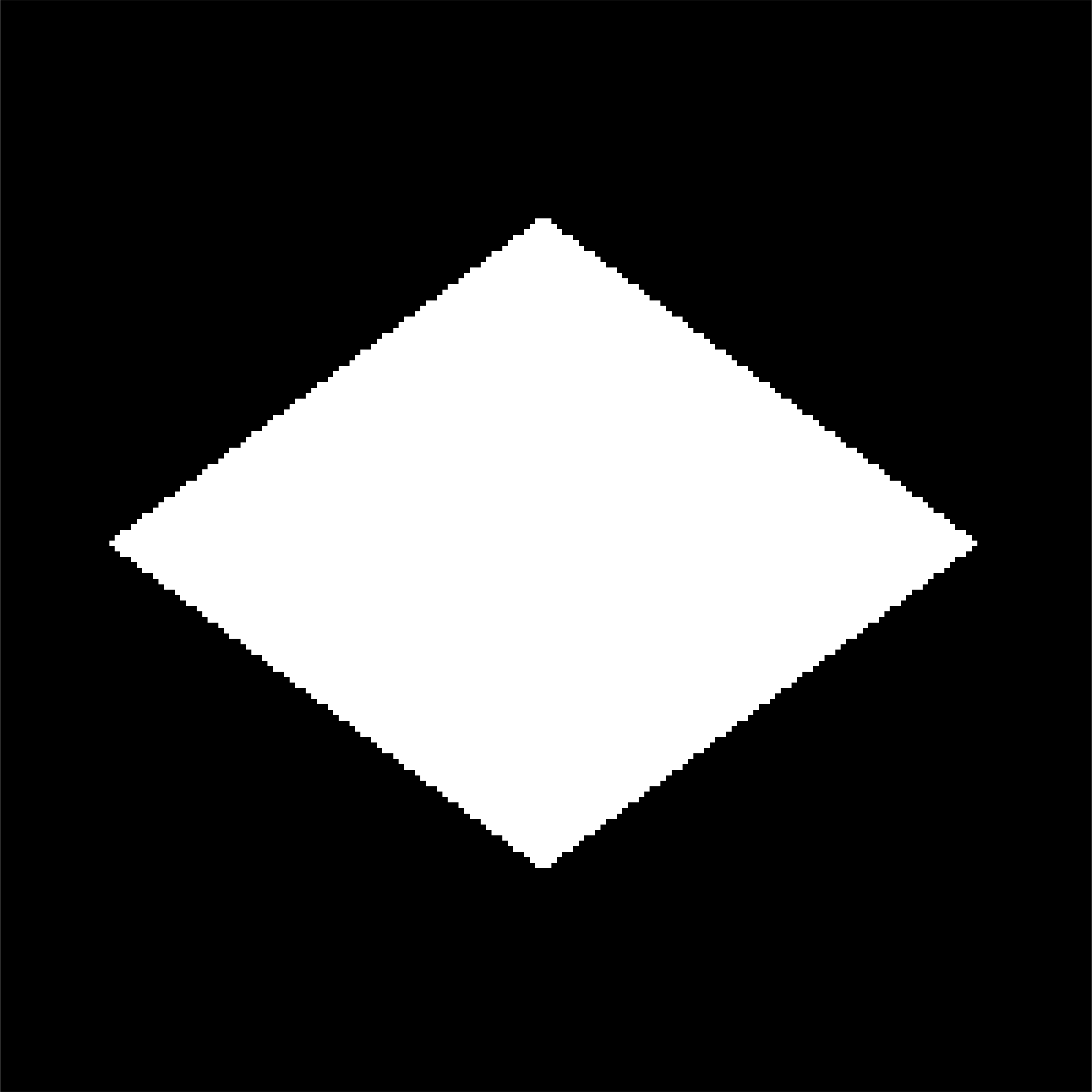}&
		\includegraphics[width=0.33\textwidth]{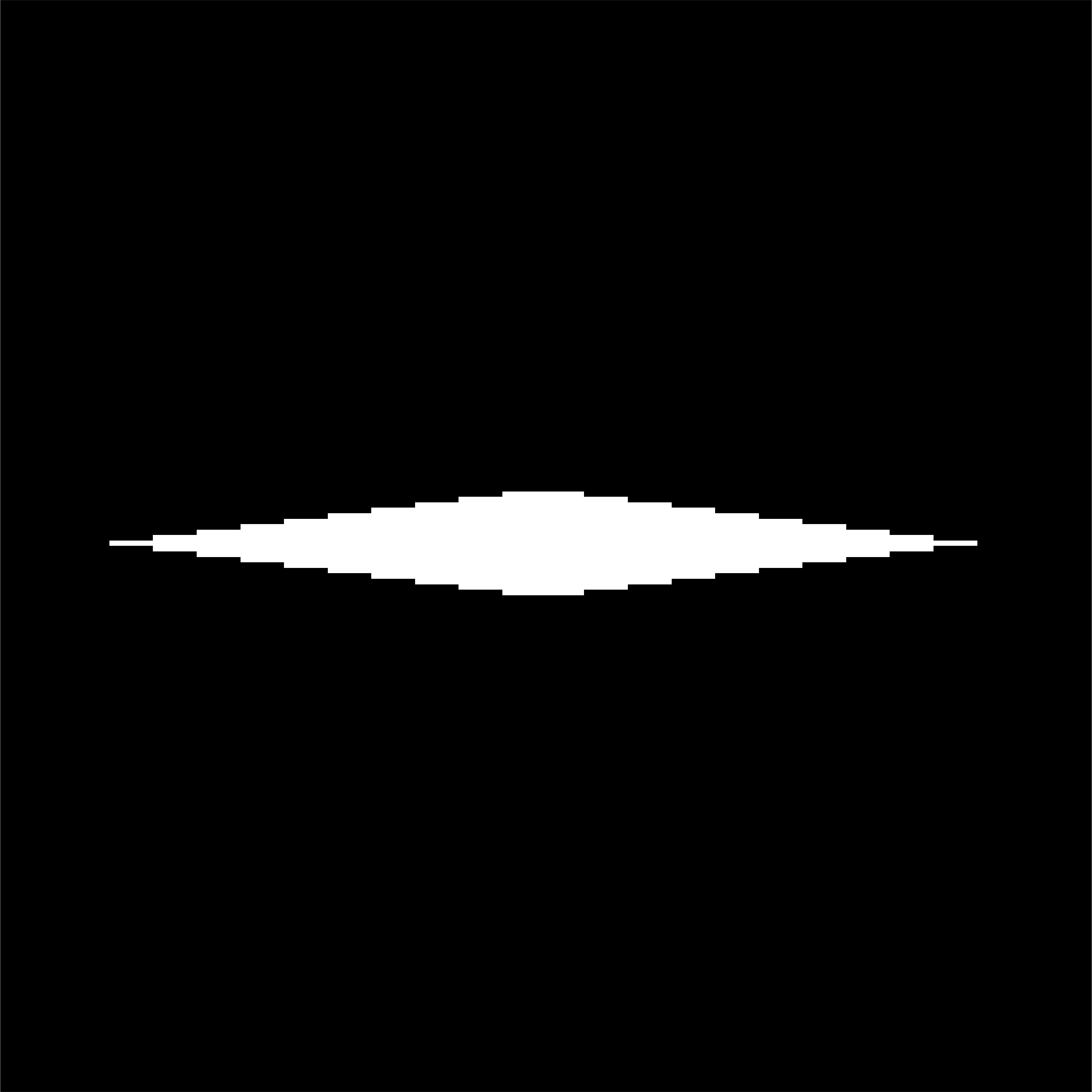}\\
		\includegraphics[width=0.33\textwidth]{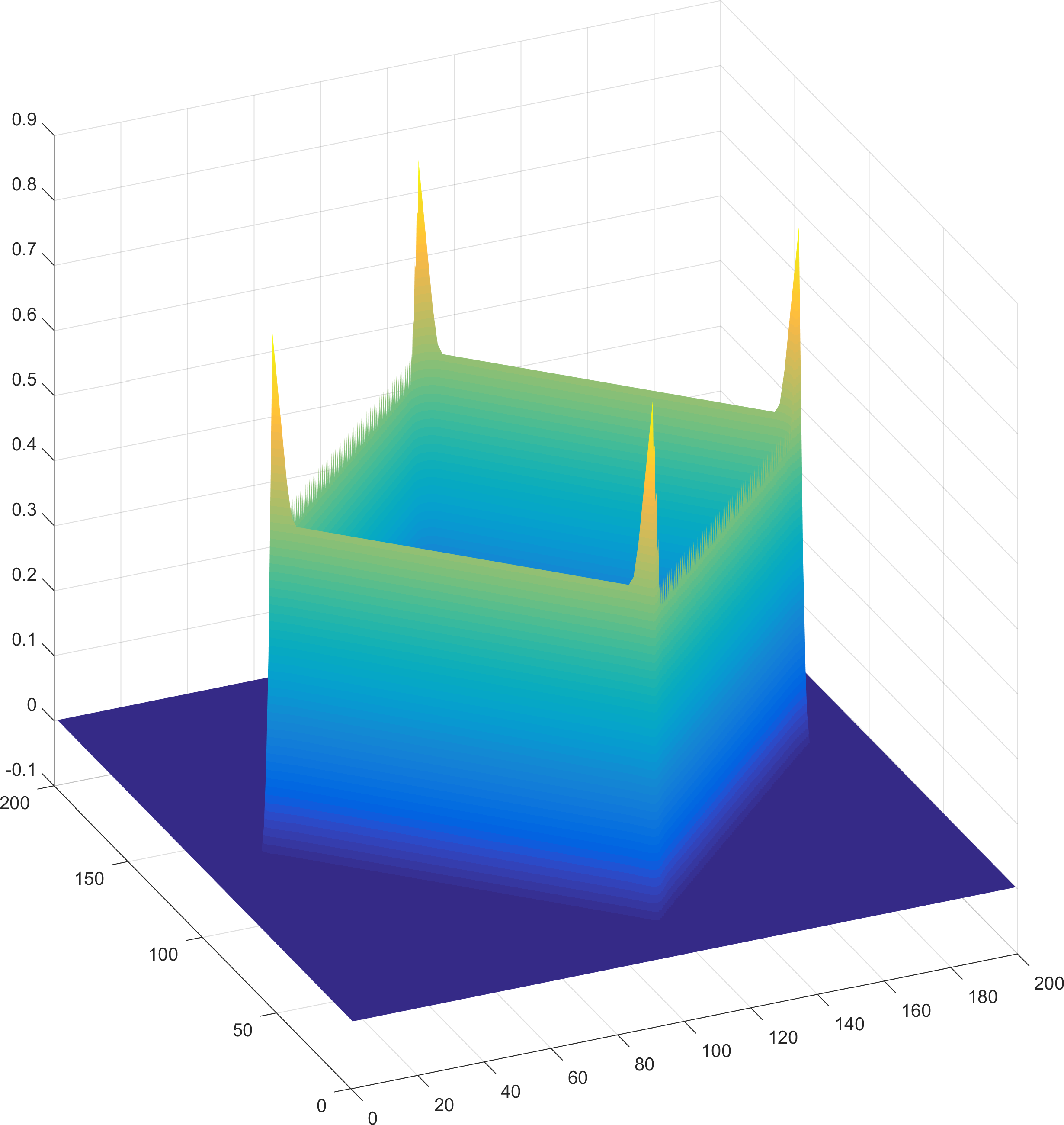}&
		\includegraphics[width=0.33\textwidth]{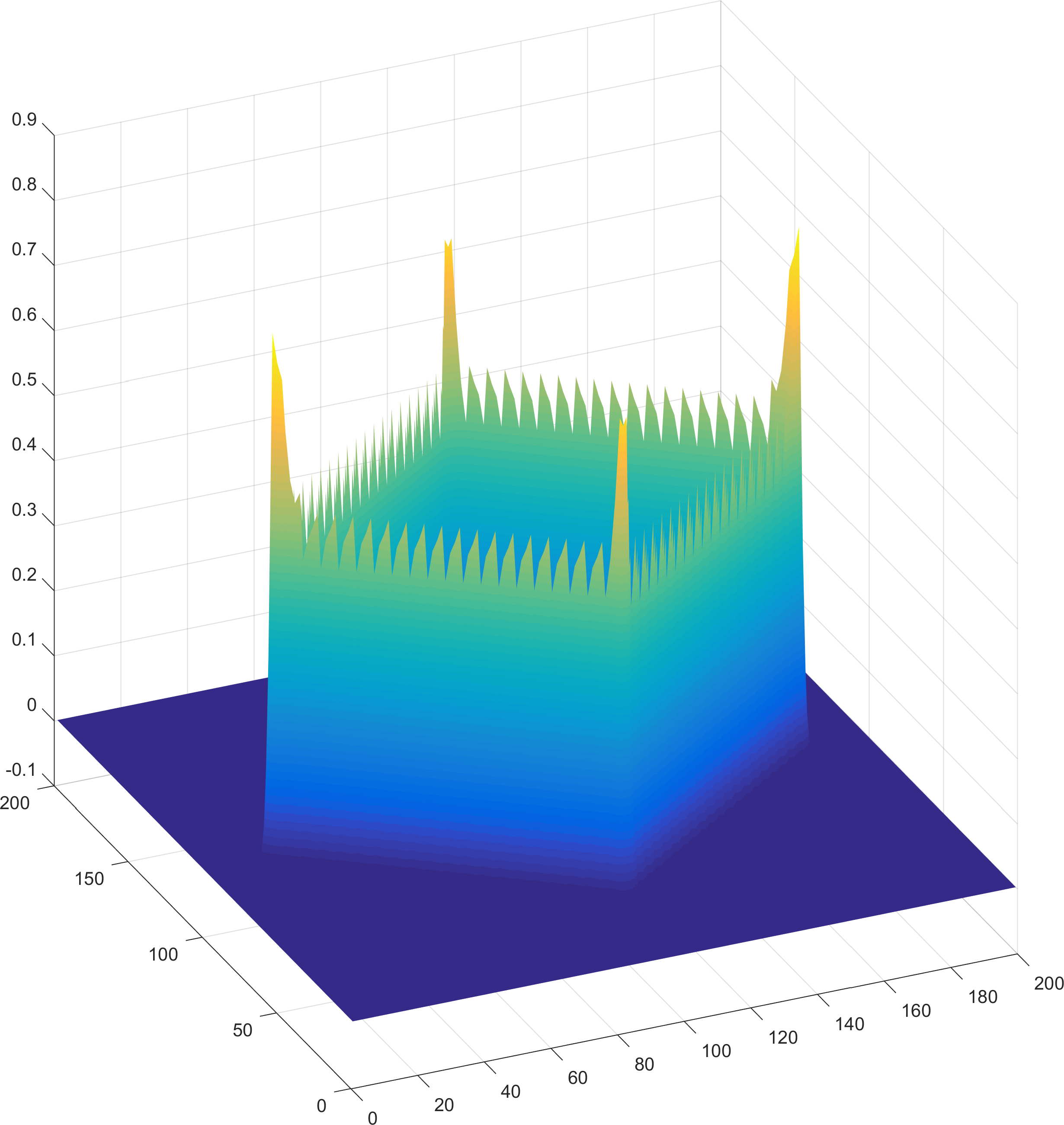}&
		\includegraphics[width=0.33\textwidth]{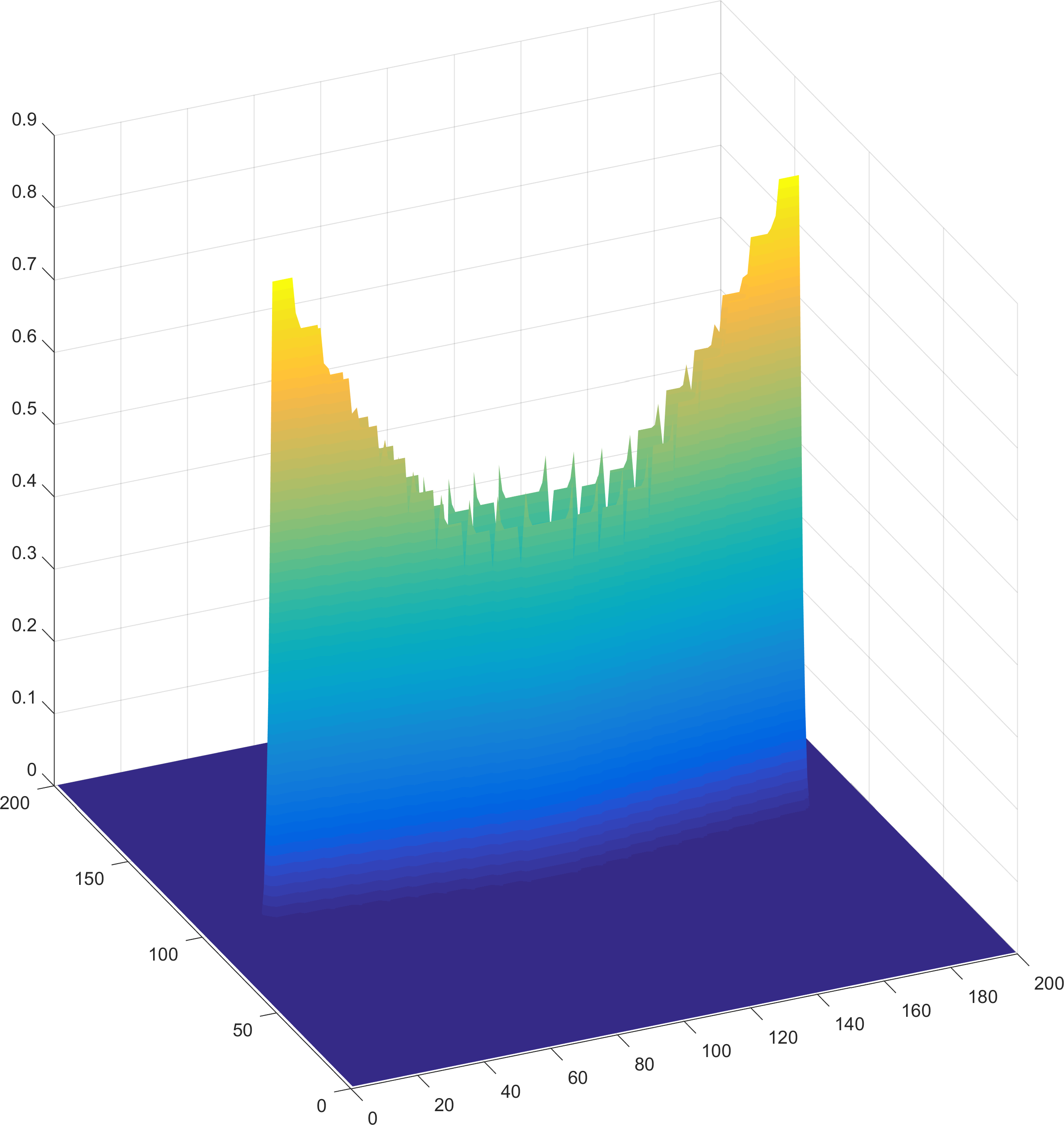}\\
		(a)&(b)&(c)
	\end{array}$
	}
\caption{\label{Sec3.Fig.CrnDetAngle}
Graph of $SR_{\lambda,\tau}(\chi_{K_a})$ for different pairs of opening angle $\theta$. 
$(a)$ $\pi/2--\pi/2$;	
$(b)$ $5\pi/12--7\pi/12$;	
$(c)$ $\pi/12--11\pi/12$.
}
\end{figure}

Let $\Omega\subset\mathbb{R}^n$ be a bounded open set with
$|\partial\Omega|=0$ (i.e. $\partial\Omega$ has zero $n-$dimensional measure) and $x\in\partial\Omega$. 
We say that the point $x\in\partial\Omega$ is a $\delta-$regular point of $\partial \Omega$
if there is an open ball $B(x_0;\delta)\subset \bar\Omega^c$, $x_0\in\Omega^c$, $\delta>0$,
such that $x\in \partial B(x_0;\delta)$ and if there is an open ball $B(x_0;\delta)\subset \Omega$, 
$x_0\in\Omega$, $\delta>0$, such that $x\in \partial B(x_0;\delta)$. 
If the point $x\in\partial\Omega$ meets only the first condition, we refer to it as exterior $\delta-$regular
point whereas if it meets only the second condition is called interior $\delta-$regular point.
Figure \ref{Fig:RegPoints} displays the different type of points of $\partial\Omega$.
The stable ridge transform
allows the characterization of such points given that if $x\in\partial\Omega$
is a $\delta-$regular point of $\Omega$ with $\delta>0$ sufficiently small, in \cite{ZOC15a} it is shown that 
there holds
\begin{equation}\label{Eq:RglPnt}
	SR_{\lambda,\tau}(\chi_{\bar\Omega})(x_0) \leq \frac{(\sqrt{\lambda+\tau}-\sqrt{\tau})^2}{\lambda}\,.
\end{equation}
As a result, we define an extractable corner point of
$\Omega$ if for at least sufficiently large $\lambda>0$ and $\tau>0$,
\begin{equation}\label{Eq:ExtrPnt}
	SR_{\lambda,\tau}(\chi_{\Omega})(x_0)>\mu_{1}(\lambda,\tau),
\end{equation}
where
\begin{equation}\label{Eq:ExtrPntmu1}
	\mu_{1}(\lambda,\tau):=\frac{(\sqrt{\lambda+\tau}-\sqrt{\tau})^2}{\lambda},
\end{equation}
is called the standard height for codimension-$1$ regular boundary points.
The analysis of the behaviour of $SR_{\lambda,\tau}(\chi_{K_a})$ in the case of the 
prototype exterior corner defined by the set $K_{a}=\{(x,y)\in\mathbb{R}^2:|y|\leq ax,a, x\geq 0\}$,
with  angle $\theta$ satisfying $a=\tan(\theta/2)$ shows that the value of 
$SR_{\lambda,\tau}(\chi_{K_a})$ at the corner tip $(0,0)$ of $K_a$ is given by 
\begin{equation}\label{Sec3.ExtCrn.Cond}
	SR_{\lambda,\tau}(\chi_{K_a})(0,0):=\mu_2(a,\lambda,\tau)=
		\left\{\begin{array}{ll}
			\displaystyle \frac{\lambda}{\lambda+(1+a^2)\tau}	& \text{if }\displaystyle a^2\leq\sqrt{\frac{\lambda+\tau}{\tau}} 	\\[1.5ex]
			\displaystyle \frac{1+a^2}{a^2}\frac{(\sqrt{\lambda+\tau}-\sqrt{\tau})^2}{\lambda}	& \text{otherwise}\,.
		\end{array}
		\right.
\end{equation}
One can then verify that  for $a>0$, and for any $\lambda,\,\tau>0$,
\[
	\mu_2(a,\lambda,\sigma)>\mu_1(\lambda,\tau)\quad\text{ and }\quad\lim_{a\to\infty}\mu_2(a,\lambda,\sigma)=\mu_1(\lambda,\sigma)\,.
\]
This result means that when the angle $\theta$ approaches $\pi$, the singularity at $(0,0)$ disappears.
Figure \ref{Sec3.Fig.CrnDetAngle} illustrates the behaviour of $SR_{\lambda,\tau}(\chi_{K_a})$
for different values of the opening angle $\theta$ and
for $\tau=\sigma\lambda$ with $\sigma=1/8$, for which the value of $a$ that separates the two conditions in \eqref{Sec3.ExtCrn.Cond} 
corresponds to $\theta=2\pi/3$.
Based on this prototype example \cite[Example 6.11]{ZOC15a}, one can therefore conclude 
that $R_{\tau}(C^u_{\lambda}(\chi_{\bar\Omega}))$ can actually detect exterior corners, whereas 
it might happen that at some $\delta$-singular points of $\partial\Omega$, $R_{\tau}(C^u_{\lambda}(\chi_{\bar\Omega}))$
takes on values lower than at the regular points of $\partial\Omega$. As a result,
a different Hausdorff stable method will be therefore needed to detect
interior corners and boundary intersections of domains.
\subsubsection{Interior corners}
\begin{figure}[H]
	\centerline{$\begin{array}{cc}
		\includegraphics[width=0.37\textwidth]{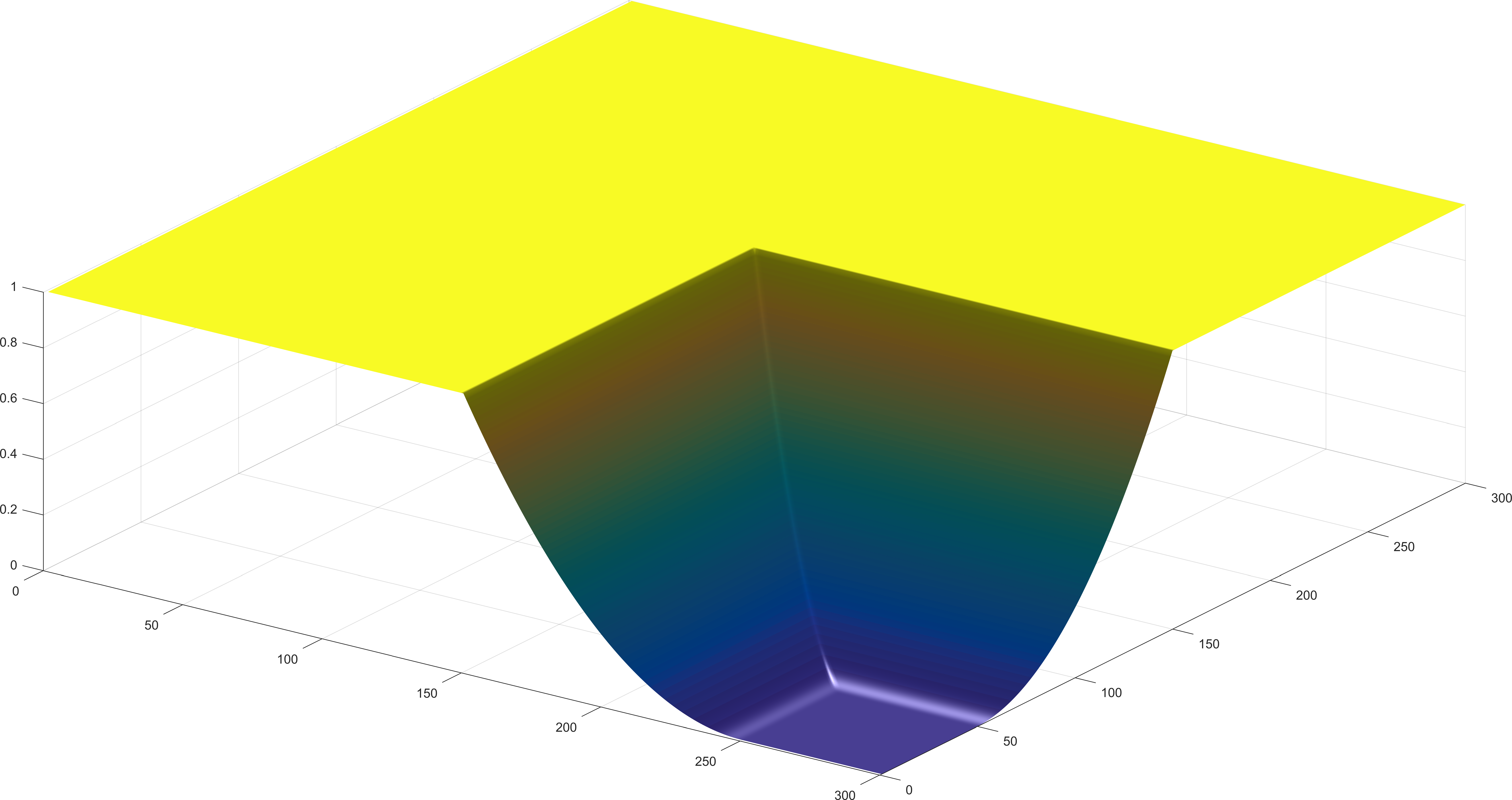}
		& \includegraphics[width=0.37\textwidth]{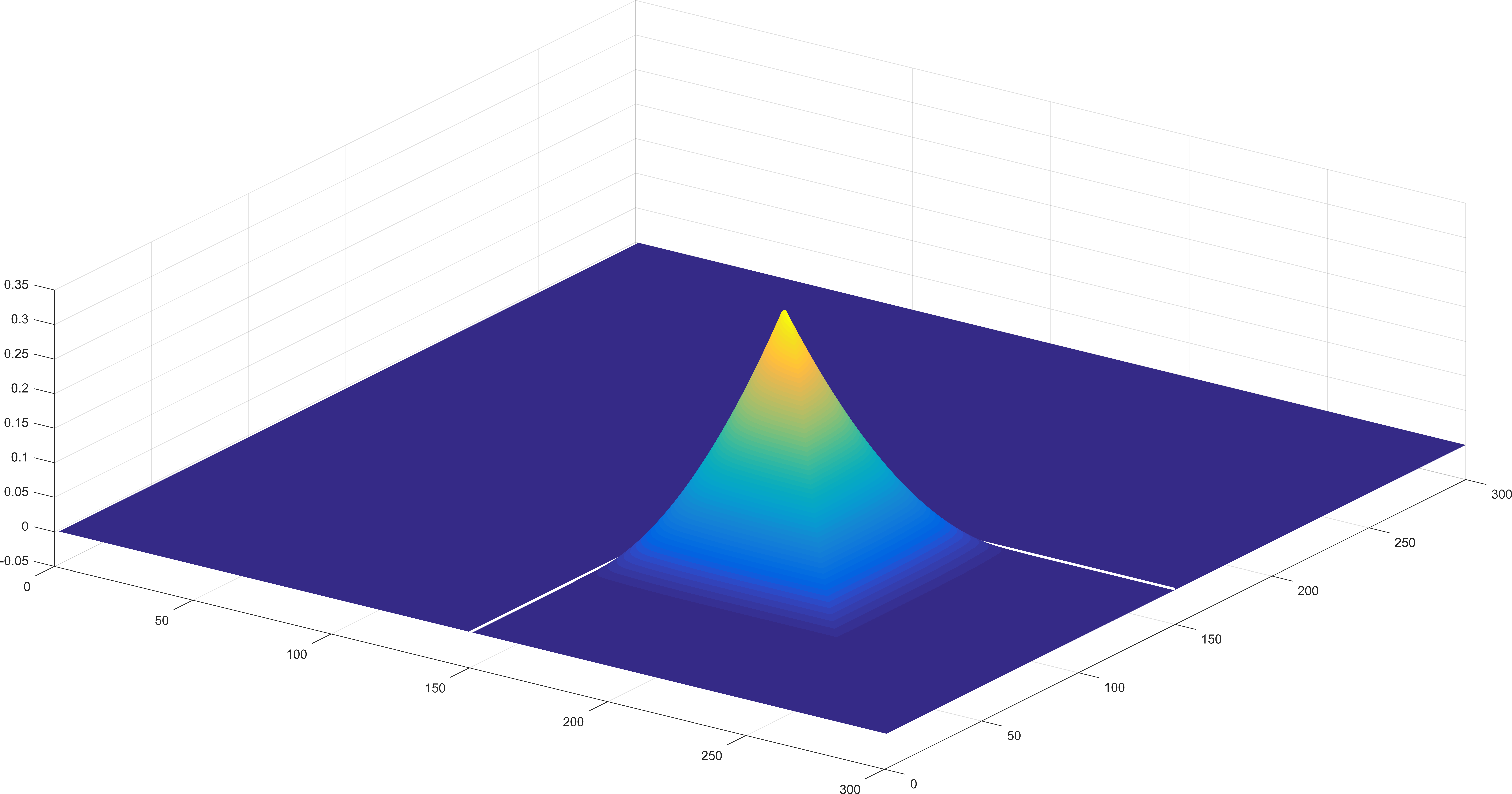}\\[1.5ex]
		(a)&(b)
		\end{array}$
		}
	\caption{\label{Fig:ProtInterCorn} Prototype of internal corner. $L-$shape domain.
	$(a)$ Graph of $D_{\lambda}^2(\cdot,\,K)$ for $\lambda=0.0001$;
	$(b)$ Graph of $V_{\lambda}^d(\cdot,\,K)$.
	}
\end{figure}

\begin{figure}[H]
\centerline{
	$\begin{array}{ccc}
		\includegraphics[width=0.30\textwidth]{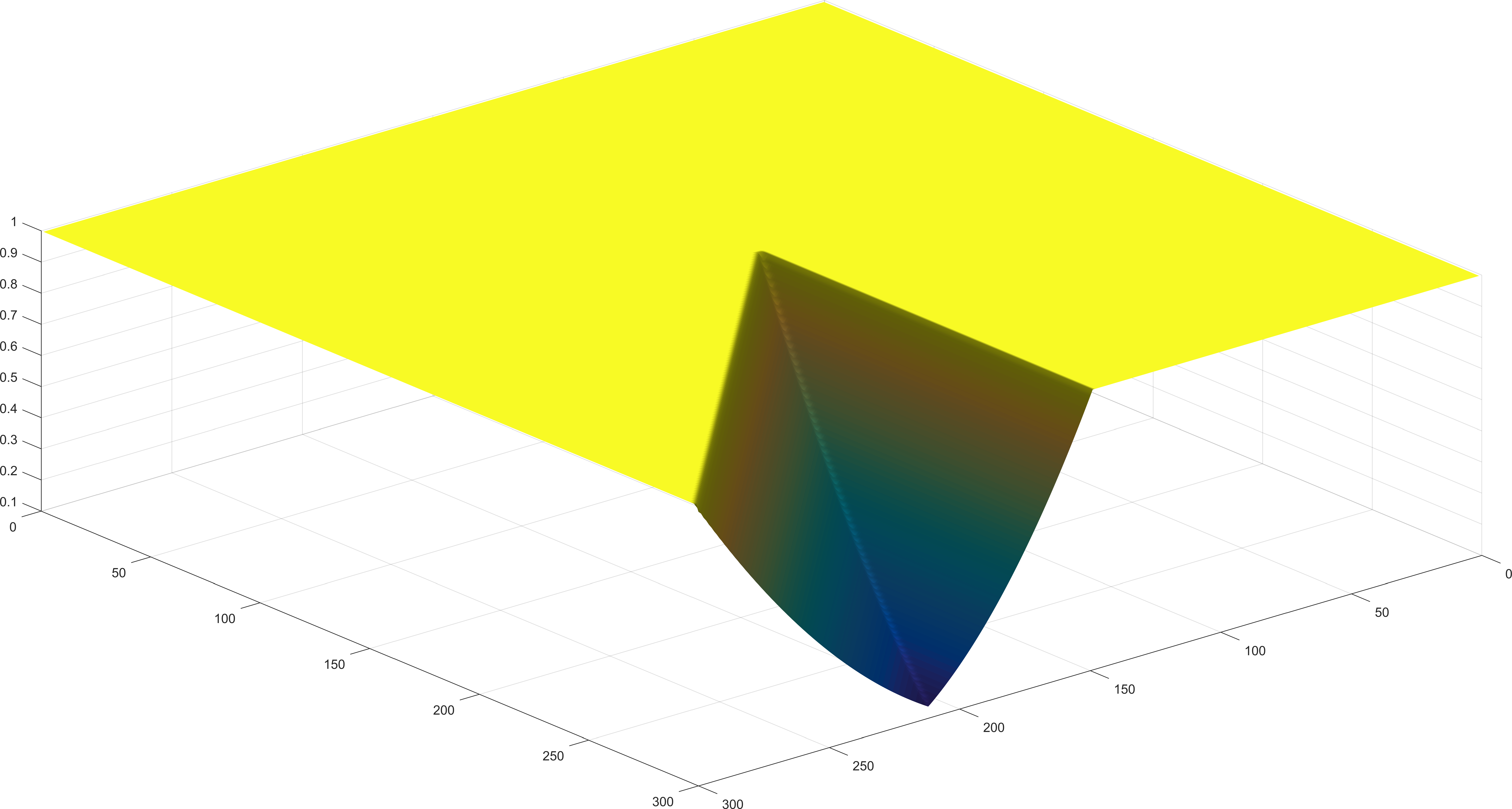}&
		\includegraphics[width=0.30\textwidth]{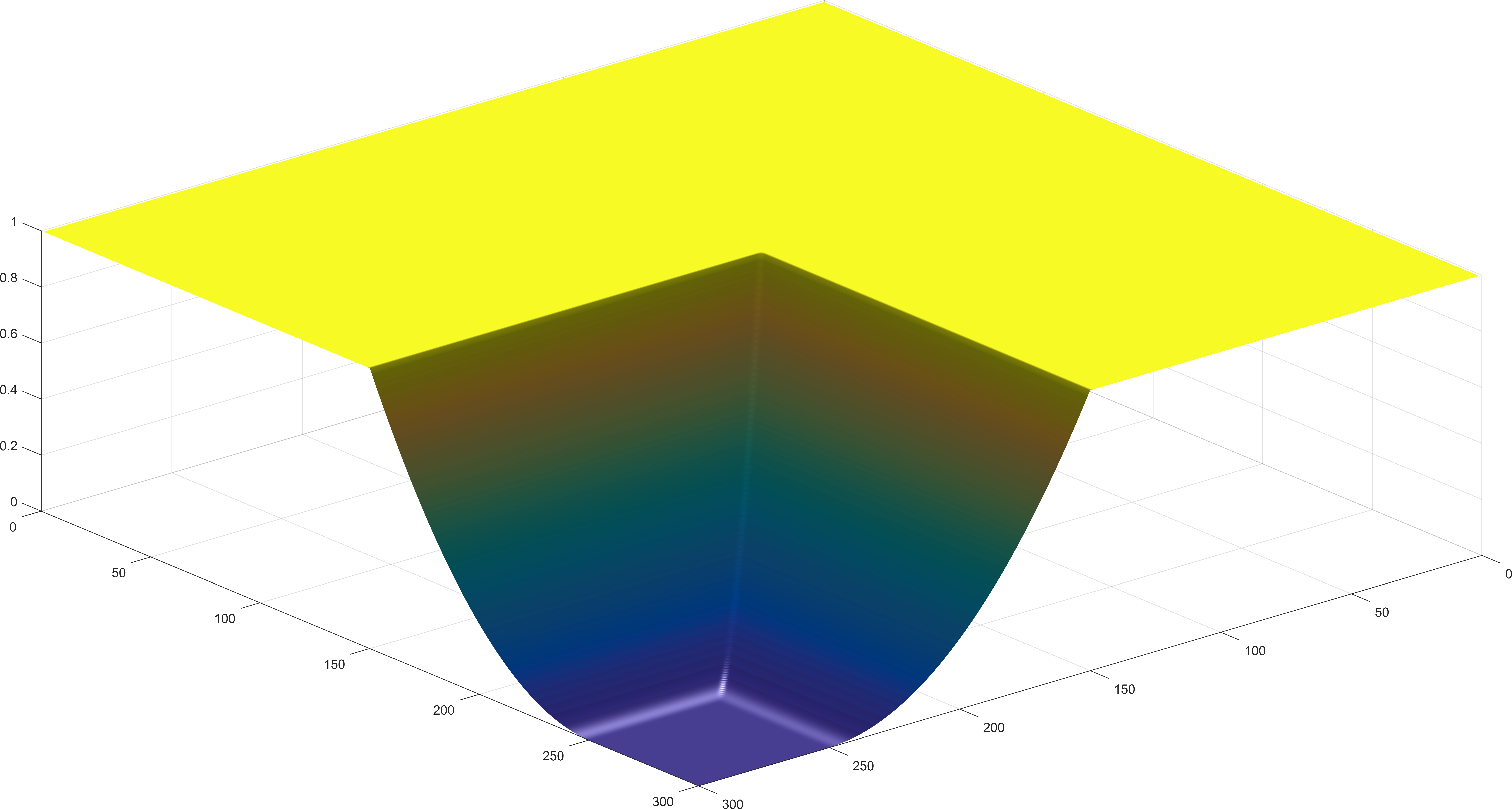}&
		\includegraphics[width=0.30\textwidth]{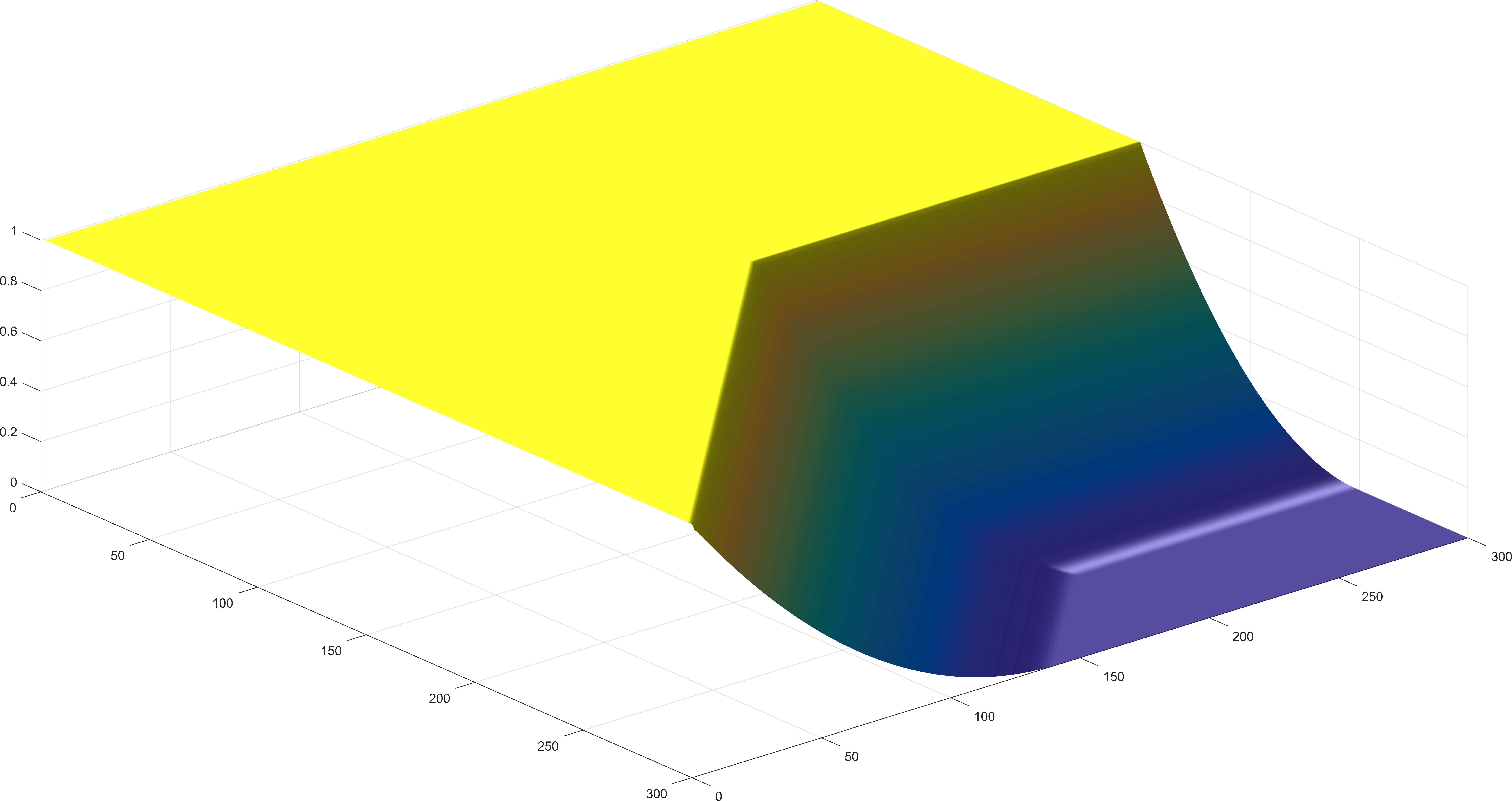}\\
		(a)&(b)&(c)\\
		\includegraphics[width=0.30\textwidth]{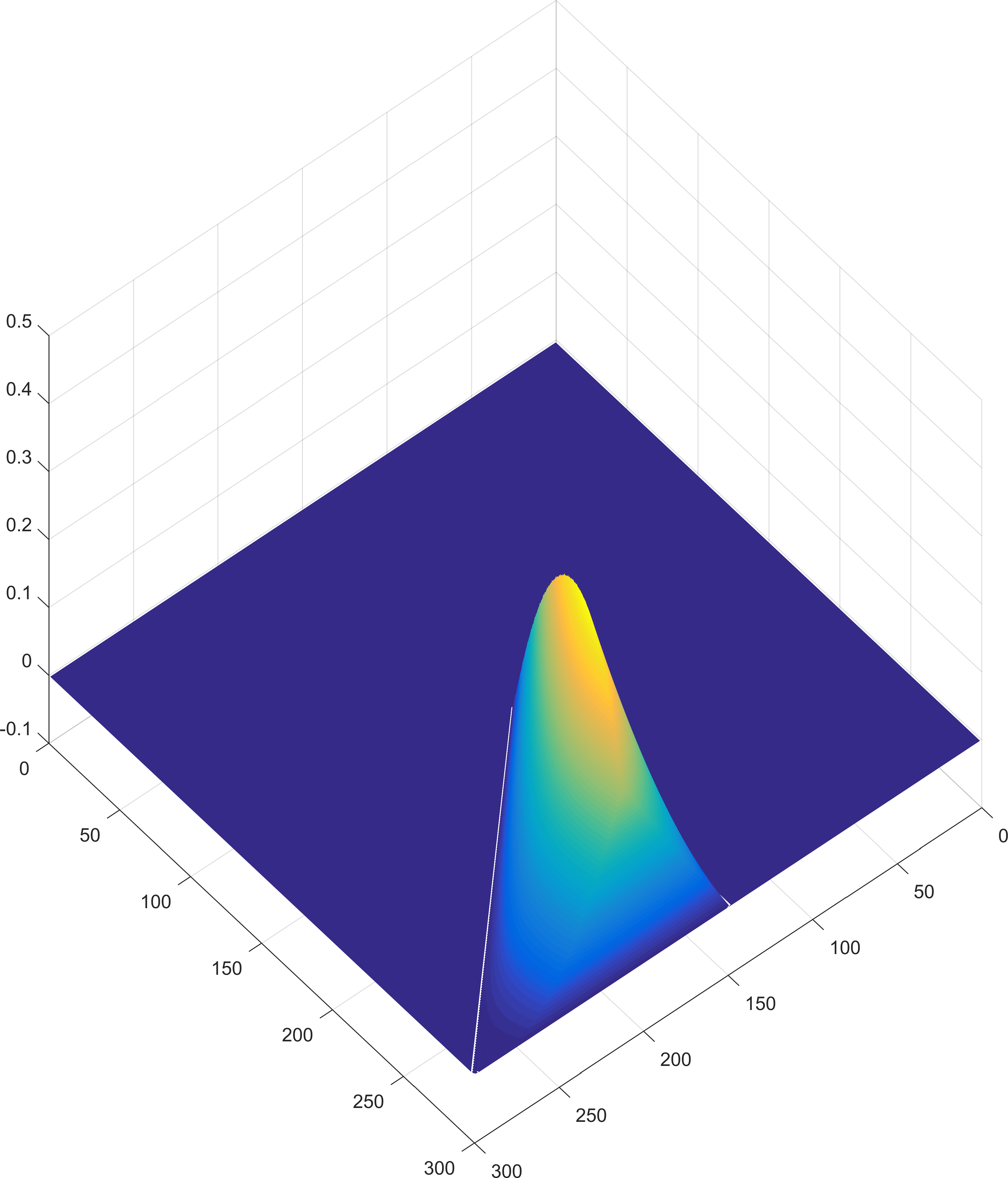}&
		\includegraphics[width=0.30\textwidth]{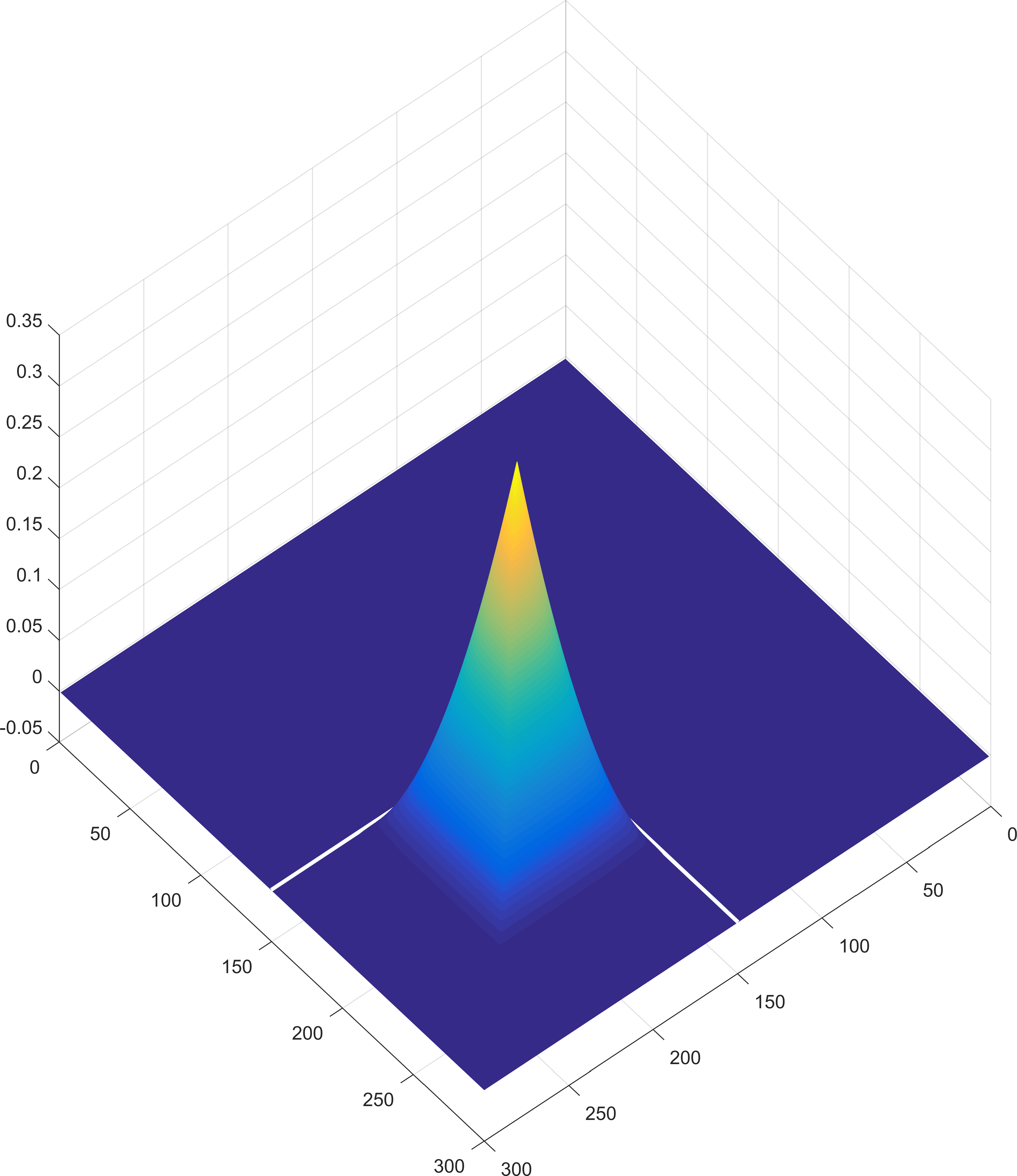}&
		\includegraphics[width=0.30\textwidth]{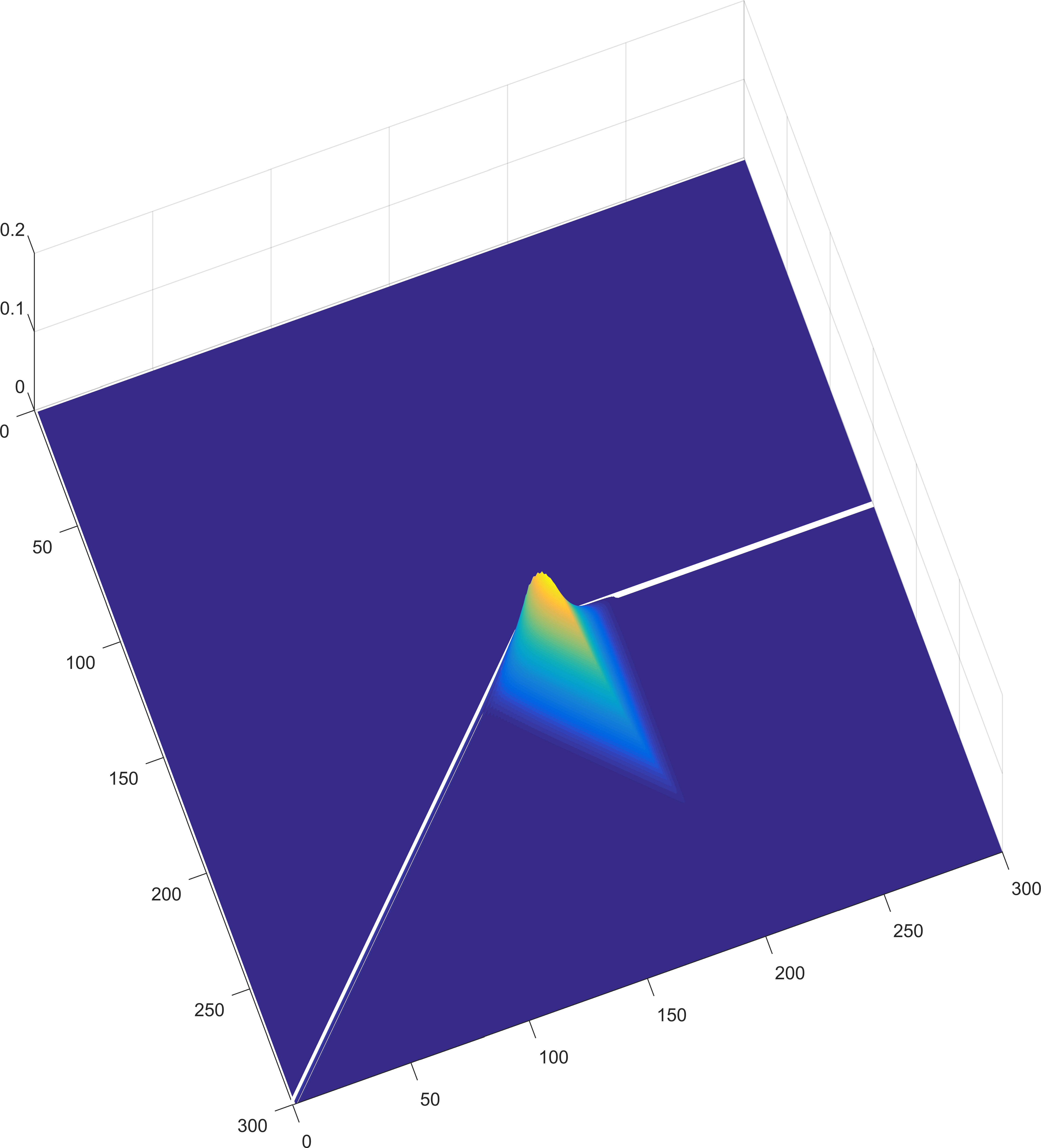}\\
		(d)&(e)&(f)\\
		\fbox{\includegraphics[width=0.30\textwidth]{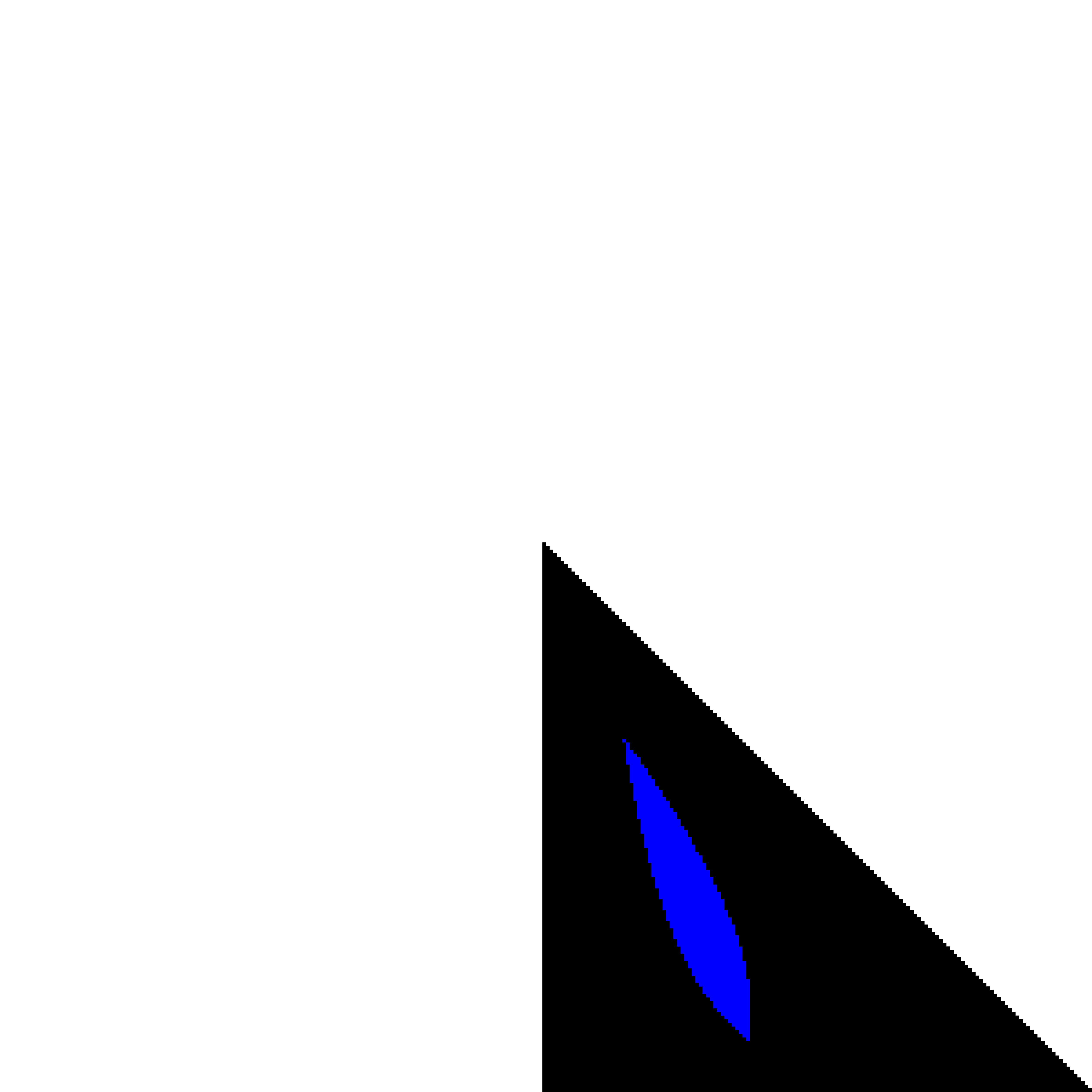}}&
		\fbox{\includegraphics[width=0.30\textwidth]{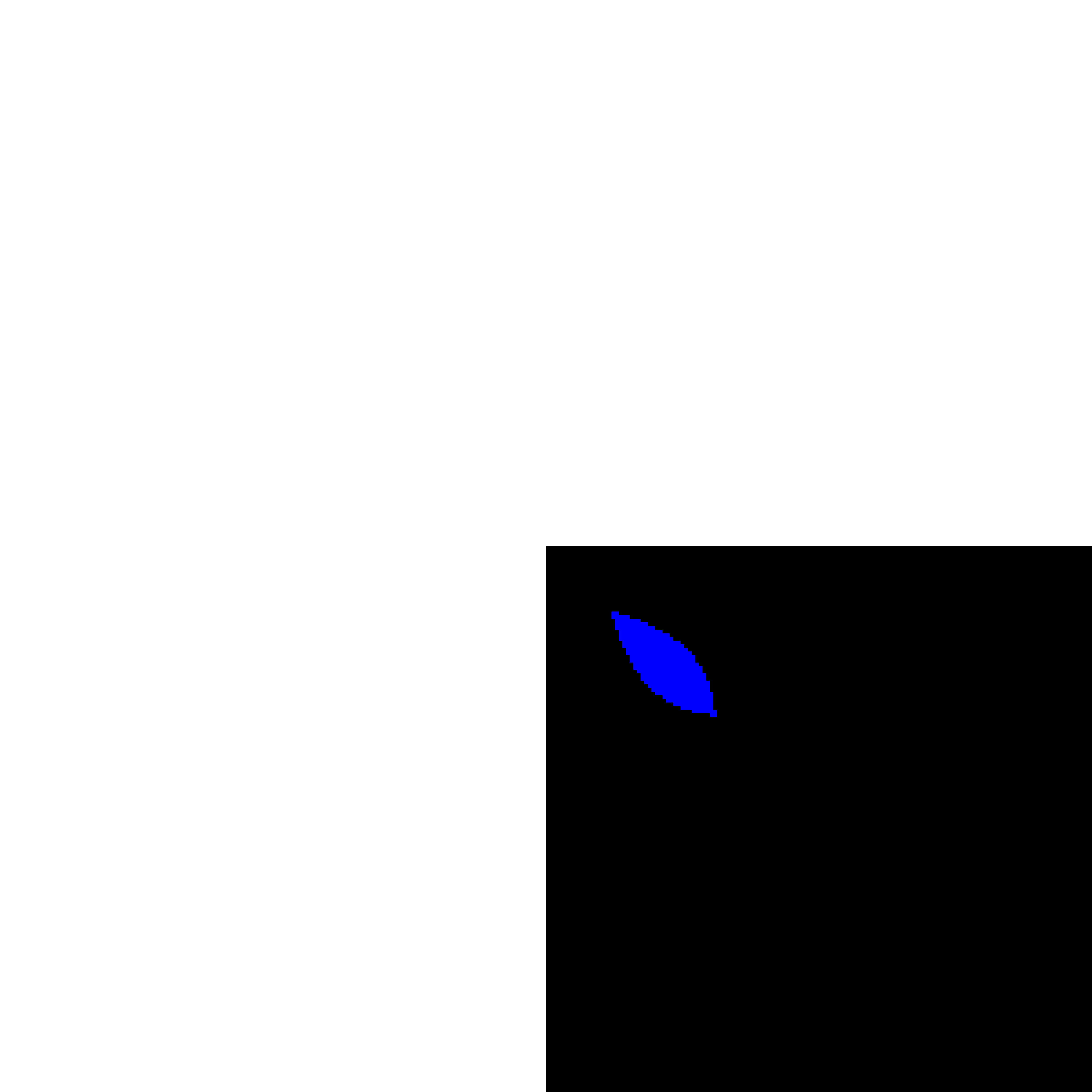}}&
		\fbox{\includegraphics[width=0.30\textwidth]{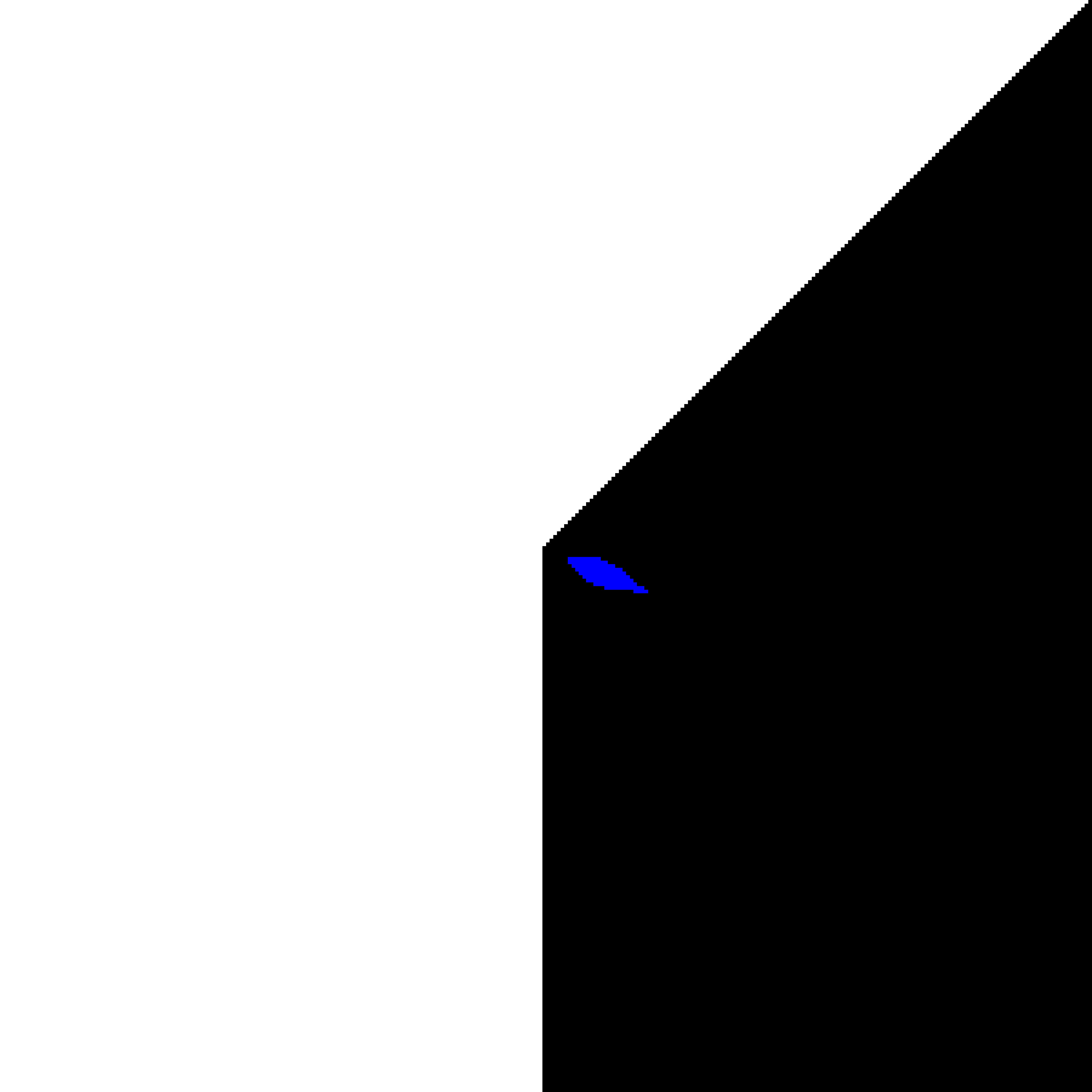}}\\
		(g)&(h)&(i)
	\end{array}$
	}
\caption{\label{Fig:InterCornVSangle}
	Graph of $D_{\lambda}^2(\cdot,\,K)$, $\lambda=0.0001$, for the three prototypes of 
	interior angle: $(a)$ acute angle; $(b)$ rectangular angle
	and $(c)$ obtuse angle. \\
	Graph of $V_{\lambda}^d(\cdot,\,K)$, $\lambda=0.0001$, for the three prototypes of interior angle:
	$(c)$ acute angle; $(d)$ rectangular angle
	and $(e)$ obtuse angle. \\
	Suplevel set of $V_{\lambda}^d(\cdot,\,K)$ with $\lambda=0.0001$ and for a level equal to 
	$\displaystyle 0.8\max_{x\in \R^2} \{V_{\lambda}^d(x,\,K)\}$ 
	for different values of the opening angle of the interior corner prototype: 
	$(g)$ Acute angle, $\max_{x\in \R^2} \{V_{\lambda}^d(x,\,K)\}=0.4137$;  
	$(h)$ Rectangular angle, $\max_{x\in \R^2} \{V_{\lambda}^d(x,\,K)\}=0.3323$;
	$(i)$ Obtuse angle, $\max_{x\in \R^2} \{V_{\lambda}^d(x,\,K)\}=0.1053$.}
\end{figure}

Since a prototype interior corner is defined as the complement of an exterior corner, one could think
of detecting interior corners of $\Omega$ by looking at the stable ridge transform of the complement of
$\Omega$ in $\R^n$. But this would not provide useful information for geometric objects
subject to finite sampling. On the other hand, traditional methods, such as Harris and Susan, as well as
other local mask based corner detection methods, would also not apply directly to such a situation.
In this case therefore we adopt an indirect approach. This consists of constructing an  
\textit{ad-hoc} geometric designed based function that is robust under sampling and is such 
that its singularities can be identified with the geomteric singularities we want to extract:
$(i)$ interior corners of a domain, and $(ii)$ intersections of smooth manifolds.
By applying one of the transforms introduced in Section \ref{Sec:BasicRVE} according to the type of singularity,
we can detect the singularity of interest. Given a non--empty closed set $K\subset\mathbb{R}^2$
with $K\not=\mathbb{R}^n$, an instance of function whose singularities capture 
the type of geometric feature of $K$ which we are interested of, is the distance--based 
function \eqref{Eq:DistTr} for $\lambda>0$, which we re-write next for ease of reference
\begin{equation}
	D_{\lambda}^2(x,\,K)=\left(\max\{0,\,1-\sqrt{\lambda}\dist(x,\, K)\}\,\right)^2,\quad x\in \mathbb{R}^n\,.
\end{equation}
Figure \ref{Fig:ProtInterCorn}$(a)$ displays the graph of $D_{\lambda}^2(x,\,K)$ for a prototype
of interior corner 
in an $L-$shape domain, and shows
that such singularity is of the valley type. 
By applyig then to $D^2_\lambda(\cdot,\, K)$ the valley transform \eqref{Eq.RVEtransf} with 
the same parameter $\lambda$ as used to compute $D^2_\lambda(\cdot,\, K)$ itself, we obtain
\begin{equation}\label{Eq:DefValIntCrn}
\begin{split}
		V^d_\lambda(x,\, K)&=-V_\lambda(D^2_\lambda(\cdot,\, K))(x)\\[1.5ex]
				   &=C_{\lambda}^u(D^2_\lambda(\cdot,\, K))(x)-D^2_\lambda(x,\, K),\quad x\in \mathbb{R}^n\,,
\end{split}
\end{equation}
whose graph is displayed in Figure \ref{Fig:ProtInterCorn}$(b)$.
We observe therefore that this transfom allows the definition of the set of interior corner 
points and intersection points of scale $1/\sqrt{\lambda}$ as the support of $V^d_\lambda(\cdot,\, K)$, that is
\begin{equation}
	I_{\lambda}(K)=\{x\in\mathbb{R}^n,\,V^d_\lambda(x,\, K)>0\}\,.
\end{equation}
In this manner we obtain a marker which is localized in the neighborhood of the 
feature. 
Figure \ref{Fig:InterCornVSangle} displays, for $\lambda=0.0001$, 
the behaviour of $D_{\lambda}^2(\cdot,\,K)$, of $V_{\lambda}^d(\cdot,\,K)$, and 
of the suplevel set of $V_{\lambda}^d(\cdot,\,K)$ for a level equal to 
$\displaystyle 0.8\max_{x\in \R^2} \{V_{\lambda}^d(x,\,K)\}$ as approximation of
$I_{\lambda}(K)$, considering different opening angles of the interior corner prototype $K$. 
As for the exterior corner, we observe that the marker reduces and the maximum of $V^d_\lambda(x,\, K)$
depends on the opening angle of the corner. The larger is
the angle, the smaller is the value of $\max\,V^d_\lambda(x,\, K)$  which agrees with what we expect given 
that the interior angle disappears and the marker vanishes. 
Finally, since $D^2_\lambda(\cdot,\, K)$ is Hausdorff-Lipschitz continuous, it is easy to see that so is 
$V^d_\lambda(x,\, K)$.

\subsection{Stable Multiscale Intersection Transform of Smooth Manifolds}\label{Sec:Filters:IT}
\begin{figure}[ht]
  \centerline{$\begin{array}{ccc}
		\includegraphics[width=0.33\textwidth]{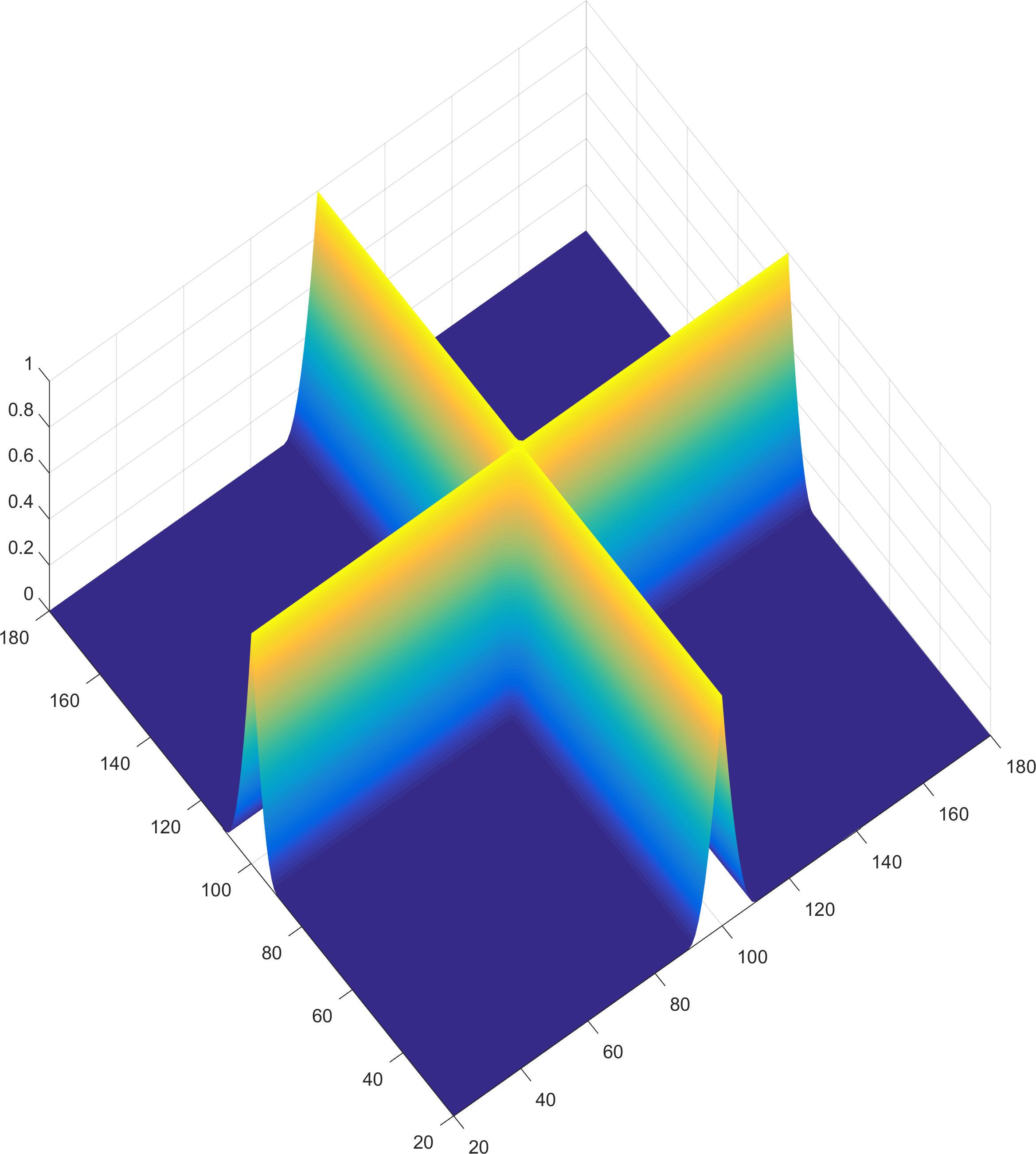}&
		\includegraphics[width=0.33\textwidth]{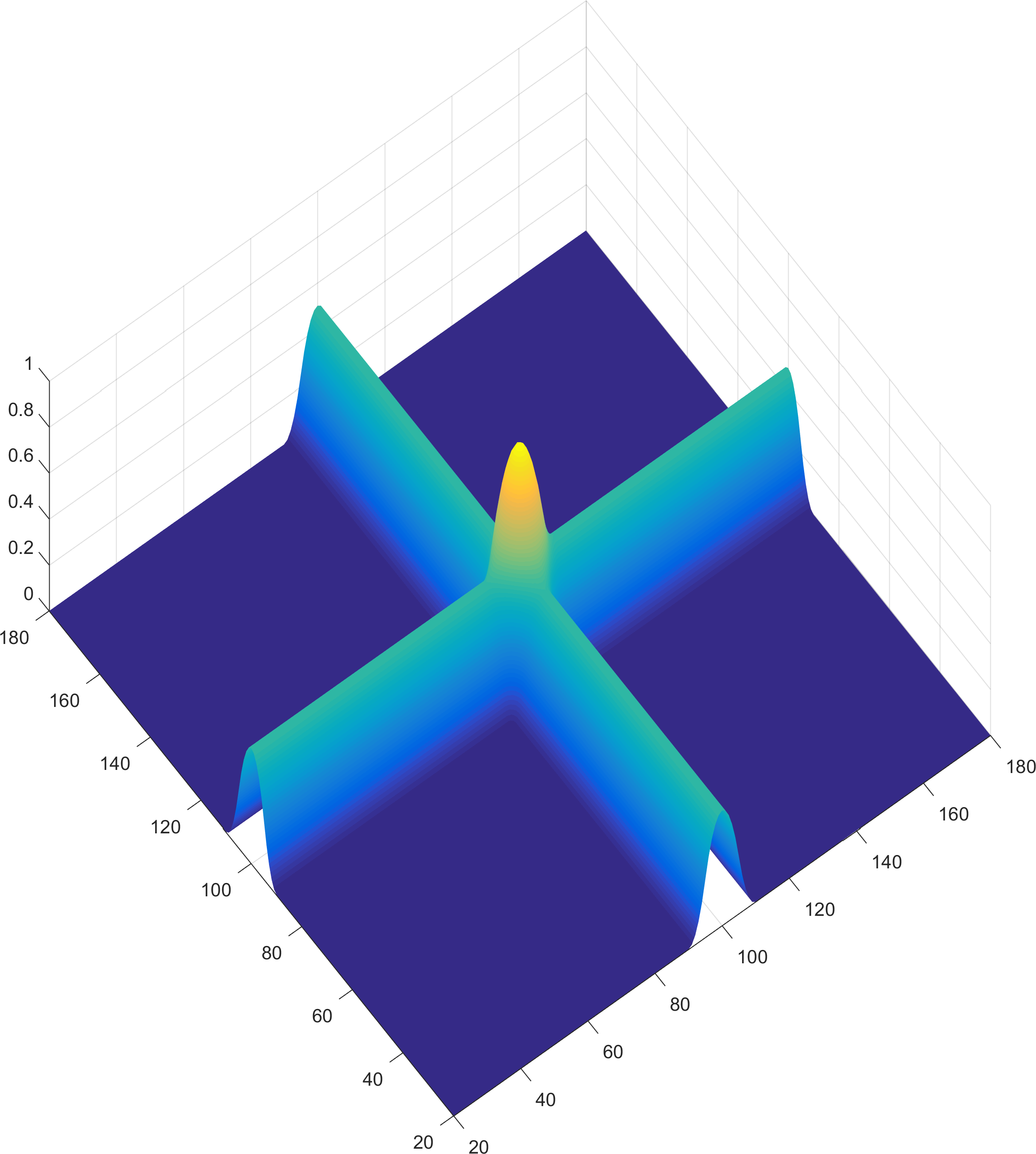}&
		\includegraphics[width=0.33\textwidth]{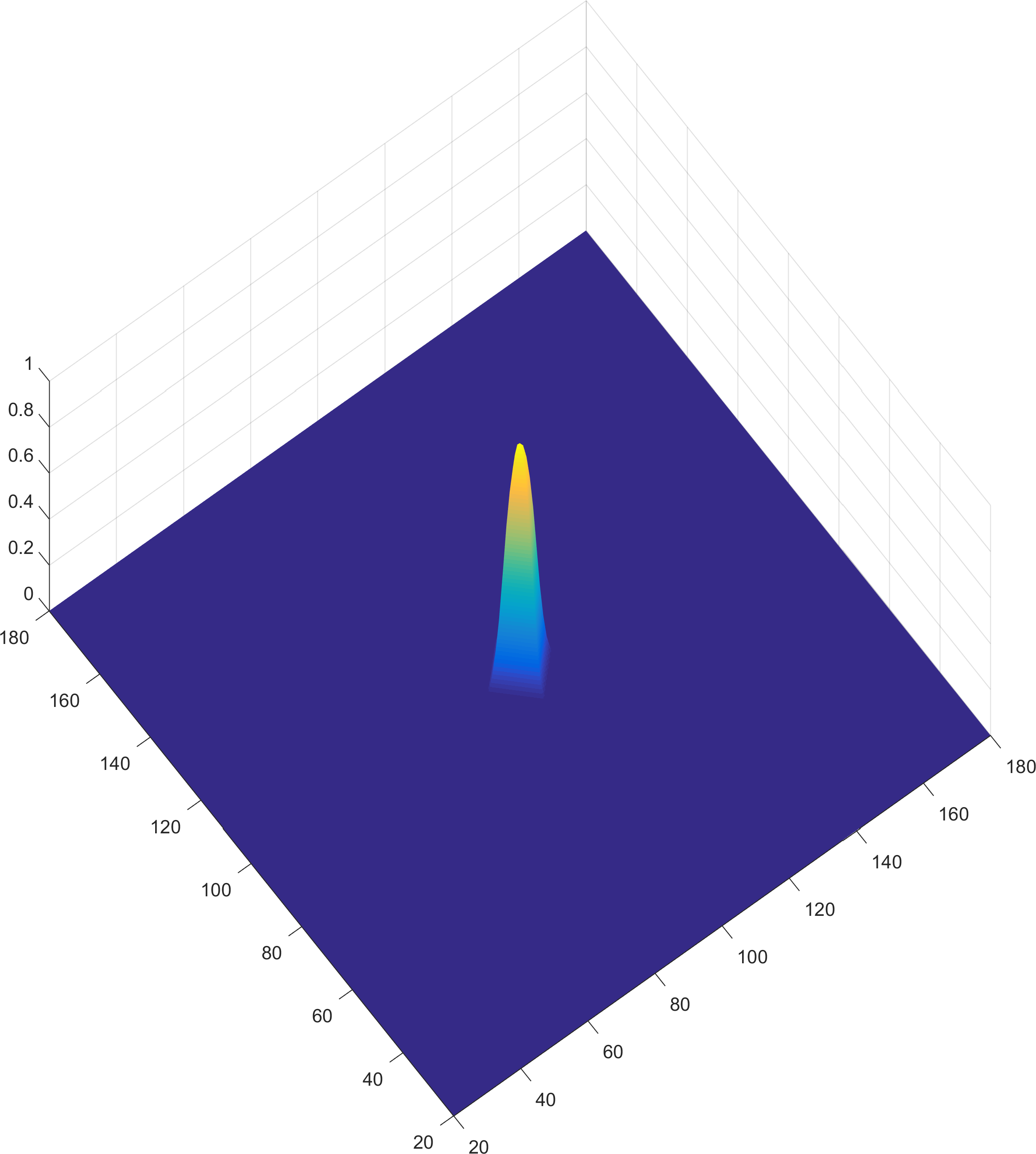}\\
		(a)&(b)&(c)
	\end{array}$}
\caption{\label{Fig:IntersFilt} Graph of:
	$(a)$ The upper transform $C^u_\lambda(\chi_{K_{\alpha=1}})(x)$ of the characteristic function of two crossing lines
	with right angle;
	$(b)$ The mixed transform $C^l_\lambda(C^u_\lambda(\chi_{K_{\alpha=1}}))(x)$;
	$(c)$ The intersection filter $I_{\lambda}(\cdot;\,K_{\alpha=1})$ together with the 
		graph of the characteristic function of $K_{\alpha=1}$ displayed as reference.}
\end{figure}

Rather than devising an ad-hoc function that embeds the geometric features as its singularities, 
one can suitably modify the landscape of the characteristic function of the object and generate singularities
which are localised in a neighborhood of the geometric feature of interest. This is for instance the rationale 
behind the transformation introduced in \cite{ZOC15a}. The objective is to obtain a Hausdorff stable multiscale 
method that is robust with respect to sampling, so that it can be applied to geometric objects represented by point clouds, 
and that is able to describe possible hierarchy of features as defined in terms of some characteristic geometric property. 
If we denote by $K\subset\mathbb{R}^n$ the union of finitely many smooth compact manifolds $M_k$, for $k=1,\ldots, m$,
in this section we are interested to extract two types of types of geometric singularities:
\begin{itemize}
	\item[$(i)$]	Transversal surface-to-surface intersections.
	\item[$(ii)$]   Boundary points shared by two smooth manifolds.
\end{itemize}
These problems are studied extensively in computer-aided geometric design under
the general terminology of shape interrogation \cite{PM02}. The traditional approach to  surface-to-surface
intersection problems is to consider parameterized
polynomial surfaces and to solve systems of algebraic equations numerically based on real algebraic
geometry \cite{PM02}. The application of these methods typically requires 
some topological information such as triangle mesh connectivity or a parameterization
of the geometrical objects, hence they are difficult to implement in the 
cases of free-form surfaces and of manifolds represented, for instance, by point clouds. For the latter case,
other types of approaches are usually used which aim at identifying, according to some criteria, the points
that are likely to belong to a neighborhood
of the sharp feature. State-of-art methods currently in use are mostly 
justified by numerical experiments, and their stability properties, under dense sampling of the set $M$, 
are not known. Let $K\subset \mathbb{R}^n$ be a non-empty compact set. By using compensated convex transforms we 
introduced the intersection extraction transform of scale $\lambda>0$ \cite{ZOC15a} by
\begin{equation}\label{Eq.Def.IE}
	I_\lambda(x;\, K)=\Big|C^u_{4\lambda}(\chi_K)(x)-2\Big(C^u_\lambda(\chi_K)(x)-C^l_\lambda(C^u_\lambda(\chi_K))(x)\Big)\Big|,
	\quad x\in\mathbb{R}^n\,.
\end{equation}
By recalling the definition  of the stable ridge transform \eqref{Sec3.Eq.StabRid} of scale $\lambda$ and $\tau$ for the
characteristic function $\chi_K$, $I_{\lambda}(x;\,K)$ can be expressed 
in terms of $\mathrm{SR}_{\lambda,\tau}(\chi_K)(x)$ for $\tau=\lambda$ as
\begin{equation}\label{Eq.DefInStb}
	I_\lambda(x;\, K)=\Big|C^u_{4\lambda}(\chi_K)(x)-2\mathrm{SR}_{\lambda,\lambda}(\chi_K)(x)\Big|,
	\quad x\in\mathbb{R}^n\,.
\end{equation}
As instance of how $I_\lambda(\cdot;\, K)$ is used to remove or filter regular points,
Figure \ref{Fig:IntersFilt} 
illustrates  the graphs of $C^u_\lambda(\chi_{K_{\alpha=1}})(x)$, 
$C^l_\lambda(C^u_\lambda(\chi_{K_{\alpha=1}}))(x)$ and of the filter 
$I_{\lambda}(\cdot;\,K_{\alpha=1})$
in the case of the intersection of two lines perpendicular to each other.
This example can be generalized 
to `regular directions' and `regular points' on manifolds $K$ and verify that $I_{\lambda}(x,\,K)=0$ at these 
points. Let $K\subset \mathbb{R}^n$ be a non-empty compact set and $e$ is a $\delta$-regular direction of $x\in K$,
then $I_\lambda(y;\,K)=0$ for $y\in [x-\delta e,\; x+\delta e]:=\{x+t\delta e,\; -1\leq t\leq 1\}$
when $\lambda\geq1/\delta^2$. In particular, we have that at the point $x$,
\begin{equation}
	C^l_\lambda(C^u_\lambda(\chi_K))(x)=1/2\,.
\end{equation}
If $K$ is a $C^1$ manifold in a neighbourhood of $x\in K$ and $x$ is a $\delta$-regular point of $K$, then
$I_\lambda(y;\,K)=0$ if  $y-x\in N_x$ and $|y-x|\leq \delta.$
Since $C_{\lambda}^{u}(\chi_K)(x)=1$ for $x\in K$, by using $I_\lambda(\cdot;\,K)$, we have that the regular 
points will be removed by the transform itself, 
leaving only points near the singular ones. In this context, for compact $C^2$ $m$-dimensional manifolds with $1\leq m\leq n-1$, 
since $I_\lambda(y;\,K)=0$ for all  $\delta-$regular points $y\in K$ when $\lambda>0$ is sufficiently large, 
the condition $I_\lambda(y;\,K)=0$ can thus be used to define singular points which can be extracted 
by $I_\lambda(\cdot;\,K)$ if there exists a
constant $c_x>0$, depending at most only on $x$, such that  $I_\lambda(x;\, K)\geq c_x>0$
for sufficiently large $\lambda>0$.

From the definition \eqref{Eq.DefInStb} of $I_\lambda(\cdot;\, K)$ in terms of the stable ridge transform and of the 
upper transform of the characteristic function of the manifold $K$, since such transforms are Hausdorff stable, 
it follows that  $I_\lambda(\cdot;\,K)$ is also Hausdorff stable, that is, for $E,\,F$ non-empty compact subsets 
of $\R^n$ and $\lambda>0$, then there holds 
\begin{equation}
	|I_\lambda(x;\,E)-I_\lambda(x;\,F)|\leq 12\sqrt{\lambda}\,\dist_{\mathcal{H}}(E,\,F)\,,\quad x\in\R^n\,.
\end{equation}

\subsection{Stable Multiscale Medial Axis Map}
The medial axis of an object is a geometric structure introduced by Blum \cite{Blu67} as a
means of providing a compact representation of a shape. 
Initially defined as the set of the shock points of a grass fire lit on the boundary and required to propagate uniformly
inside the object. Closely related definitions of skeleton \cite{CH68} and cut-locus \cite{Wol93}
have since been proposed, and have 
served for the study of its topological properties \cite{Alb14,ACNS13,CCM91,Lie04,Mat88,SPW96}, 
its stability \cite{ChoiS04,CS04a} and for the development of fast and efficient algorithms for its computation 
\cite{AAAHJR09,AM97,AL13,KSKB95,OK95}. 
Hereafter we refer to the definition given in \cite{Lie04}.
For a given non-empty closed set $K\subset \mathbb{R}^n$, with $K \neq \R^n$, we define the 
medial axis $M_K$ of $K$ as  the set of points  $x\in\mathbb{R}^n\setminus K$ such that
$x\in M_K$ if  and only if there are at least two different points $y_1,\, y_2\in K$,
satisfying $\dist(x;\,K)=|x-y_1|=|x-y_2|$, whereas for a non-empty bounded open set 
$\Omega\subset \mathbb{R}^n$, the medial axis  of $\Omega$ is defined by $M_\Omega:=\Omega\cap M_{\partial\Omega}$. 

The application of the lower transform to study the medial axis $M_K$ of a set $K$ is motivated by 
the identification of the medial axis with 
the singularity set of the Euclidean distance function and by the geometric structure 
of this set \cite{ACNS13,CP01,MM03}. However, for our setting, it is more convenient to consider 
the squared distance function and to use the identification of the singular set of the squred distance function 
with the set of points where the squared distance function fails to be locally $C^{1,1}$. Since
the lower compensated convex transform to the Euclidean squared-distance function gives 
a smooth ($C^{1,1}$) tight approximation outside a neighbourhood of the closure of the medial axis, 
in \cite{ZCO15} the quadratic multiscale medial axis map with scale $\lambda>0$
is defined as a scaled difference between the squared-distance function and its lower transform, that is,
\begin{equation}\label{Eq.Def.MMA}
	M_{\lambda}(x;\,K):=(1+\lambda)R_\lambda(\dist^2(\cdot;\,K))(x)
			=(1+\lambda)\Big(\dist^2(x;\,K))-C_{\lambda}^l(\dist^2(\cdot;\,K)))(x)\Big)\,,
\end{equation}
whereas for a bounded open set $\Omega\subset \mathbb{R}^n$
with boundary $\partial\Omega$, the quadratic multiscale medial axis map of $\Omega$ 
with scale $\lambda>0$ is defined by
\begin{equation*}
	M_{\lambda}(x;\,\Omega):=M_{\lambda}(x;\,\partial\Omega)\qquad x\in\Omega.
\end{equation*}

A direct consequence of the definition of $M_{\lambda}(x,\,K)$ is that 
for $x\in\mathbb{R}^n\setminus M_K$ we have 
\begin{equation}\label{Eq:LimMMAM}
	\lim_{\lambda\to\infty} M_{\lambda}(x,\,K)=0\,,
\end{equation}
and the limit map $M_{\infty}(x,\,K)$ presents well
separated values, in the sense that they are zero outside the medial axis and 
remain strictly positive on it. 
To gain an insight of the geometric structure of $M_{\lambda}(x;\,K)$,
for $x\in M_K$, \cite{ZCO15} makes use of the separation angle $\theta_x$ introduced in \cite{Lie04}. 
Let $K(x)$ denote the set of points of $\partial K$ that realise the
distance of $x$ to $K$ and by $\angle[y_1-x,y_2-x]$ the angle between the two nonzero vectors
$y_1-x$ and  $y_2-x$ for $y_1,y_2\in K(x)$, then
\begin{equation}
	\theta_x=\max\{\angle[y_1-x,y_2-x]\,,\quad y_1,y_2\in K(x)\}.
\end{equation}
By means of this geometric parameter, it was shown in \cite{ZCO15} that for every $\lambda>0$ and $x\in M_K$ that
\begin{equation}
	\sin^2(\theta_x/2)\,\dist^2(x,\,K)\leq M_{\lambda}(x,\,K)\leq \dist^2(x,\,K)\,.
\end{equation}
This result along with the examination of prototype examples ensures that the multiscale medial axis map of 
scale $\lambda$ keeps a constant height
along the part of the medial axis generated by a two-point subset, with the value of the height depending on
the distance between the two generating points. Such values can, therefore, be used to define a 
hierarchy between different parts of the medial axis and one can thus select the relevant parts through 
simple thresholding, that is, by taking suplevel sets of the multiscale medial axis map, justifying the  
the word "multiscale" in its definition. 
For each branch of the medial axis, the multiscale medial axis map automatically defines a scale associated 
with it. In other words, a given branch has a strength which depends on some geometric features of the part 
of the set that generates that branch.

An inherent drawback of the medial axis $M_K$ is in fact its sensitivity to boundary details, in the sense that 
small perturbations of the object (with respect to the Hausdorff distance) can produce huge variations 
of the corresponding medial axis. This does not occur in the case of the quadratic multiscale medial axis map,
given that \cite{ZCO15} shifts somehow the focus from the support of $M_{\lambda}(\cdot,K)$ to the whole map.
Let $K,\, L\subset \R^n$ denote non-empty compact sets and $\mu:=\dist_{\mathcal{H}}(K,\,L)$,
it was shown in \cite{ZCO15}  that for $x\in \mathbb{R}^n$, we have
\begin{equation}\label{Eq:HAUSMMAM} 
	\Big|M_{\lambda}(x;\,K)-M_{\lambda}(x;\,L)\Big|\leq \mu(1+\lambda)
		\Big((\dist(x;\,K)+\mu)^2+2\dist(x;\,K)+2\mu+1\Big)\,.
\end{equation}
While the medial axis of $K$ is not a stable structure with respect to the Hausdorff distance, 
its medial axis map $M_{\lambda}(x;\,K)$ is by contrast  a stable structure. This result
complies with \eqref{Eq:HAUSMMAM} which shows that as $\lambda$ becomes large, 
the bound in \eqref{Eq:HAUSMMAM} becomes large. 

\begin{figure}[htbp]
\centerline{
	$\begin{array}{ccc}
		\includegraphics[width=0.25\textwidth]{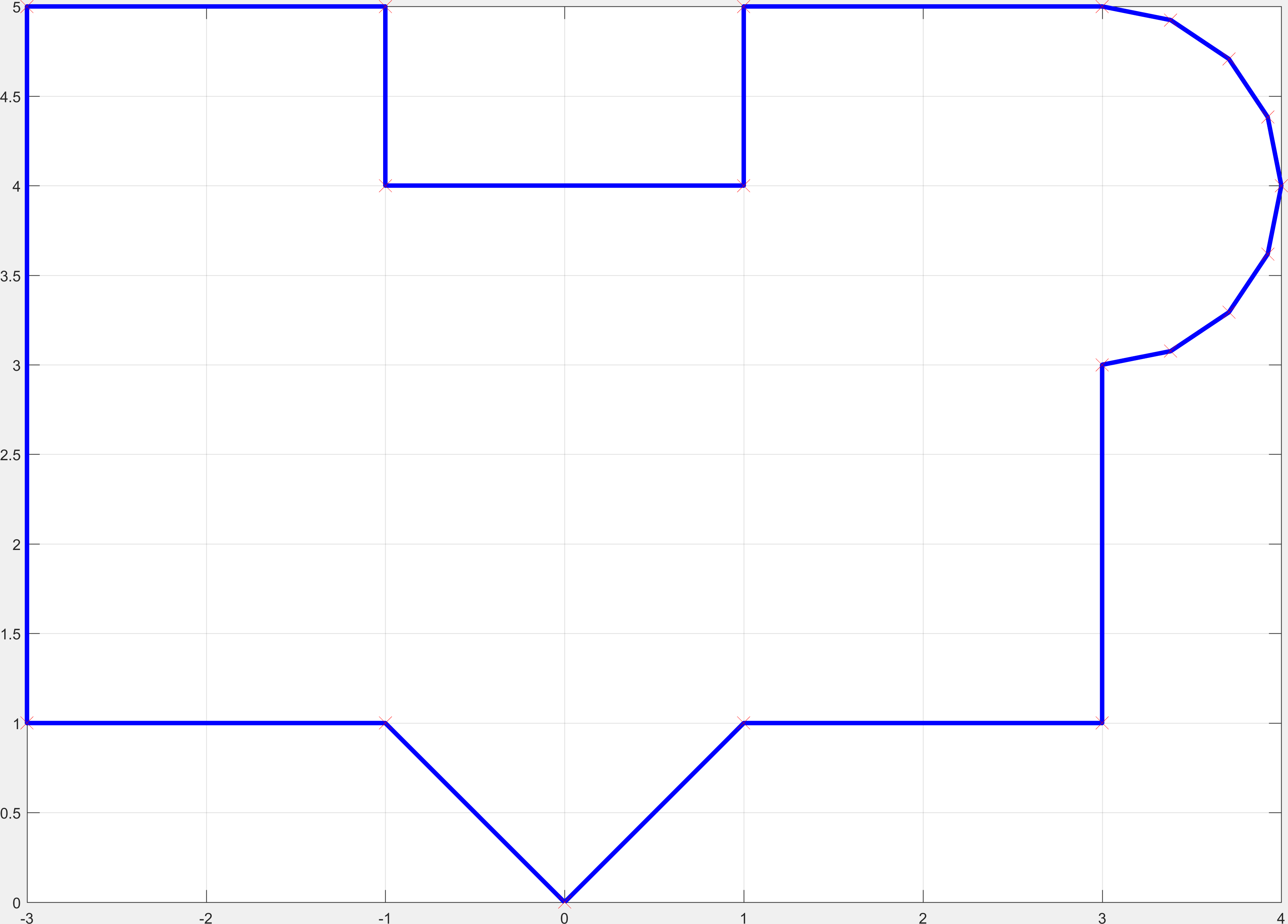}&
		\includegraphics[width=0.32\textwidth]{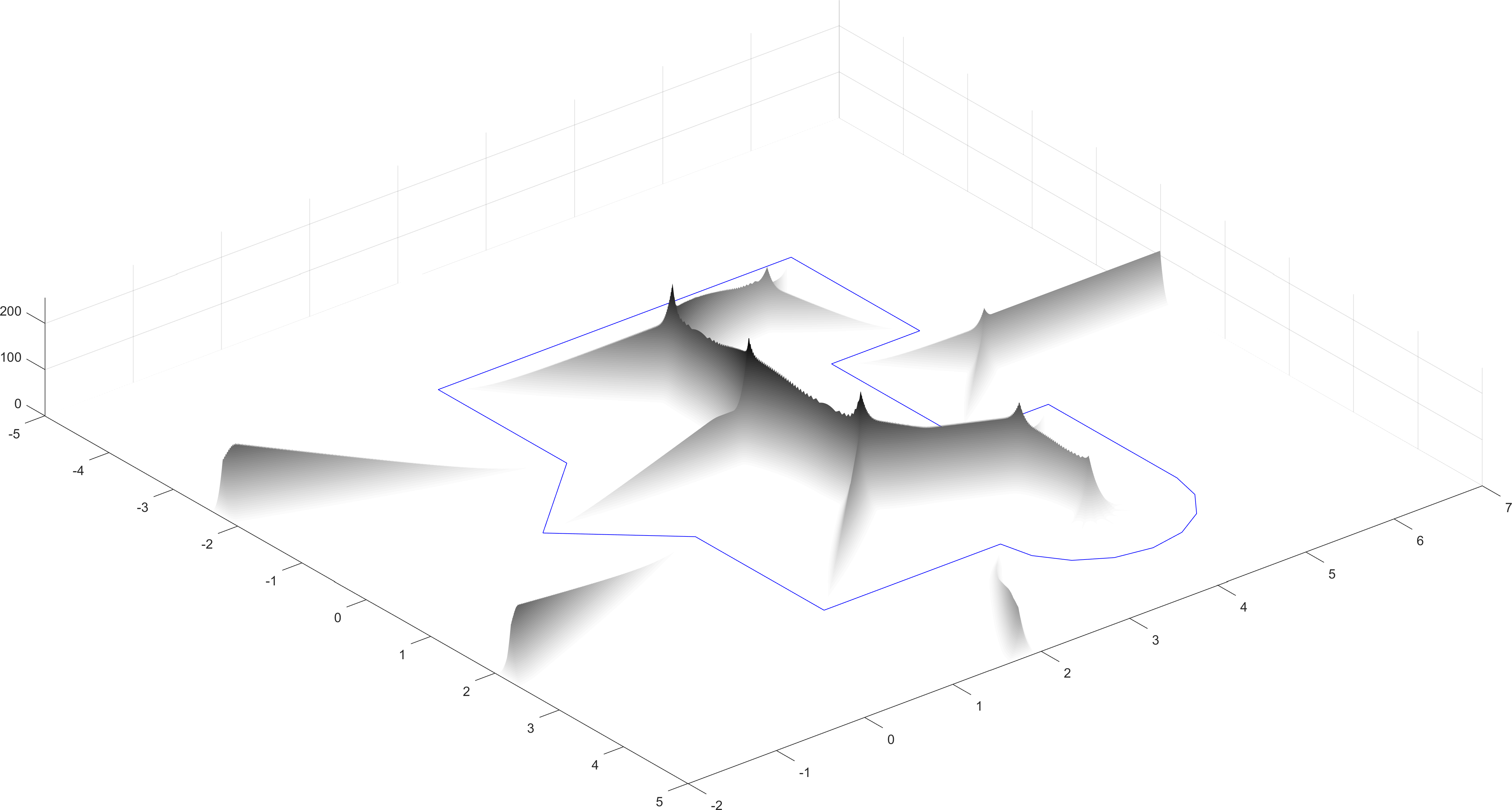}&
		\includegraphics[width=0.32\textwidth]{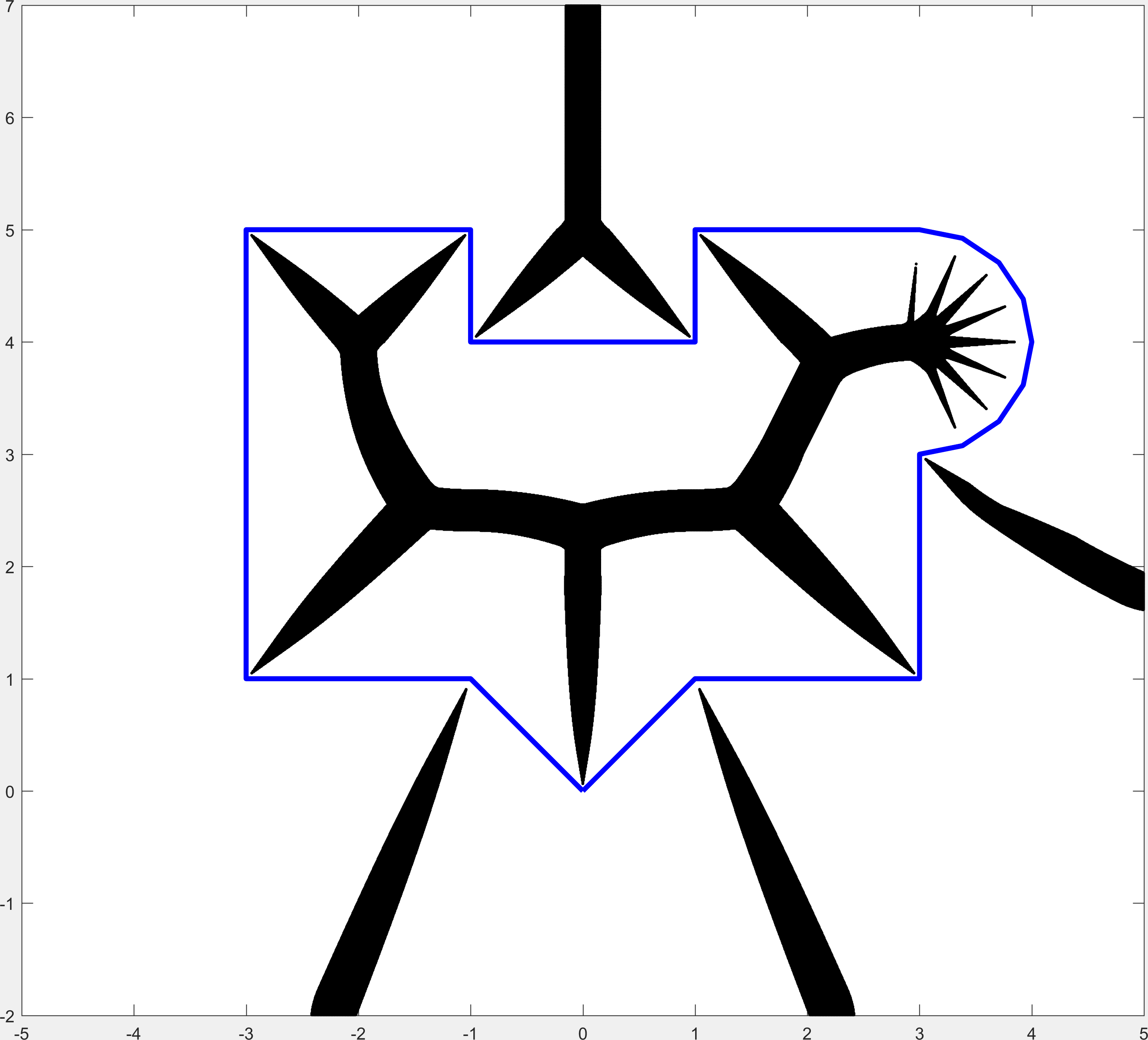}\\
				(a) & (b) & (c)\\
		\includegraphics[width=0.32\textwidth]{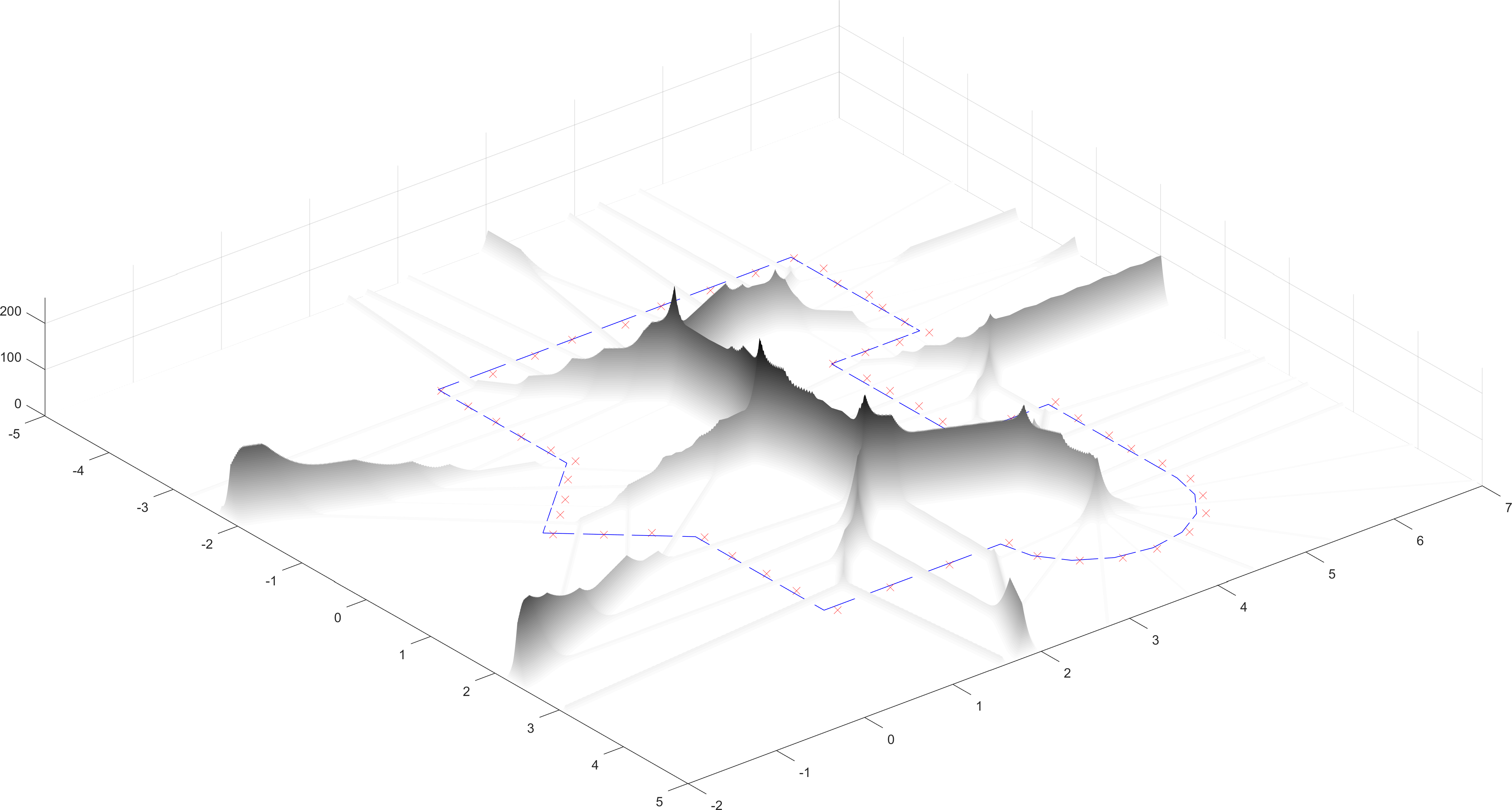}&
		\includegraphics[width=0.32\textwidth]{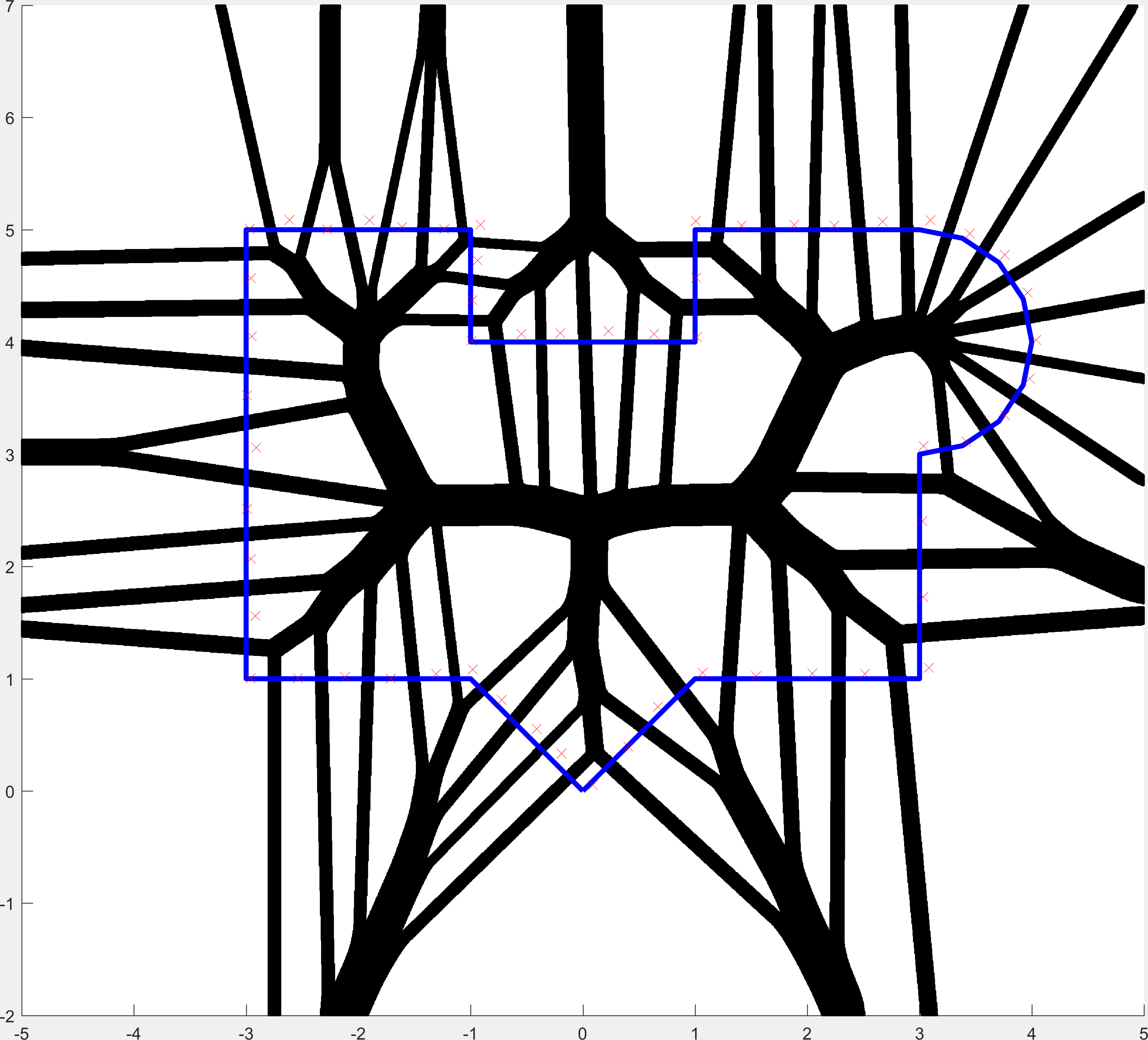}&
		\includegraphics[width=0.32\textwidth]{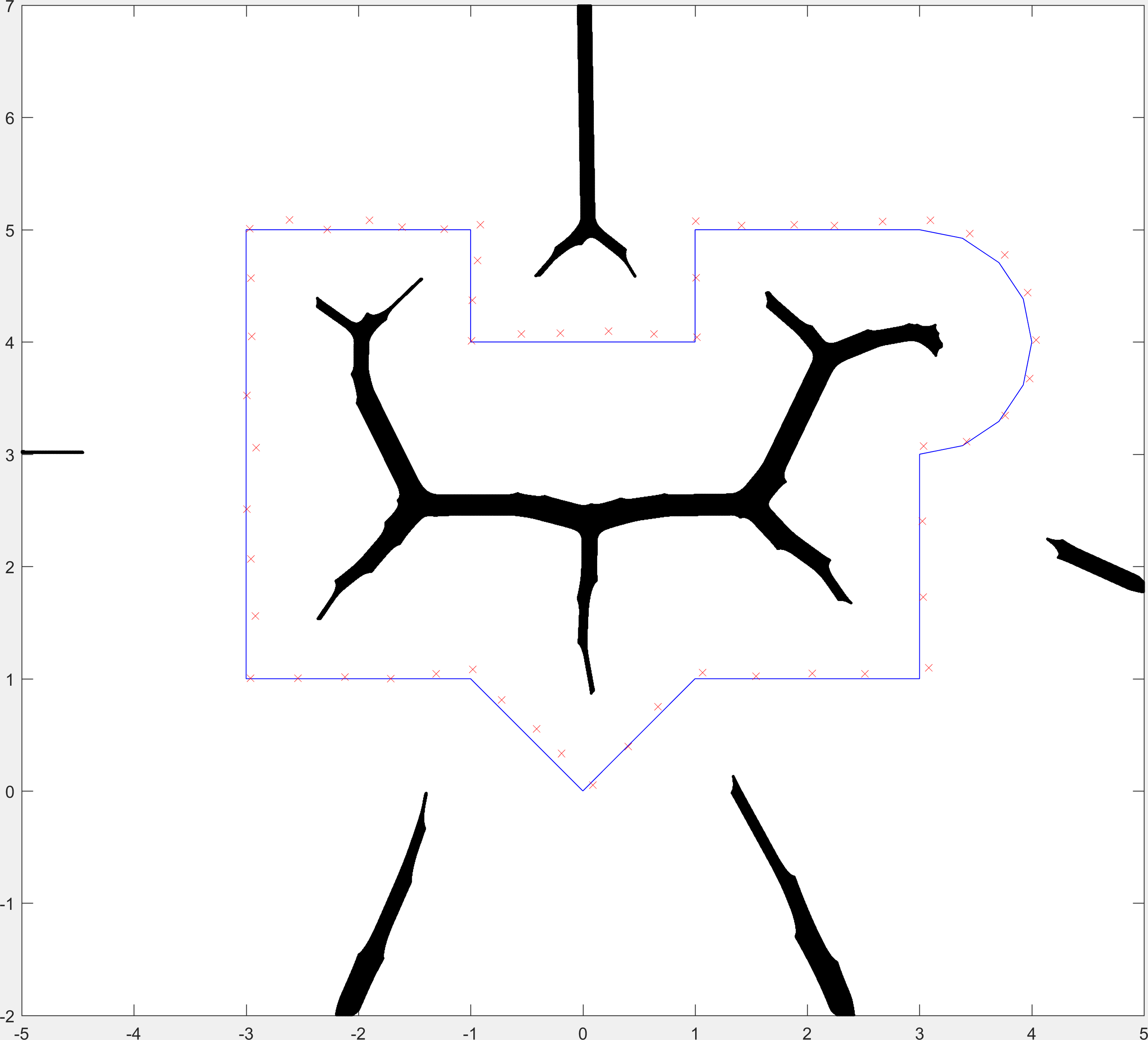}\\
		(d) & (e) & (f)\\
	\end{array}$
}
\caption{\label{Fig:Haus} 
	Multiscale Medial Axis Map of a nonconvex domain $\Omega$ 
	and of an $\epsilon$-sample $K_{\epsilon}$ of its boundary.
	$(a)$ Nonconvex domain $\Omega$;
	$(b)$ Graph of $M_{\lambda}(\cdot;\,\partial\Omega)$ for $\lambda=2.5$;
	$(c)$ Support of $M_{\lambda}(\cdot;\,\partial\Omega)$;
	$(d)$ Graph of $M_{\lambda}(\cdot;\,K_{\epsilon})$; 
	$(e)$ Support of $M_{\lambda}(\cdot;\,\Omega)$;
	$(f)$ Suplevel set of $M_{\lambda}(x;\,K_{\epsilon})$ for a threshold equal 
	to $\displaystyle 0.15\max_{x\in \R^2} \{M_{\lambda}(x;\,K_{\epsilon})\}$.
	}
\end{figure}

With the aim of giving insights into the implications of the Hausdorff stability of $M_{\lambda}(x;\,\partial\Omega)$, 
we display in Figure \ref{Fig:Haus} the graph of the
multiscale medial axis map of a non-convex domain $\Omega$ and of an $\epsilon$-sample $K_{\epsilon}$
of its boundary.
An inspection of the graph of $M_{\lambda}(x;\,\partial\Omega)$ and $M_{\lambda}(x;\,K_{\epsilon})$, 
displayed in Figure \ref{Fig:Haus}$(a)$ and Figure \ref{Fig:Haus}$(b)$, reveals
that both functions take comparable values along the main branches of $M_{\Omega}$. Also,
$M_{\lambda}(x;\,K_{\epsilon})$ takes small values along the secondary branches, generated by the 
sampling of the boundary of $\Omega$. These values can therefore be filtered out
by a simple thresholding so that a stable approximation of the medial axis of $\Omega$ can be
computed. This can be appreciated by looking at Figure \ref{Fig:Haus}$(d)$, which displays a suplevel
set of $M_{\lambda}(x;\,K_{\epsilon})$ that appears to be a reasonable approximation of the support of
$M_{\lambda}(x;\,\partial\Omega)$ shown in Figure \ref{Fig:Haus}$(c)$.

A relevant implication of \eqref{Eq:HAUSMMAM} concerns with the continuous approximation
of the medial axis of a shape starting from subsets of the Voronoi 
diagram of a sample of the shape boundary which is pertinent for 
shape reconstruction from point clouds. Let us consider an $\epsilon$-sample $K_\epsilon$ 
of $\partial\Omega$, that is, a discrete set of points such that 
$\dist_{\mathcal{H}}(\partial\Omega,K_\epsilon)\leq \epsilon $.
Since the medial axis of $K_{\epsilon}$ is the Voronoi diagram of $K_{\epsilon}$, 
if we denote by $V_{\epsilon}$ the set of all the vertices
of the Voronoi diagram $\mathcal{V}or(K_{\epsilon})$ of $K_{\epsilon}$,  and denote by  
$P_{\epsilon}$ the subset of $V_{\epsilon}$ formed by the `poles' of $\mathcal{V}or(K_{\epsilon})$ 
introduced in \cite{AB99}, (i.e. those
vertices of $\mathcal{V}or(K_{\epsilon})$ that converge to the medial axis of $\Omega$ as the sample 
density approaches infinity), 
then, for $\lambda>0$, it was established in \cite{ZCO15} that 
\begin{equation*}
	\lim_{\epsilon\to 0+}\,M_{\lambda}(x_{\epsilon};	\,K_{\epsilon})=0
	\quad\text{for }x_{\epsilon}\in V_{\epsilon}\setminus P_{\epsilon}\,.
\end{equation*}
Since as $\epsilon \to 0+$, $K_{\epsilon}\to \partial \Omega$,
and knowing that 
$P_{\epsilon}\to M_{\Omega}$ \cite{ACK01b,BC00}, 
then on the vertices of 
$\mathcal{V}or(K_{\epsilon})$ that do not tend to $M_{\Omega}$, $M_{\lambda}(x_{\epsilon};\,K_{\epsilon})$ 
must approach zero in the limit because of \eqref{Eq:LimMMAM}.
As a result, in the context of the methods of approximating the medial axis starting from the Voronoi diagram of a sample set
(such as those described in \cite{ACK01b,Dey06,DZ04,SP08}), the use of the multiscale medial axis map offers 
an alternative and much easier 
tool to construct continuous approximations to the medial axis with guaranteed convergence as $\epsilon \to 0+$.

We conclude this topic by showing how compensated convex transform 
is used to obtain a fine result of geometric measure theory. Let us introduce the set $V_{\lambda,K}$ defined as
\begin{equation}
	V_{\lambda,\,K}= \left\{x\in \mathbb{R}^n: \, \lambda\dist(x;\, M_K)\leq \dist(x;\,K)\right\}\,,	
\end{equation}
which represents a neighborhoof of $\overline{M}_K$. From the property of the tight approximation of the lower transform of 
the squared-distance function, it was shown in \cite{ZCO15}  that 
\begin{equation}
	\dist^2(\cdot;\,K)\in C^{1,1}(\mathbb{R}^n\setminus V_{\lambda,K})\,,
\end{equation} 
and a sharp estimate for the Lipschitz constant of $D\dist^2(\cdot,K)$ was also obtained. This result can be viewed 
as a weak Lusin type theorem for the squared-distance function which extends regularity results
of the squared-distance function to any closed non-empty subset of $\R^n$.

\subsection{Approximation Transform}
The theory of compensated convex transforms can also be applied to define Lipschitz continuous and 
smooth geometric approximations and interpolations for bounded real-valued functions 
sampled from either a compact set $K$ in $\mathbb{R}^n$ or the
complement of a bounded open set $\Omega$, i.e. $K=\mathbb{R}^n\setminus \Omega$.
The former is motivated by approximating or interpolating sparse data and/or contour lines whereas
the latter by the so-called inpainting problem in image processing \cite{CS05}, where some parts 
of the image content are missing. The aim of `inpainting' is to use other information from parts of the image to repair 
or reconstruct the missing parts. 

Let $f:\R^n\mapsto \R$ denote the underlying function to be approximated, $f_K:K\subset \R^n\to \mathbb{R}$ 
the sampled function defined by $f_K(x) = f(x)$ for $x \in K$,  and $\Gamma_{f_K}:=\{(x,f_K(x)),\, x\in K\}$ its graph, the setting
for the application of the compensated convex transforms to obtain an approximation transform is the following.
Given $M>0$, we define first two functions extending $f_K$ to $\R^n\setminus K$, namely
\begin{equation}\label{Eq.Def.ExtFnct}
\begin{array}{lll}
	\displaystyle f^{-M}_K(x) & \displaystyle = f(x)\chi_K(x)-M\chi_{\R^n\setminus K}
				& \displaystyle =
		\left\{\begin{array}{ll}
				f_K(x),	& x\in K,\\[1.5ex]
				-M,	& x\in\mathbb{R}^n\setminus K\,;
			\end{array}\right.\\[2.5ex]
	\displaystyle f^{M}_K(x) & \displaystyle = f(x)\chi_K(x)+M\chi_{\R^n\setminus K}
				& \displaystyle =
		\left\{\begin{array}{ll}
			f_K(x),	& x\in K,\\[1.5ex]
			M,	& x\in\mathbb{R}^n\setminus K\,,
		\end{array}\right.
\end{array}
\end{equation}
\noindent where  $\chi_G$ denotes the characteristic
function of a set $G$. We then compute the arithmetic average of the proximal hull of $f^{M}_K(x)$
and the upper proximal hull of $f^{-M}_K$ as follows,
\begin{equation}\label{Eq.Def.AvAprx}
	A^{M}_{\lambda}(f_K)(x)= \frac{1}{2}
		\left(C^l_\lambda(f^{M}_K)(x)+C^u_\lambda(f^{-M}_K)(x)\right),
		\quad x\in \mathbb{R}^n\,,
\end{equation}
which we refer to as the average compensated convex approximation transform of $f_K$ of scale $\lambda$ and level $M$  \cite{ZCO16a}.

In the case that $K\subset\mathbb{R}^n$ is a compact set and 
$f:\mathbb{R}^n\mapsto \mathbb{R}$ is bounded and uniformly continuous,
error estimates are available for $M\to \infty$ and for $x\in\co[K]$.
If for $x\in\co[K]\setminus K$ we denote by $r_c(x)$ the convex density radius 
as the smallest radius of a closed ball $\bar{B}(x;\,r_c(x))$ such that $x$
is in the convex hull of $K\cap \bar{B}(x;\,r_c(x))$, then
for $\lambda>0$ and all $x\in \co[K]$ there holds
\begin{equation}\label{Eq:ErrEstInftyKcmpct}
	|A^{\infty}_\lambda(f_{K})(x)-f(x)|\leq
	\omega\left(r_c(x)+\frac{a}{\lambda}+\sqrt{\frac{2b}{\lambda}}\right)\,,
\end{equation}
where $\omega=\omega(t)$ is the least concave majorant of the modulus of continuity 
$\omega_f$ of $f$ and $a\geq 0$, $b\geq 0$ are such that $\omega(t)\leq at+b$ for $t\geq 0$.
Error estimates are also available for a finite $M>0$ under the extra restriction that 
$f(x)=c_0$ for $|x|\geq r$ where $c_0\in\mathbb{R}$ and $r>0$ are constants. In this case,
for $R>r$, we extend $f_K$ to be equal to $c_0$ outside
a large ball $B(0;\,R)$ containing $K$ and define $K_R=K\cup B^c(0;\,R)$. Thus we obtain
similar error estimate to \eqref{Eq:ErrEstInftyKcmpct} for $A^M_\lambda(f_{K_R})(x)$.
Furthermore, we have that when $M>0$ is sufficiently large, $A^M_\lambda(f_K)$ approaches $f_K$ in $K$ 
as $\lambda \to \infty$, whereas if $f$ is a $C^{1,1}$ function and $\lambda>0$ is large enough, 
$A^M_\lambda(f_K)$ is an interpolation of $f$ in the convex hull $\co[K]$ of $K$.
In the special case of a finite set $K$, the average approximation
$A_{\lambda}^M(f_K)$ defines an approximation  for the scattered data
$\Gamma_{f_K}=\{(x,f_K(x)),\; x\in K\}$.

If the closed set $K$ is the complement of a non-empty bounded open 
set $\Omega\subset \mathbb{R}^n$, we can also obtain 
estimates that are similar to \eqref{Eq:ErrEstInftyKcmpct}. Clearly,
$\co[K]=\R^n$ for such a $K$, thus if $f:\mathbb{R}^n\mapsto \mathbb{R}$ is bounded and 
uniformly continuous, satisfying $|f(x)|\leq A_0$ for
some constant $A_0>0$ and for all $x\in \R^n$ and $d_\Omega$ denotes the diameter of $\Omega$,
then for $\lambda>0$,  $M>A_0+\lambda d_\Omega^2$ and  all $x\in \mathbb{R}^n$, we have
\begin{equation}\label{Eq:EstCompSet}
	|A^M_\lambda(f_{K})(x)-f(x)|\leq \omega\left(r_c(x)+\frac{a}{\lambda}+
	\sqrt{\frac{2b}{\lambda}}\right)\,,
\end{equation}
where, as for \eqref{Eq:ErrEstInftyKcmpct}, the constants $a\geq 0$ and $b\geq 0$ are such that 
$\omega(t)\leq at+b$ for $t\geq 0$ with $\omega=\omega(t)$ the least concave majorant of the modulus
of continuity $\omega_f$ of $f$.

Both the estimates \eqref{Eq:ErrEstInftyKcmpct} and \eqref{Eq:EstCompSet} can be improved 
for Lipschitz functions and for $C^{1,1}$ functions.

Another natural and practical question in data approximation and interpolation is  
the stability of a given method. For approximations and interpolations of sampled
functions, we would like to know, for two sample sets which are `close' to each other,
say, under the Hausdorff distance \cite{AT04}, whether the corresponding approximations are close
to each other. It is easy to see that differentiation and integration based approximation
methods are not Hausdorff stable because continuous functions can be sampled over a finite dense set.
One of the advantages of the compensated convex approximation is that for a bounded 
uniformly continuous function $f$,
and for fixed $M>0$ and $\lambda>0$, the mapping $K\mapsto A^M_\lambda(f_K)(x)$ is continuous
with respect to the Hausdorff distance
for compact sets $K$, and the continuity is uniform with respect to $x\in \mathbb{R}^n$.
This means that if another sampled subset $E\subset \mathbb{R}^n$ (finite or compact) is close
to $K$, then the output  $A^M_\lambda(f_E)(x)$ is close to $A^M_\lambda(f_K)(x)$ uniformly with
respect to $x\in \mathbb{R}^n$.
As far as we know, not many known interpolation/approximation methods  share
such a property.

By using the mixed compensated convex transforms \cite{Zha08a}, it is possible to define a mixed 
average compensated convex approximation with scales $\lambda>0$
and $\tau>0$ for the sampled function $f_K:K \to \R$ by
\begin{equation}
	(SA)^{M}_{\tau,\lambda}(f_K)(x)=
	\frac{1}{2}( C^u_\tau(C^l_\lambda(f_K^M))(x)+C^l_\tau(C^u_\lambda(f_K^{-M}))(x)\,,
	\quad x\in \mathbb{R}^n\,.
\end{equation}
Since the mixed compensated convex transforms are $C^{1,1}$ functions \cite[Theorem 2.1(iv) and Theorem 4.1(ii)]{Zha08a},
the mixed average approximation $(SA)^{M}_{\tau,\lambda}$ is a smooth version of our average approximation.
Also, for a bounded function $f:\mathbb{R}^n\mapsto\mathbb{R}$,
satisfying $|f(x)|\leq M$, $x\in\mathbb{R}^n$ for some constant $M>0$,
we have the following estimates \cite[Theorem 3.13]{ZOC15a}
\begin{equation*}
	0\leq C^u_\tau(C^l_\lambda(f))(x)-C^l_\lambda(f)(x)\leq \frac{16M\lambda}{\tau},\quad
	0\leq C^u_\lambda(f)(x)-C^l_\tau(C^u_\lambda(f))(x)\leq \frac{16M\lambda}{\tau}
\end{equation*}
for all $x\in\mathbb{R}^n$, $\lambda>0$ and $\tau>0$, and hence can easily show that for any closed set $K\subset \R^n$, 
\begin{equation*}
	|(SA)^M_{\tau,\lambda}(f_K)(x)-A_\lambda^M(f_K)(x)|\leq \frac{16M\lambda}{\tau},
	\quad  x\in\R^n\,.
\end{equation*}
This  implies that for given $\lambda>0$ and $M>0$,  the mixed approximation $(SA)^M_{\tau,\lambda}(f_K)$
converges to the basic average approximation $A_\lambda^M(f_K)$ uniformly in $\mathbb{R}^n$ as $\tau\to\infty$,
with rate of convergence $16M \lambda/\tau$.

\setcounter{equation}{0}
\section{Numerical Algorithms}\label{Sec:NumAlg}
\begin{figure}[H]
\centerline{
	$\includegraphics[width=0.48\textwidth]{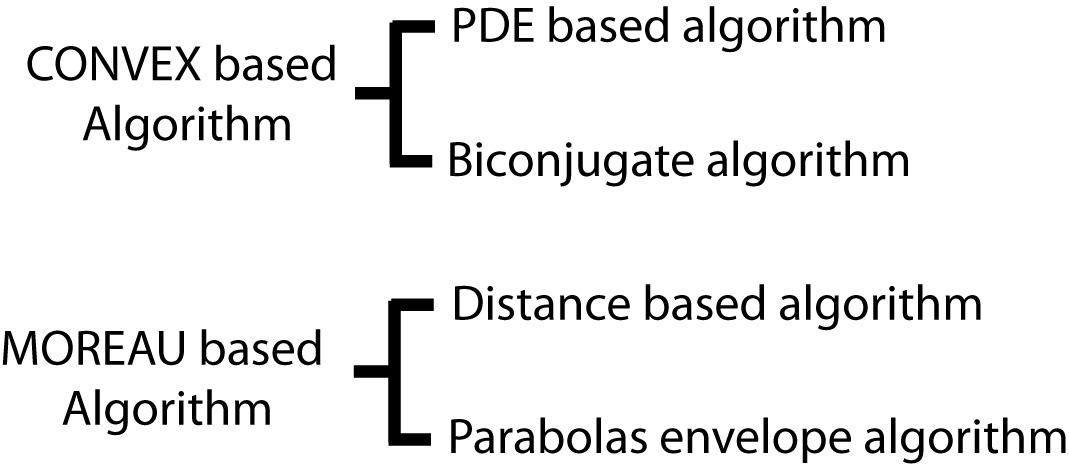}$
} \caption{\label{Ex01.NumSch}
Different approaches for computing the Lower Compensated Convex transform $C_{\lambda}^{l}(f)$.
}
\end{figure}

The numerical realisation of the convex transforms introduced in Section \ref{Sec:Filters} relies on the availability
of numerical schemes for computing the upper and lower transforms of a given function. Because of the relation 
\eqref{Eq:UpTr} between the upper and lower transform, the computation of the above transforms  
ultimately boils down to the evaluation of the lower compensated convex transform. 
As a result, without loss of generality, in the following we refer just to the actual implementation
of $C_{\lambda}^l(f)$. With this respect, we can proceed in two different ways according to whether 
we use definition \eqref{Eq:LwTr}
in terms of the convex envelope or the characterization \eqref{Eq:CmpTrMo} as proximity hull of the function and use 
its definition in terms of the Moreau envelopes. In the following, we describe some algorithms that can be 
used successfully for the computation of $C_{\lambda}^l(f)$ and discuss their relative merits. 
Figure \ref{Ex01.NumSch} summarizes the different approaches considered in this paper.

\subsection{Convex based algorithms}\label{Sec:NumAlg:CB}
Algorithms to compute convex hull such as the ones given in \cite{Cha96,BDH96} are more suitable 
for discrete set of points and their complexity is related to the cardinality of the set. 
An adaptation of these methods to our case, with the set to convexify given by the epigraph 
of $f+\lambda|\cdot|^2$, does not appear to be very effective, especially for functions 
defined in subsets of $\mathbb{R}^n$ for $n\geq 2$, compared to the methods that (directly) 
compute the convex envelope of a function \cite{Ves99,BC94,Obe08,CEV15}. 

Of particular interest for applications to image processing, where functions involved are defined on
grid of pixels, is the characterization of the convex envelope
as the viscosity solution of a nonlinear obstacle problem \cite{Obe08}. 
An approximated solution is then obtained by using centered finite differences 
along directions defined by an associated stencil to approximate the first eigenvalue of 
the Hessian matrix at the grid point. A generalization of the scheme introduced in \cite{Obe08} 
in terms of the number of convex combinations of the function values at the grid points 
of the stencil, is briefly summarized in Algorithm \ref{Algo:CnvxEnvOBE} and described below.
Given a uniform grid of points $x_k\in\mathbb{R}^n$, equally spaced with grid size $h$,  denote by
$S_{x_k}$ the $d-$point stencil of $\mathbb{R}^n$ with center at $x_k$ defined as 
$S_{x_k}=\{x_k+hr, |r|_{\infty}\leq 1, r\in\mathbb{Z}^n\}$ with $|\cdot|_{\infty}$ the $\ell^{\infty}$-norm of 
$r\in\mathbb{Z}^n$ and $d=\#(S)$, cardinality of the finite set $S$. At each grid point $x_k$ we compute an 
approximation of the 
convex envelope of $f$ at $x_k$ by an iterative scheme where each iteration step $m$ is given by
\[
	(\co f)_m(x_k)=\min\Big\{ 
		f(x_k),\,\sum\lambda_i (\co f)_{m-1}(x_i):\,\,\,
		\sum \lambda_i=1,\,\lambda_i\geq 0,\,x_i\in S_{x_k}\Big\}
\]
with the minimum taken between $f(x_k)$ and only some convex combinations of $(\co f)_{m-1}$ 
at the stencil grid points $x_i$ of $S_{x_k}$. It is then not difficult to show that the scheme 
is monotone, thus convergent. However, there is no estimate of the rate of convergence which, 
in actual applications, appears  to be quite slow. 
Furthermore, results are biased by the type of underlying stencil.

\begin{algorithm}[H]
\begin{algorithmic}[1]
	\STATE{Set $\displaystyle m=1,\,(\co f)_{0}=f,\,\,tol$}
	\STATE{$\displaystyle \epsilon=\|f\|_{L^2}$}
	\WHILE{$\displaystyle \epsilon>tol$}
	\STATE{$\displaystyle \forall x_k, \quad (\co f)_{m}(x_k)=\min\Big\{ 
		f(x_k),\,\sum\lambda_i (\co f)_{m-1}(x_i):\,\,\,
		\sum \lambda_i=1,\,\lambda_i\geq 0,\,x_i\in S_{x_k}\Big\}$
		}
	\STATE{$\displaystyle \epsilon=\|(\co f)_m-(\co f)_{m-1}\|_{L^2}$}
	\STATE{$m\leftarrow m+1$}
	\ENDWHILE
\end{algorithmic}
\caption{\label{Algo:CnvxEnvOBE} Computation of the convex envelope of $f$ according to \cite{Obe08} 
}
\end{algorithm}

Based on the characterization of the convex envelope of $f$ in terms of the biconjugate $(f^{\ast})^{\ast}$ of $f$ \cite{BC17,HL01,Roc70},
where $f^{\ast}$ is the Legendre-Fenchel transform of $f$,  
we can approximate the convex envelope by computing twice the discrete Legendre-Fenchel transform.
We can thus improve speed efficiency with respect to a brute force algorithm, which computes $(f^{\ast})^{\ast}$ with 
complexity $O(N^2)$ with $N$ the number of grid points, if we have an efficient scheme 
to compute the discrete Legendre-Fenchel transform of a function.  
For functions $f: X\to\mathbb{R}$ defined on cartesian sets of the type $X=\prod_{i=1}^nX_i$ with 
$X_i$ intervals of $\mathbb{R}$, $i=1,\ldots,n$, the Legendre-Fenchel transform of $f$ can be reduced to the
iterate evaluation of the Legendre-Fenchel transform of functions dependent only on one varable as follows
\begin{equation}\label{Eq:LFDcmp}
\begin{split}
	(\xi_1,\xi_2,\ldots,\xi_n)&\in\mathbb{R}^n\to 
			f^{\ast}(\xi_1,\ldots,\xi_n)=\sup_{x\in X}\,\{
			\xi\cdot x-f(x)
				\}\\[1.5ex]
			&=\sup_{x_1,\ldots,x_{n-1}\in \prod_{i=1}^{n-1}X_i}\,\Big\{
		x_1\xi_1+\ldots+x_{n-1}\xi_{n-1}-\sup_{x_n\in X_n}\{x_n\xi_n-f(x_1,\ldots,x_{n-1},x_n)
	\}\Big\}\,.
\end{split}
\end{equation}
As a result, one can improve the complexity of the computation of $f^{\ast}$ if one has an efficient scheme
to compute the Legendre-Fenchel transform of functions of only one variable. 
For instance,
the algorithm described in \cite{Luc97,HM11}, which exploits an idea of \cite{Bre89} and improves the implementation 
of \cite{Cor96}, computes the discrete Legendre-Fenchel transform in linear time, that is, with complexity 
$O(N)$. If $g_h$ denote the grid values of a function of one variable, the key idea of \cite{Bre89,Cor96} 
is to compute $(g_h)^{\ast}$ as approximation of $g^{\ast}$ using the following result
\begin{equation}
	(g_h)^{\ast}(\xi)=\left(\co[\Pi f_h]\right)^{\ast}(\xi)\,,\quad\quad\xi\in\mathbb{R}
\end{equation}
where $\Pi g_h$ denotes the continuous piecewise affine interpolation of the grid values $g_h$.
Therefore by applying an algorithm with linear complexity, for instance the beneath-beyond algorithm 
\cite{PS85}, to compute the convex envelope $\co[\Pi g_h]$, followed by the use of analytical expressions for the Legendre-Fenchel transform of a convex piecewise affine function
yields an efficient method to compute $(g_h)^{\ast}$ \cite{Luc97}. For functions defined in a bounded domain, In \cite{Luc97} it was
recommended to increment the size of the 
domain for a better precision of the computation of the Legendre-Fenchel transform. 
The work \cite{HM11} avoids this by elaborating the exact expression of the 
Legendre-Fenchel transform of a convex piecewise affine function defined in a bounded domain which is equal to 
infinity in $\R\setminus X$ or it has an affine variation. In this manner, they can avoid bounday effects.
For ease of reference, we report next the analytical expression of $g^{\ast}$ in the case where $g:\R\to\overline\R$ is
convex piecewise affine. Without loss of generality, let $x_1<\ldots<x_N$ be a grid of points of $\R$,
$c_1<\ldots<c_N$ and assume $g:\R\to\overline\R$ to be defined as follows:
\begin{equation}
	g:x\in\R\to\left\{\begin{array}{ll}
			\displaystyle +\infty		&\displaystyle \text{if }x\leq x_1\\[1.5ex]
			\displaystyle g_i+c_i(x_i-x)	&\displaystyle \text{if }x_i\leq x\leq x_{i+1},\quad i=1,\ldots,N-1\\[1.5ex]
			\displaystyle g_N+c_N(x_N-x)		&\displaystyle \text{if }x\geq x_N
			\end{array}\right.
\end{equation}
where $g_i=g(x_i)$ and $c_i$, for $i=1,\ldots,N$, represent the slopes of each affine piece of $g$. It is  not
difficult to verify that the analytical expression of $g^{\ast}$ is given by \cite{HM11}
\begin{equation}
	g^{\ast}:\xi\in\R\to\left\{\begin{array}{ll}
			\displaystyle x_1\xi-g_1	 &\displaystyle \text{if }\xi\leq c_1\\[1.5ex]
			\displaystyle x_{i+1}\xi-g_{i+1} &\displaystyle \text{if }c_i\leq \xi\leq c_{i+1},\quad i=1,\ldots,N-2\\[1.5ex]
			\displaystyle +\infty		 &\displaystyle \text{if }\xi\geq c_N\,.
			\end{array}\right.
\end{equation}
Once we know $g^{\ast}$, using the decomposition \eqref{Eq:LFDcmp}, we can compute $f^{\ast}$ and thus the biconjugate $f^{\ast\ast}$.

\subsection{A Moreau envelope based algorithm}\label{Sec:NumAlg:MB}
The computation of the Moreau envelope is an established task in the field of computational convex analysis
\cite{Luc10} that has been tackled by various different approaches aimed at reducing the quadratic complexity 
of a direct brute force implementation of the transform. 
Such reduction is achieved, one way or another,  by a dimensional reduction. 
The fundamental idea of the scheme presented in \cite{ZOC20}, for instance, is the generalization of the 
Euclidean distance transform of binary images, by replacing the binary image by an arbitrary function on a grid.
The decomposition of the structuring element which yields the exact Euclidean distance transform \cite{SM92}
into basic ones, leads to a simple and fast algorithm  where the discrete lower Moreau envelope can be 
computed by a sequence of local operations, using one-dimensional neighborhoods. 
Unless otherwise stated, in the following, $i,\,j,\,k,\, r,\,s,\, p,\, q\in \mathbb{Z}$ denote 
integers whereas $m,\, n\in\mathbb{N}$ are non-negative integers. 
Given $n\geq 1$, we introduce grid of points of the space $\mathbb{R}^n$ with regular spacing $h>0$
denoted by $x_k\in\R^n$, $k\in\mathbb{Z}$ and define the discrete lower Moreau envelope  
at $x_k\in\mathbb{R}^n$ as
\begin{equation}\label{Sec4.Def.DiscrMor} 
	M_\lambda^h(f)(x_k)=
	\inf\{ f(x_k+rh)+ \lambda h^2|r|^2,\; r\in \mathbb{Z}^n\}\,.
\end{equation}
By taking the infimum in \eqref{Sec4.Def.DiscrMor}
over a finite number $m\geq 1$ of directions, we obtain the $m-$th approximation of the 
discrete Moreau lower envelope $M_\lambda^h(f)(x_k)$ which can be evaluated by 
taking the values  $f_m(x_k)$ given by Algorithm \ref{Algo:MoreauEnv}.
For the convergence analysis and converegence rate we refer to \cite{ZOC20} where it is shown that the
scheme has a linear convergence rate with respect to  $h$.

\begin{algorithm}[H]
\begin{algorithmic}[1]
	\STATE{Set $\displaystyle i=1,\, m\in\mathbb{N}$}
	\STATE{$\displaystyle \forall x_k,\, f_0(x_k)=f(x_k)$}
	\WHILE{$\displaystyle i<m$}
	\STATE{$\displaystyle \tau_i=2i-1$
		}
	\STATE{$\displaystyle f_{i}(x_k)=\min\{ f_{i-1}(x_k+rh)+\lambda h^2|r|^2\tau_{i}: \,
		r\in\mathbb{Z}^n,\,|r|_{\infty}\leq 1\}$}
	\STATE{$i\leftarrow i+1$}
	\ENDWHILE
\end{algorithmic}
\caption{\label{Algo:MoreauEnv} Computation of $f_m(x_k)$ at the points $x_k$
of the grid of $\R^n$ of size $h$ for given $m\geq 1$.}
\end{algorithm}

Likewise the computation of the Legendre-Fenchel transform, in the scheme proposed by \cite{FH12},
the authors apply the dimensional reduction directly to the computation of the Moreau envelope which 
is factored by $n$
one dimensional Moreau envelope. For instance, in the case of $n=2$,
let $\Omega=X\times Y$, with $X,Y\subset\mathbb{R}$, 
and $(\xi_1,\xi_2)\in \Omega=X\times Y$, for any $x=(x_1,x_2)\in\mathbb{R}^2$, we have
\begin{equation}
\begin{split}
	M_{\lambda}(f)(x_1,x_2)&=\inf_{(\xi_1,\xi_2)\in\Omega}\,\{\lambda|(x_1,x_2)-(\xi_1,\xi_2)|^2+f(\xi_1,\xi_2)\}\\[1.5ex]
			       &=\inf_{\xi_1\in X}\,\Big\{\lambda|x_1-\xi_1|^2+
					\inf_{\xi_2\in Y}\,\{\lambda|x_2-\xi_2|^2+  
				       f(\xi_1,\xi_2)\}\Big\}\,.
\end{split}
\end{equation}
For the computation of $M_{\lambda}(f)$ with $f$ function of one variable, if we  
denote by $\mathcal{F}$ the family of parabolas with given 
curvature $\lambda$ of the following type
\[
	\mathbcal{p}_{q}:\,x\in\mathbb{R}\,\to\, \mathbcal{p}_{q}(x)=\lambda|x-q|^2+f(q)\,,
\]
parameterized by $q\in\Omega\subset\mathbb{R}$, we have that 
\begin{equation}\label{Eq:MorEnvParab}
	M_{\lambda}(f)(x)=\inf\,\{\mathbcal{p}_{q}(x):\,\,\mathbcal{p}_{q}\in\mathcal{F}\}\,,
\end{equation}
that is, the Moreau envelope of a function of one variable is reduced to the computation of the 
envelope of parabolas of given curvature $\lambda$. 
The computation of such envelope is realised by \cite{FH12} 
in two steps. In the first one, they compute the envelope 
by adding the parabolas one at time which is done in linear time, 
and comparing each parabola to the parabolas that realise the envelope, which is done in constant time,  
whereas in the second step they compute the value of the envelope at the given point $x\in\mathbb{R}$.
The key points of the scheme result from two observations. The first one is that 
given any two parabolas of $\mathcal{F}$ parameterized by $q,\,r\in\Omega$, their interesection 
occurs ony at one point with coordinate
\[
	x_s=\frac{(f(q)-f(r))+\lambda(q^2-r^2)}{2\lambda(q-r)}\,,
\]
whereas the second one regards the relation between the parabolas so that if $q<r$, then $\mathbcal{p}_{q}(x)\leq \mathbcal{p}_{r}(x)$ 
for $x<x_s$ and $\mathbcal{p}_{q}(x)\geq \mathbcal{p}_{r}(x)$  for $x>x_s$. 
This scheme allows the evaluation of 
$M_{\lambda}(f)(x)$ for any $x\in\mathbb{R}^n$ even if $f$ is defined only on a bounded open set $\Omega$, 
without any consideration on how to extend $f$ on $\mathbb{R}^n\setminus\Omega$. 
We will refer next to this scheme as the parabola envelope scheme.

By using the link between the Moreau envelope and the Legendre-Fenchel transform given by \cite{RW98,Luc06} 
\begin{equation}\label{Eq:MorLF}
	M_{\lambda}(f)(x)=\lambda|x|^2-2\lambda\left(\frac{f}{2\lambda}+\frac{|\cdot|^2}{2}\right)^{\ast}(x)\,,
\end{equation}
it is possible to design another scheme to calculate the Moreau envelope by computing 
the Legendre-Fenchel transform of the augmented function that appears in \eqref{Eq:MorLF} \cite{Luc06}. 
In this case, however, special considerations must be taken about the primary domain, 
where the Moreau envelope is defined, and the dual domain, which is the one where the 
Legendre-Fenchel transform is defined.

\setcounter{equation}{0}
\section{Numerical Examples}\label{Sec:NumExmpl}
In this section we present some illustrative numerical examples of implementation of the transforms 
introduced in Section \ref{Sec:Filters}. We preceed this discussion by the computation
of a two-dimensional prototype example with analytical expression of $C_{\lambda}^u(\chi_K)$ which 
we use to select the most suitable numerical scheme out of 
those described in Section \ref{Sec:NumAlg} for the computation of the compensated convex transforms.
\subsection{Prototype Example: Upper transform of a singleton set of $\R^2$}
Given the singleton set $K=\{0\}\subset\R^2$, the 
analytical expression of $C_{\lambda}^u(\chi_K)$ established in \cite[Example 1.2]{ZOC15b} is given by
\begin{equation}
	C_{\lambda}^u(\chi_K)(x)=\left\{\begin{array}{ll}
		\displaystyle	0\,,				  &\displaystyle\text{if }|x|> 1/\sqrt{\lambda}\,,	\\[1.5ex]
		\displaystyle	\lambda(1/\sqrt{\lambda}-|x|)^2\,,&\displaystyle\text{if }|x|\leq 1/\sqrt{\lambda}\,.	
	\end{array}\right.
\end{equation}
We compute then $C_{\lambda}^u(\chi_K)$ by applying the convex based algorithms, i.e. 
Algorithm \ref{Algo:CnvxEnvOBE} \cite{Obe08} and the biconjugate based scheme 
(shorted as $BS$ hereafter) \cite{Luc97,HM11}, 
and  the Moreau based algorithms, i.e. Algorithm \ref{Algo:MoreauEnv} and the parabola envelope
scheme (shorted as $PES$ hereafter)\cite{FH12}. To compare the accuracy of the schemes,
we will consider: $(i)$ the Hausdorff distance between the support of the exact and the computed upper 
transform, 
\[
	e_{\mathcal{H}}=\dist_{\mathcal{H}}\,\left(\overline B(0; 1/\sqrt{\lambda}),\,
		\mathsf{sprt}\left(C_{\lambda}^{u,h}(\chi_K)\right)\right)
\]
with $C_{\lambda}^{u,h}(\chi_K)$ the computed upper compensated transform; $(ii)$ the relative 
$L^{\infty}$~error norm given by
\[
	e_{L^{\infty}}=\frac{\max_{x\in\R^2}|C_{\lambda}^{u,h}(\chi_K)(x)-C_{\lambda}^{u}(\chi_K)(x)|}
				{\max_{x\in\R^2}|C_{\lambda}^{u}(\chi_K)(x)|}
\]
and $(iii)$ the execution time $t_c$ in seconds by a PC with processor Intel\textregistered\, 
Core\texttrademark\, i7-4510U CPU@2.00 GHz and 8GB of memory RAM. 
\begin{figure}[H]
\begin{center}
	$\begin{array}{cccc}
		\includegraphics[width=0.22\textwidth]{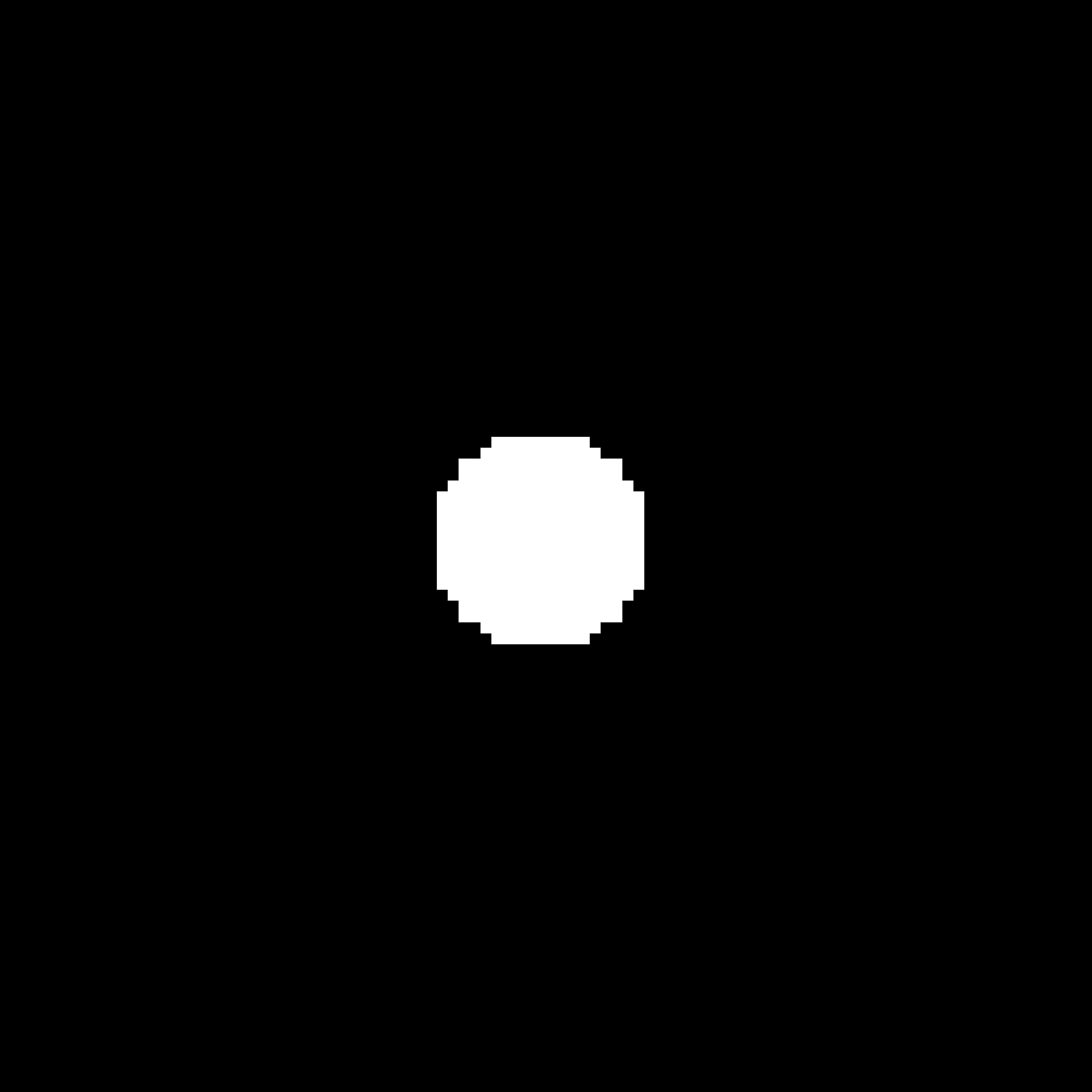}&
		\includegraphics[width=0.22\textwidth]{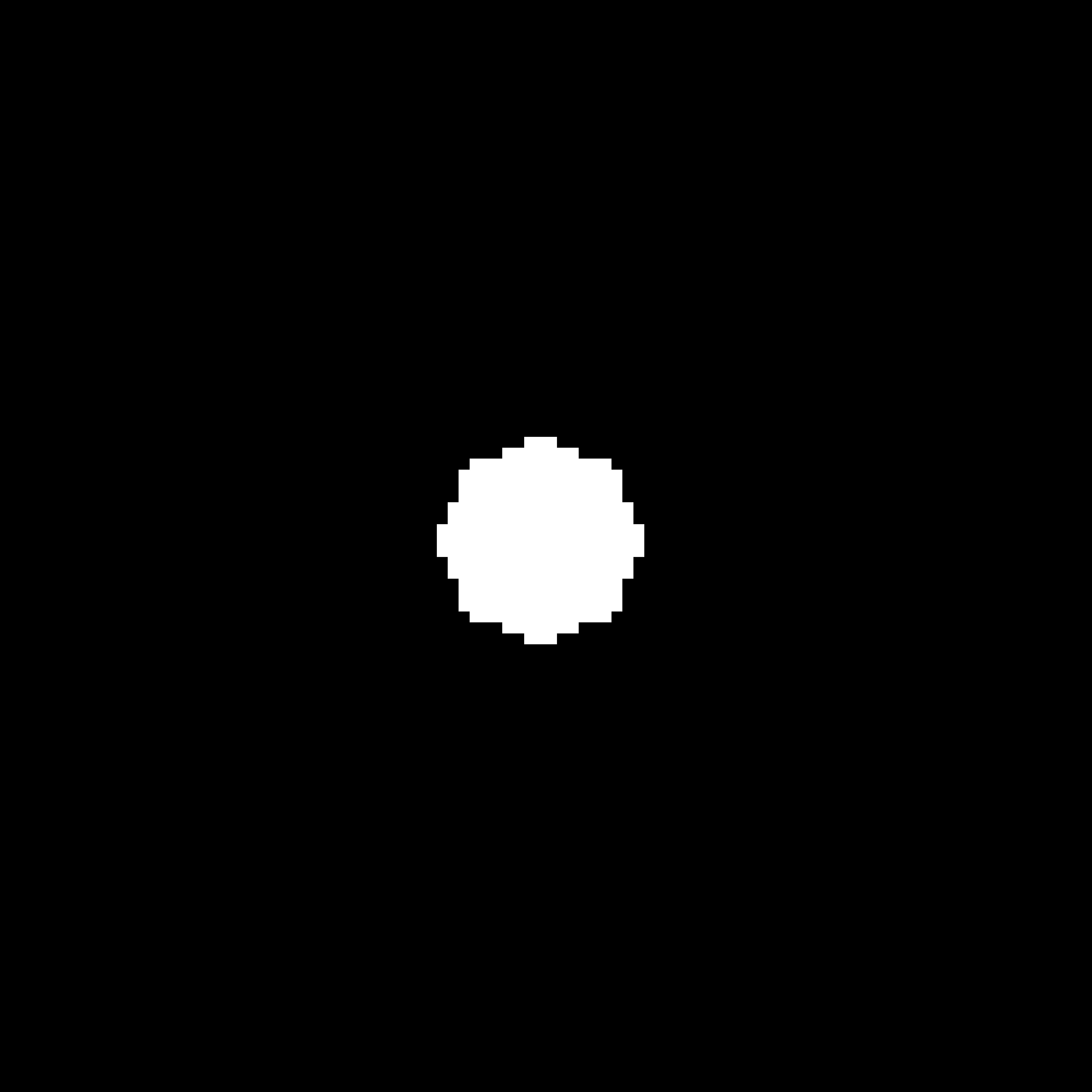}&
		\includegraphics[width=0.22\textwidth]{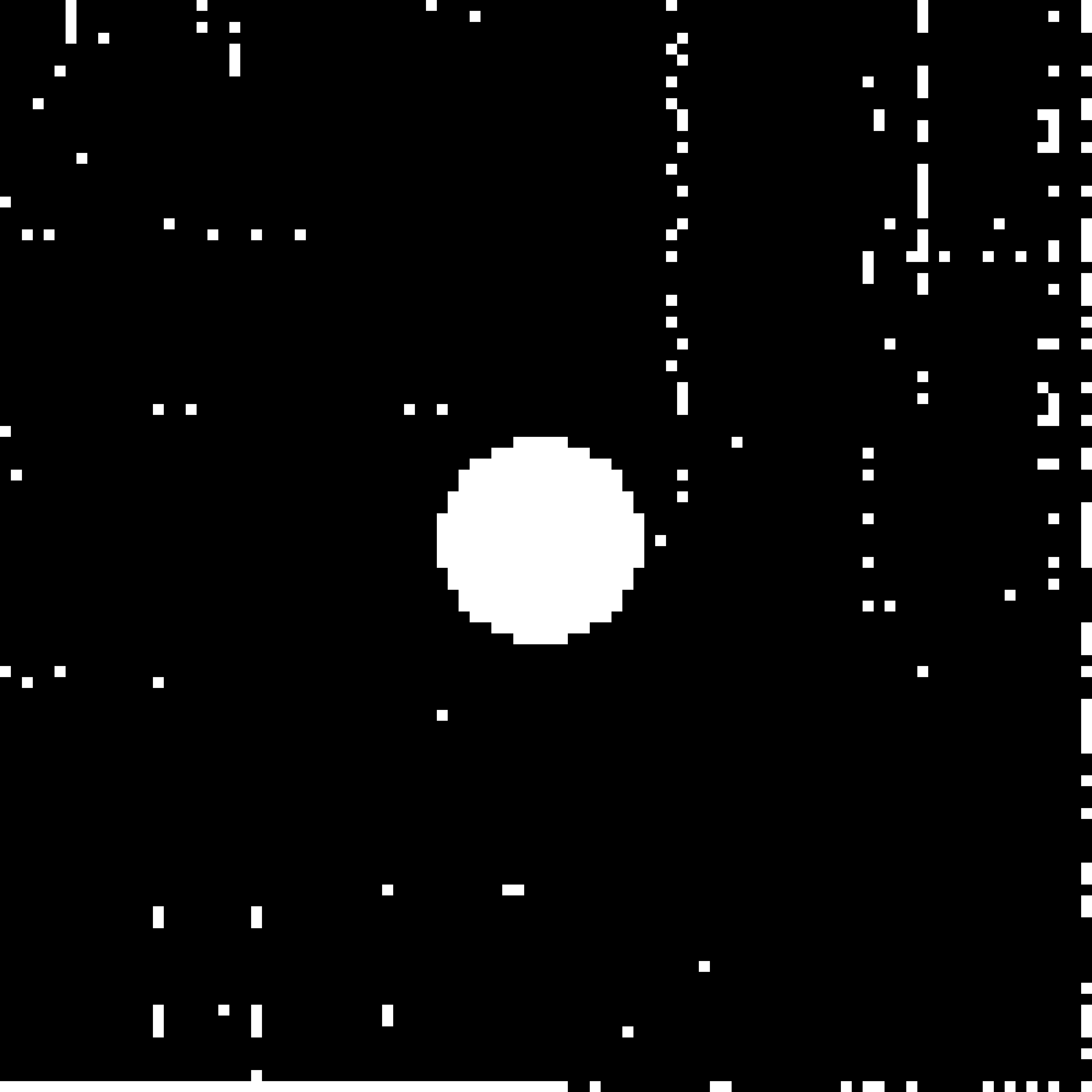}&
		\includegraphics[width=0.22\textwidth]{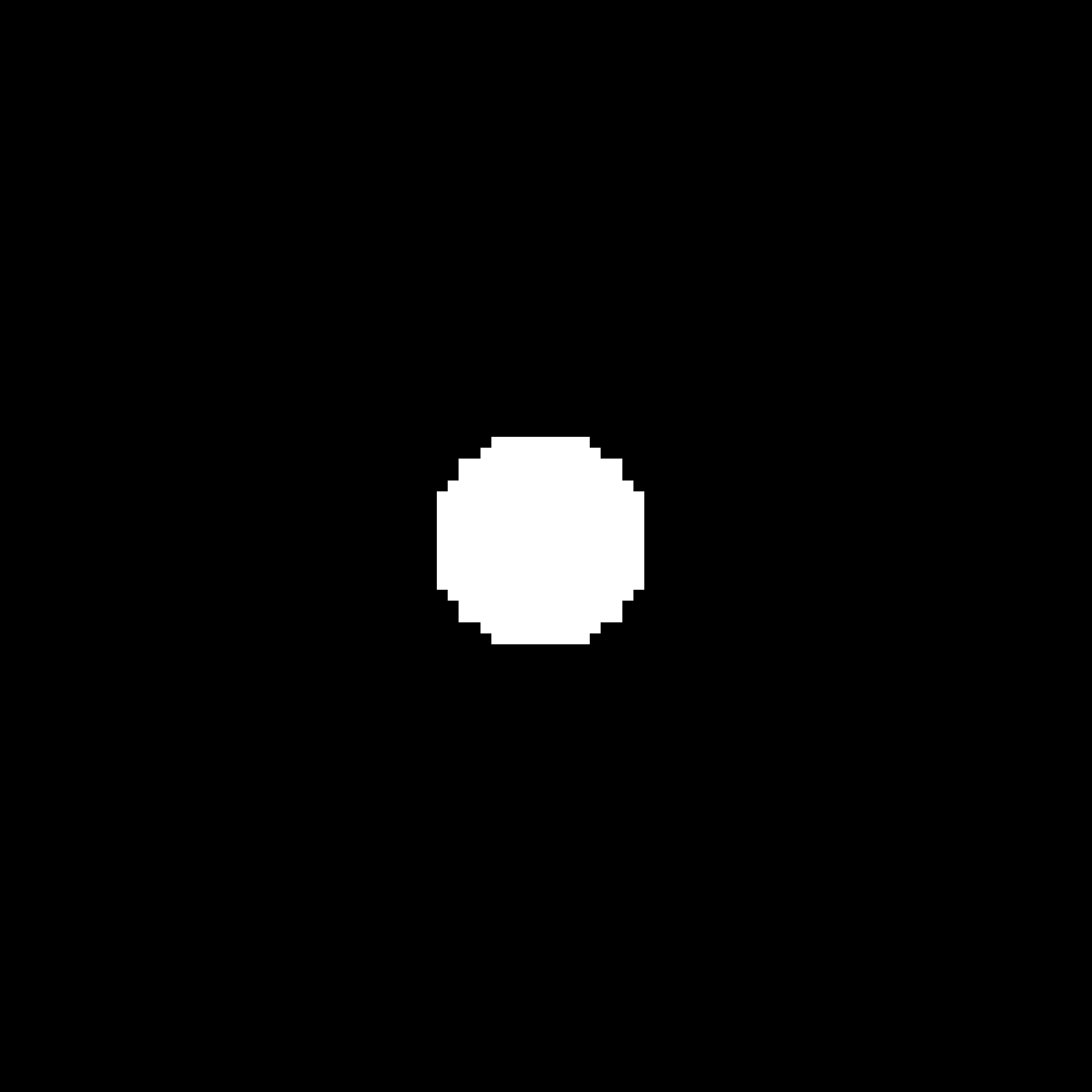}\\[1.5ex]
		(a)&(b)&(c)&(d)
	\end{array}$
\end{center}
\caption{\label{Ex:SingletonUpTr} 
Supports of the exact and computed upper compensated transform of the characteristic function of a
singleton set of $\R^2$ by the different numerical schemes. 
$(a)$ Exact support given by $\overline{B}(0; 1/\sqrt{\lambda})$ for $\lambda=0.01$; 
$(b)$ Support of $C_{\lambda}^{u,h}(\chi_K)$ computed by Algorithm \ref{Algo:CnvxEnvOBE} \cite{Obe08};
$(c)$ Support of $C_{\lambda}^{u,h}(\chi_K)$ computed by the biconjugate based scheme \cite{Luc97,HM11} for
$h_d=0.001$;
$(d)$ Support of $C_{\lambda}^{u,h}(\chi_K)$ computed by Algorithm \ref{Algo:MoreauEnv} \cite{ZOC20} 
which coincides with the one computed using the parabola envelope scheme \cite{FH12}.
}
\end{figure}
Figure \ref{Ex:SingletonUpTr} displays the support of $C_{\lambda}^{u}(\chi_K)$ given by 
$\overline{B}(0; 1/\sqrt{\lambda})$ and of $C_{\lambda}^{u,h}(\chi_K)$ computed by the numerical schemes mentioned above. 
Algorithm \ref{Algo:MoreauEnv} and the parabola envelope algorithm yield the same results, 
thus Figure \ref{Ex:SingletonUpTr} displays the support as computed by only one of the two schemes. 
In this case we observe that the support coincides with the exact one. This does not happen 
for the support computed by the other two schemes. The application of Algorithm \ref{Algo:CnvxEnvOBE}
evidences the bias of the scheme with the underlying stencil whereas by applying the biconjugate based scheme
we note some small error all over the domain. The spread of this error depends on the dual mesh grid size 
$h_d$. Table \ref{Table:UpTrSing} reports the values of $t_c$, $e_{L^{\infty}}$ and 
$d_{\mathcal{H}}$ for the different schemes. For the biconjugate based scheme, we have different results 
according to the parameter $h_d$ that controls the uniform discretization of the dual mesh.
The value $h_d=1$ means that we are considering the same grid size as the grid of the input function $\chi_K$
whereas lower values for $h_d$ means that we are computing on a finer dual mesh compared to the primal one. 
The results given in Table \ref{Table:UpTrSing} show that in terms of the values of $C_{\lambda}^u(\chi_K)$
the biconjugate based scheme is the one that produces the best results (compare the values of $e_{L^{\infty}}$),
but this occurs at the fraction of cost of reducing $h_d$ which means to increase the number of the dual 
grid nodes and consequently the computational time. The issue of the choice of the dual grid on the accuracy 
of the computation of the convex envelope by the conjugate has been also tackled and recognized in \cite{CEV15}.
However, as already pointed out in the analysis of Figure \ref{Ex:SingletonUpTr}, the support of 
$C_{\lambda}^{u,h}(\chi_K)$ computed by the biconjugate scheme is the one to yield the worst value for 
$e_{\mathcal{H}}$.


\begin{table}[tbhp]
\centerline{
\begin{tabular}{ccc|c|c|c|}
\cline{4-6}
			& &	    & $t_c$      & $e_{L^{\infty}}$	& $e_{\mathcal{H}}$\\ \hline
\multicolumn{1}{ |c }{\multirow{5}{*}{\rotatebox[origin=c]{90}{\parbox{2cm}{\centering \small{Convex based schemes}}}}}
			& \multicolumn{2}{ |c| }{Algorithm \ref{Algo:CnvxEnvOBE}} & $1.9791$   & $0.0390$	&  $1.7321$		   \\ \cline{2-6}
\multicolumn{1}{ |c }{}	& \multicolumn{1}{ |c }{\multirow{4}{*}{\rotatebox[origin=c]{90}{\parbox{2cm}{\centering \small{biconjugate scheme}}}}}
							& \multicolumn{1}{ |l| }{$h_d=1$ }	& $0.1575$  & $48$	& $9.4999$		\\ \cline{3-6}
\multicolumn{1}{ |c }{}	& \multicolumn{1}{ |c }{}	& \multicolumn{1}{ |l| }{$h_d=0.1$ }	& $0.2157$  & $0.2400$	& $9$		\\ \cline{3-6}
\multicolumn{1}{ |c }{}	& \multicolumn{1}{ |c }{}	& \multicolumn{1}{ |l| }{$h_d=0.01$ }	& $0.5935$  & $0.0142$   & $7.6158$ \\ \cline{3-6}
\multicolumn{1}{ |c }{}	& \multicolumn{1}{ |c }{}	& \multicolumn{1}{ |l| }{$h_d=0.001$ }	& $16.6603$   & $0.0032$   & $7.5498$ \\ \hline
\multicolumn{1}{ |c }{\multirow{2}{*}{\parbox{2cm}{\centering \footnotesize{Moreau based schemes}}}}			
			&\multicolumn{2}{ |c| }{Algorithm \ref{Algo:MoreauEnv}} &  $0.1246$	&	 $0.0249$	&  $0$	   \\ \cline{2-6}
\multicolumn{1}{ |c }{}	&\multicolumn{2}{ |c| }{PE scheme}			&  $0.2553$	&	 $0.0249$	&  $0$		   \\ \hline
\end {tabular}
}
\caption{\label{Table:UpTrSing} Comparison between the different numerical schemes for the computation of 
$C_{\lambda}^u(\chi_{K})$ for $\lambda=0.01$. The symbol $h_d$ refers to the dual mesh size of the scheme that computes 
the convex envelope via the biconjugate.
	 }
\end{table}

\subsection{Intersection of Sampled Smooth Manifolds}
\begin{figure}[H]
  \centerline{$\begin{array}{ccc}
     \includegraphics[width=0.33\textwidth]{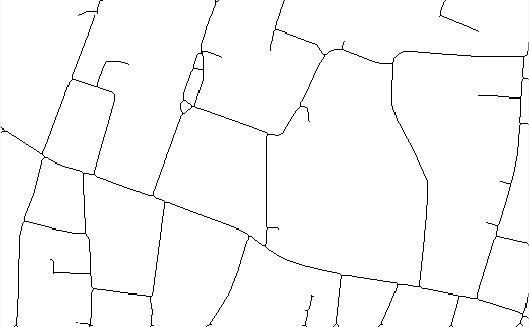}&
     \includegraphics[width=0.33\textwidth]{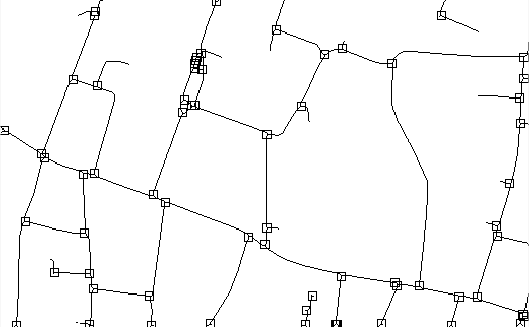}&
     \includegraphics[width=0.33\textwidth]{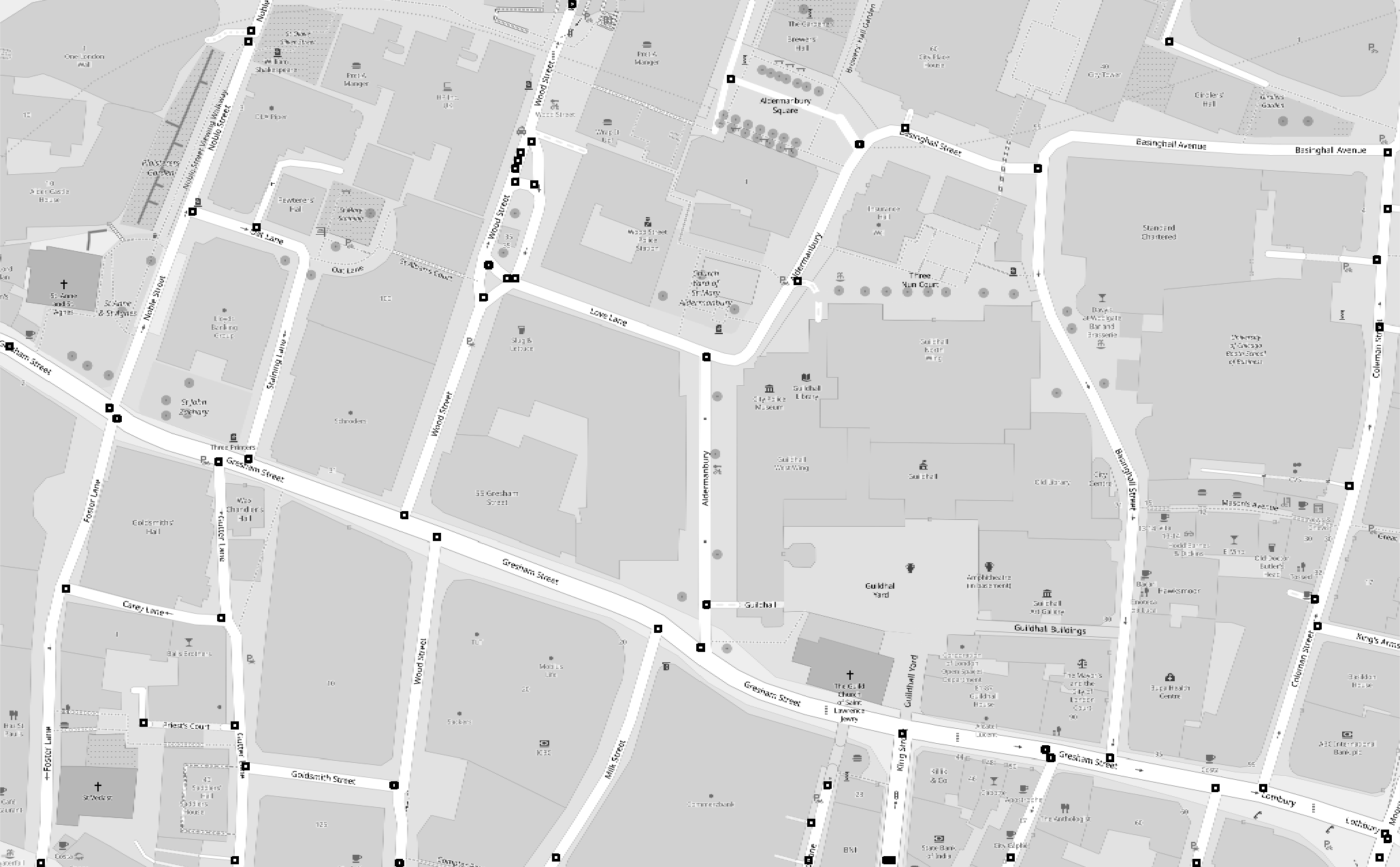}\\
     (a)&(b)&(c)
    \end{array}$
}
   \caption{\label{Fig.IntPointsRoadNetwork} 
	$(a)$ Medial axis of the road network; 
	$(b)$ Location of the intersection points;
	$(c)$ Map of the road network and location of the intersection points shown in 
	$(b)$.
   }
\end{figure}

In the following numerical experiments we verify the effectiveness of
the filter $I_{\lambda}(\cdot;\,K)$ introduced in Section \ref{Sec:Filters:IT} and its Hausdorff stability property.
We will consider both $2d-$ and $3d-$geometries. The geometry is digitized and 
input as an image, but also other computer representations of the geometry can clearly be handled.
This depends finally on
the representation of the input geometry for the numerical scheme that is used to compute the compensated transforms.
Figure \ref{Fig.IntPointsRoadNetwork} displays a road network extract from a map of teh city of London, 
and represents a set of $2d$ curves which
intersect to each other in different manner.
The Figure shows the position of the local maxima of $I_{\lambda}(\cdot;\,K)$ which
are seen to coincide with all the crossing and turning points of the given curves.
We also have some false positive due to the digitization of the road network.

\begin{figure}[H]
  \centerline{$\begin{array}{ccc}
		\includegraphics[width=0.30\textwidth]{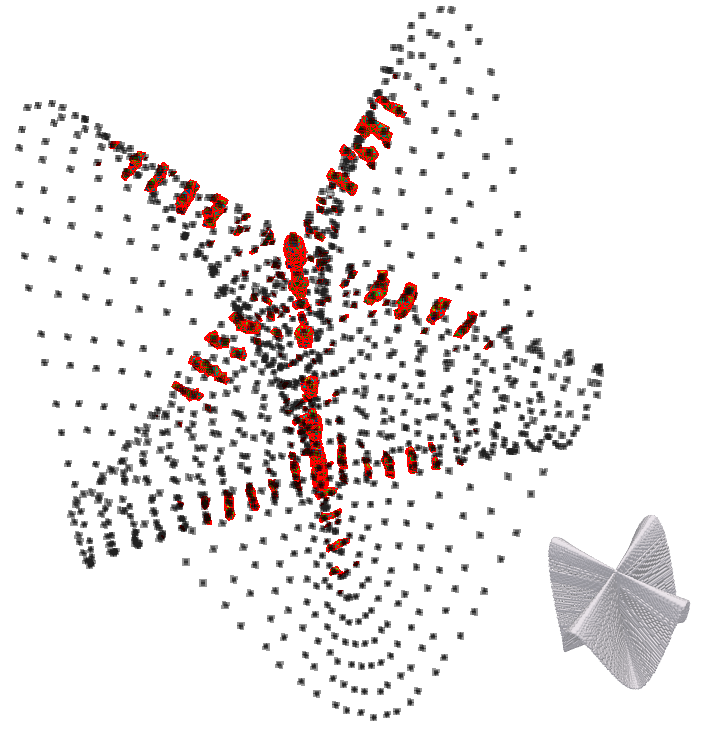}&
		\includegraphics[width=0.33\textwidth]{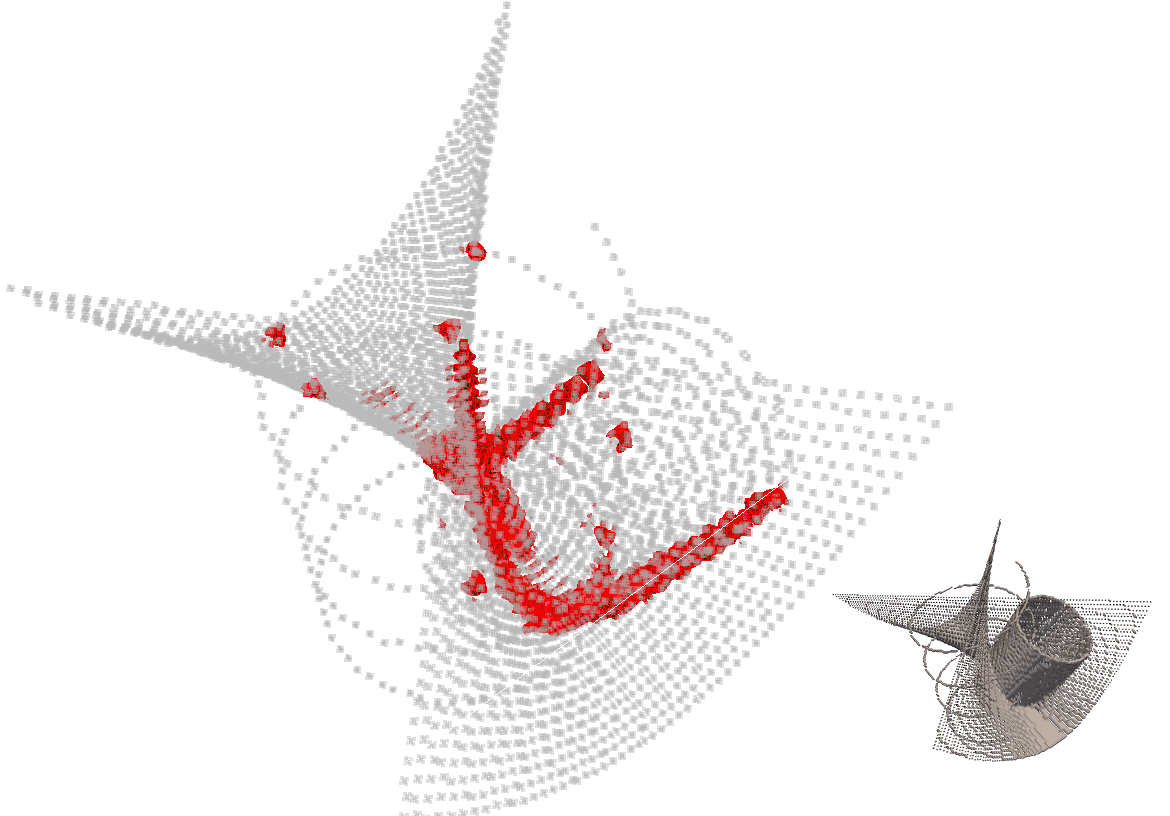}&
		\includegraphics[width=0.33\textwidth]{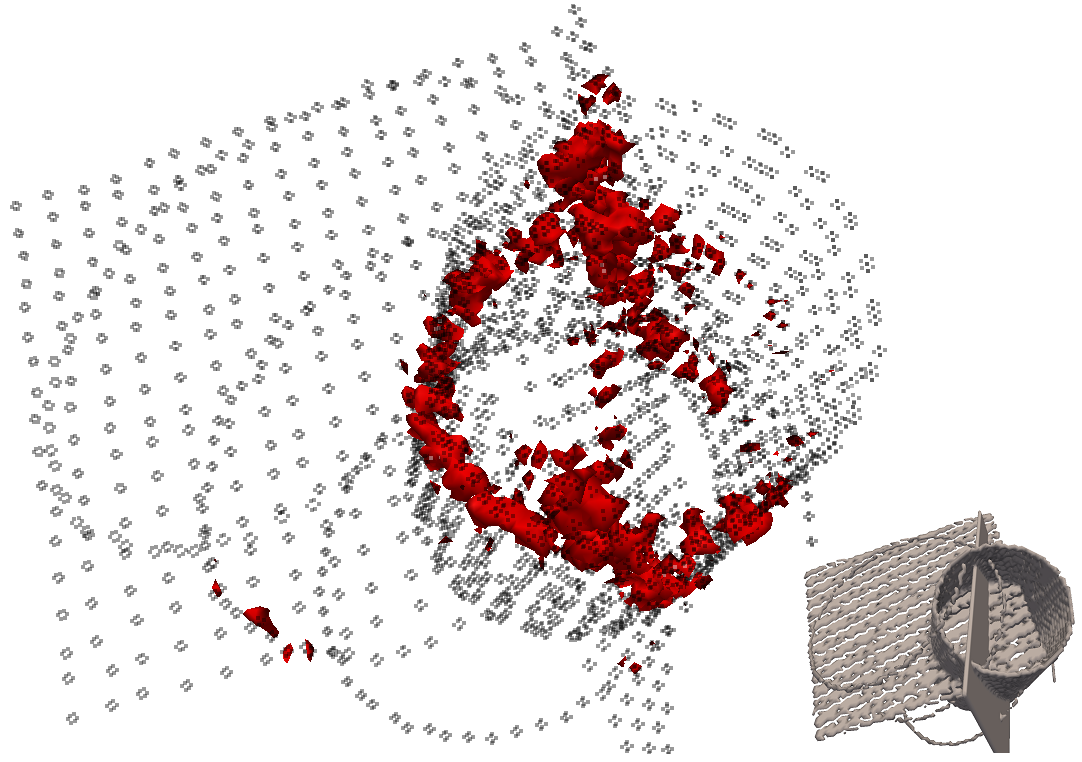}\\
		(a)&(b)&(c)
	\end{array}$
	}
   \caption{\label{Fig.3dGeo1}
	    $(a)$ Pl\"ucker surface with identification of its singular lines and surface parts of higher curvaturs;
	    $(b)$ Intersections of the Whitney surface of equation $x^2=y^2z$ with an helix and a cylinder;
	    $(c)$ Intersections of planes with a cylinder and an helix.
	    }
\end{figure}

Figure \ref{Fig.3dGeo1} displays the results of the application of the filter
$I_{\lambda}(\cdot;\,K)$ to $3d$ geometries represented by
point clouds. Figure \ref{Fig.3dGeo1}$(a)$ displays 
the Pl\"ucker's conoid of parametric equation
\[
	x=v\cos u,\quad	y=v\,\sin u,\quad z=sin 4u\quad\text{for }u\in[0,\,2\pi[,\,\,v\in[-1,\,1]\,,
\]
with the location of its singular lines and the parts of surface with higher curvature.
Figure  \ref{Fig.3dGeo1}$(b)$ depicts the intersections between manifolds of different dimensions,
namely, in the Figure, we have the Whitney umbrella
of the implicit equation $x^2=y^2z$, a cylinder and an helix, with the location of their mutual 
intersections and also of where the Whitney surface intersects itself; finally, Figure 
\ref{Fig.3dGeo1}$(c)$ displays the intersection between a cylinder, planes
and an helix.

The intersection of the line with the plane for the geometry shown in Figure \ref{Fig.3dGeo1} is weaker than the
geometric singularities of the surfaces. With this meaning, the values of the local maxima
of $I_{\lambda}(\cdot;\,K)$ determine a scale between the different type of intersections present in the manifold $K$
and represents the multiscale nature of the filter $I_{\lambda}(\cdot;\,K)$.

\begin{figure}[H]
	\centerline{$\begin{array}{cc}
		\includegraphics[width=0.45\textwidth]{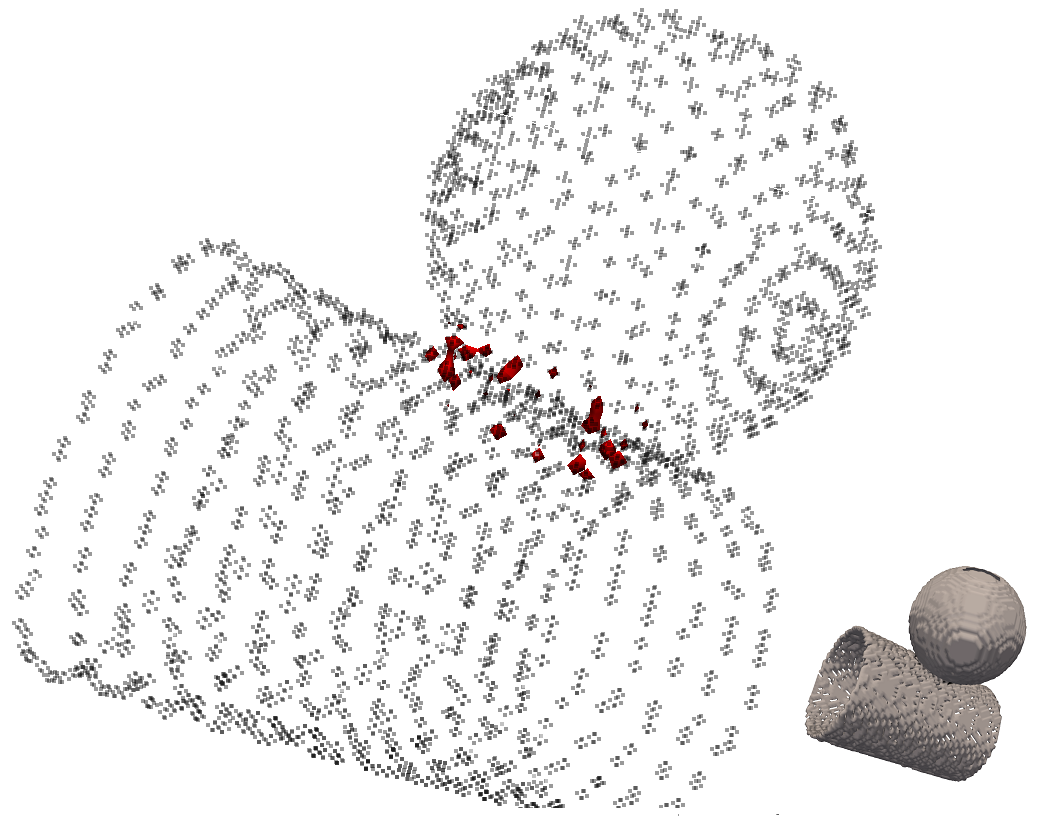}&
		\includegraphics[width=0.45\textwidth]{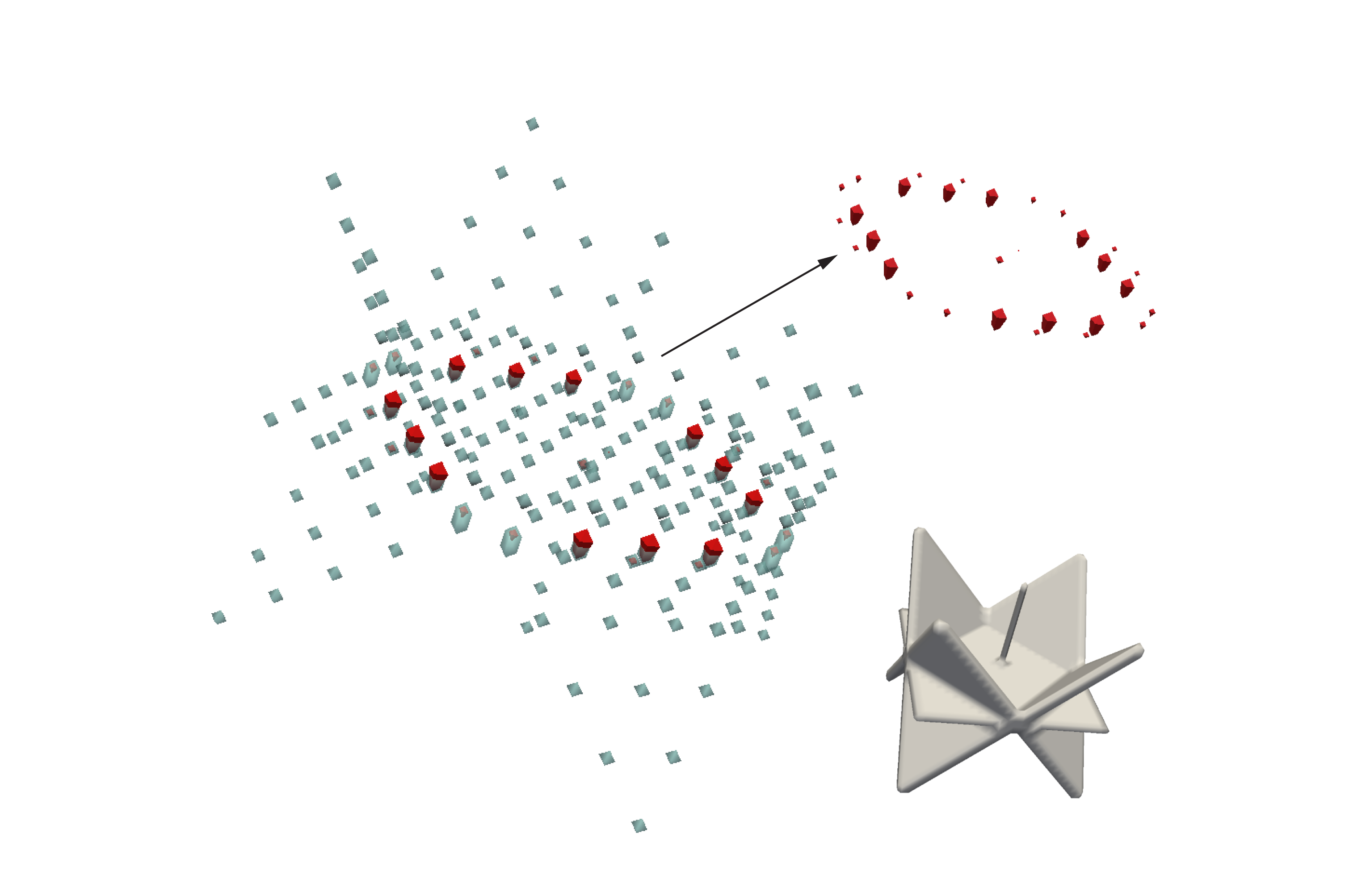}\\
			(a)&(b)
	\end{array}$
	}
	\caption{\label{Fig.3dGeo2}
		$(a)$ Tangential intersection of a sampled sphere and cylinder 
		which are `almost' tangentially intersected, and indication of the intersection marker. 
		$(b)$ Intersection markers for the intersection among loosely sampled piecewise affine 
		surfaces of equation
		$||10x-75|-|10y-75|+|10z-75|-45|$=0, the circle of equation $(10x-75)^2+(10z-75)^2\leq 45^2$
		on the plane of equation $y=75$ and the line of equation $x=75, z=75$.
		}
\end{figure}

Finally, the numerical experiments displayed in Figure \ref{Fig.3dGeo2} 
refer to critical conditions that are not directly
covered by the theoretical results we have obtained. Figure \ref{Fig.3dGeo2}$(a)$
shows the result of the application of $I_{\lambda}(\cdot;\,K)$ to a sphere and a cylinder that are 
`almost' tangentially
intersecting each other,
whereas Figure \ref{Fig.3dGeo2}$(b)$ illustrates the results of the application of the filter
to detect the intersection between loosely sampled piecewise affine functions, a plane and a line.

\subsection{Approximation Transform}
We report here on applications of the average approximation compensated convex transform developed in 
\cite{ZCO16a,ZCO18} to three class of problems. These include $(i)$ surface reconstruction from 
real world data using level lines and single points; $(ii)$ Salt \& Pepper noise restoration
and $(iii)$ image inpainting.

\subsubsection{Level set reconstruction}
%

\begin{figure}[H]
  \centerline{$\begin{array}{ccc}
	\includegraphics[width=0.50\textwidth]{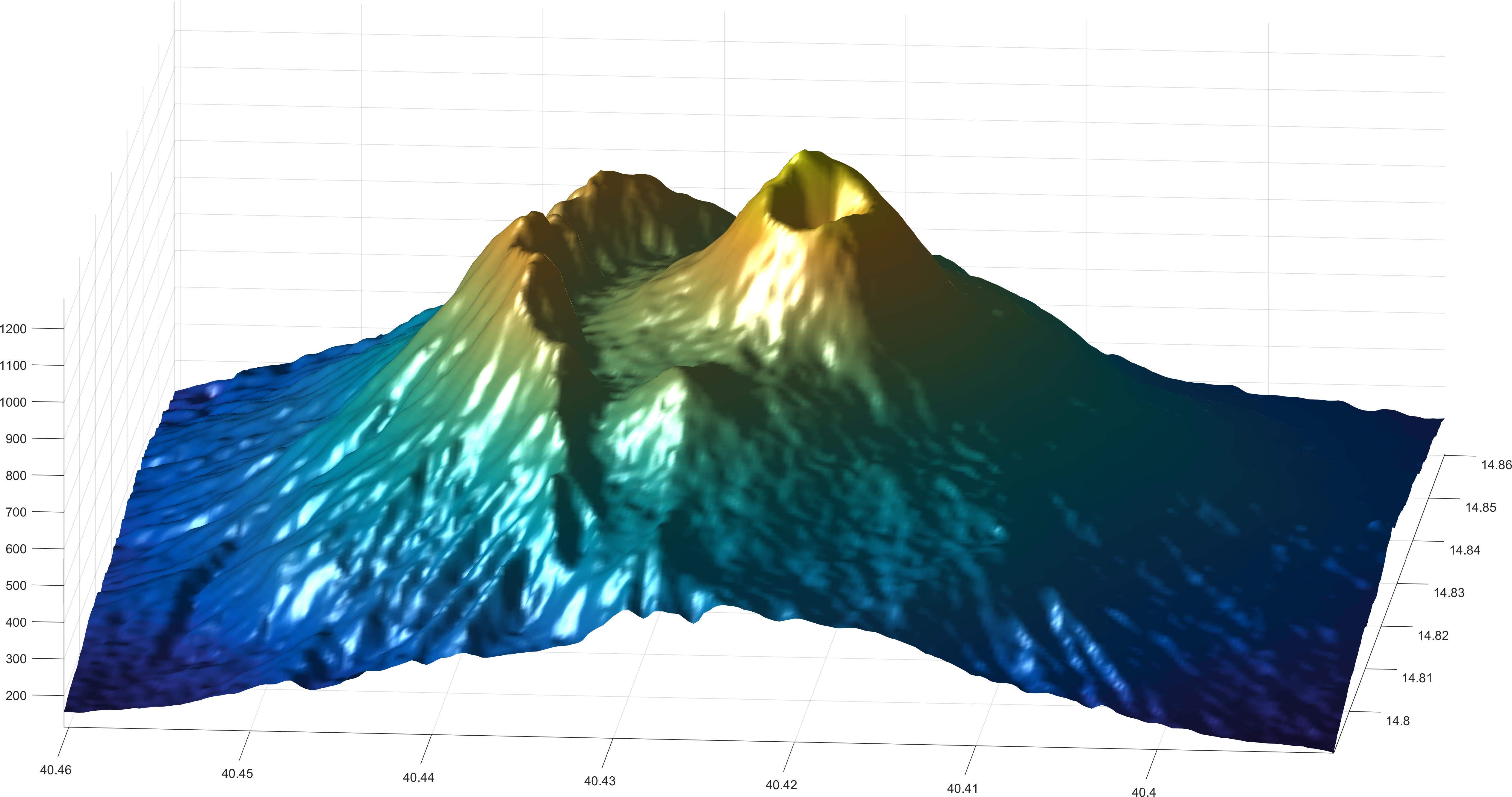}
	&\includegraphics[width=0.25\textwidth]{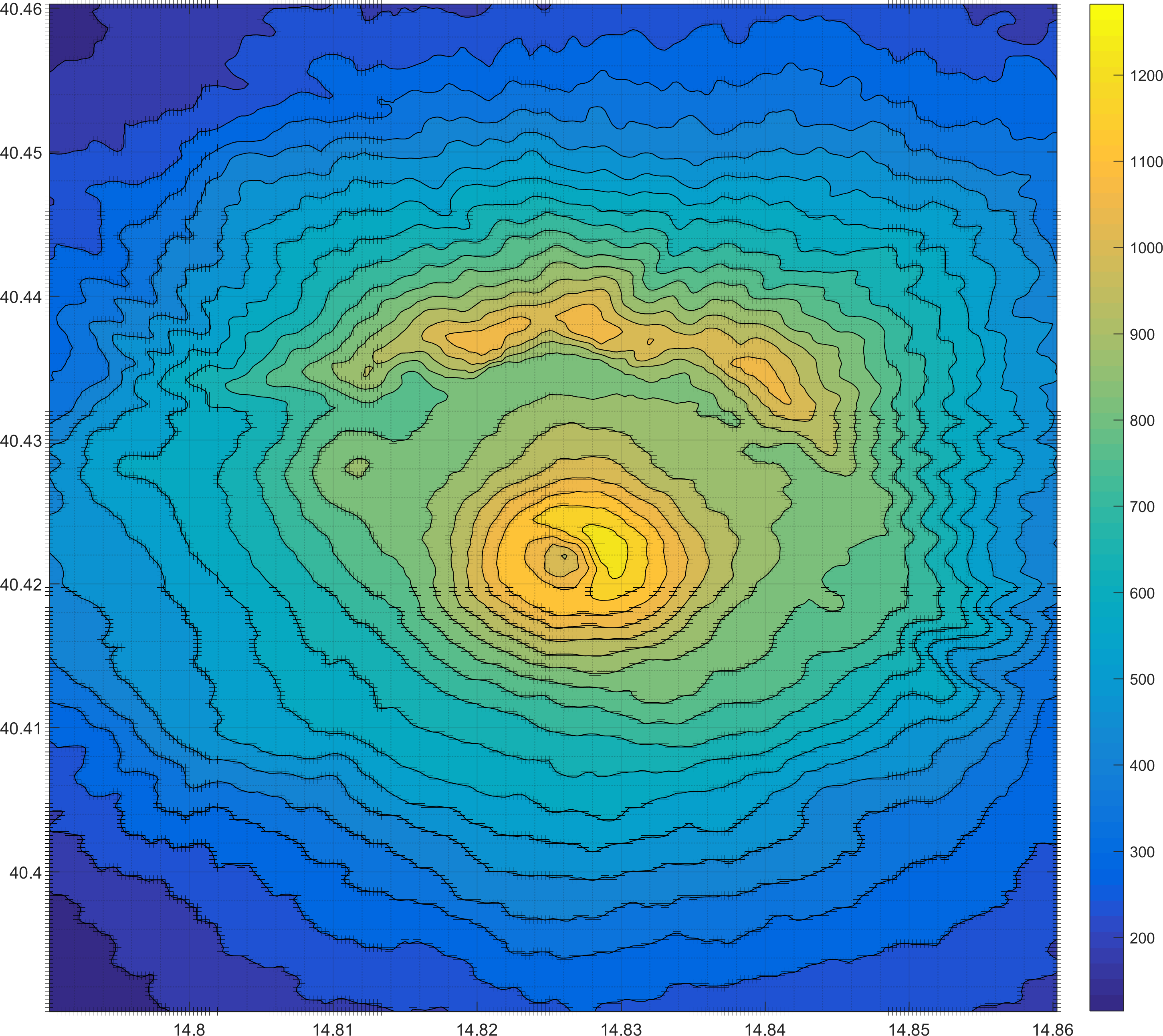}
	&\includegraphics[width=0.25\textwidth]{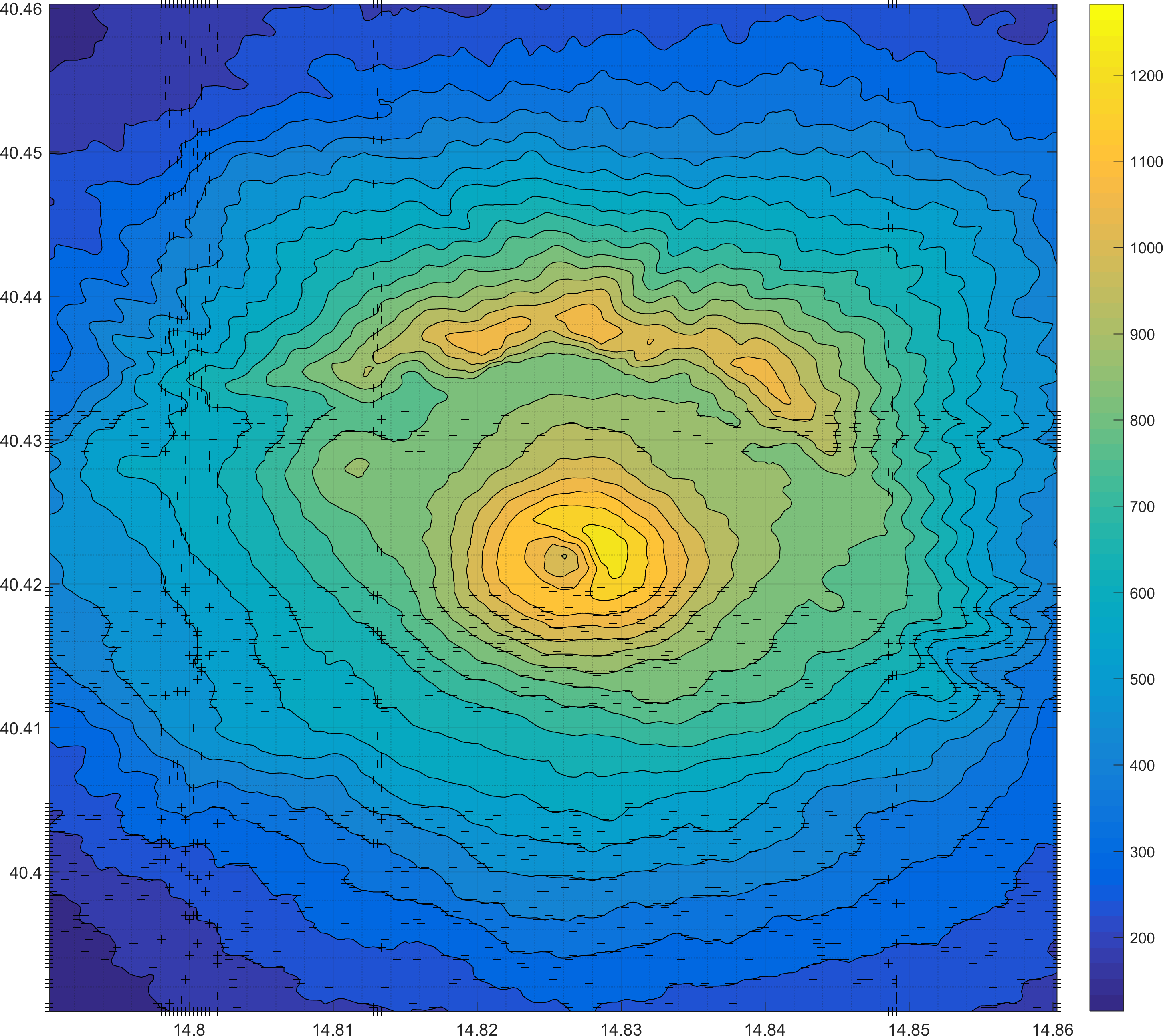}\\
	(a) & (b) & (c)
    \end{array}$
}
   \caption{\label{Fig.NumEx7.2.4a} Reconstruction of real-world digital elevation maps.
	 $(a)$ Ground truth model from USGS-STRM1 data relative to the area with geographical coordinates;
	 $[\mathrm{N}\,40^{\circ}23'25'',\,\mathrm{N}\,40^{\circ}27'37'']\times[\mathrm{E}\,14^{\circ}47'25'',\,\mathrm{E}\,14^{\circ}51'37'']$. 
	 $(b)$ Sample set $K_1$ formed by only level lines at regular height interval of $58.35\,\mathrm{m}$. 
		The set $K_1$ contains $14\%$ of the ground truth points.
	 $(c)$ Sample set $K_2$ formed by taking randomly $30\%$ of the
		points belonging to the level lines of the set $K_1$ and scattered points 
		corresponding to $5\%$ density. The sample set $K_2$ contains $7\%$ of the 
		ground truth points.
	}
\end{figure}

We consider here the problem of producing a Digital Elevation Map from a sample of the 
the NASA SRTM global digital 
elevation model of Earth land. The data provided by the 
National Elevation Dataset \cite{GEMHC09} contain geographical coordinates (latitude, longitude and elevation)
of points sampled at one arc-second intervals in latitude and longitude. For our experiments,
we choose the region defined by the coordinates 
$[\mathrm{N}\,40^{\circ}23'25'',\,\mathrm{N}\,40^{\circ}27'37'']\times[\mathrm{E}\,14^{\circ}47'25'',\,\mathrm{E}\,14^{\circ}51'37'']$
extracted from the SRTM1 cell $N40E014.hgt$ \cite{SRTM1}. Such region consists of an area 
with extension $7.413\,\mathrm{km}\times 5.844\,\mathrm{km}$ and height varying between $115\,\mathrm{m}$ and 
$1282\,\mathrm{m}$, 
with variegated topography features. In the digitization by the US Geological Survey, 
each pixel represents a $30\,\mathrm{m} \times 30\,\mathrm{m}$ patch.
Figure \ref{Fig.NumEx7.2.4a}$(a)$ displays the elevation model from the SRTM1 data
which we refer in the following to as the ground truth model. 
We will take a sample $f_K$ of such data, make the reconstruction using the $A_{\lambda}^M(f_K)$
computed with Algorithm \ref{Algo:CnvxEnvOBE}
and the AMLE interpolant \cite{ACGR02,CMS98} using the MatLab\textregistered\, code described in \cite{PS16}, 
and compare them with the ground truth model.

In the numerical experiments, we consider two sample data, characterized by different 
data density and typo of information. The first, which we refer to as sample set $K_1$, 
consists only of level lines at regular height interval of $658.35\,\mathrm{m}$ and contains 
the $14\%$ of the ground truth real digital data.
The second sample set, denoted by $K_2$, has been formed by taking randomly the $30\%$ of the
points belonging to the level lines of the set $K_1$ and scattered points corresponding to
$5\%$ density so that the sample set $K_2$ amounts to about $7\%$ of the ground truth points.
The two sample sets $K_1$ and $K_2$ are shown in Figure \ref{Fig.NumEx7.2.4a}$(b)$ and 
Figure \ref{Fig.NumEx7.2.4a}$(c)$, respectively.

\begin{figure}[H]
  \centerline{$\begin{array}{cc}
	\includegraphics[width=0.50\textwidth]{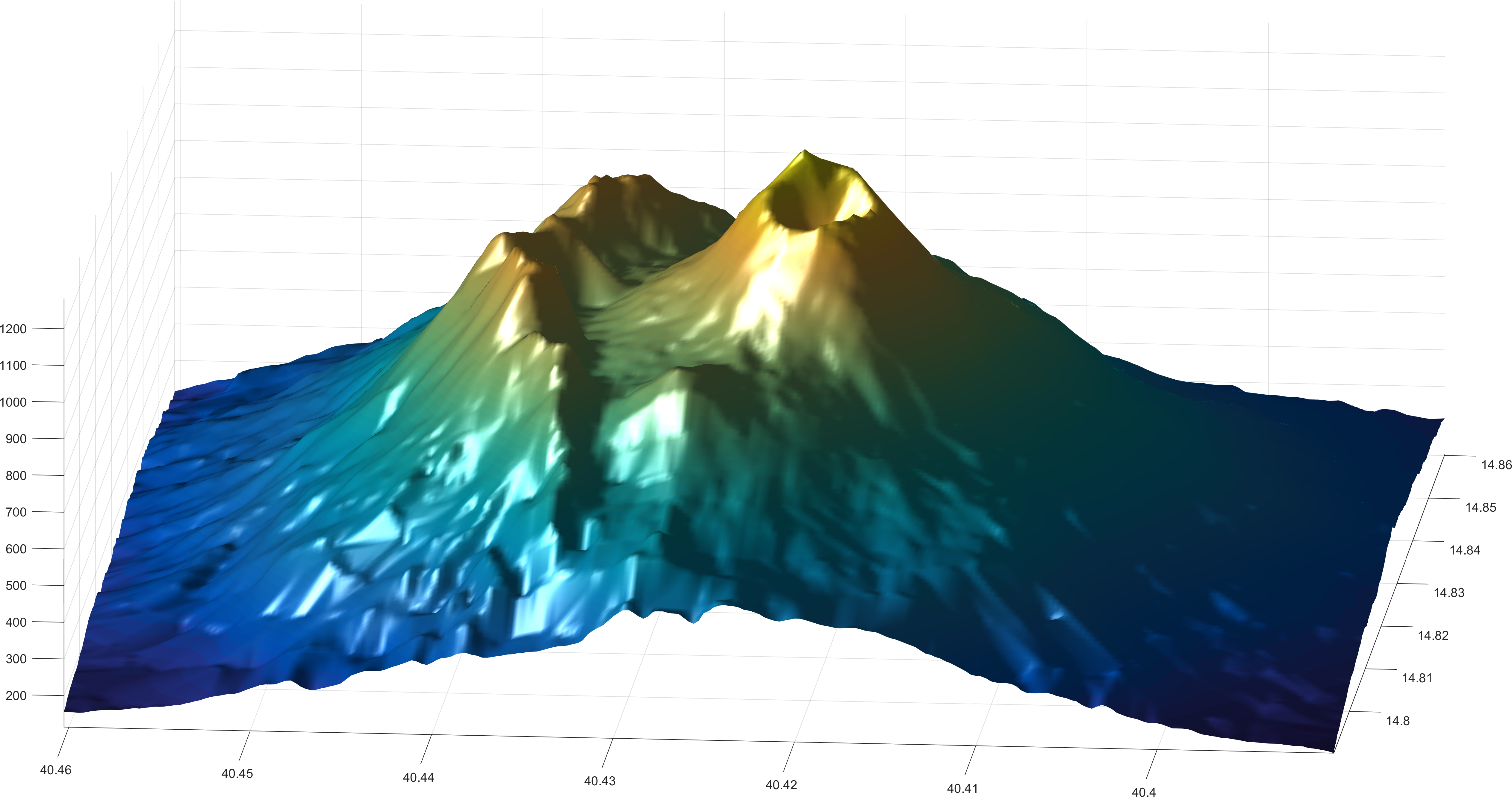}
	&\includegraphics[width=0.50\textwidth]{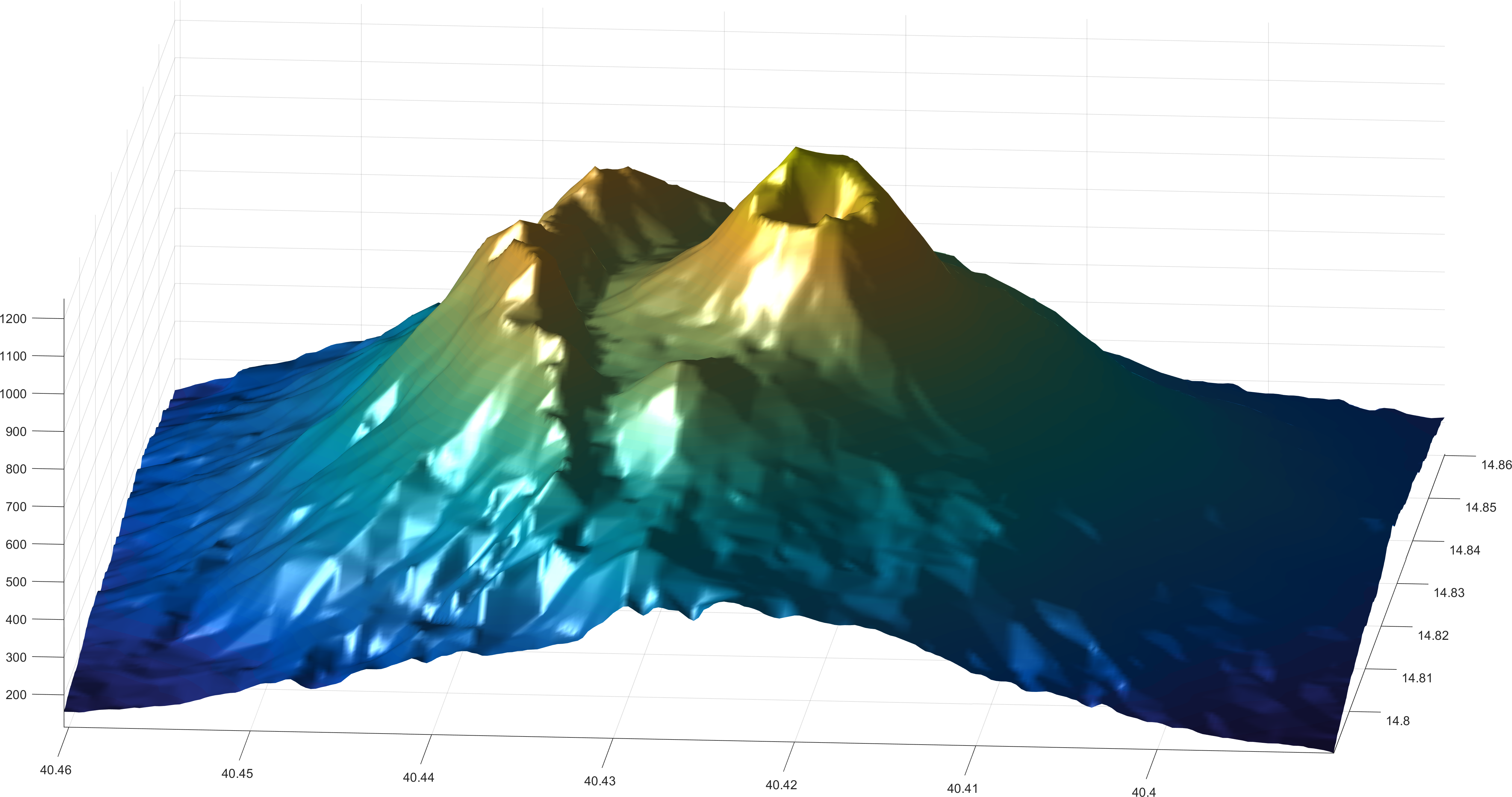}\\
	(a) & (b)\\
	\includegraphics[width=0.25\textwidth]{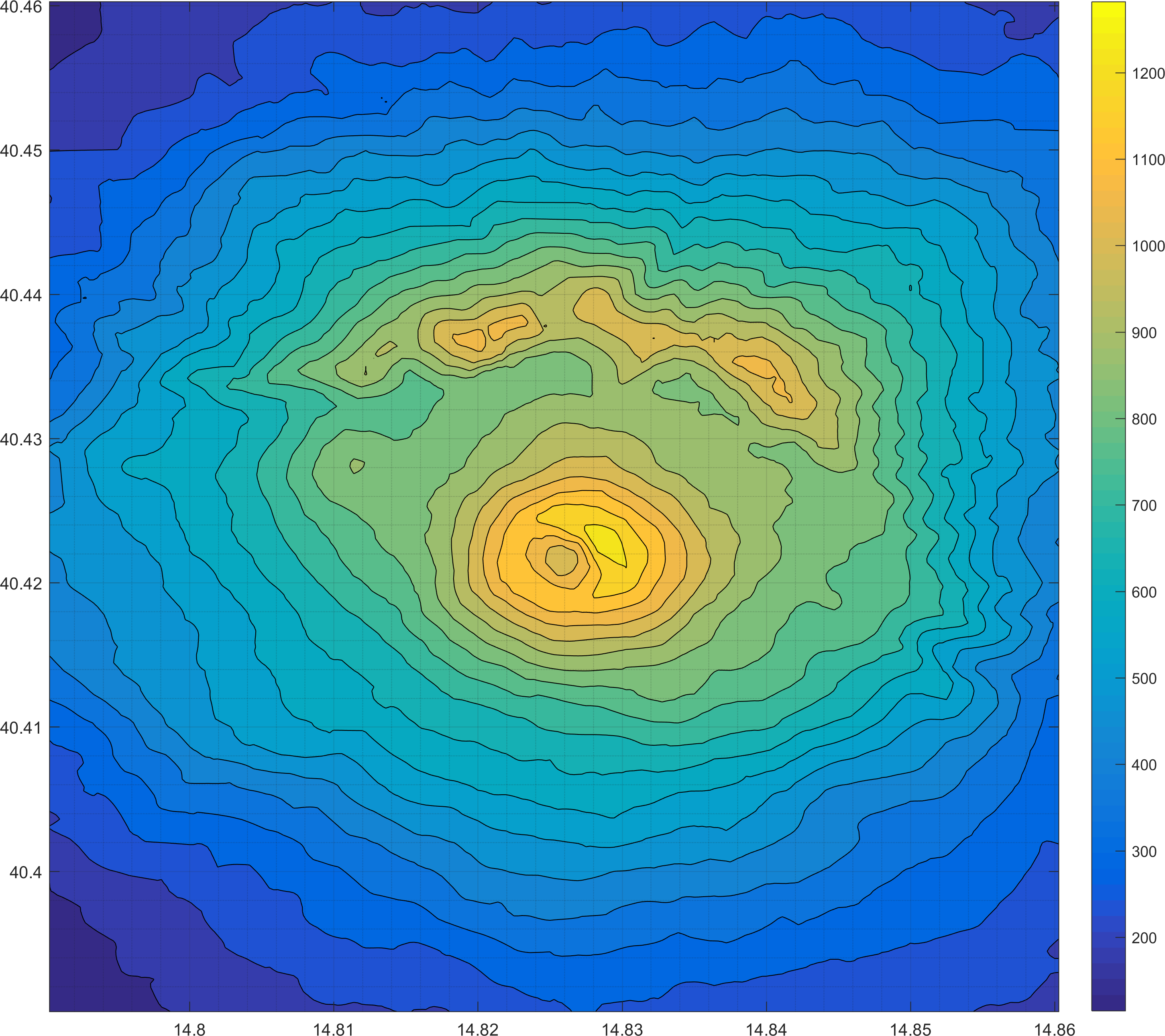}
	&\includegraphics[width=0.25\textwidth]{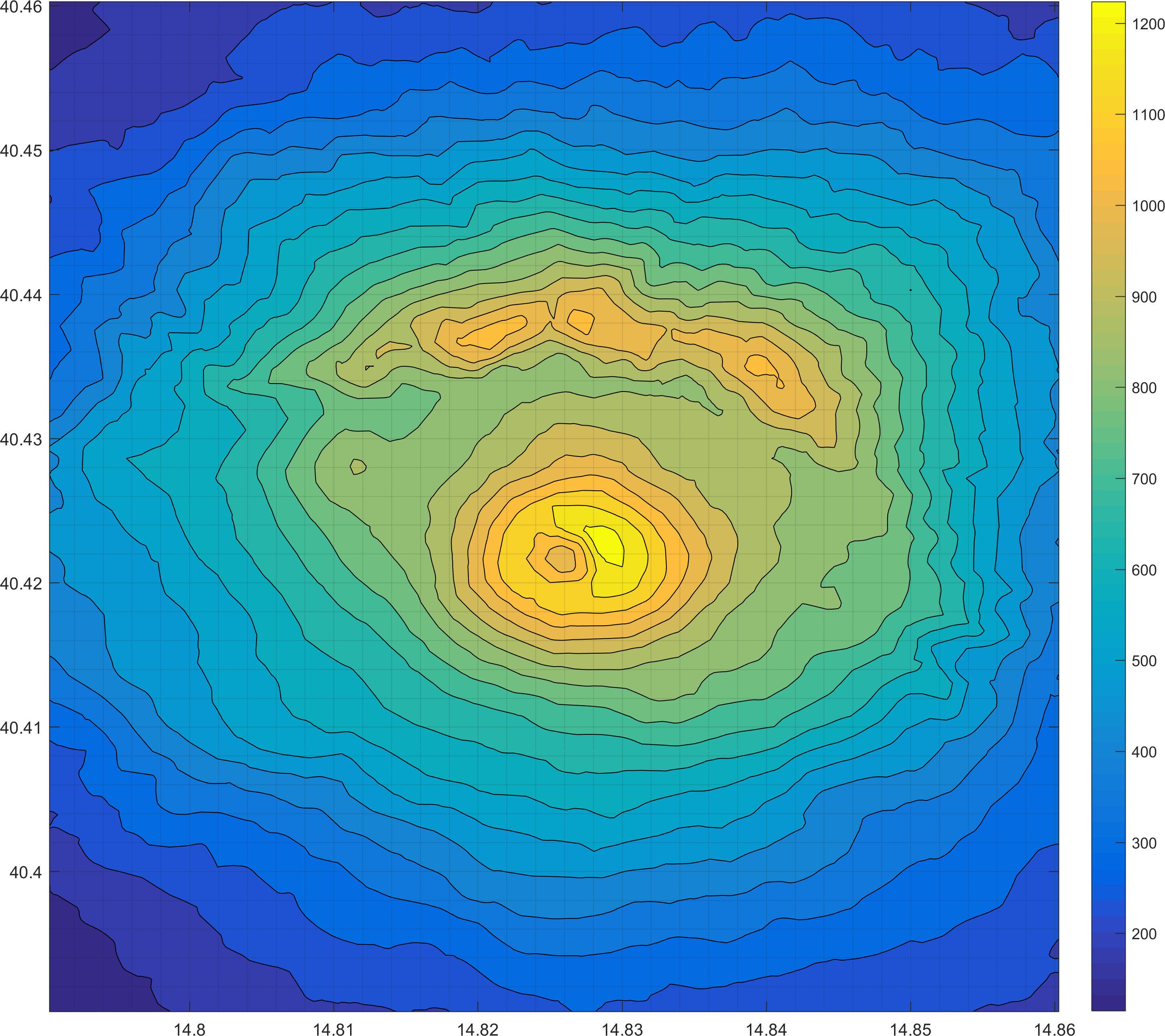}\\
	(c) & (d)
    \end{array}$
}
   \caption{\label{Fig.NumEx7.2.4b} 
   Reconstruction of real-world digital elevation maps.
	 $(a)$ Graph of $A_{\lambda}^M(f_K)$ for sample set $K_1$. 
		Relative $L^2$-Errors: $\epsilon=0.0118$, $\epsilon_K=0$.
		Parameters: $\lambda=2\cdot 10^3$, $M=1\cdot 10^6$.   
		Total number of iterations: $3818$.
	 $(b)$ Graph of $A_{\lambda}^M(f_K)$ for sample set $K_2$. 
		Relative $L^2$-Errors: $\epsilon=0.0109$, $\epsilon_K=0$.
		Parameters: $\lambda=2\cdot 10^3$, $M=1\cdot 10^6$.   
		Total number of iterations: $1662$.
	 $(c)$ Isolines of $A_{\lambda}^M(f_K)$ from sample set $K_1$
		at regular heights of $58.35\,\mathrm{m}$.
	 $(d)$ Isolines of $A_{\lambda}^M(f_K)$ from sample set $K_2$
		 at regular heights of $58.35\,\mathrm{m}$.
	}
\end{figure}

\begin{figure}[H]
  \centerline{$\begin{array}{cc}
	\includegraphics[width=0.50\textwidth]{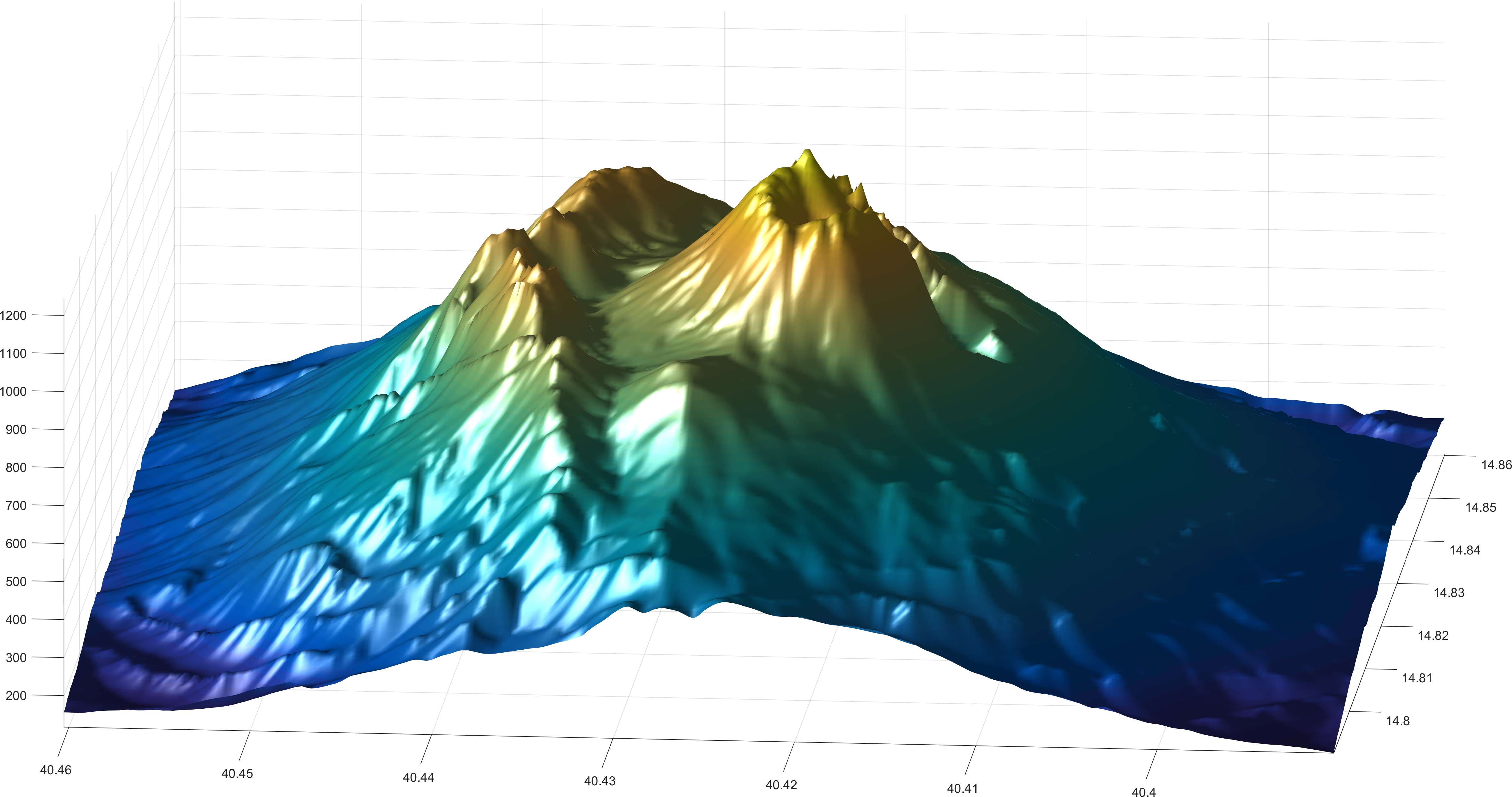}
	&\includegraphics[width=0.50\textwidth]{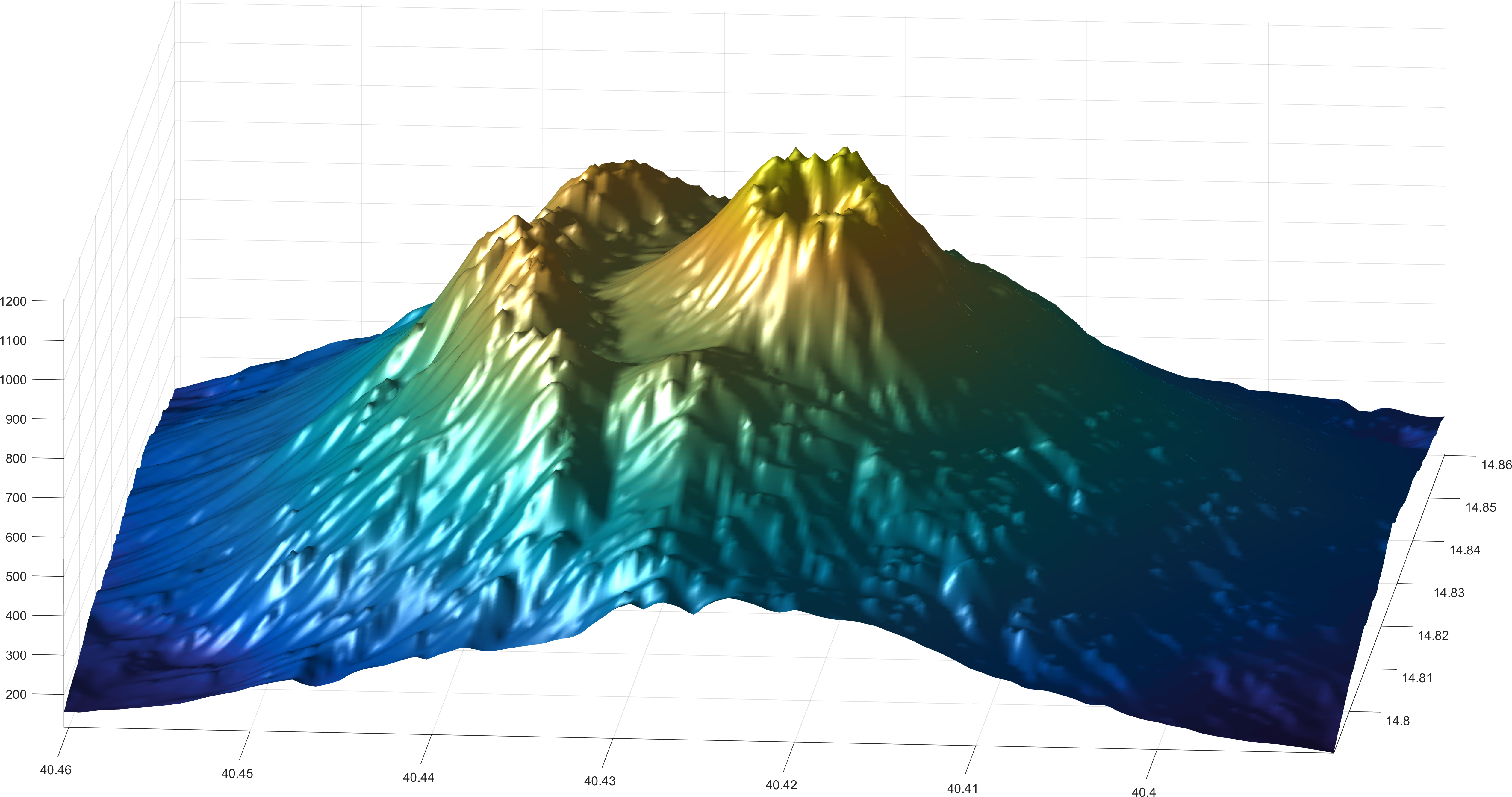}\\
	(a) & (b)\\
	\includegraphics[width=0.25\textwidth]{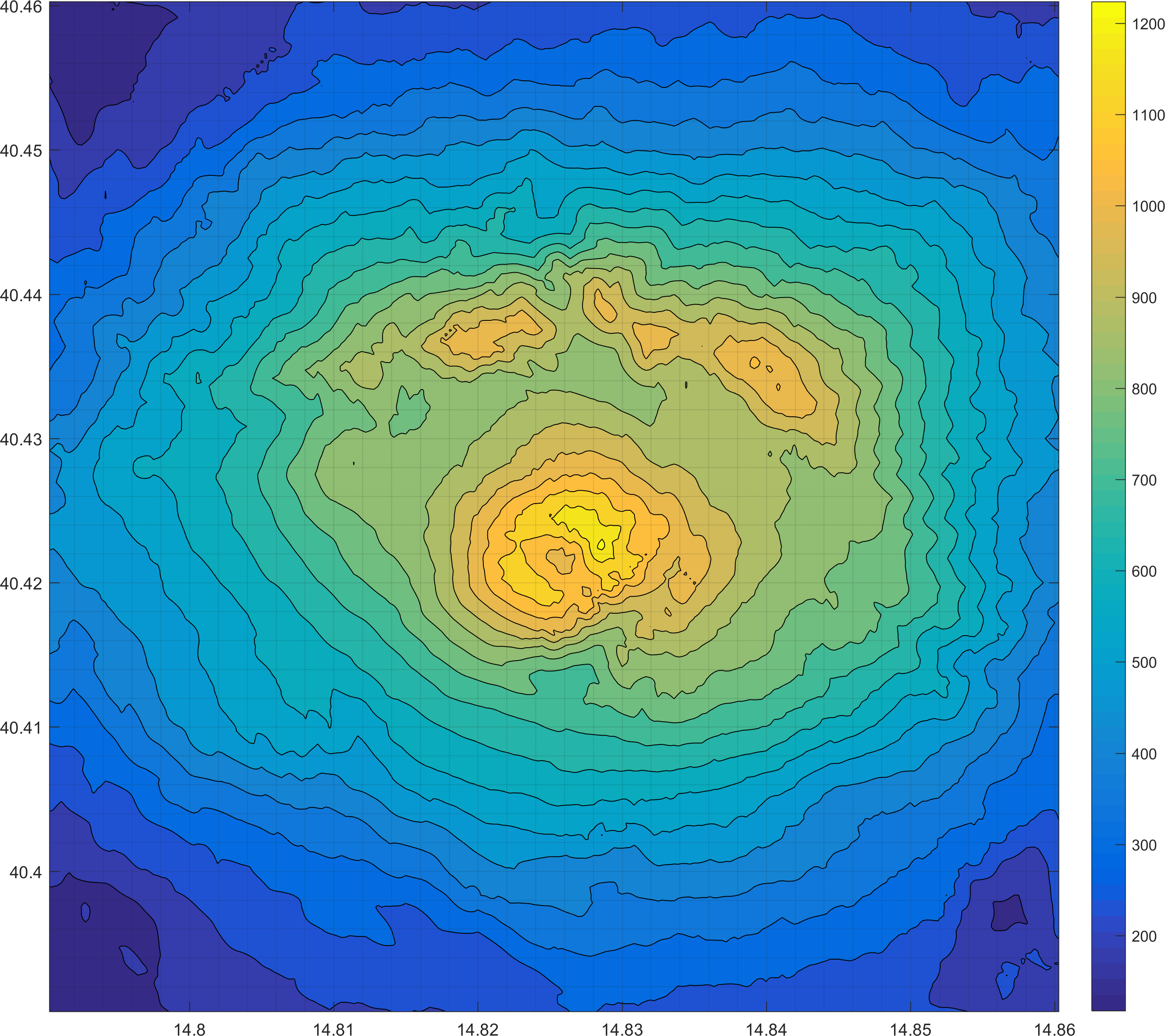}
	&\includegraphics[width=0.25\textwidth]{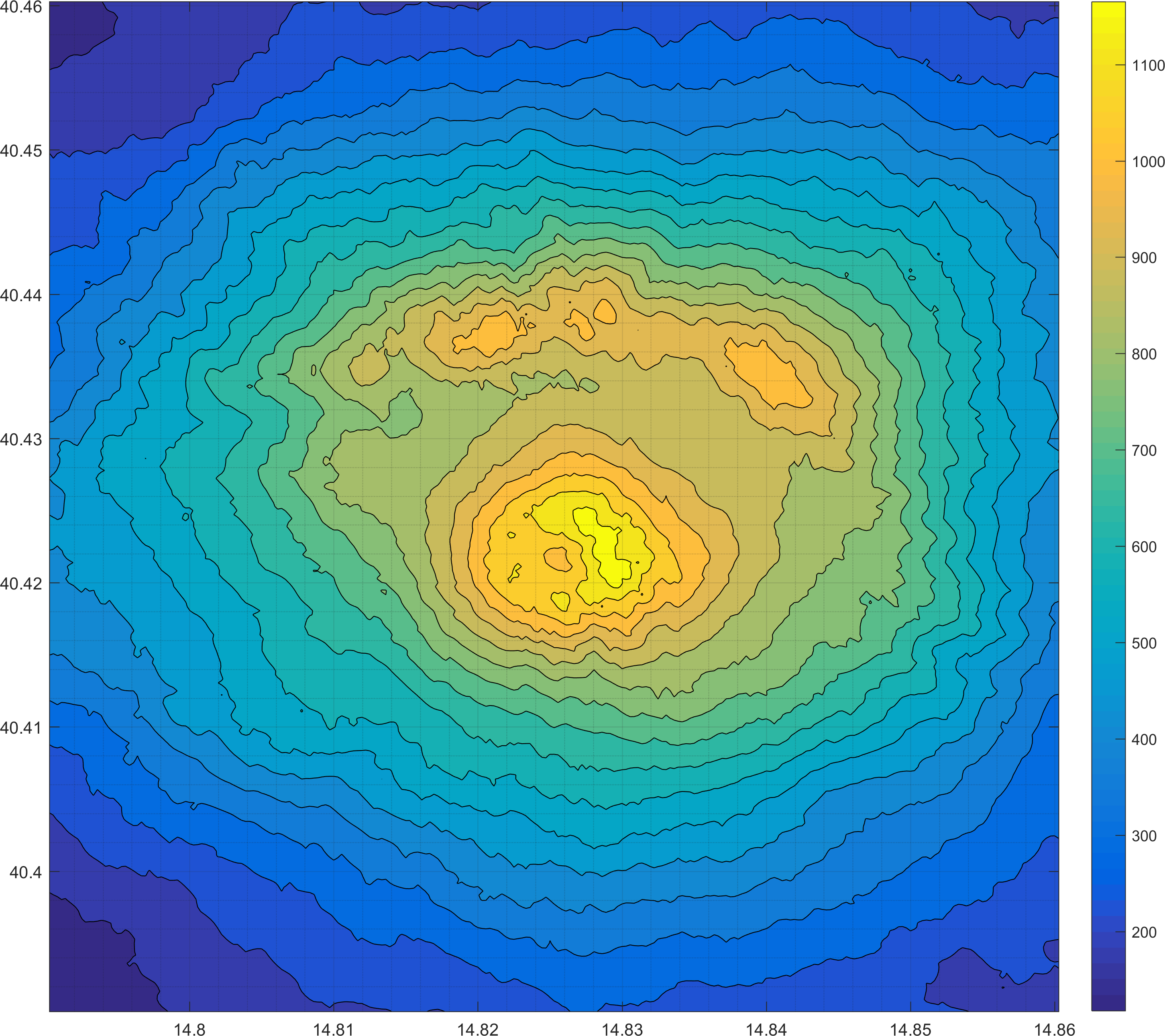}\\
	(c) & (d)
    \end{array}$
}
   \caption{\label{Fig.NumEx7.2.4c}
   Reconstruction of real-world digital elevation maps.
	 $(a)$ Graph of the AMLE Interpolant from set $K_1$.   
		Relative $L^2$-Error: $\epsilon=0.0410$,
		$\epsilon_K=0.0110$.
		Total number of iterations: $11542$.
	 $(b)$ Graph of the AMLE Interpolant from set $K_2$.   
		Relative $L^2$-Error: $\epsilon=0.02863$,
		$\epsilon_K=0.0109$.
		Total number of iterations: $12457$.
	 $(c)$ Isolines of the AMLE Interpolant from sample set $K_1$
		at regular heights of $58.35\,\mathrm{m}$.
	 $(d)$ Isolines of the AMLE Interpolant from sample set $K_2$
		 at regular heights of $58.35\,\mathrm{m}$. 
	}
\end{figure}

The graph of the $A_{\lambda}^M(f_K)$ 
interpolant and of the $AMLE$ interpolant for the two sample sets along with the respective 
isolines at equally spaced heighs equal to $58.35\,\mathrm{m}$, are displayed in 
Figure \ref{Fig.NumEx7.2.4b} and Figure \ref{Fig.NumEx7.2.4c}, respectively, 
whereas Table \ref{Tab:RealDat} contains the values of the 
relative $L^2$-error $\epsilon$  on $\Omega$ and $\epsilon_K$ on the sample set $K$
between such interpolants and the ground truth model,
given by, respectively,
\begin{equation}\label{Eq.RelErr}
	\epsilon=\frac{\|f-A_\lambda^M(f_K)\|_{L^2(\Omega)}}{\|f\|_{L^2(\Omega)}}\quad\text{and}
	\quad
	\epsilon_K=\frac{\|f_K-A_\lambda^M(f_K)\|_{L^2(K)}}{\|f_K\|_{L^2(K)}}
	\,,
\end{equation}
where $f$ is the ground truth model and $A_\lambda^M(f_K)$ is the average approximation
of the sample $f_K$ of $f$ over $K$. We observe that while $A_{\lambda}^M(f_K)$ yields an exact interpolation 
of $f_K$ over $\Omega$, this is not the case for the AMLE approximation.


\begin{table}[H]
\centerline{
\begin{tabular}{c|c|c|c|c|}
\cline{2-5}
    & \multicolumn{2}{c|}{ $\epsilon$}& \multicolumn{2}{c|}{ $\epsilon_K$} \\ \hline
\multicolumn{1}{ |c| }{Sample set}	& $A_{\lambda}^M(f_K)$  &  AMLE		& $A_{\lambda}^M(f_K)$  &  AMLE     \\ \hline
\multicolumn{1}{ |c| }{$K_1$}		& $0.0118$ & $0.0410$			& $0$ & $0.0110$	\\ \hline
\multicolumn{1}{ |c| }{$K_2$}		& $0.0109$ & $0.0286$			& $0$ & $0.0109$\\ \hline
\end {tabular}
}
\caption{\label{Tab:RealDat} Relative $L^2$-error for the DEM Reconstruction from the two sample sets
using the $A_{\lambda}^M(f_K)$ and the AMLE interpolant. The realization of $\epsilon_K=0$
for $A_{\lambda}^M(f_K)$ says that $A_{\lambda}^M(f_K)$ yields an exact interpolation 
of $f_K$ over $\Omega$, unlike the AMLE approximation.
}
\end{table}


Though both reconstructions are comparable visually to the ground truth model, 
a closer inspection of the pictures show that the reconstruction from the synthetic data, 
the AMLE interpolant does not reconstruct correctly the mountains peaks, which appear to be smoothed,
and introduce artificial ridges along the slopes of the mountains. In contrast, the 
$A_{\lambda}^M(f_K)$ interpolant appears to better for capturing features of the ground truth model.
Finally, we also note that though the sample set $K_1$ contains a number 
of ground truth points higher than the sample set $K_2$, the reconstruction from $K_2$ appears 
to be better than the one obtained from $K_1$. This behaviour was found for both interpolations,
though it is more notable in the case of the $A_{\lambda}^M(f_K)$ interpolant. By taking scattered data,
we are able to get a better characterization of irregular surfaces, compared
to the one obtained from a structured representation such as provided by the level lines.

\subsubsection{Salt \& Pepper Noise Removal}
\begin{figure}[H]
	\centerline{$\begin{array}{cc}
		\includegraphics[width=0.35\textwidth]{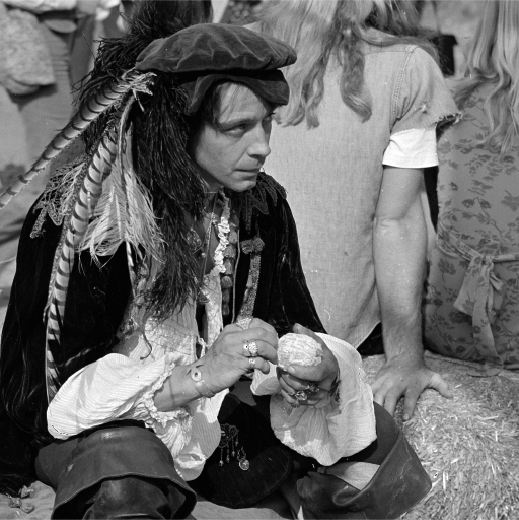}&
		\includegraphics[width=0.35\textwidth]{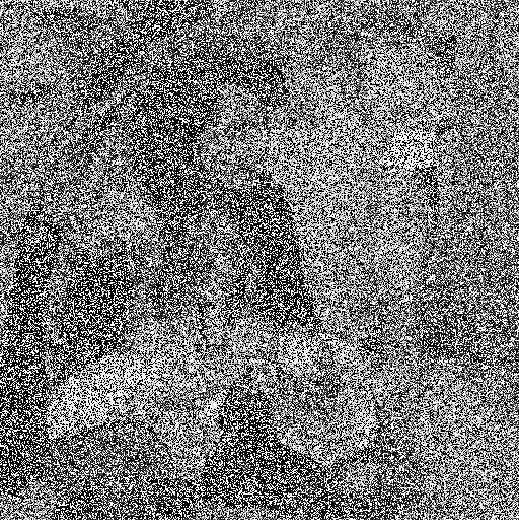}\\
		(a)&(b)\\
		\includegraphics[width=0.35\textwidth]{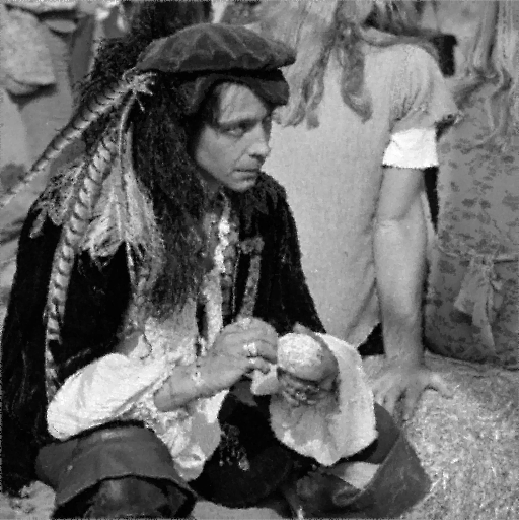}&
		\includegraphics[width=0.35\textwidth]{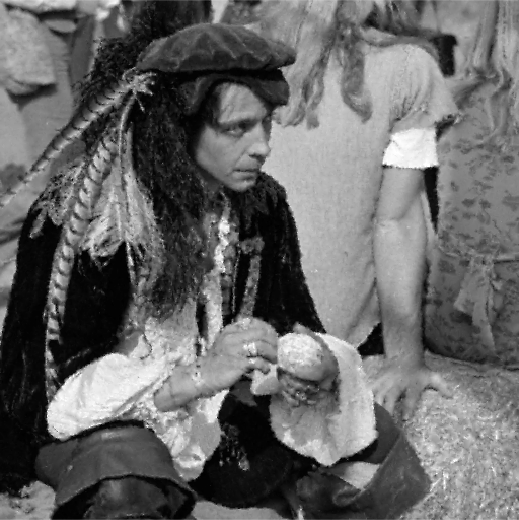}\\
		(c)&(d)\\
		\includegraphics[width=0.35\textwidth]{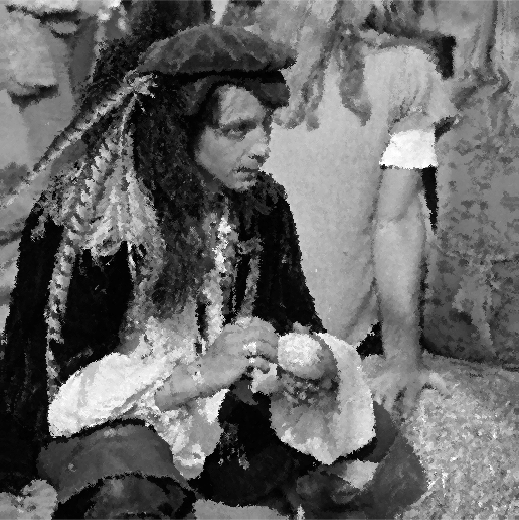}&
		\includegraphics[width=0.35\textwidth]{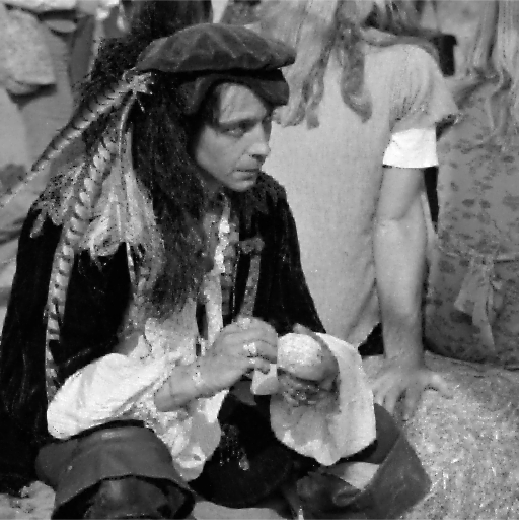}\\
		 (e)&(f)
	\end{array}$}
	\caption{\label{Fig.NumEx7.2.5} 
		$(a)$ Original image;
		$(b)$ Original image covered by a salt \& pepper noise density of $70\%$. 
			$\mathrm{PSNR}=6.426\,\mathrm{dB}$; 
		$(c)$ Restored image $A_{\lambda}^M(f_K)$ by Moreau based scheme (Algorithm \ref{Algo:MoreauEnv})
			with the set $K$ padded by two pixels. $\mathrm{PSNR}=26.020\,\mathrm{dB}$.
			$\lambda=20$, $M=1E13$. Total number of iterations: $21$. 
		$(d)$ Restored image $A_{\lambda}^M(f_K)$ by Convex based scheme (Algorithm \ref{Algo:CnvxEnvOBE})
			with the set $K$ padded by two pixels. $\mathrm{PSNR}=26.642\,\mathrm{dB}$.
			$\lambda=20$, $M=1E13$. Total number of iterations: $1865$. 
		$(e)$ Restored image by the Adaptive Median filter \cite{HH95} used as starting guess for the 
			two-stage TV-based method described in \cite{CCM07,CHN05}. Window size $w=33$ pixels. 	
			$\mathrm{PSNR}=22.519\,\mathrm{dB}$.
		$(f)$ Restored image by the two-stage TV-based method described in \cite{CCM07,CHN05} with
			the set $K$ padded by two pixels.
			$\mathrm{PSNR}=26.475\,\mathrm{dB}$. Total number of iterations: $3853$.
	}
\end{figure}

As an application of  scattered data approximation to image processing, 
we consider here the restoration of 
an image corrupted by salt \& pepper noise. This is an impulse type noise that is caused, for instance, by 
malfunctioning pixels in camera sensors or faulty memory locations
in hardware, so that information is lost at the faulty pixels and the 
corrupted pixels are set alternatively
to the minumum or to the maximum value of the range of the image values. 
When the noise density  is low, about less than $40\%$, the median filter \cite{AK97} or its improved 
adaptive median filter \cite{HH95}, is quite effective for restoring the 
image. However, this filter loses  its denoising power for higher noise density given that 
details and features of the original image are smeared out. In those cases, other techniques must be 
applied; one possibility is the two-stage TV-based method proposed in \cite{CHN05}
which consists of applying first an adaptive median filter to identify the pixels that are likely to
contain noise and construct thus a starting guess which is used in the second stage  
for the minimization of a functional of the form
\[
	F(u,y)=\Psi(u,y)+\alpha\Phi(u)
\]
where $y$ denotes the noisy image, $\Psi$ is a data-fidelty term and $\Phi$
is a regularization term, with $\alpha>0$ a parameter. 
In the following numerical experiments,  we consider the image displayed in Figure \ref{Fig.NumEx7.2.5}$(a)$ 
with size $512\times 512$ pixels, damaged by $70\%$ salt \& pepper noise. The resulting corrupted image is displayed in 
Figure \ref{Fig.NumEx7.2.5}$(b)$ where on average only  $78643$ pixels out of the total $262144$ pixels 
carry true information.
The true image values represent our sample function $f_K$ whereas the set of
the true pixels forms our sample set $K$. 
To assess the restoration performance we use 
the peak signal-to-noise ratio ($\mathrm{PSNR}$) which is expressed in the units of $\mathrm{dB}$ and, 
for an $8-$bit image, i.e. with values in the range $[0,\,255]$,
is defined by
\begin{equation}
	\mathrm{PSNR}=10\log_{10}\displaystyle \frac{255^2}{\frac{1}{mn}\sum_{i,j}|f_{i,j}-r_{i,j}|^2}
\end{equation}
where $f_{i,j}$ and $r_{i,j}$ denote the pixels values of the original and restored image, respectively,
and $m,\,n$ denote the size of the image $f$. 
In our numerical experiments, we have considered the following cases. 
The first one assumes the set $K$ to be given by the noise-free interior pixels of the corrupted image together with the boundary pixels of the original image.
In the second case,  $K$ is just the set of the noise-free pixels of the corrupted image, without any special 
consideration on the image boundary pixels.  
In analysing this second case, to reduce the boundary effects produced by the application of 
Algorithm \ref{Algo:CnvxEnvOBE} and 
Algorithm \ref{Algo:MoreauEnv}, 
we have applied our method to an enlarged image and then 
restricted the resulting restored image to the original domain. The enlarged image has been obtained by padding a 
fixed number of pixels before the first image element and after the last image element along each dimension, making 
mirror reflections with respect to the boundary. The values used for padding are all from the corrupted image.
In our examples, we have considered two versions of enlarged images, 
obtained by padding the corrupted image with 2 pixels and 10 pixels, respectively. 
Table \ref{Table:PSNR:BndFree}, Table \ref{Table:PSNR:Ext2} and Table \ref{Table:PSNR:Ext10} compare 
the values of the $\mathrm{PSNR}$ of the restored images by our method  and the 
TV-based method applied to the corrupted image with noise-free boundary and to the two versions of the enlarged images 
with the boundary values of the enlarged images given by the padded noisy image data. 
We observe that there are no important variations in the denoising result between the different methods of treating the image boundary.
This is also reflected by the close value of the $\mathrm{PSNR}$ of the resulting restored images.
For $70\%$ salt \& pepper noise, Figure \ref{Fig.NumEx7.2.5}$(c)$  and Figure \ref{Fig.NumEx7.2.5}$(d)$ 
display the restored image $A_{\lambda}^M(f_K)$ 
by Algorithm \ref{Algo:CnvxEnvOBE} and Algorithm \ref{Algo:MoreauEnv}, respectively,
with $K$ equal to the true 
set that has been enlarged by two pixels, whereas Figure \ref{Fig.NumEx7.2.5}$(e)$ and 
Figure \ref{Fig.NumEx7.2.5}$(f)$ show the restored image 
by the Adaptive median Filter and the TV-based method \cite{CCM07,CHN05} using the same set $K$. 
Although the visual quality of the images restored from $70\%$ noise corruption is comparable between our method 
and the TV-based method, the $\mathrm{PSNR}$ using our method with Algorith \ref{Algo:CnvxEnvOBE}
is higher than that for the TV-based method in all 
of the experiments reported in Table \ref{Table:PSNR:BndFree}, Table \ref{Table:PSNR:Ext2} 
and Table  \ref{Table:PSNR:Ext10}.
An additional advantage of our method is its speed. 
Our method does not require initialisation which is in contrast with the two-stage TV-based method, 
for which the initialisation, for instance, 
is given by the restored image using an adaptive median filter.


\begin{table}[H]
\centerline{
\begin{tabular}{c|c|c|c|}
	\cline{2-4}			& \multicolumn{3}{c|}{$\mathrm{PSNR}$} \\ \cline{2-4}
	\cline{2-4}			& \multicolumn{3}{c|}{$K$ with noise-free boundary} \\ \cline{2-4}
	\cline{2-3}					& \multicolumn{2}{c|}{$A_{\lambda}^M(f)$} &\multirow{2}{*}{TV}\\ \cline{1-3}
\multicolumn{1}{|c|}{Noise Density}			& \multicolumn{1}{c|}{Algorithm \ref{Algo:CnvxEnvOBE}} & \multicolumn{1}{c|}{Algorithm \ref{Algo:MoreauEnv}} &  \\ \hline
\multicolumn{1}{|c|}{$70\%$ ($6.426\,\mathrm{dB}$)}	& $26.674\,\mathrm{dB}$	& $26.634\,\mathrm{dB}$	& $26.506\,\mathrm{dB}$	\\ \hline
\multicolumn{1}{|c|}{$90\%$ ($5.371\,\mathrm{dB}$)}	& $23.117\,\mathrm{dB}$	& $22.968\,\mathrm{dB}$	& $22.521\,\mathrm{dB}$	\\ \hline
\multicolumn{1}{|c|}{$99\%$ ($4.938\,\mathrm{dB}$)}	& $18.424\,\mathrm{dB}$	& $18.357\,\mathrm{dB}$	& $17.420\,\mathrm{dB}$	\\ \hline
\end {tabular}
}
\caption{\label{Table:PSNR:BndFree} 
	Comparison of $\mathrm{PSNR}$ of the restored images by the compensated convexity based method 
	($A_{\lambda}^M(f_K)$) 
	by applying the Moreau based scheme (Algorithm \ref{Algo:CnvxEnvOBE}) and the convex based scheme 
	(Algorithm \ref{Algo:MoreauEnv}),  
	and by the two-stage TV-based method (TV), with the set $K$ with noise--free boundary.
}
\end{table}

\begin{table}[H]
\centerline{
\begin{tabular}{c|c|c|c|}
	\cline{2-4}			& \multicolumn{3}{c|}{$\mathrm{PSNR}$} \\ \cline{2-4}
	\cline{2-4}			& \multicolumn{3}{c|}{$K$ padded by two pixels} \\ \cline{2-4}
	\cline{2-3}					& \multicolumn{2}{c|}{$A_{\lambda}^M(f)$} &\multirow{2}{*}{TV}\\ \cline{1-3}
\multicolumn{1}{|c|}{Noise Density}			& \multicolumn{1}{c|}{Algorithm \ref{Algo:CnvxEnvOBE}} & \multicolumn{1}{c|}{Algorithm \ref{Algo:MoreauEnv}} &  \\ \hline
\multicolumn{1}{|c|}{$70\%$ ($6.426\,\mathrm{dB}$)}	& $26.642\,\mathrm{dB}$	& $26.020\,\mathrm{dB}$	& $26.475\,\mathrm{dB}$	\\ \hline
\multicolumn{1}{|c|}{$90\%$ ($5.371\,\mathrm{dB}$)}	& $23.078\,\mathrm{dB}$	& $22.654\,\mathrm{dB}$	& $22.459\,\mathrm{dB}$	\\ \hline
\multicolumn{1}{|c|}{$99\%$ ($4.938\,\mathrm{dB}$)}	& $18.240\,\mathrm{dB}$	& $18.026\,\mathrm{dB}$	& $17.314\,\mathrm{dB}$	\\ \hline
\end {tabular}
}
\caption{\label{Table:PSNR:Ext2} 
	Comparison of $\mathrm{PSNR}$ of the restored images by the compensated convexity based method 
	($A_{\lambda}^M(f_K)$) 
	by applying the Moreau based scheme (Algorithm \ref{Algo:CnvxEnvOBE}) and the convex based scheme (Algorithm \ref{Algo:MoreauEnv}),  
	and by the two-stage TV-based method (TV), with the set $K$ padded by two pixels.
}
\end{table}

\begin{table}[H]
\centerline{
\begin{tabular}{c|c|c|c|}
	\cline{2-4}			& \multicolumn{3}{c|}{$\mathrm{PSNR}$} \\ \cline{2-4}
	\cline{2-4}			& \multicolumn{3}{c|}{$K$ padded by ten pixels} \\ \cline{2-4}
	\cline{2-3}					& \multicolumn{2}{c|}{$A_{\lambda}^M(f)$} &\multirow{2}{*}{TV}\\ \cline{1-3}
\multicolumn{1}{|c|}{Noise Density}			& \multicolumn{1}{c|}{Algorithm \ref{Algo:CnvxEnvOBE}} & \multicolumn{1}{c|}{Algorithm \ref{Algo:MoreauEnv}} &  \\ \hline
\multicolumn{1}{|c|}{$70\%$ ($6.426\,\mathrm{dB}$)}	& $26.640\,\mathrm{dB}$	& $26.020\,\mathrm{dB}$	& $26.468\,\mathrm{dB}$	\\ \hline
\multicolumn{1}{|c|}{$90\%$ ($5.371\,\mathrm{dB}$)}	& $23.068\,\mathrm{dB}$	& $22.654\,\mathrm{dB}$	& $22.446\,\mathrm{dB}$	\\ \hline
\multicolumn{1}{|c|}{$99\%$ ($4.938\,\mathrm{dB}$)}	& $18.342\,\mathrm{dB}$	& $18.026\,\mathrm{dB}$	& $17.330\,\mathrm{dB}$	\\ \hline
\end {tabular}
}
\caption{\label{Table:PSNR:Ext10} 
	Comparison of $\mathrm{PSNR}$ of the restored images by the compensated convexity based method 
	($A_{\lambda}^M(f_K)$) 
	by applying the Moreau based scheme (Algorithm \ref{Algo:CnvxEnvOBE}) and the convex based scheme (Algorithm \ref{Algo:MoreauEnv}),  
	and by the two-stage TV-based method (TV), with the set $K$ padded by ten pixels.
}
\end{table}

\begin{figure}[H]
	\centerline{$\begin{array}{ccc}
		\includegraphics[width=0.33\textwidth]{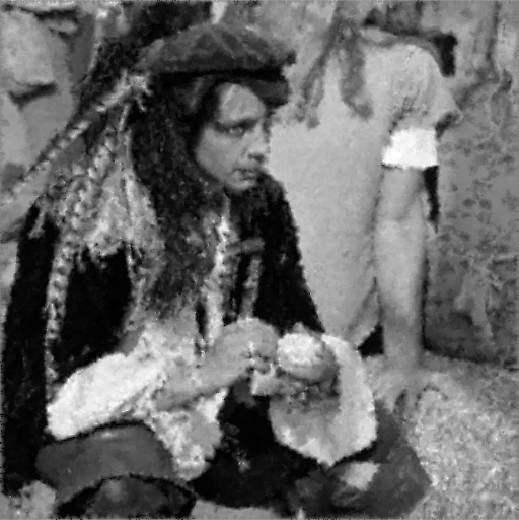}&
		\includegraphics[width=0.33\textwidth]{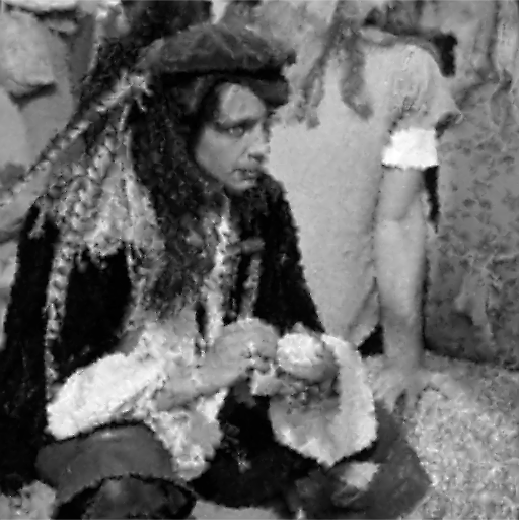}&
		\includegraphics[width=0.33\textwidth]{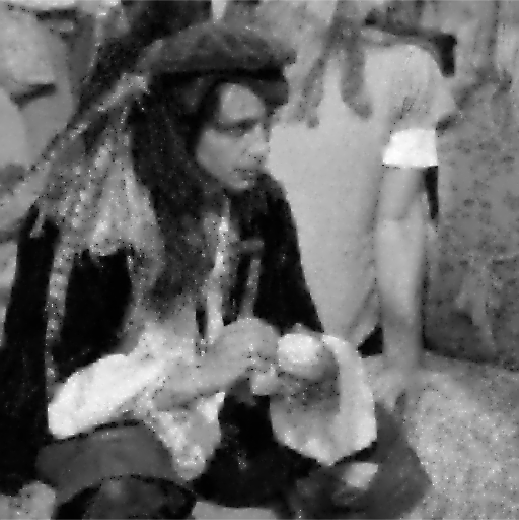}\\
		(a)&(b)&(c)\\
		\includegraphics[width=0.33\textwidth]{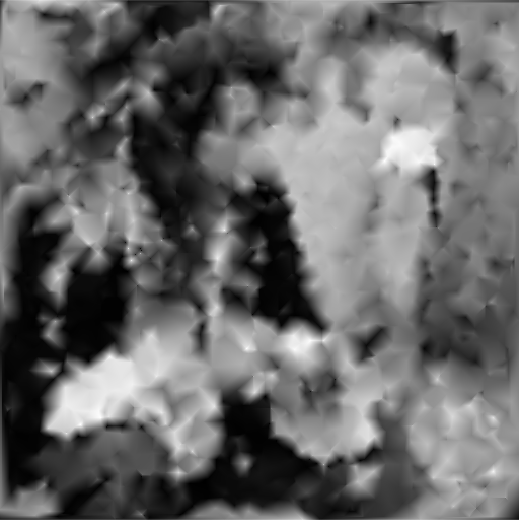}&
		\includegraphics[width=0.33\textwidth]{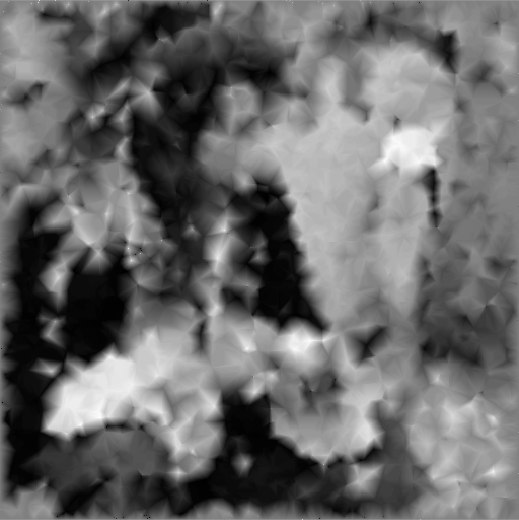}&
		\includegraphics[width=0.33\textwidth]{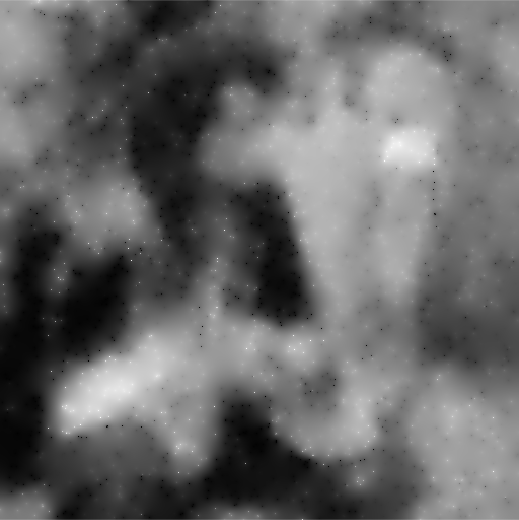}\\
		 (d)&(e)&(f)
	\end{array}$}
	\caption{\label{Fig.NumEx7.2.5a} 
		  Restoration of $90\%$ corrupted image ($\mathrm{PSNR}=5.372\,\mathrm{dB}$) with the set 
		  $K$ padded by two pixels.	
			$(a)$ Restored image $A_{\lambda}^M(f_K)$ by Moreau based scheme 
				(Algorithm \ref{Algo:MoreauEnv}). $\mathrm{PSNR}=22.654\,\mathrm{dB}$.
				$\lambda=10$, $M=1e13$. Total number of iterations equal to $32$
			$(b)$ Restored image $A_{\lambda}^M(f_K)$ by Convex based scheme 
				(Algorithm \ref{Algo:CnvxEnvOBE}). $\mathrm{PSNR}=23.078\,\mathrm{dB}$.
				$\lambda=10$, $M=1e13$. Total number of iterations equal to $10445$
			$(c)$ Restored image by the two-stage TV-based method described in \cite{CCM07,CHN05}.
				$\mathrm{PSNR}=22.459\,\mathrm{dB}$. Total number of iterations: $2679$.\\
			Restoration of $99\%$ corrupted image ($\mathrm{PSNR}=4.938\,\mathrm{dB}$),
			with the set $K$ padded by ten pixels.	
			$(d)$ Restored image $A_{\lambda}^M(f_K)$ by Moreau based scheme 
				(Algorithm \ref{Algo:MoreauEnv}).
				$\mathrm{PSNR}=18.026\,\mathrm{dB}$.
				$\lambda=2$, $M=1e13$. Total number of iterations equal to $78$
			$(e)$ Restored image $A_{\lambda}^M(f_K)$ by Convex based scheme 
				(Algorithm \ref{Algo:CnvxEnvOBE}). $\mathrm{PSNR}=18.342\,\mathrm{dB}$.
				$\lambda=2$, $M=1e13$. Total number of iterations equal to $54823$
			$(f)$ Restored image by the two-stage TV-based method described in \cite{CCM07,CHN05}.
				$\mathrm{PSNR}=17.330\,\mathrm{dB}$. Total number of iterations: $13125$.}
\end{figure}

Finally, to demonstrate the performance of our method in some extreme cases of very sparse data, we consider 
cases of noise density equal to $90\%$ and $99\%$.
Figure \ref{Fig.NumEx7.2.5a} displays the restored image by the compensated convexity based method and by 
the TV-based method for the case where 
$K$ is padded by two pixels and ten pixels for $90\%$ and $99\%$ noise level, respectively. 
As far as the visual quality of the restored 
images is concerned, and to the extent that such judgement can make sense given the high level of noise density,
the inspection of Figure \ref{Fig.NumEx7.2.5a} seems to indicate that 
$A_{\lambda}^M(f_K)$ gives a better approximation of details than the TV-based restored image. This is also reflected by the values
of the $\mathrm{PSNR}$ index in the Table \ref{Table:PSNR:BndFree}, Table \ref{Table:PSNR:Ext2} 
and Table  \ref{Table:PSNR:Ext10}.

\subsubsection{Inpainting}
Inpainting is the problem where 
we are given an image that is damaged in some parts and we want to reconstruct the values in the damaged part
on the basis of the known values of the image. 
This topic has attracted lot of interest especially as an application of 
TV related models \cite{CS05,Sch15}. The main motvation is that functions 
of bounded variations provide the appropriate functional setting given that such functions are allowed to 
have jump discontinuities \cite{AFP00}.
These authors usually argue that continuous functions cannot
be used to model digital image related functions as functions representing images may have jumps \cite{CS05},
which are associated with the image features. 
However, from the human vision perspective, it is hard to distinguish between a jump discontinuity, where values
change abruptly, and a continuous function with sharp changes within a very small transition layer. By the application
of our compensated convex based average transforms we are adopting the latter point of view. A comprehensive study of 
this theory applied to image inpainting can be found in \cite{ZCO16a,ZCO18} where we also establish error estimates for
our inpainting method and compare with the error analysis for image inpainting discussed in \cite{CK06}.
We note that for the relaxed Dirichlet problem of the minimal graph \cite{CK06}
or of the TV model used in \cite{CK06}, as the boundary value of the solution does not have to agree
with the original boundary value, 
extra jumps can be introduced along the boundary. By comparison, since our average 
approximation is continuous, it will not introduce such a jump discontinuity at the boundary.  

\medskip

\begin{figure}[H]
  \centerline{$\begin{array}{cc}
	\includegraphics[width=0.45\textwidth]{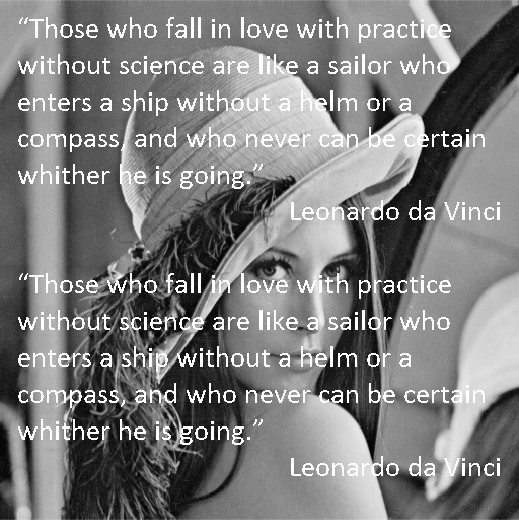}&
	\includegraphics[width=0.45\textwidth]{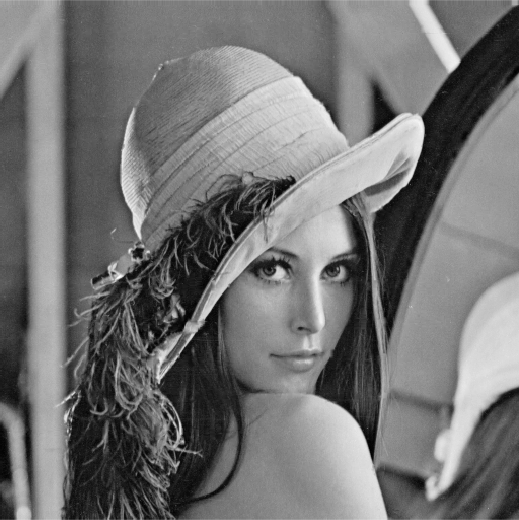}\\
	(a)&(b)\\
	\includegraphics[width=0.45\textwidth]{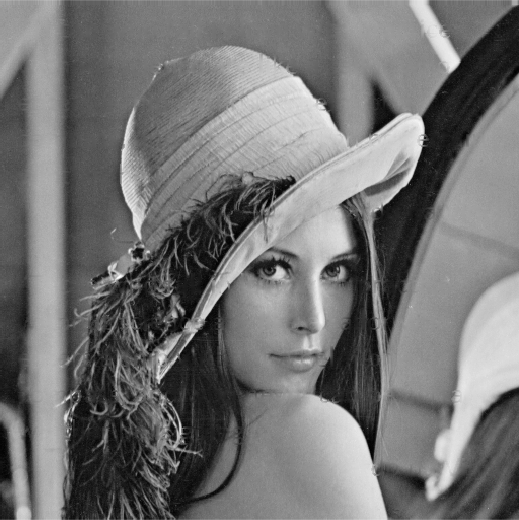}&
	\includegraphics[width=0.45\textwidth]{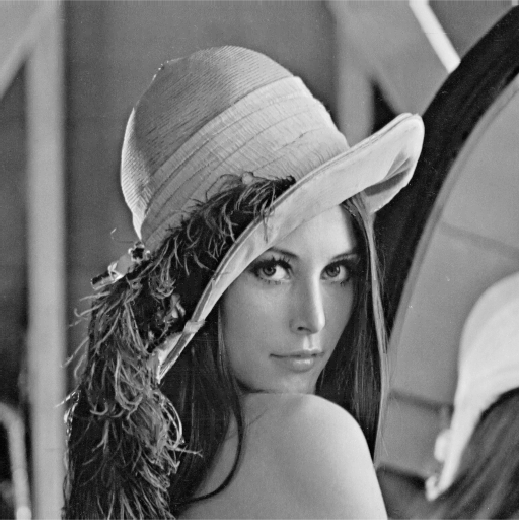}\\
	(c)&(d)
    \end{array}$
}
   \caption{\label{Fig.NumEx7.3a}
	Inpainting of a text overprinted on an image: 
	$(a)$ Input image.
	$(b)$ Restored image $A_{\lambda}^M(f_K)$ using Algorithm \ref{Algo:MoreauEnv}.
		$\mathrm{PSNR}=39.122 \,\mathrm{dB}$.
		Parameters: $\lambda=18$ and $M=1\cdot 10^5$. Total number of iterations: $19$.
	$(c)$ Restored image by the AMLE method described in \cite{Sch15,PS16}. 
		$\mathrm{PSNR}=36.406 \,\mathrm{dB}$. Total number of iterations: $5247$.
	$(d)$ Restored image by the Split Bregman inpainting method described in \cite{Get12}. 
		$\mathrm{PSNR}=39.0712\,\mathrm{dB}$.
		Total number of iterations: $19$.
	}
\end{figure}

\begin{figure}[H]
  \centerline{$\begin{array}{c}
	  \begin{array}{cc}
		\includegraphics[width=0.45\textwidth]{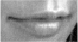}&
		\includegraphics[width=0.45\textwidth]{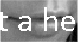}\\
		(a)&(b)
	  \end{array}\\
	  \begin{array}{ccc}
		\includegraphics[width=0.33\textwidth]{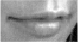}&
		\includegraphics[width=0.33\textwidth]{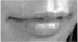}&
		\includegraphics[width=0.33\textwidth]{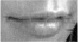}\\
		(c)&(d)&(e)
	   \end{array}
    \end{array}$
}
   \caption{\label{Fig.NumEx7.3b}
	Comparison of a detail of the original image with the corresponding detail of the
	restored images according to the compensated convexity method and the TV-based method. 
	Lips detail of the original: image $(a)$ without and $(b)$ with  overprinted text.
	Lips detail of the: $(c)$  restored image $A_{\lambda}^M(f_K)$ using Algorithm \ref{Algo:MoreauEnv}; 
	$(d)$ AMLE-based restored image; $(d)$ TV-based restored image.
	}
\end{figure}

To assess the performance of our reconstruction compared to state-of-art inpainting methods, 
we consider synthetic example where we are given an image $f$ and we overprint some text on it.
The problem is then removing the text overprinted on the image displayed 
in Figure \ref{Fig.NumEx7.3a}$(a)$ and how close we can get to the original image $f$. 
If we denote by $P$ the set of pixels containing the overprinted text, and 
by $\Omega$ the domain of the whole image, then $K=\Omega\setminus P$ is the set of the true pixels and 
the inpainting problem is in fact the problem of reconstructing the image over $P$ from knowing $f_K$, if we denote by 
$f$ the original image values. 
we compare our method 
with the total variation based image inpainting method solved by the split Bregman method described in \cite{Get12}
and with the AMLE inpainting reported in \cite{Sch15}.
The restored image $A_{\lambda}^M(f_K)$ obtained by our compensated convexity method is displayed in 
Figure \ref{Fig.NumEx7.3a}$(b)$, the restored image by the AMLE method is shown 
in Figure \ref{Fig.NumEx7.3a}$(d)$ whereas \ref{Fig.NumEx7.3a}$(c)$ presents 
the restored image by the the split Bregman inpainting method. 
All the restored images look visually quite good. However, if we use 
the $\mathrm{PSNR}$ as a measure of the quality of the restoration, we find that $A_{\lambda}^M(f_K)$
has a value of $\mathrm{PSNR}$ equal to $ 39.122\,\mathrm{dB}$,  the split Bregman inpainting
restored image gives a value for $\mathrm{PSNR}=39.071\,\mathrm{dB}$,
whereas the AMLE restored image has $\mathrm{PSNR}$ equal to $ 36.406\,\mathrm{dB}$.
To assess how well $A_{\lambda}^M(f_K)$ is able to preserve image details 
and not to introduce unintended effects such as image blurring and staircase effects, Figure \ref{Fig.NumEx7.3b}
displays details of the original image and of the restored images by the three methods. 
Once again, the good performance of $A_{\lambda}^M(f_K)$ can be appreciated visually.

\begin{figure}[H]
  \centerline{$\begin{array}{cc}
	\includegraphics[width=0.45\textwidth]{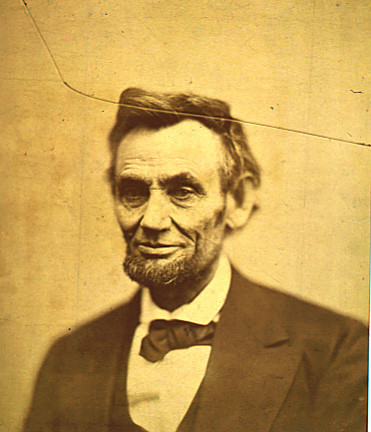}&
	\includegraphics[width=0.45\textwidth]{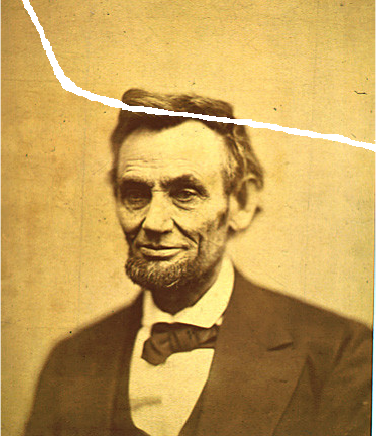}\\
	(a)&(b)\\
	\includegraphics[width=0.45\textwidth]{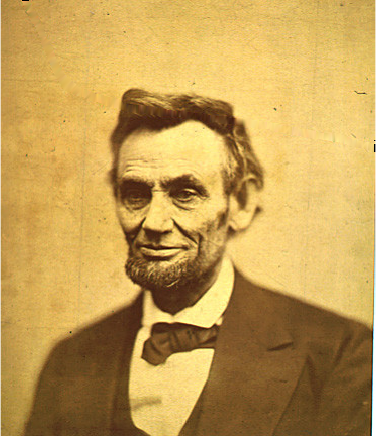}&
	\includegraphics[width=0.45\textwidth]{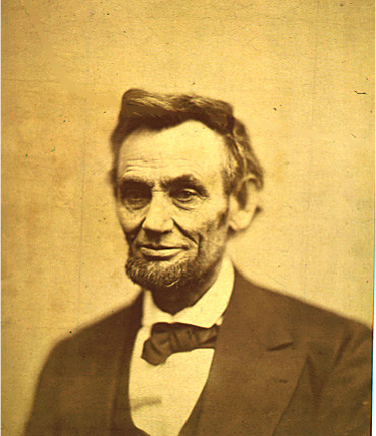}\\
	(c)&(d)
    \end{array}$
}
   \caption{\label{Fig.NumEx7.3c} Restoration of an old image. 
	$(a)$ Input image with the scratch.
	$(b)$ Input image with manual definition of the mask, given by the domain to respair.
	$(c)$ Restored image $A_{\lambda}^M(f_K)$ with $\lambda=15$, $M=10^6$.
	$(d)$ TV-based restored image.
	}
\end{figure}

We conclud ethis section with two real--world applications, where we actaully do not know the true background 
pciture $f$, thus the assessment of the inpainting must simply rely on the visual quality 
of the approximation. Figure \ref{Fig.NumEx7.3c} compares the results of the Average compensated approximation 
and of the TV-based approximation in the case of the restoration of an image containing a scratch,
whereas Figure \ref{Fig.NumEx7.3d} refers to the removal of an unwanted thin object from a picture.
For both the examples, the two approximations yield qualitatively good results.

\begin{figure}[H]
  \centerline{$\begin{array}{cc}
	\includegraphics[width=0.45\textwidth]{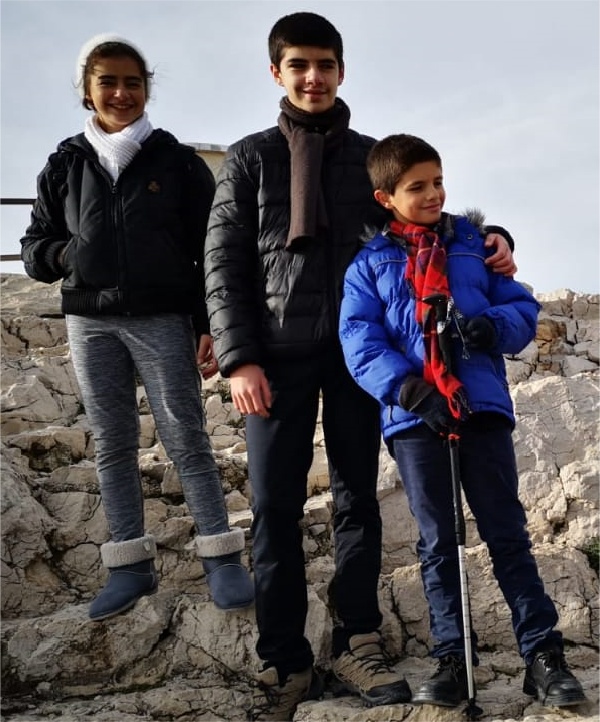}&
	\includegraphics[width=0.45\textwidth]{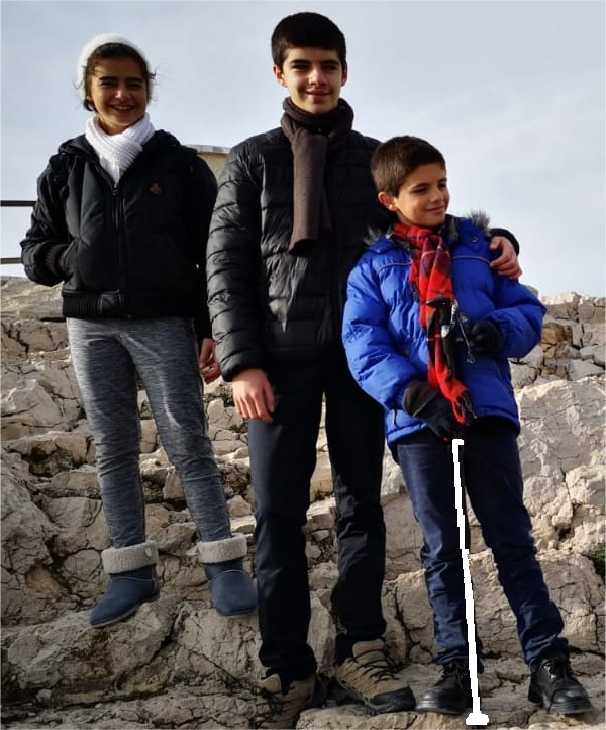}\\
	(a)&(b)\\
	\includegraphics[width=0.45\textwidth]{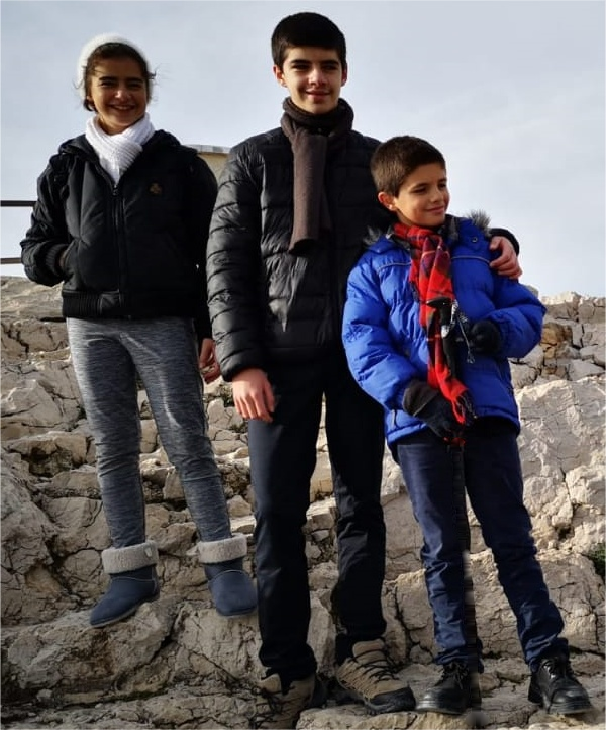}&
	\includegraphics[width=0.45\textwidth]{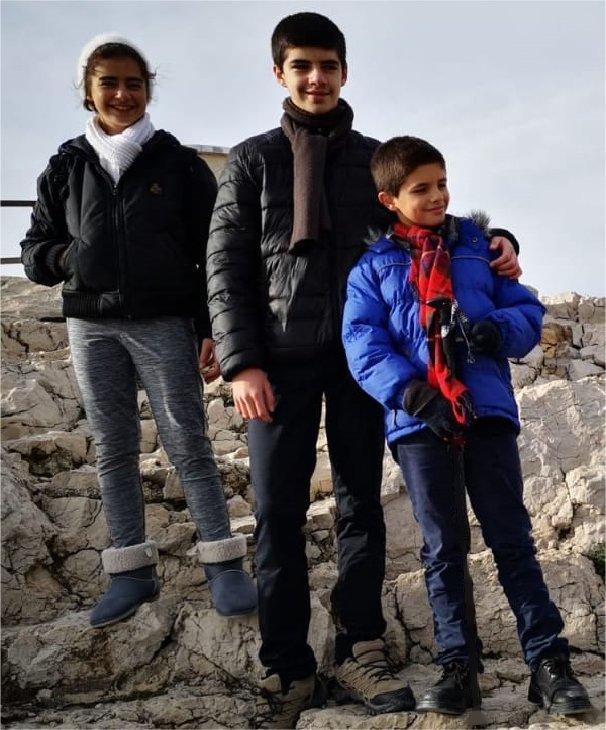}
	\\
	(c)&(d)
    \end{array}$
}
   \caption{\label{Fig.NumEx7.4c} Removal of a thin object from a picture.
	$(a)$ Input image.
	$(b)$ Input image with manual definition of the mask, given by the domain to be inpainted.
	$(c)$ Restored image $A_{\lambda}^M(f_K)$ with $\lambda=15$, $M=10^6$.
	$(d)$ TV-based restored image.
	}
\end{figure}


\section*{Acknowledgements}

AO acknowledges the partial financial support of the Argentinian Research Council (CONICET)
through the project PIP 11220170100100CO,
the National University of Tucum\'{a}n through the project PIUNT CX-E625 and the FonCyT 
through the project PICT 2016 201-0105 Prestamo Bid. 
EC is grateful for the financial support of the College of Science, Swansea University,
and KZ wishes to thank University of Nottingham for its support.


\end{document}